\documentclass[useAMS,usenatbib]{mn2e}
\usepackage{amssymb, latexsym, amsmath, graphicx, url, lscape,tikz}

\usepackage{wasysym}

\title{LOFAR Observations of X-ray Cavity Systems}
\author[L.~B\^{\i}rzan et al.]{L.~B\^{\i}rzan$^{1}$\footnote{E-mail address: lbirzan@hs.uni-hamburg.de},
D.~A.~Rafferty$^{1}$, M. Br\"{u}ggen$^{1}$, and A.~Botteon$^{2,3}$, G.~Brunetti$^{3}$,
\newauthor
V. Cuciti$^{1}$, A.~C. Edge$^{4}$, R. Morganti$^{5,6}$, H.~J.~A. R\"{o}ttgering$^{2}$, T.~W.~Shimwell$^{5}$
 \\
$^{1}$Hamburger Sternwarte, Universit\"{a}t Hamburg, Gojenbergsweg 112, 21029, Hamburg, Germany\\
$^{2}$Leiden Observatory, Leiden University, Oort Gebouw, P.O. Box 9513, 2300 RA Leiden, The Netherlands\\
$^{3}$INAF-Instituto di Radioastronomia, via P. Gobetti, 101, I-40129, Bologna, Italy\\
$^{4}$Institute for Computational Cosmology, Department of Physics, Durham University, Durham, DH1 3LE, UK\\
$^{5}$ASTRON, the Netherlands Institute for Radio Astronomy, Oude Hoogeveensedijk 4, 7991 PD Dwingeloo, The Netherlands\\
$^{6}$Kapteyn Astronomical Institute, University of Groningen, P.O. Box 800,
9700 AV Groningen, The Netherlands.}

\begin{document}

\maketitle

\begin{abstract}

We present LOFAR observations at 120-168 MHz of 42 systems with possible X-ray
cavities in their hot atmosphere, of which 17 are groups or ellipticals, 19 are
nearby clusters ($z<0.3$), and six are higher-redshift clusters ($z>0.3$). The
X-ray cavities, formed by the radio lobes of the central active galactic nucleus
(AGN), are evidence of radio-mode AGN feedback. In the groups and ellipticals
sample, half of the systems have X-ray cavities for which no
associated lobe emission was detected. Conversely, we report the discovery
of large radio lobes in NGC 6338,
extending far beyond the emission reported previously in the literature. In the
case of the nearby clusters, our observations show that there is little
low-frequency radio emission that extends beyond the cavities (e.g., MS
0735.6+7421 and A2052).
For the first time, we report secure cavity-radio associations
in 2A 0335+096, ZwCl 2701, and ZwCl 8276 that strengthens their interpretation
as AGN-created cavities. However, in some known cavity systems (e.g., A1795 and
ZwCl 3146) we report the lack of detectable low-frequency radio emission
associated with the cavities. Our sample of higher-redshift systems is small,
and unfortunately the present LOFAR observations are not able to resolve the
lobes in many of them. Nevertheless, our sample represents one of the best
available for investigating the connection between radio and jet power in
radio-mode AGN feedback.

\end{abstract}


\section{Introduction}\label{S:intro}

The AGN feedback paradigm in galaxy clusters posits that the central active
galactic nucleus (AGN) is connected in a feedback loop to the cooling
intracluster medium (ICM) in which the AGN resides \citep[see the reviews,][]{mcna07,fabi12}. This feedback is generally
negative, such that when the cooling increases the AGN heating increases to
compensate, reducing the cooling. AGN feedback has been observed in systems
ranging from massive clusters to isolated ellipticals
\citep[e.g.,][]{raff08,cava08,voit08,hoga15,puli18,lakh18,baby19}.
Sophisticated AGN feedback simulations, when they account for both  negative AGN
feedback
\citep[e.g.,][]{gasp13,gasp17c,pras15,pras17,wang19,li14,yang16,meec17,mart19}
and positive AGN feedback \citep[e.g., feedback that enhances the cooling and
star formation activity;][]{wagn12,gaib12,silk13,wagn16,vale20}, have
demonstrated its importance to galaxy formation and evolution. For example, AGN
feedback provides a mechanism to truncate cooling in massive galaxies
\citep{crot06,alex12,sija15,crot16,wyle16,deke19}, to reconcile the star
formation (SF) history of the largest elliptical galaxies with those predicted
from hierarchical clustering through dry mergers \citep{fabe07}, and to prevent
overcooling of the ICM in the cooling-flow clusters \citep[the cooling flow
problem,][]{fabi94}.

Direct observational evidence for AGN feedback comes from
high-angular-resolution \emph{Chandra} X-ray observations of giant elliptical
galaxies, groups and clusters that contain large amounts of hot plasma. These
observations show X-ray cavities in the hot atmospheres, filled with radio
emission from the lobes of the central radio source associated with the
brightest cluster galaxy (BCG). To date, \emph{Chandra} has discovered $\sim 100$ such systems
\citep[see cavity
samples,][]{birz04,birz12,dunn04,raff06,dunn06,nuls09,dunn10,dong10,cava10,
osul11,hlav12,shin16,hlav15,birz17}. The X-ray cavities are direct evidence of a
strong coupling between the AGN jets and the hot atmospheres \citep[see the
reviews of][]{mcna07,catt09,mcna12,fabi12,voit15b}. This feedback mode is known
in the literature as the \emph{maintenance-mode} or \emph{radio-mode} feedback,
to distinguish it from the radiatively dominated quasar-mode feedback.

In \emph{radio-mode AGN feedback}, the heating is thought to be mainly done by
the buoyantly rising cavities created by the AGN, along with the weak shocks
\citep{nuls05b,rand15,form17}, sound waves \citep{fabi03,tang17,fabi17},
subsonic turbulence through gravity waves, g-modes \citep{reyn15,bamb18}, mixing
of the inflated cavity's contents with the ICM \citep{brug02n,hill17}, shocks
and turbulent mixing \citep{yang16}, internal waves and turbulence mixing
\citep{kim03,gasp14,gasp15,zhur14,zhur18,zhan18}, cosmic rays
\citep[CR;][]{guo08,pfro13,jaco17,rusz17}, and uplifting of the cool, central gas by
the expanding jets and rising cavities
\citep{pete06,reva08,pope10,li14,kirkp15,brig15,mcna16,gend17,voit17}. It is not
yet established which of these processes is the dominant source of heat, but
there is a consensus that the heating is self-regulated in a gentle process, as
the entropy increases continuously from the center to the cluster outskirts
\citep[e.g.,][]{voit17}. Furthermore, it is also important to connect all these
processes responsible for AGN feedback and feeding from the smallest scales
(micro scales) to the largest \citep[meso and macro scales,][]{gasp20}.
However, proper modeling
of the multiphase nature of the cooling gas that occurs on small scales is
computationally challenging, and next-generation simulations are likely needed
to detangle this problem \citep{jian18,ogiy18,mart19}.

The X-ray cavities seen in groups and clusters are not only direct evidence of
the interplay between the radio source and the ICM, they also allow one to
systematically quantify the bulk of the energy injected by the AGN into the
cluster atmosphere by measuring the work done by the buoyantly rising cavities
\citep{birz04,dunn06,raff06,birz12,hlav12}. Until recently, cavities were
detected at redshifts up to $z=0.544$ \citep[e.g; MACS
J1423.9+2404,][]{raff06,hlav12}, but with the advent of the Sunyaev-Zel'dovich
selected samples of clusters \citep[e.g., SPT, ACT,
Planck;][]{reic13,marr11,plan13}, there are now cavity candidates up to
$z=1.132$ \citep[e.g., SPT-CL J2106-5845;][]{birz17}. However, at these high
redshifts the details of the AGN feedback process are even less well understood,
with some evidence that the primary mode of feedback transitions from a
mechanically dominated mode to a radiatively dominanted one, e.g., from
low-excitation radio galaxies (LERGs) to high-excitation radio galaxies
(HERGs)\footnote{LERG vs. HERG dichotomy is based on the presence of weak,
narrow low-ionization lines \citep{hine79,hard06,hard07}.}, or radio mode to
quasar mode feedback \citep{chur05,russ13,hlav13,birz17,pint18,mcdo18}.

An important result from X-ray cavity studies is the determination of scaling
relations between the cavity power and the radio power
\citep{birz04,merl07,birz08,cava10,osul11,heck14}. There is a large range of
(logarithmic) slopes found in these scaling relations \citep[e.g., from 0.35 to
0.75 in the case of monochromatic relations at 1.4
GHz;][]{birz08,cava10,osul11}, with the latter relation spanning over seven
orders of magnitude in radio and jet power. At 325 MHz, the best-fit relation
has a slope of $\approx 0.5$ \citep{birz08,koko17}. There is also a difference
in the above scaling relation slopes if we some information on the spectral age
of the lobe emission, through the break frequency of the synchrotron spectrum
\cite[e.g., for the scaling relations of the cavity power vs. the bolometric
radio luminosity a slope of 0.5 or 0.6 was found for the total source or lobes
only vs. 0.7 when the break frequency information is included; ][]{birz08}. With
the break frequency included, the scatter about the best-fit relation is reduced
by $\sim 50\%$ \citep{birz08}.

These scaling relations have been used for a variety of purposes by a number of
authors. e.g.: for studies of how jet-mode heating balances cooling for large
samples of galaxies \citep[e.g.,][]{best06,best07,magl07,hart09,ma13,best14},
for studies of the cosmic evolution of AGN feedback to higher redshifts
\citep[e.g.,][]{catt09b,smol09,dani12,simp13,best14,smol15,prac16,smol17,hard19}, and for studies of the accretion mechanism and accretion rates
\citep{sun09,saba19}. While fairly uncertain, for a sample with a wide range in
luminosities a slope of $\sim$ 0.7 is widely used \citep[see discussion
in][]{best06,catt09,heck14,smol17,hard19}.

Additionally, there are also theoretical models for the scalings derived from
Fanaroff-Riley type II \citep[FRII;][]{fana74} expansion models \citep{will99,daly12,ines17}. From theoretical
considerations, the slope is expected to range from 0.5 according to buoyancy arguments
\citep{godf16} to 0.8 according to FRII modeling \citep{will99,daly12}. As a result, it
is important to better constrain and understand the observed slopes and to
understand whether they are in conflict with the theoretical ones. This point is
especially true for Fanaroff-Riley type I (FRI) sources which constitute the majority of radio sources observed in cluster centers \citep[see also the FRI source model,][]{luo10}. It
is also important to understand whether there are variations in the slope that
depend on environment and redshift; for example, for radio sources in
groups/elliptical category, in order to explain the lower average kinetic power
of the jets, deceleration of the jets by mass entrainment was invoked
\citep{bick84,peru14b,lain14}. This deceleration may have an effect on the
spectral age of the source, which could in turn affect the slope of the radio to
jet-power scaling. Our goal is to use LOFAR observations to better constrain the
scaling relations at lower frequencies, adding information regarding the
spectral shape, and to increase the sample at lower luminosities and higher
redshifts.

In this paper, we present LOFAR observations at 120-168 MHz for 42 systems with
likely cavities, ranging from ellipticals to massive clusters. Our goal is to
supplement our previous sample \citep{birz08} with additional systems, and we
particularly focus on groups and ellipticals and higher redshift clusters
($z>0.3$), which had very little representation in \citet{birz08} sample. We
also expand the  lower-redshift clusters sample ($z<0.3$), since we want to
ensure that we have a wide distribution of halo masses.  We will present an
analysis of the low-frequency jet power to radio power scaling relations in a
subsequent paper. The paper is organized as follows: the sample is presented in section 2, details of the X-ray (\emph{Chandra}) and radio (LOFAR) data analysis are presented in
section 3, our results and discussion in section 4 and 5, respectively, and our conclusions in section 6.

 \section{The Sample}\label{S:sample}

Our sample consists of 42 systems with possible X-ray cavities observed with LOFAR (see Table \ref{Xray_table}, Table \ref{LOFAR_table} and Table~\ref{summary_table}), based on the cavity sample of \citet{birz08}
(called the B08 sample henceforth) of systems with multifrequency Very Large Array (VLA) radio data at four
frequencies (327 MHz, 1.4 GHz, 4.5 GHz and 8.5 GHz) and at high angular
resolution (e.g., $\approx 1.0\arcsec \times 1.0\arcsec$ at 1.4 GHz, A array).
These systems are highlighted in bold in Table \ref{Xray_table}, Table
\ref{LOFAR_table} and Table~\ref{summary_table}. The B08 sample consists of 5 groups and ellipticals,
two high redshift clusters,
and 17 nearby clusters.
In the B08 sample we were able to separate the lobe vs. core radio emission for a subsample of 12 systems (4 groups and 8 nearby clusters).
For the remaining systems, the lobe break frequency could not be well constrained, either because the
data did not sample the emission at low enough frequencies or because the lobes
could only be detected and resolved at one frequency (e.g., A1835).

From the original B08 sample, we imaged the majority of the systems that could
be observed with LOFAR (those situated at $\delta_{2000}>+0^{\circ}$).\footnote{From the B08 sample, two groups (Centaurus and HCG 62) and five nearby clusters (A133, Hydra A, Sersic 159/03, A2597, A4059) lie at $\delta_{2000}<+0^{\circ}$. The
B08 systems missing from our sample that that are situated at $\delta_{2000}>+0^{\circ}$ are Perseus, M84, and M87, for which the
LOFAR reduction is nontrivial due to the presence of very bright sources,
RBS~797, MACS J0423.8+2402, and A1835 (all works in progress), and Cygnus~A, the LOFAR observations of which were already published by \citet{mcKe16}.}
More recently, \citet{koko17} imaged many of these systems
with LOFAR at 140 MHz, but at a resolution of only $\approx 23 \arcsec \times 23
\arcsec $. The main goal of this paper is to expand the low-frequency imaging of
\citet{koko17} to higher resolutions and to systems at higher redshifts and with
lower X-ray luminosities (i.e., groups and ellipticals).

In order to expand the B08 sample to lower-luminosity systems, we identified
known groups and ellipticals with X-ray cavities in the literature
\citep[e.g.,][]{raff06,dunn10,cava10,dong10,osul11} that are accessible to
LOFAR. To this end, we limited our sample to systems that lie at
$\delta_{2000}>+0^{\circ}$, but in principle even lower declinations are
accessible, although LOFAR's sensitivity declines as the projected area of the
stations decreases. Also, we added  a number of ellipticals which likely harbor
cavities and where significant H$\alpha$ emission
is present \citep[e.g., NGC
499, NGC 410;][]{lakh18}.

Additionally, we also expanded the lower-redshift ($z<0.3$) cluster sample. This
was done by adding a number of clusters found recently to have cavities that are
not present in the B08 sample, e.g., 4C+55.16 \citep{raff06}, ZwCl 8276
\citep{etto13}, A2390 \citep{savi19}, RX J0820.9+0752 \citep{vant19}, A1361,
ZwCl 0235, RX J0352.9+1941, MS 0839.9+2938 \citep{shin16}. We also included some
cooling flow clusters which might harbor cavities and where significant
H$\alpha$ emission is present \citep[e.g., A1668, ZwCl 0808;][]{craw99}. To
expand the sample to higher redshifts ($z > 0.3$), we used the sample of
\citet{hlav12} supplemented with some putative cavity systems from
\citet{shin16}, e.g., MACS J1621.3+3810.

In the tables (Table 1 - Table 3), systems are grouped into three categories:
groups and ellipticals, nearby clusters ($z<0.3$), and higher-redshift clusters
($z>0.3$), all of them ordered from lower to higher redshift. However, it is
important to mention that there is an overlap between these categories, due to
the range of mass, radio power and redshift. Furthermore, at the end of each
category, we list the systems for which LOFAR observations from this paper
failed to find radio emission filing the reported X-ray cavities (e.g; A1795,
MACS J1359.8+6231, NGC 3608, NGC 777). Throughout this paper we use the term
\emph{radio-filled cavities}; the radio association is interpreted as clear
evidence for radio-mode AGN feedback (see the summary table, Table
\ref{summary_table}).

Lastly, in Table \ref{summary_table}, we also added
available information from the literature regarding the presence of H$\alpha$ filaments or molecular gas and evidence for sloshing.
The H$\alpha$ filaments and the molecular gas imaged in nearby groups and clusters  are interpreted as the end product of the cooling of the X-ray gas \citep[e.g., the chaotic cold accretion mechanism;][]{gasp13}, and they have a diverse range of morphologies \citep[e.g., discs, filaments, etc.;][]{hame16}. Sloshing, in which the BCG oscillates around the cluster center, is thought to be due to a perturbation of the gravitational potential of the cluster that follows an off-axis minor merger \citep{mark00,mark07}. It has been postulated that such sloshing might produce enough heating to balance the cooling of the inner regions \citep[$r\lesssim 30$ kpc;][]{zuho10}. Information on the presence of sloshing,
H$\alpha$ filaments, and molecular gas can be used to understand whether heating by sloshing and the presence of molecular gas and H$\alpha$ filaments are common in systems with X-ray cavities.

\section{Data Analysis}\label{S:analysis}

\subsection{LOFAR Data}\label{S:radio_analysis}

All systems were observed with the High-Band Array (HBA) of LOFAR at frequencies
of 120-168 MHz (for observational details see Table \ref{LOFAR_table}). Most
systems were observed as part of LoTSS, the LOFAR Two-meter Sky Survey
\citep{shim19}, and for 8 hours of total integration time, except for the lower
declination systems that were observed for 4 hours (see Table
\ref{LOFAR_table})\footnote{In the case of not target on source observations,
such as the  pointings of the LoTSS, there will be a lower effective integration
time because of the primary beam attenuation ($\approx 4-8$ h effective
integration time).}. The \textsc{prefactor}\footnote{Available at
\url{https://github.com/lofar-astron/prefactor}} and \textsc{factor} pipelines\footnote{Available at \url{https://github.com/lofar-astron/factor}} were used to
calibrate and image the data using the facet-calibration scheme described in
\citet{vanw16}, following the process detailed in \citet{birz19}.
Version 2.0.2 of \textsc{prefactor} and version 1.3 of
\textsc{factor} were used. A
conservative systematic uncertainty of 15\% was adopted on all LOFAR flux densities
throughout our analysis, as done in previous LOFAR-HBA work (see Table
\ref{LOFAR_table} for the total flux density, the rms noise, and the resolution
of the final image).

\subsection{X-ray Data}\label{S:Xray_analysis}
Table \ref{Xray_table} lists information on the \emph{Chandra} X-ray
observations used in this work, such as the observation IDs, the total
integration time on source after reprocessing and the presence of cavities as
reported in the literature. The X-ray data were reprocessed with \textsc{CIAO}
4.9\footnote{See \url{cxc.harvard.edu/ciao/index.html}.} using \textsc{CALDB}
4.7.3\footnote{See \url{cxc.harvard.edu/caldb/index.html}.} and used to make
exposure-corrected X-ray images and residual maps, following the steps detailed in
\citet{raff13}. To make the residual maps, a model of the extended X-ray emission is
subtracted from the corresponding exposure-corrected image. The model was found
using the multi-Gaussian expansion technique of \citet{capp06}.

The X-ray and radio images for the systems in our sample are shown in Figure 1
through Figure 6 in three panels (\emph{left panels}: LOFAR images; \emph{middle
panels}: overlays of LOFAR contours and the X-ray residual maps; \emph{right
panels}: smoothed X-ray images).  The systems in the figures (Figure 1 $-$ Figure
6) are shown in the same order as in the tables, starting with the groups and ellipticals with clear cavities (Figure 1), groups and ellipticals without clear cavities (Figure 2), nearby clusters with clear cavities (Figure 3), nearby clusters without clear cavities (Figure 4), high-redshift clusters with clear cavities (Figure 5), and high-redshift clusters without clear cavities (Figure 6).

\section{Results and Comparison with Previous Observations}\label{S:results}

\subsection{LOFAR Images for B08 Sample}\label{B08_sample}

The systems from the  B08 sample are highlighted in bold in Table
\ref{Xray_table}, Table \ref{LOFAR_table} and Table \ref{summary_table} (see Section \ref{S:sample} for a summary of the B08 sample).
The LOFAR and X-ray
images for the B08 systems present in our sample are displayed in Figure~1 (A262),
Figure~3 (A2199, 2A0335+096, A2052, MKW3S, A478, ZwCl 2701, MS 0735+0721),
and Figure~4 (A1795, ZwCl 3146). Below, we describe our new observations
for the B08 sample:\footnote{For more references
of the presence of the X-ray cavities see Table 1 and more references for  the
presence of the central radio source see Table 2.}\\

\begin{description}
\item[{  $\bullet$}] The LOFAR observation of A262 (see Figure 1) detected the
western lobe at higher significance than previous VLA and GMRT observations \citep[e.g., 327 MHz VLA and 610 MHz GMRT;][]{birz08,clar09}. The full extent of the eastern lobe, as seen by LOFAR, is similar to that seen in the previous observations (e.g., VLA and GMRT).
\end{description}

\begin{description}
\item[{  $\bullet$}] The LOFAR observation of 2A0335+096 (see Figure 3) is
significantly more sensitive than previous VLA images \citep{birz08}, where the
327 MHz (B array) and 1.4 GHz (C array)  detected only hints of the lobes. In
the LOFAR image the lobes are clearly seen, with the north lobe filling in the
X-ray cavity visible in the \emph{Chandra} image.\footnote{The emission seen beyond the western lobe to the north-west is due to a head-tail radio galaxy seen in previous VLA images.}
\end{description}

\begin{description}
\item[{  $\bullet$}] The LOFAR observation of ZwCl 2701 (see Figure 3) shows
that lobe emission likely fills the X-ray cavities, in contrast to the previous
VLA images \citep{birz08} where we did not detect any lobe emission.
\end{description}

\begin{description}
\item[{  $\bullet$}] The LOFAR image for ZwCl 3146 (see Figure
4) shows for the first time a central radio source with well-resolved lobes. The
X-ray residual image shows spiral structure probably created by gas sloshing
\citep[see the references in Table \ref{summary_table}, e.g.,][]{form02}, which
suggests that the cluster may be going through a minor merger
\citep[see][]{zuho10,zuho16}. Additionally, there is no direct evidence of
cavities at the lobe locations. Because of the complexity of the X-ray
morphology we do not include this system in our follow-up cavity sample.
\end{description}

\begin{description}
\item[{  $\bullet$}] In the case of A1795 there is no apparent association
between the central radio source and the large NW cavity which is further out \citep{walk14n}: the central radio
source is extended NE-SW, the same orientation as  the emission seen at higher resolution with the VLA at 1.4 GHz \citep{ge93,birz08}. Our result is consistent
with previous GMRT observation from \citet{koko18}. Since the central radio
source and the X-ray cavity appear to have no association, we will not consider
this system in our final cavity sample \citep[for more discussion
see][]{koko18}.
\end{description}

\begin{description}
\item[{  $\bullet$}] The LOFAR image of A478 is from \citet{savi19} and does not
resolve the small scales of the X-ray cavities as the previous VLA observation
does \citep[e.g., VLA at 1.4 GHz, A array;][]{birz08}.
\end{description}

\begin{description}
\item[{  $\bullet$}] The LOFAR observations of A2199, A2052 and
MS 0735.6+7421 (Figure 3) show resolved central radio sources that fill the
X-ray cavities, similar to previous VLA observations \citep{birz08}.
\end{description}

\begin{description}
\item[{  $\bullet$}] The LOFAR observation of MKW3S (see Figure 3) detects the
emission seen with the VLA and GMRT  \citep{mazz02,giac07,birz08}. For the southern lobe there is a corresponding X-ray cavity in the \emph{Chandra} image, but no corresponding X-ray cavity for the northern lobe has been identified \citep[see][]{mazz02,birz04,dunn04,raff06}.
\end{description}

\subsection{Groups and Ellipticals Sample}\label{groups_sample}

In the groups and ellipticals sample, besides A262 (see Figure 1) that is part
of the B08 sample, there are other clear cavity systems (we list them in the
same order as in Table 1--3):

\begin{description}
\item[{  $\bullet$}]  For NGC 5846, X-ray cavities are reported in
\citet{dunn10} and \citet{mach11}, and radio images at multiple frequencies have
been published for this system (see the references in Table 3): from VLA data at
5 GHz and 1.4 GHz \citep{mach11} and GMRT data at 610 MHz \citep{giac11b}. The
LOFAR observation of NGC 5846 detects the central radio source, but because of
the low declination of this source ($\delta_{2000} \sim+01^{\circ}$), the
sensitivity of the observation is quite low.
\end{description}

\begin{description}
\item[{  $\bullet$}] For NGC 5813, X-ray cavities are reported in
\citet{rand15}. As with NGC 5846, because of the low declination of the source
($\delta_{2000} \sim+01^{\circ}$), the sensitivity of the LOFAR observation is
quite low. As a result, the LOFAR image detects only emission associated with
the inner lobes, whereas previous 235 MHz GMRT observations show radio emission
associated with both the inner and outer cavities \citep{giac11b}.
\end{description}

\begin{description}
\item[{  $\bullet$}] NGC 193 has clear X-ray cavities, as presented in
\citet{bogd14}. The radio lobes, seen also with LOFAR, were imaged
previously with the VLA \citep{lain11} and GMRT \citep{giac11b}. However, the
radio-cavity association is complex, and there might be two generations of AGN outbursts  \citep[see][]{bogd14}.
\end{description}

\begin{description}
\item[{  $\bullet$}] NGC 6338 is an interesting system that is undergoing a
merger and has possible cavities  \citep{pand12,wang19,osul19}. Previous radio
observations at 1.4 GHz  \citep{wang19,osul19} did not reveal the large lobes
seen with LOFAR. These lobes are spatially coincident with the H$\alpha$
emission \citep[see][]{pand12}, as if the radio lobes dragged the cold gas
further in the cluster atmosphere \citep[see][]{mcna16}. In a
reasonably deep \emph{Chandra} observation ($\sim$ 300 ks), there is some evidence of an
X-ray cavities at the location of the eastern lobe \citep{osul19}, but not at that of the western lobe.
\end{description}

\begin{description}
\item[{  $\bullet$}] Another spectacular elliptical in our sample is IC1262,
with clear X-ray cavities in the \emph{Chandra} image \citep{dong10,pand19} filled by the radio emission.
The two large radio lobes seen in the LOFAR image were reported first in \citet{rudn09}, in WENSS images at 327 MHz, and more
recently by \citet{pand19}, who used GMRT observations at 325 MHz. Additionally, \citet{pand19} used VLA observations at 1.4 GHz to image the inner lobes of the central radio source, which appear to fill the inner X-ray cavities. They also reported that the outer lobes (seen in our LOFAR observations) have a steep spectral index and that the southern lobe fills a ghost cavity visible in the \emph{Chandra} image. However, there is no visible X-ray cavity associated with the northern lobe (as is also the case in MKW3S). Lastly, \citet{pand19} reported the presence of a phoenix radio source embedded in the southern lobe. This phoenix emission is also visible in the LOFAR image.

\end{description}

\begin{description}
\item[{  $\bullet$}] In the case of NGC 6269, the \emph{Chandra} image does not
show clear X-ray cavities\footnote{The putative cavities in NGC 6269 are graded as 'C' in \citet{cava10}; but the X-ray data are not sufficiently deep for a detailed analysis \citep[see also][]{bald09b}.}, but there is a central radio source with well
resolved lobes in the LOFAR image, with structure on a similar scale as other
radio observations taken previously \citep[VLA 1.4 GHz and GMRT 235
MHz;][]{bald09b,giac11b}.
\end{description}

\begin{description}
\item[{  $\bullet$}] NGC 5098 has clear X-ray cavities and 1.4 GHz radio
emission associated with them \citep[see][]{rand09}. The LOFAR image resolves
the radio lobes on scales similar to the VLA image.
\end{description}

Among the remaining groups and ellipticals, NGC 741 is a complicated system,
where the radio emission observed with LOFAR, and previously with the VLA and
GMRT \citep{jeth08,giac11b}, is probably associated with the nearby galaxy NGC
742, with which NGC 741 is undergoing a merger. However, for NGC 3608 and NGC
777 from \citet{cava10}, NGC 2300, UGC 5088, and RX J1159.8+5531 from
\citet{dong10}, NGC 4104 from \citet{shin16}, and NGC 499 and NGC 410, the LOFAR
observations do not detect any lobe emission that fill the reported X-ray
cavities.   Additionally, NGC 2300 might be merging with the nearby galaxy NGC
2276. The LOFAR image shows that the spiral-shaped radio emission due to SF
activity in NGC 2276 extends toward NGC 2300  (see \emph{left} panel of Figure 7).  In the case of
NGC 4104 the putative X-ray cavity reported  in \citet{shin16} is centered on the X-ray core (see Figure 2), which also corresponds to the core of the BCG, a very unusual location for an AGN cavity, since most AGN cavities are located at a distance of approximately two cavity radii from the core. The LOFAR image detects a central radio source, but does not resolve any lobe emission. Furthermore, NGC 4104 shows
 diffuse emission on scale of $\sim$ 100 kpc (see \emph{right} panel of Figure 7), which could be an old AGN lobe from a previous AGN outburst. This emission will be investigated in an upcoming
paper.

\subsection{Nearby Clusters Sample ($z<0.3$)}\label{nearby_cl_sample}

In this section we elaborate on the nearby cluster category ($z<0.3$) that are
not present in B08 sample (in the same order as they are listed in Table 1 -
Table 3):

\begin{description}
\item[{  $\bullet$}] The LOFAR image of A1668  shows large radio lobes which
extend far into the ICM \citep[see also][]{hoga15}, but the X-ray image is not deep enough to confirm the
presence of  X-ray cavities coincident with the lobes.
\end{description}

\begin{description}
\item[{  $\bullet$}] In the case of ZwCl 8276, the LOFAR image shows a
well-resolved central radio source that fills in the cavities reported by
\citet{etto13}. On the other hand, previous VLA 1.4 GHz DnA-array radio observations
from \citet{giac14} did not resolve the lobes (e.g., beam size $10.7 \arcsec \times 9.7 \arcsec$), and as a result the nature of the
extended emission was unclear at that time.
\end{description}

\begin{description}
\item[{  $\bullet$}] For A1361, the \emph{Chandra} data were severely affected
by flares and only 1.0 ks out of the initial 8.33 ks were used to make the image
in Figure 3. The X-ray residual map image shows some evidence of
depressions at the location of the radio lobes, but deeper \emph{Chandra} data
would be needed to confirm them. This source was previously imaged with VLA, A
array at 1.4 GHz \citep{owen97}, and 4.5 GHz \citep{hoga15}, which show a
two-sided lobe morphology.
\end{description}

\begin{description}
\item[{  $\bullet$}] For ZwCl 0808 there is no clear evidence for X-ray cavities at
the location of the extended radio emission \citep[see also][]{hoga15}, but this has be be further confirmed with
deeper \emph{Chandra} data.
\end{description}

\begin{description}
\item[{  $\bullet$}] For A2390, the presence of X-ray cavities was reported by \citet{sonk15} and \citet{shin16}. However, it wasn't until the LOFAR observations of \citet{savi19} that a central radio source with large lobes was detected.\footnote{The LOFAR image of A2390 used in this paper is from \citet{savi19}.}
A2390 is a good example of the lobes of the central radio source filling the X-ray cavities.
\end{description}

\begin{description}
\item[{  $\bullet$}] For 4C+55.16, LOFAR confirms the presence of radio emission
filling the X-ray cavities \citep[e.g.,][]{raff06,hlav11,xu95}.
\end{description}

Next, based on our LOFAR observations, we discuss the systems where the X-ray
cavities reported in the literature are not filled with radio emission (the
systems are presented in the same order as in Table 1- Table 3):

\begin{description}
\item[{  $\bullet$}]  LOFAR observation of ZwCl 0235 (see Figure 4) shows a
central radio source with small lobes associated with the BCG. However, there
are no evident X-ray cavities at the lobe location.\footnote{For ZwCl 0235 there
were putative cavities reported in the literature \citep{shin16}. However, we do
not know the size and location of these reported cavities.} This source might be
similar to ZwCl 3146, since  shows clear evidence for a spiral residual pattern
in the ICM, often found to be associated with sloshing \citep[see][]{zuho10}.
\end{description}

\begin{description}
\item[{  $\bullet$}] LOFAR image of RX J0352.9+1941 shows a point-like central
radio source, associated with the BCG. As in the case of ZwCl 0235 (see footnote) putative X-ray cavities were reported in the literature by \citet{shin16}. However, there does not seem to be any clear X-ray depression in this system and no cental radio source with lobes either.
\end{description}

\begin{description}
\item[{  $\bullet$}] For RX J0820.9+0752, the LOFAR image does not show radio
lobes filling the putative X-ray cavity \citep{vant19}. The reported cavity is
larger and further out in the ICM than the location of the central radio source imaged with LOFAR
\citep{vant19}.
\end{description}

\begin{description}
\item[{  $\bullet$}] The LOFAR image of MS 0839.9+2938 confirms the presence of
a central radio source with small lobes, previously reported by \citet{giac17}
using VLA B and C arrays at 1.4 GHz. However, there are no clear corresponding
X-ray cavities at the location of the lobes. Low-significance cavities were
reported by \citet{shin16}, but we do not know their size and location relative
to the radio lobes seen in the LOFAR image.
\end{description}

\subsection{High-Redshift Clusters Sample ($z>0.3$)}\label{highz_sample}

In the high redshift sample, there are two clear cavity systems and one possible
cavity system:

\begin{description}
\item[{  $\bullet$}]  For MACS J1532.9+3021, where the X-ray cavities are
reported in \citet{hlav12} and \citet{hlav13b}, LOFAR did not resolve the lobes.
The interpretation in the literature is that the radio emission is a mini-halo
\citep[see][]{kale13,hlav13b,giac14}.
\end{description}

\begin{description}
\item[{  $\bullet$}] For IRAS 09104+4109, X-ray cavities were reported in
\citet{osul12} and \citet{hlav12}. The morphology of the radio emission in the  LOFAR image is similar to the GMRT and
VLA images presented in \citet{osul12}.
\end{description}

\begin{description}
\item[{  $\bullet$}] MACS J1621.3+3810 was previously imaged at 365 MHz as part
of WENSS, the Westerbork Northern Sky Survey \citep[see][]{edge03}. The LOFAR
image shows an extended central radio source, but better resolution is required
to resolve any lobes.
\end{description}

For the remaining higher-redshift systems, MACS J2245.0+2637, MACS J1359.8+6231,
and MACS J1720.2+3536, there is no detected LOFAR radio emission in the X-ray
cavities reported in \citet{hlav12}. In these systems, the reported cavities are
far beyond the extent of the central radio sources imaged with LOFAR. On a much smaller scale
than the cavities reported in \citet{hlav12}, there might be hints of lobe emission for
the central radio source in MACS J2245.0+263, however the source is not well
resolved in the present LOFAR images.\footnote{In many of the higher-redshift
systems, use of the LOFAR international stations will be required to achieve the
arcsecond resolution needed to resolve any emission in the cavities identified in the X-ray images, as the size of the cavities is below the LOFAR resolution limit when international stations are not used. The development of techniques
to use the international stations is a work in progress
\citep[e.g.,][]{vare15,vare16,mora16}.}

\section{Discussion}\label{S:discuss}

This work presents LOFAR HBA observations at 143 MHz for a sample of clusters,
groups and ellipticals with previously reported X-ray cavities. We separated the
sample into three subsamples: groups and ellipticals, nearby ($z<0.3$) clusters,
and higher-redshift ($z>0.3$) clusters.

\subsection{Group and ellipticals subsample}

For the group and elliptical subsample, in addition to A262 that was present in
B08, we observed candidate cavity systems from \citet{cava10},
\citet{osul11}, and \citet{dong10}. We found that only 6 out of 17 systems are
good AGN feedback candidates (see Table \ref{summary_table}), and the two best
cavity systems, NGC 5813 and NGC 5846, unfortunately have relatively low radio
flux densities and are located at $\delta_{2000} \approx +0^{\circ}$. Therefore,
the sensitivity of the LOFAR observations of these systems is not sufficient to
image the full extent of the radio lobes. For NGC 6338, the LOFAR observations
reveal  previously unknown extended emission, with the eastern lobe being coincident with a putative X-ray cavity \citep{osul19}.

In 9 of the 17 group and elliptical cavity system candidates we did not detect
lobes (e.g; NGC 777, NGC 3608, NGC 2300), so the construction of a sample of
radio-filled cavities systems in the groups and ellipticals category has been so
far a difficult problem. Additionally, in X-rays the study of AGN feedback in
group and ellipticals is limited by \emph{Chandra's} capability to image the
diffuse gas in such systems. Nevertheless, although we do not have a large
sample from which to draw firm conclusions, the established picture is expected
to hold, in which mechanical AGN feedback in elliptical galaxies is less
powerful and efficient than in clusters \citep{gasp12b}, with an average duty
cycle of $\sim$ 1/3 \citep{osul17}. The duty cycle may increase with the size of
the system \citep{nuls09}, and generally the reservoir of cold gas has a major
influence in the AGN feedback process \citep{gasp12b,li14b,vale15}.
Additionally, another complication with the lower X-ray luminosity systems is
that one cannot assume that any undetected X-ray cavities are well traced by the
radio lobes, since such systems often host high-power radio sources whose lobes
extend far beyond the dense atmospheres \citep[e.g., NGC 4261, IC4296, IC1459,
NGC 1600, NGC 5090, UGC11294,
ARP308;][]{dieh08a,sun09,cava10,duta15,kolo18,ruff19,gros19}.

Table \ref{summary_table} shows that H$\alpha$ filaments are mostly found
in groups and ellipticals where the radio emission is filling the X-ray cavities
(the exceptions are NGC 499, NGC 410 and NGC 4104, see
Table \ref{summary_table}). This result is broadly consistent with the study of
\citet{lakh18}, where in a sample of 49 nearby elliptical galaxies they found a
hint of a trend between the presence of H$\alpha$ emission and the AGN jet power
\citep[see also][]{baby19}.

\subsection{Nearby and higher-redshift cluster subsamples}

For the nearby cluster sample, our LOFAR observations have sufficient sensitivity and spatial resolution to
detect the radio lobes present at the center of most nearby cooling flow
clusters (for 17 out of 19 systems we detected radio lobes, the exceptions being RX J0352.9+1941 and RX J0820.9+0752). In some such systems with known X-ray cavities, e.g., 2A0335+096
\citep{mazz03}, ZwCl 2701 \citep{raff06} and ZwCl 8276 \citep{etto13}, the
cavity-radio association was not clear from previous radio images. The LOFAR
observations of these systems show us that the X-ray cavities are indeed filled
with low-frequency radio emission.

On the other hand, the LOFAR images of A1795, ZwCl 3146
 and ZwCl 235 do not show diffuse radio emission\footnote{Lower-frequency LOFAR LBA observations might provide further constraints on the presence of even older electron populations.}
associated with the cavities \citep{koko18,raff06,shin16}, and as a result the  nature of the cavities is
still unclear. In particular, A1795, besides showing evidence for sloshing
activity \cite[see Table \ref{summary_table} for references,][]{ghiz10}, may be
going through a merging process, with the H$\alpha$ filaments
\citep{craw05,mcdo09, mitt15,trem15} being dragged along by the ``flying''
cluster core \citep{ehle15}.
Furthermore, RX J0352.9+1941 and RX J0820.9+0752 do not have lobe-like central
radio emission, and as a result we cannot confirm a X-ray/radio association
\citep{shin16,vant19}, but they do show H$\alpha$ filaments and molecular gas
\citep{baye02,hame16,vant19}.

In addition to the AGN heating trough X-ray cavities, the
ICM heating can occur through other means, e.g.,
uplifting of the cold gas from the center of the cluster through sloshing motions \citep[e.g., Fornax cluster, A1068;][]{su17b,mcna04}, by the central radio source and/or radio bubbles \citep{pete06,reva08,kirkp15,mcna16,hame16}, by the flying cluster core \citep[e.g., A1795,][]{craw05,ehle15} or even by major merger with another cluster or subcluster \citep[e.g., A2146;][]{cann12}\footnote{In some of the nearby and high redshift  clusters there is additional evidence for sloshing motion or minor merging activity, such as a displacement between the X-ray peak, the H$\alpha$ peak and the BCG \citep[e.g;][]{craw05,ehle15,hame16,mcdo16}.}.
Also, all nearby cooling flow systems, regardless of whether or not they have X-ray cavities, show evidence of sloshing activity and possess H$\alpha$ filaments (see Table \ref{summary_table}), as if the heating done by cavities and sloshing goes together in some cases \citep[e.g., Fornax Cluster, Perseus;][]{su17b,walk18}. It would be important to understand if such heating is more critical for the cooling flow clusters without cavities and without a central radio source with lobes (e.g., A1068).

Additionally, some systems in our sample are likely in a cooling stage. We know from studying complete samples that the duty cycle of radio-mode feedback is $\approx$ 70$\%$ \citep{birz12} and that the cooling stage is not always clearly separated from the heating stage. Also, the detectability of a cavity depends on its location, orientation, angular size and the depth of the X-ray  observations \citep{enss02,dieh08,brug09}.
Also, it is important to note that there is an evolution to any X-ray cavity
and we will tend to observe X-ray cavities in only a fraction of clusters where
the system is in the middle stage of the cycle with well inflated cavities that
are well filled with the energetic electrons (the lobes seen in radio). As a
result, some systems might be in the early stages of their current activity, and
what we see in LOFAR images might be from previous radio activity (e.g., A2390).

Our LOFAR observations generally show that the low-frequency radio-emitting
plasma does not appear to extend much beyond the cavity edges, e.g., in MS
0735.6+7421 and A2052 \citep[see also M87,][]{deGa12}. This finding has
important implications for simulations of the X-ray cavities, such as the
interaction of radio lobes with the ICM and the magnetic field configuration
inside the cavities \citep[see][]{pfro13}. Observationally, it was found that
the energy content of the cavities is not dominated by the radio-emitting
electrons \citep[e.g.,][]{morg88,dunn04,birz08,cros08,cros18}, and there are
observational constraints on the amount of hot gas filling the cavities
\citep{abdu19}. As a result, the most promising candidate for pressure support
of the cavities is CR protons. There are some constraints on the confinement
time of CR protons from the nondetection with $\gamma$-ray telescopes
\citep{prok17}. There are also a large number of simulations that investigate
the heating of the ICM with CRs
\citep{guo08,shar10,pfro13,wien13,rusz17,wein17,ehle18,thom19,yang19}, or by the
mixing of the bubble contents, which can be either hot gas and/or CR protons,
with the ICM. The latter process is thought to happen at the bubble surface
through vortices \citep{brug02n,brug09,yang16,hill17}. Also, the cavities could have  pressure support from very hot thermal plasma, for example in the case of Perseus, half of the cavity volume could be filled with 50 keV thermal gas \citep{sand07}.

However, not all the X-ray cavities are as well defined
as in A2052 or MS 0735.6+7421. But, the LOFAR
observations show that, in the case of nearby clusters, the radio lobes
generally appear to be well confined. Thus we can postulate that  the radio
lobes generally fill the X-ray cavities with no major evidence of CR electrons leaking.
However, it is important to remember that the interaction of the radio source
with these rich and dynamic clusters environments is complicated, since sloshing
and other cluster weather is also often present (e.g., in A2199 the western
lobes is curved and appears to be moving back in the direction of cluster
motion). As a result, the FRI sources that tend to exist in rich cluster
environments might be different than the FRI sources that are more common in
poor cluster and group environments \citep[see also][]{cros18}, since sloshing
motions can provide some re-acceleration of the existing electron population \citep[e.g.,][]{deGa17b}.
Additionally, in the case of nearby and higher-redshift  clusters, Table \ref{summary_table} shows that,  even if the H$\alpha$
emission is present in systems with  and without a good correlation between the X-ray cavities and the lobe radio emission, the systems with
cavities filled by the radio lobe emission tend to host more powerful radio sources than those without \citep[see
also][]{hoga15}.

 \section{Conclusions}\label{S:summary}

The goal of this paper is to search for diffuse radio emission at low
frequencies in a sample of systems that have possible cavities in X-ray images.
To this end, we have imaged a total of 42 such systems with LOFAR, of which 17 are nearby groups and ellipticals, 19
are nearby massive clusters ($z<0.3$), and
6 are higher-redshift clusters ($z>0.3$).

Based on the presence of low-frequency radio emission that fills the X-ray
cavities, we conclude that only 11/19 of the nearby massive clusters show clear
evidence for radio mode AGN feedback, where the cavities and the central radio
source are well correlated (the associations for A1668 and ZwCl 0808
need to be further confirmed). Additionally, 3/6 high redshift clusters and 7/17
nearby groups and ellipticals show such evidence (NGC 6269
need to be further confirmed; see Table~\ref{summary_table}).
As a result, building a large, statistically significant
sample of low-frequency observations of systems with cavities in each of the
three categories will require the use of other telescopes (e.g; the VLA and
GMRT) to add systems situated at $\delta_{2000} < 0^{\circ}$ and the use of the
LOFAR international stations since, generally, the LOFAR observations of systems
at higher redshift are limited due to the lack of resolution. In particular, the typical resolution achievable without the international stations ($\approx 5-10\arcsec$) implies a limiting physical scale of $\sim 20-40$ kpc at redshifts of $\sim 0.3$, whereas typical cavities observed in  such systems have sizes of $\sim 10-20$ kpc, \citep[e.g., RBS 797, MACS J1423.8+2404, and MACS
J1532.9+3021,][]{raff06,hlav13b}.

\begin{figure*} \begin{tabular}{@{}cc}
\includegraphics[width=54mm]{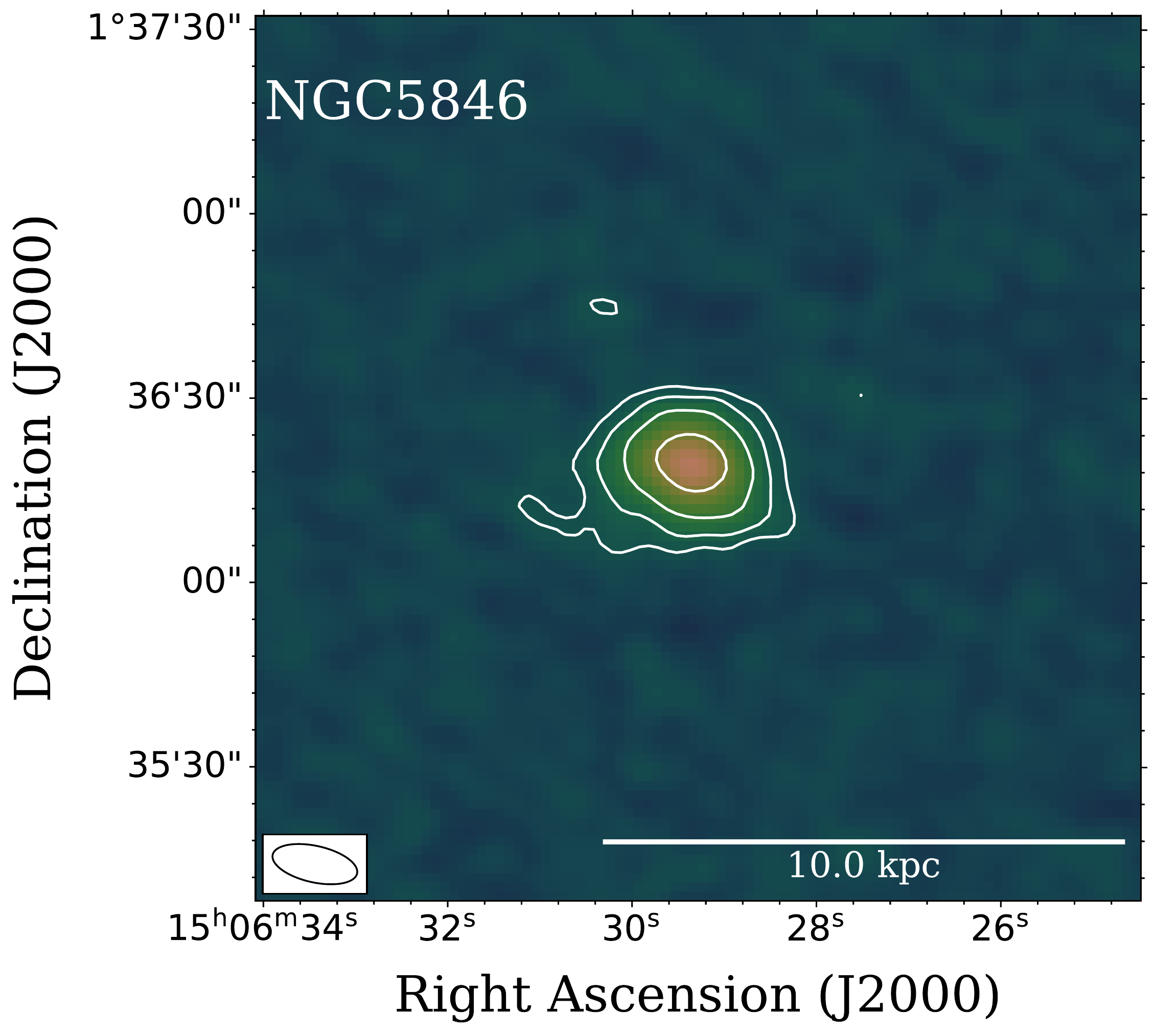} &
\includegraphics[width=54mm]{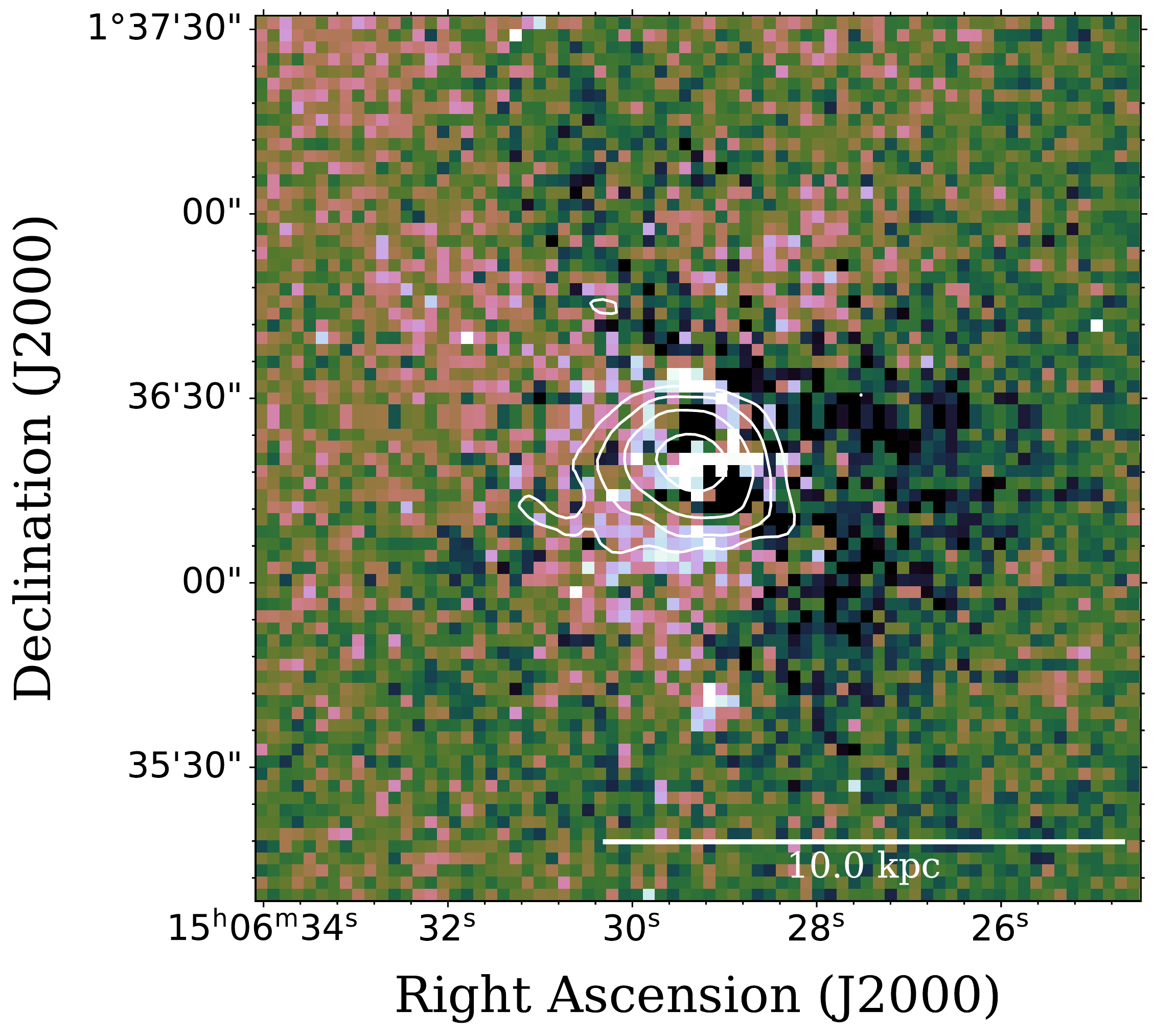}
\includegraphics[width=54mm]{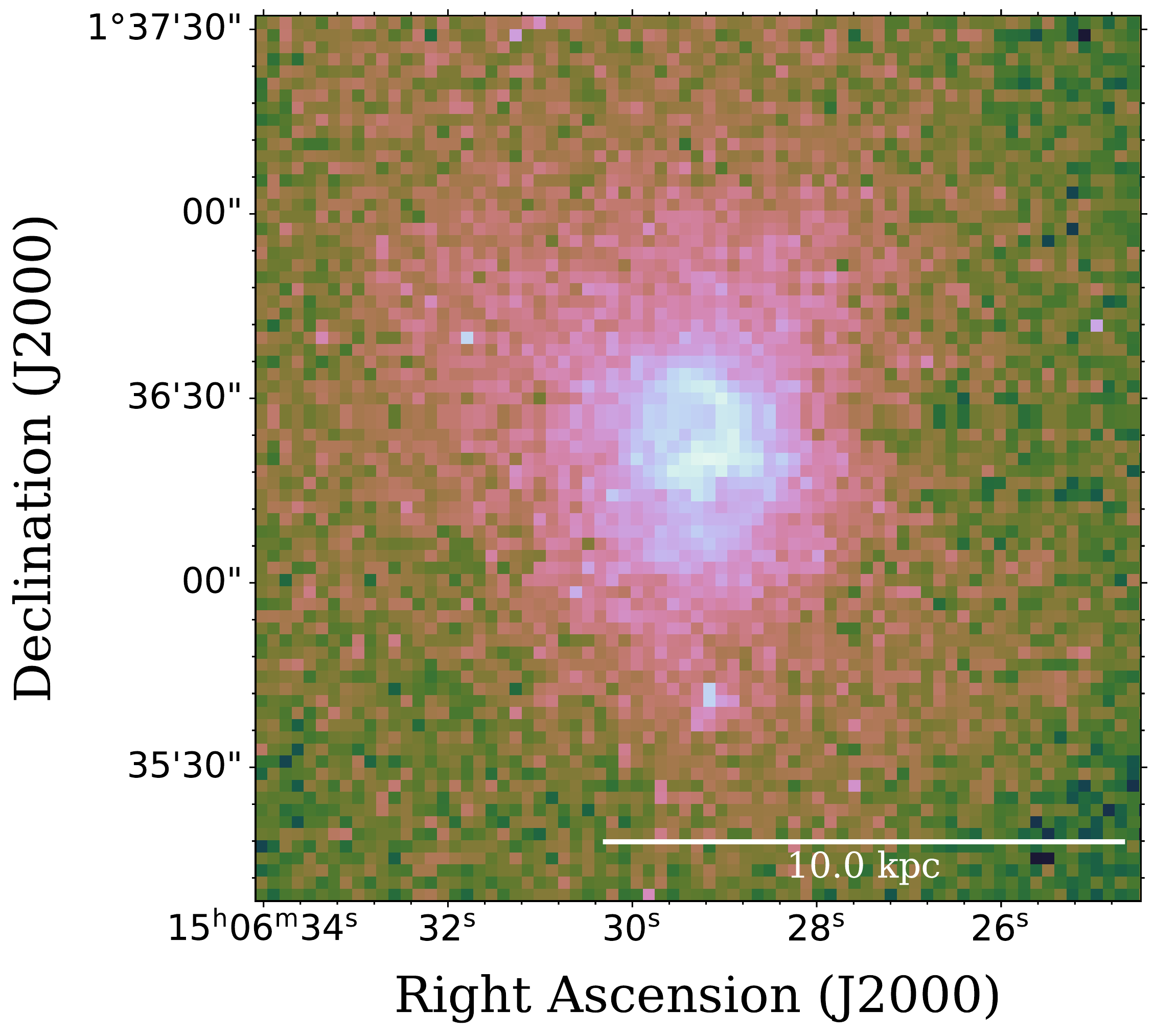}  \\
\includegraphics[width=54mm]{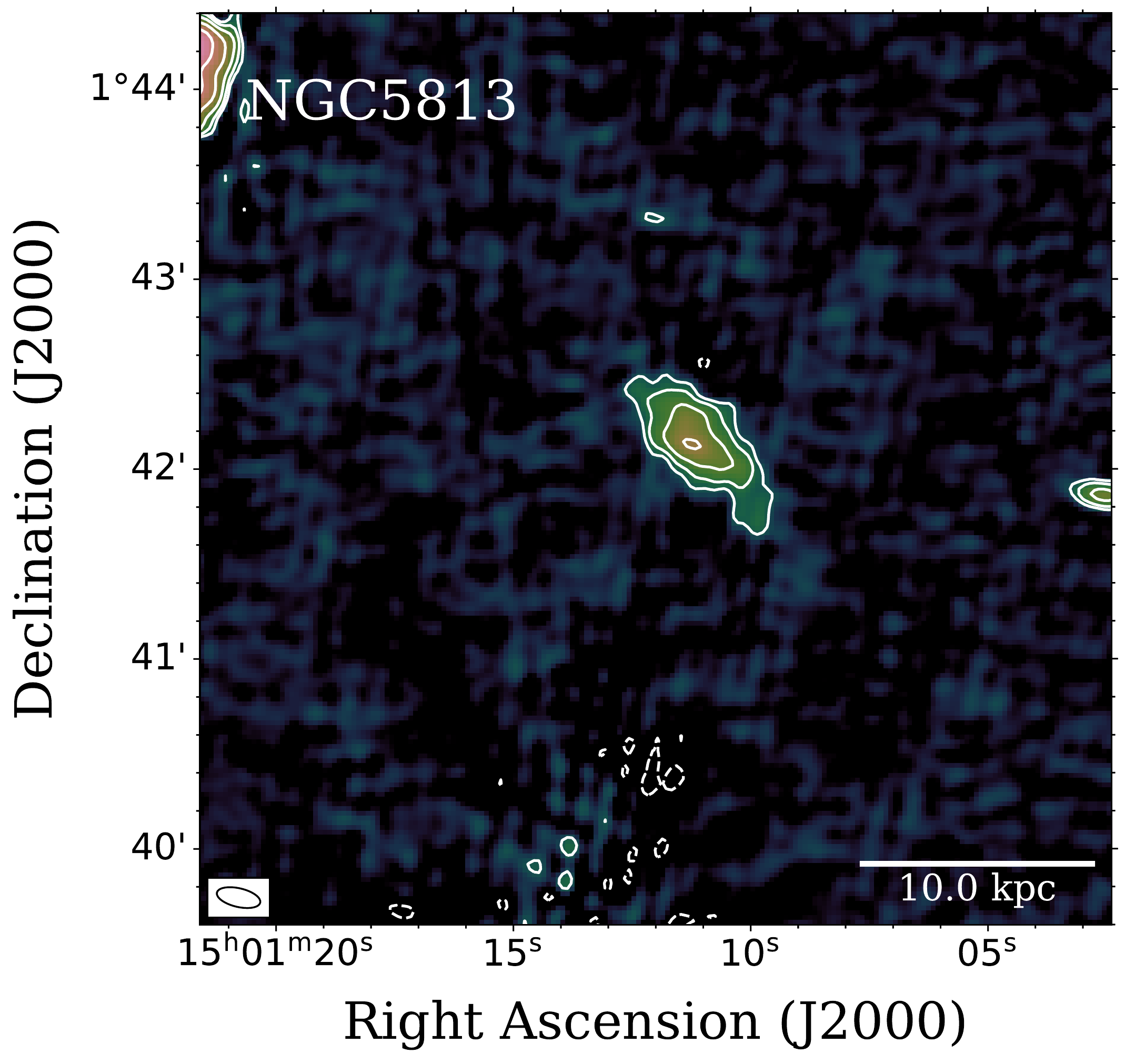} &
\includegraphics[width=54mm]{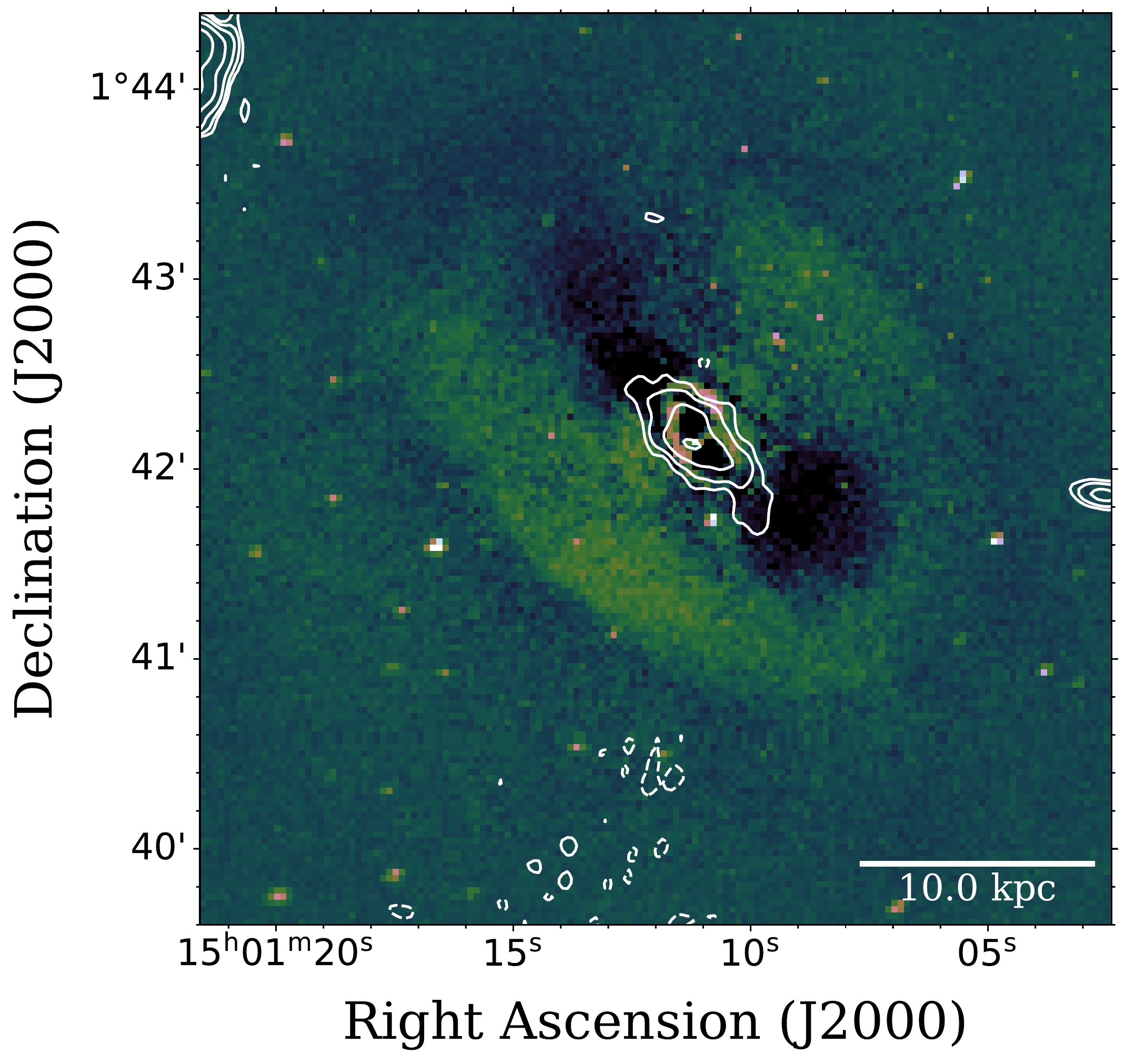}
\includegraphics[width=54mm]{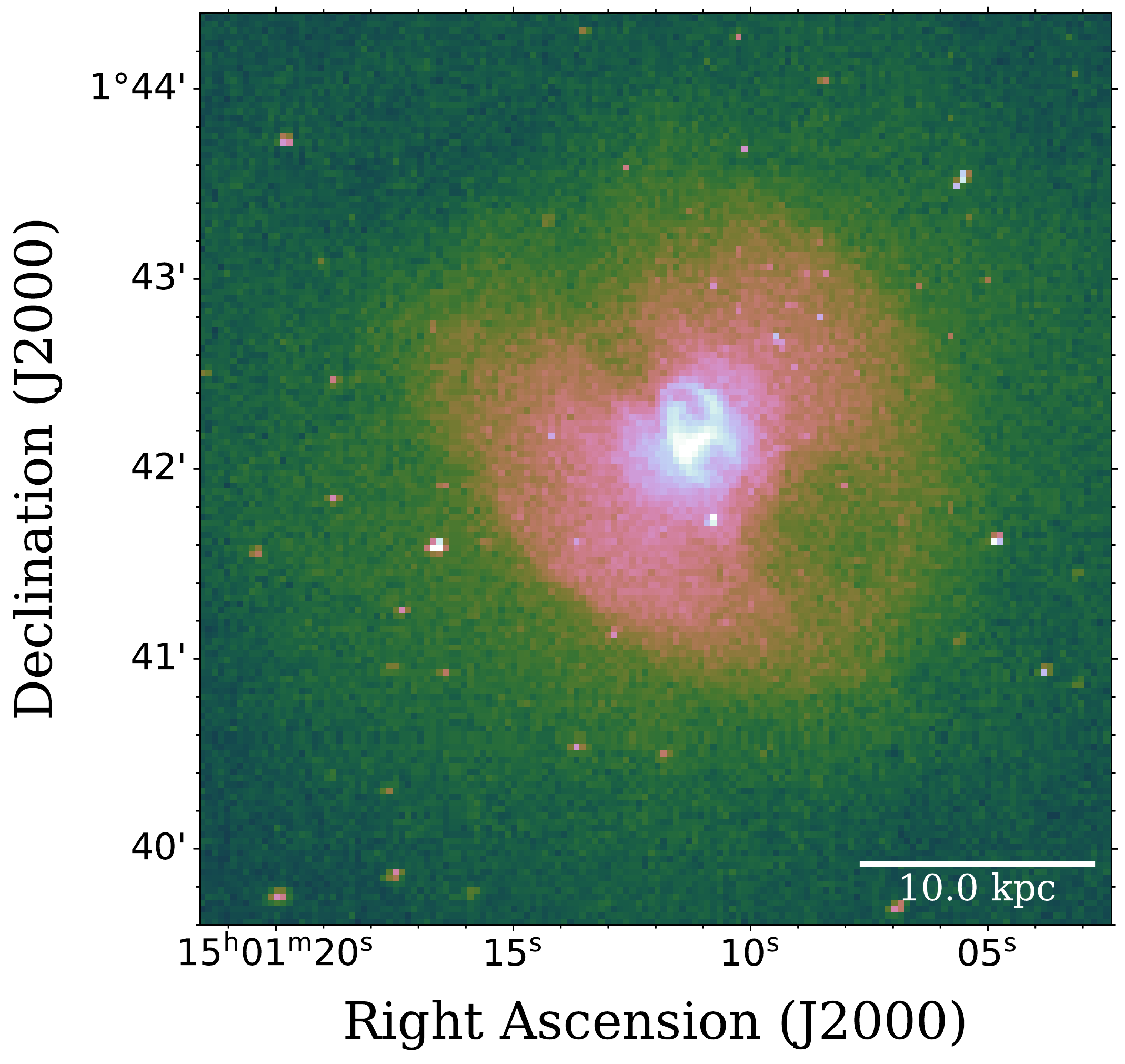} \\
\includegraphics[width=54mm]{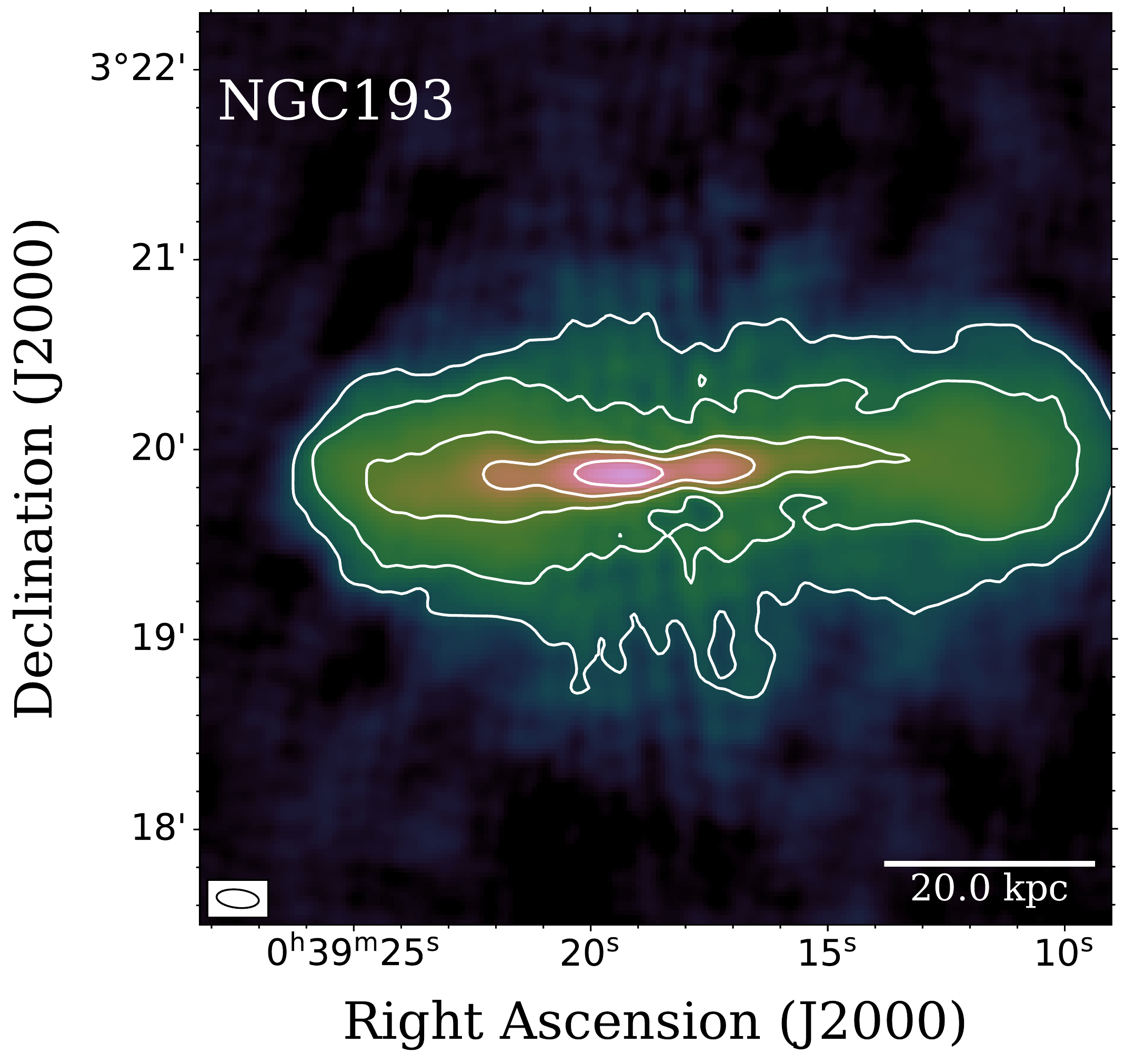} &
\includegraphics[width=54mm]{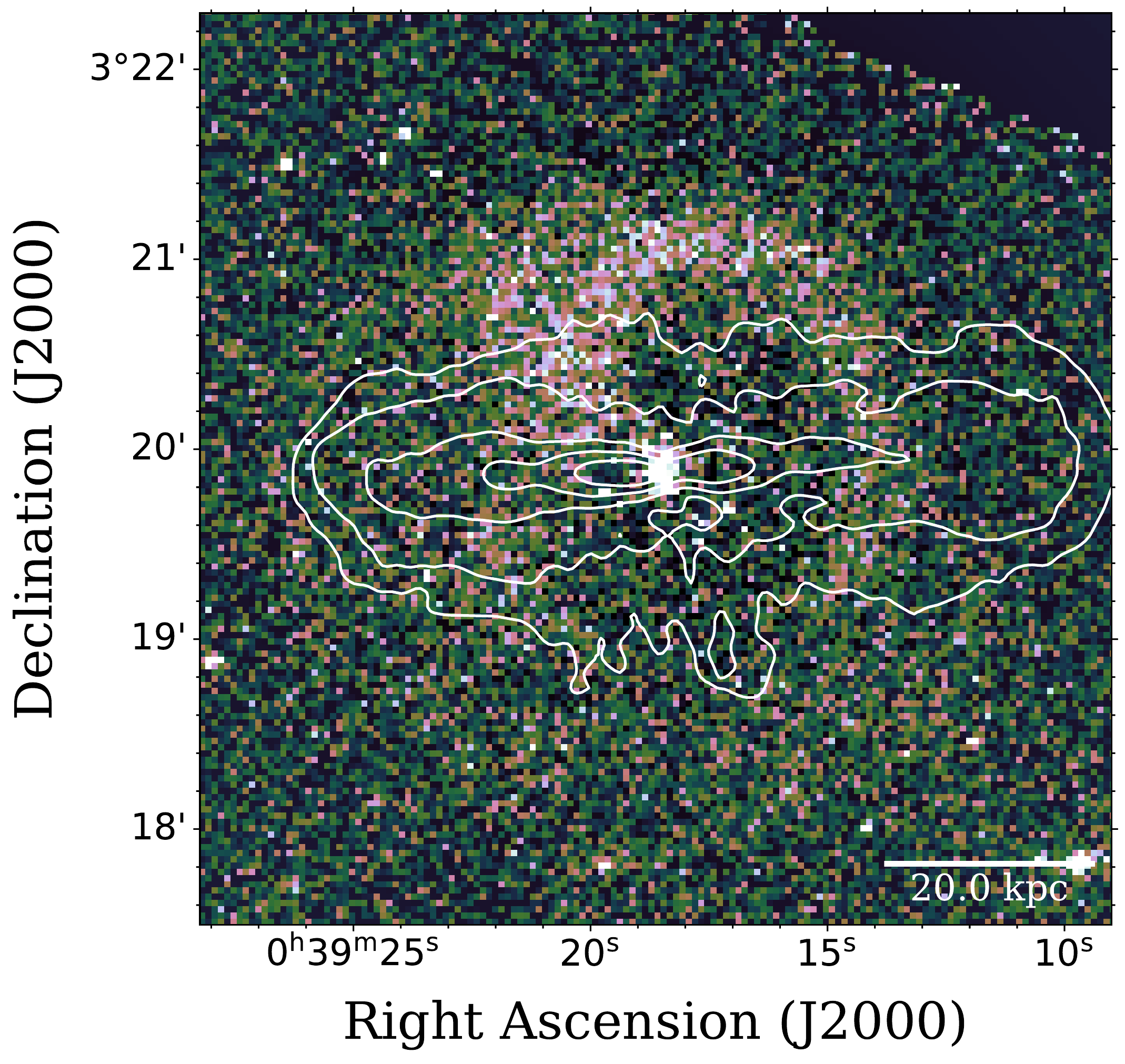}
\includegraphics[width=54mm]{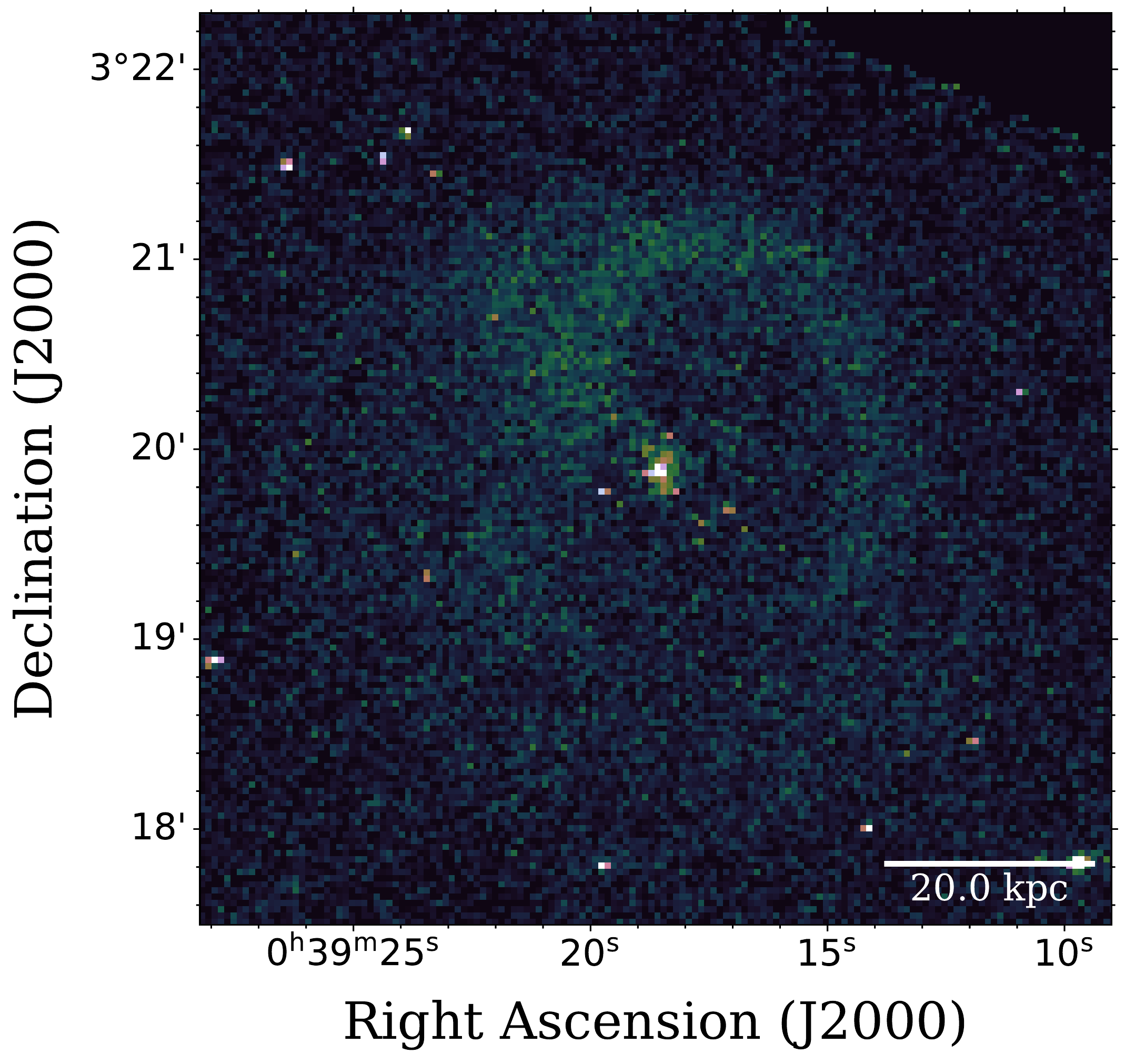}  \\
\includegraphics[width=54mm]{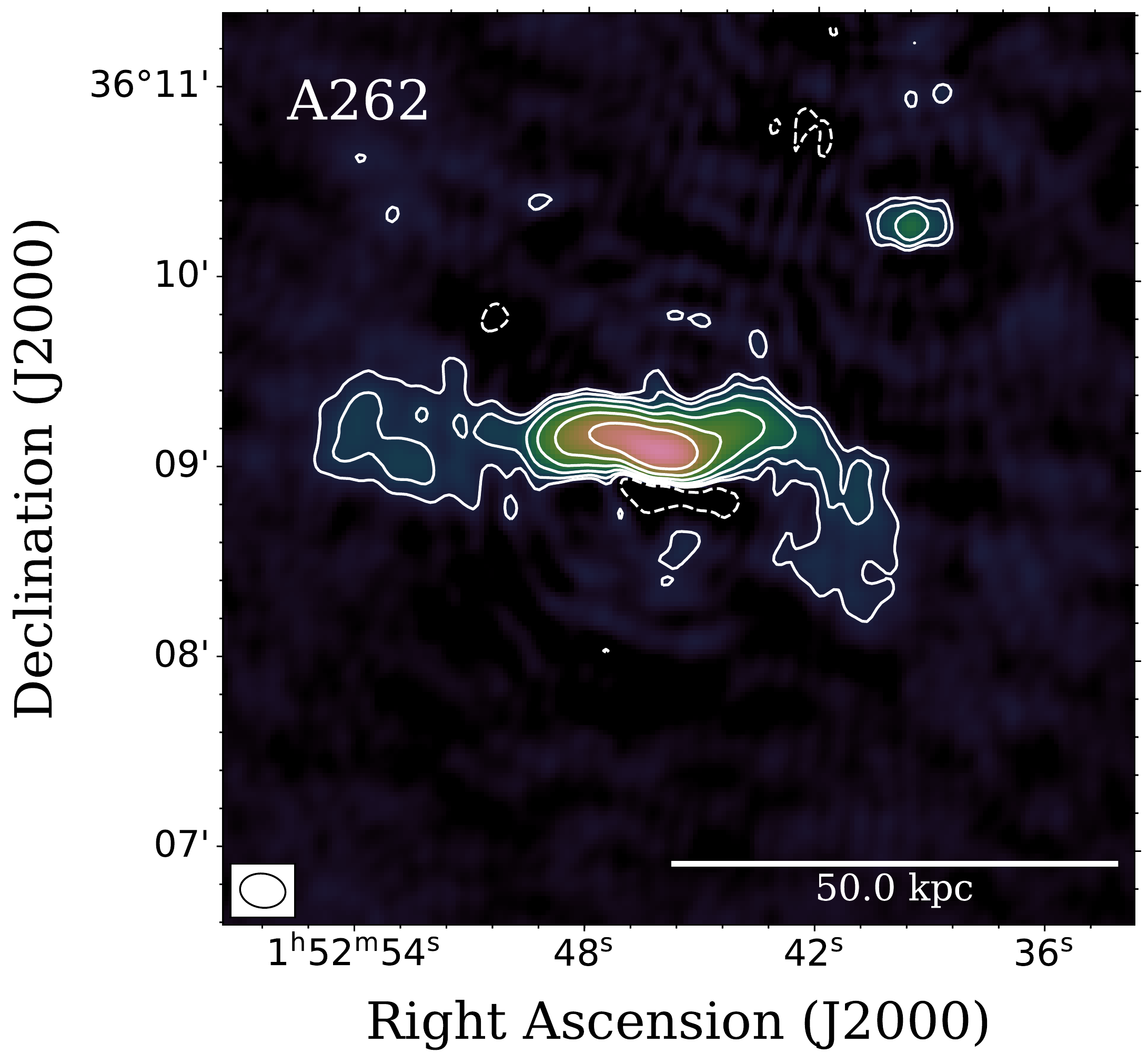} &
\includegraphics[width=54mm]{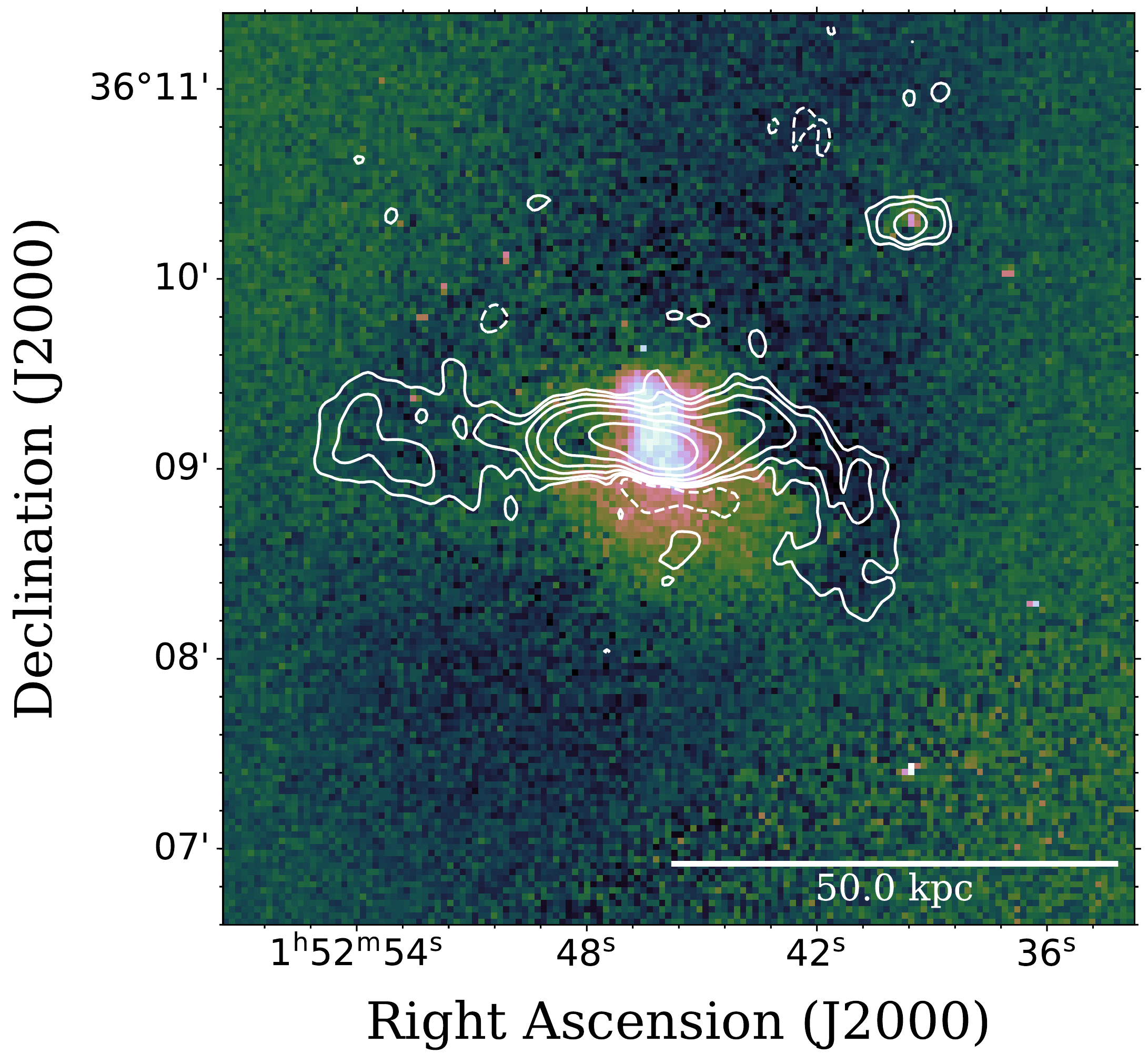}
\includegraphics[width=54mm]{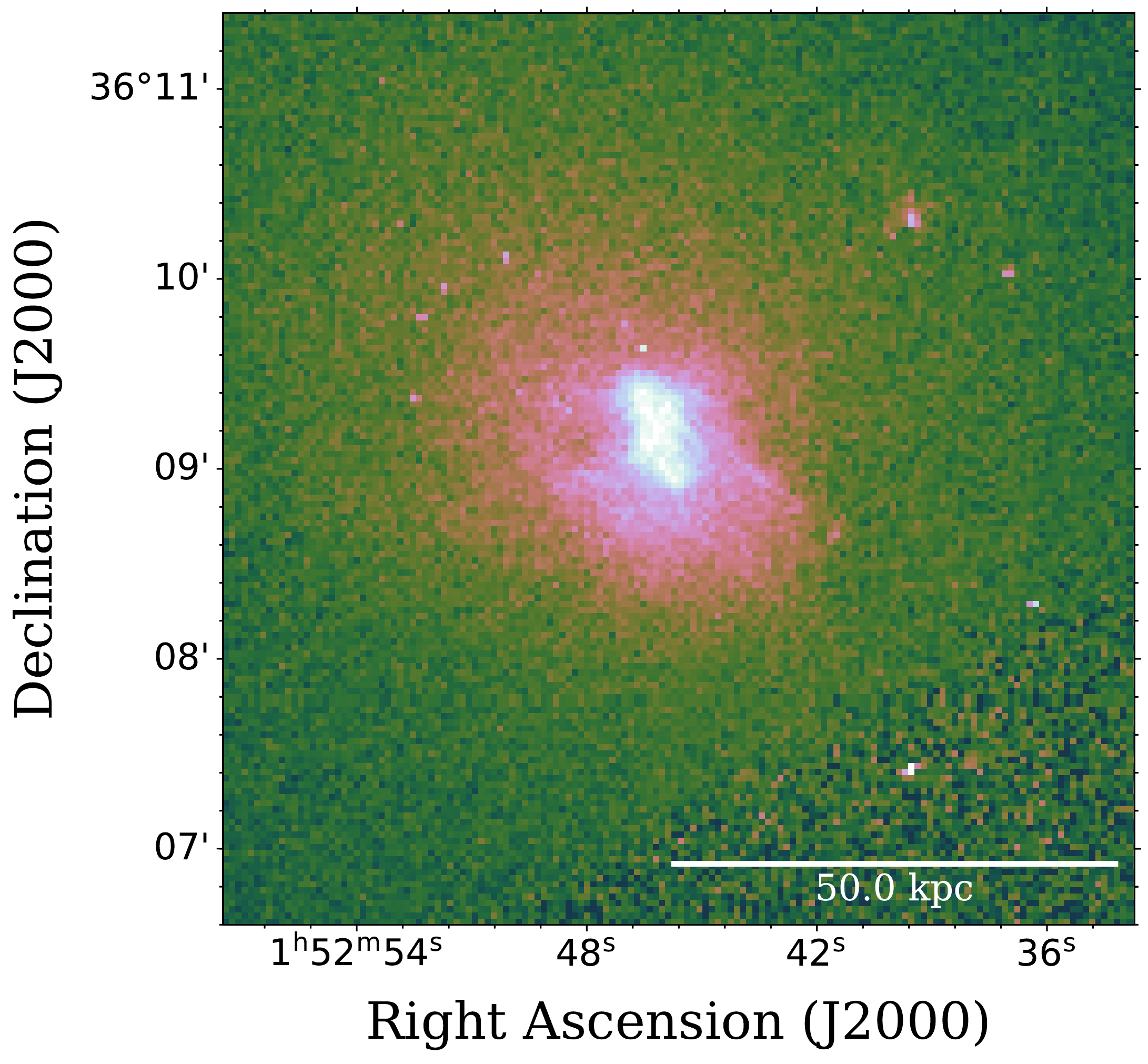} \\ \end{tabular}
\caption{\emph{Chandra} and LOFAR images for the groups and ellipticals with evident X-ray cavities, in the same order as in the tables (NGC~5846, NGC 5813, NGC 193 and A262 are shown above, and the others are shown in  Figure 1-continued).
\emph{Left panels}: LOFAR images at 143 MHz, the first contour is at 0.0026 mJy beam$^{-1}$ (NGC~5846), 0.0018 mJy beam$^{-1}$ (NGC 5813), 0.009 mJy beam$^{-1}$ (NGC 193), 0.003 mJy beam$^{-1}$ (A262), and each contour increases by a factor of two (the beam size is shown in the lower left-hand corner); \emph{middle panels}: overlays of LOFAR contours and the X-ray residual maps; \emph{right panels}: smoothed X-ray images.}
\label{F:images_1}
\end{figure*}

\begin{figure*} \begin{tabular}{@{}cc}
\includegraphics[width=54mm]{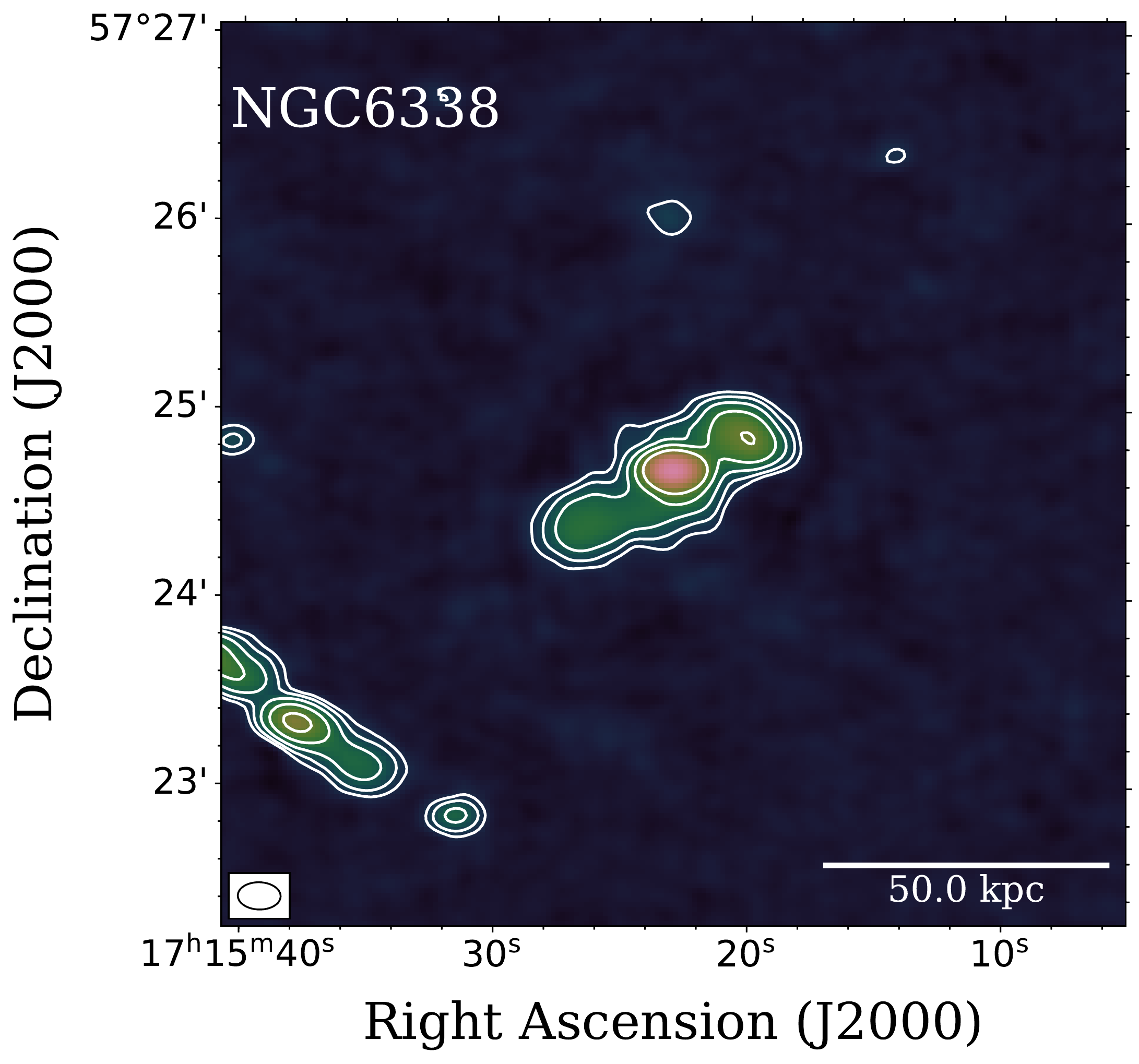} &
\includegraphics[width=54mm]{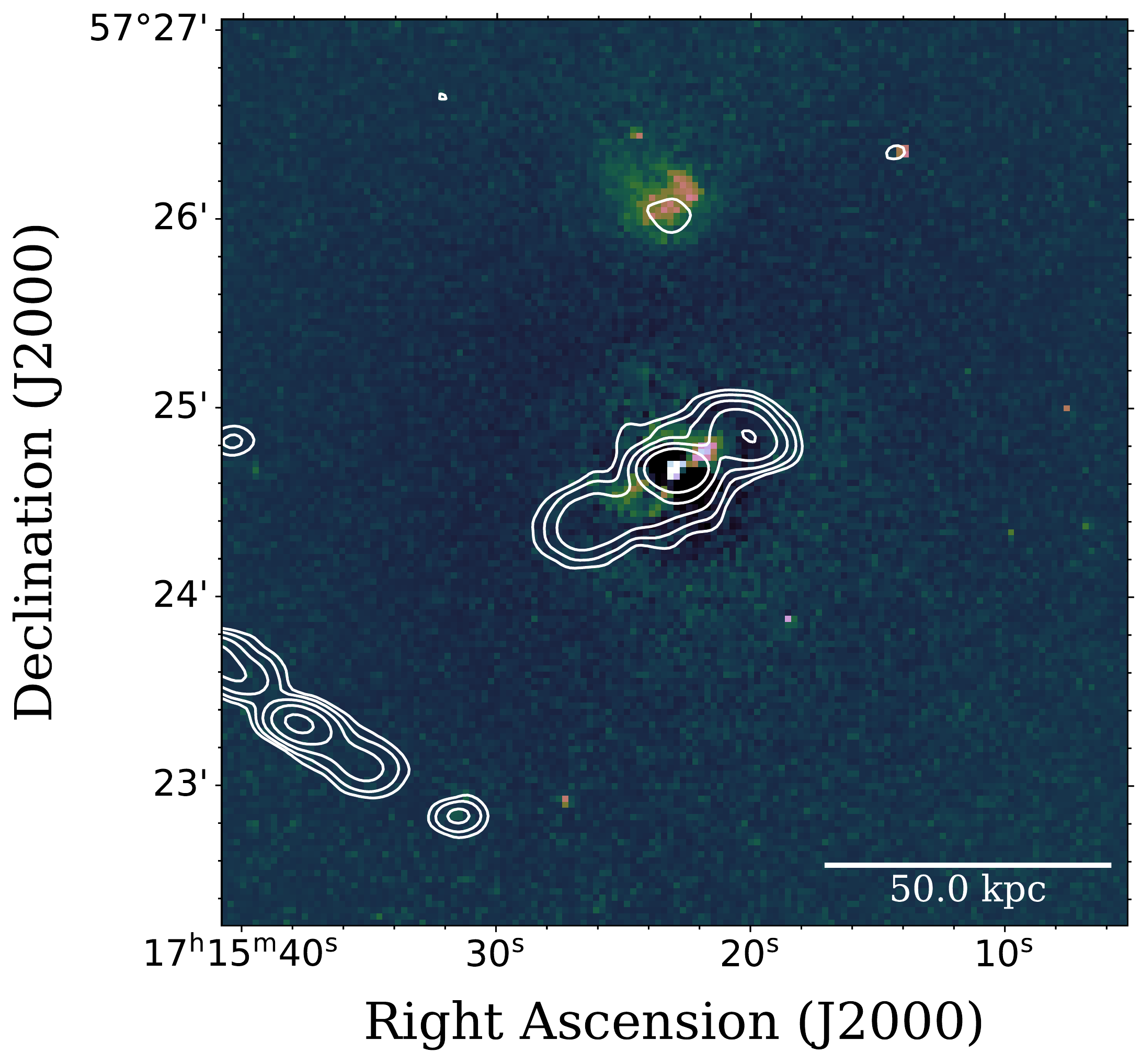}
\includegraphics[width=54mm]{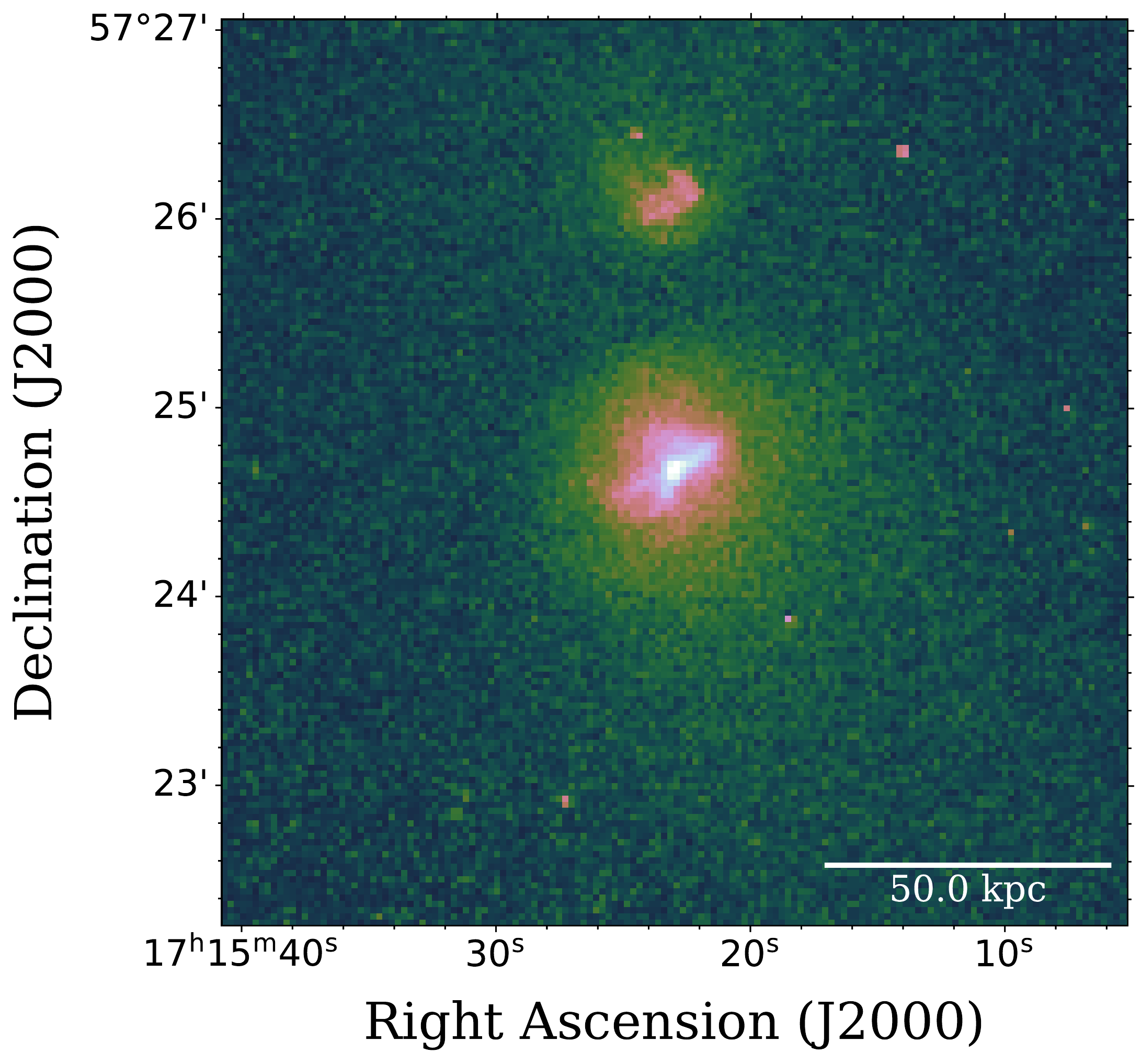} \\
\includegraphics[width=54mm]{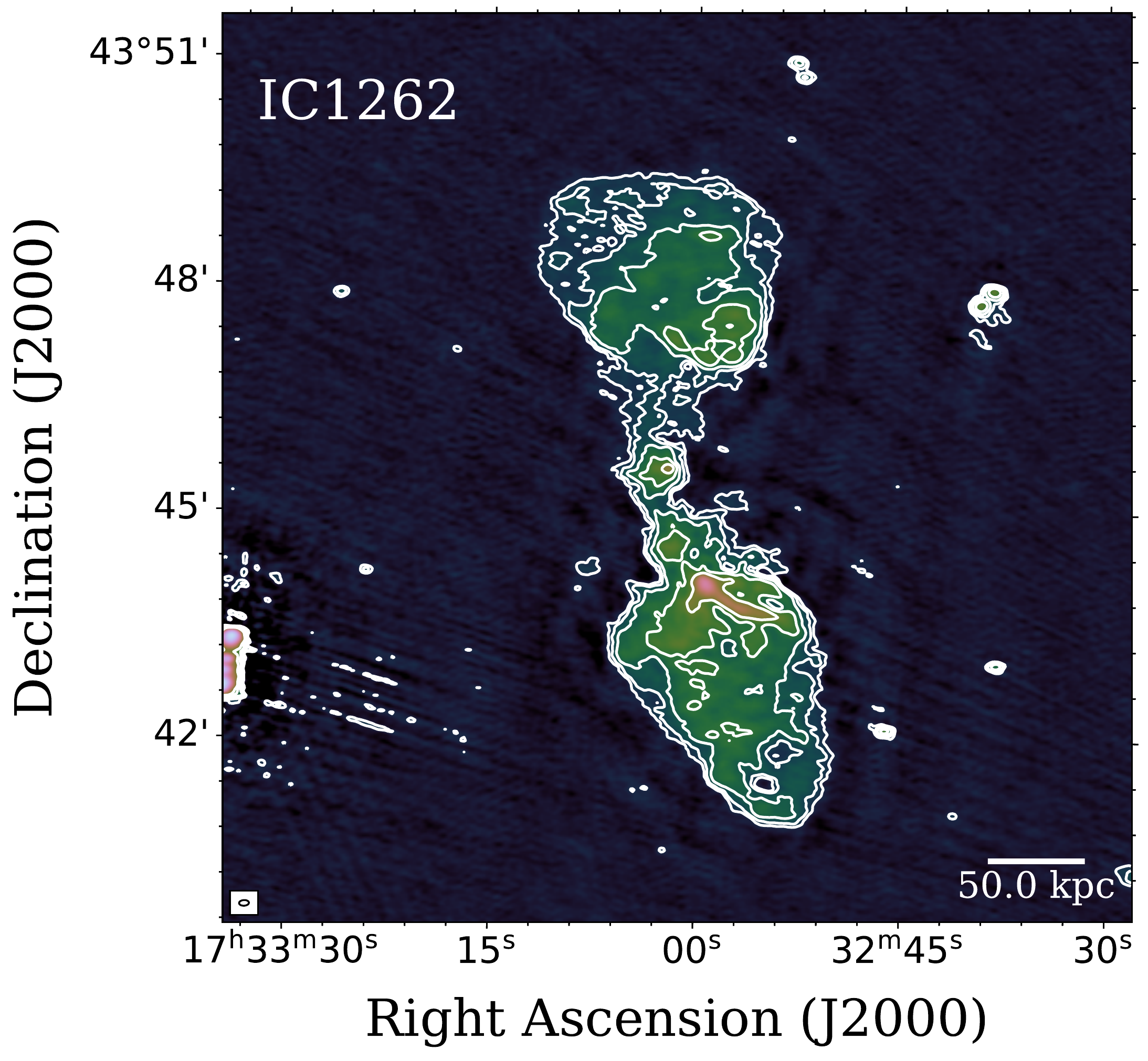} &
\includegraphics[width=54mm]{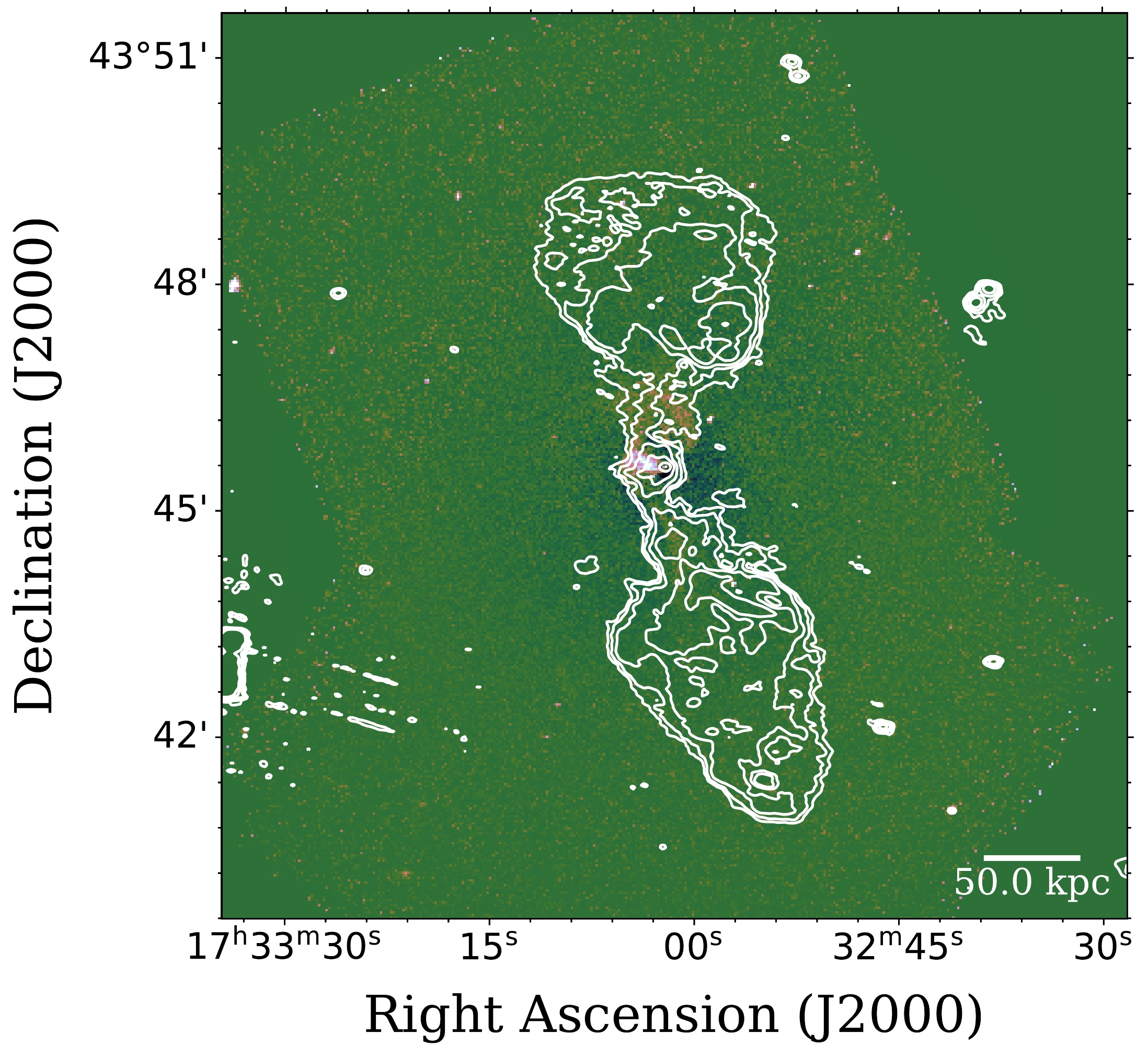}
\includegraphics[width=54mm]{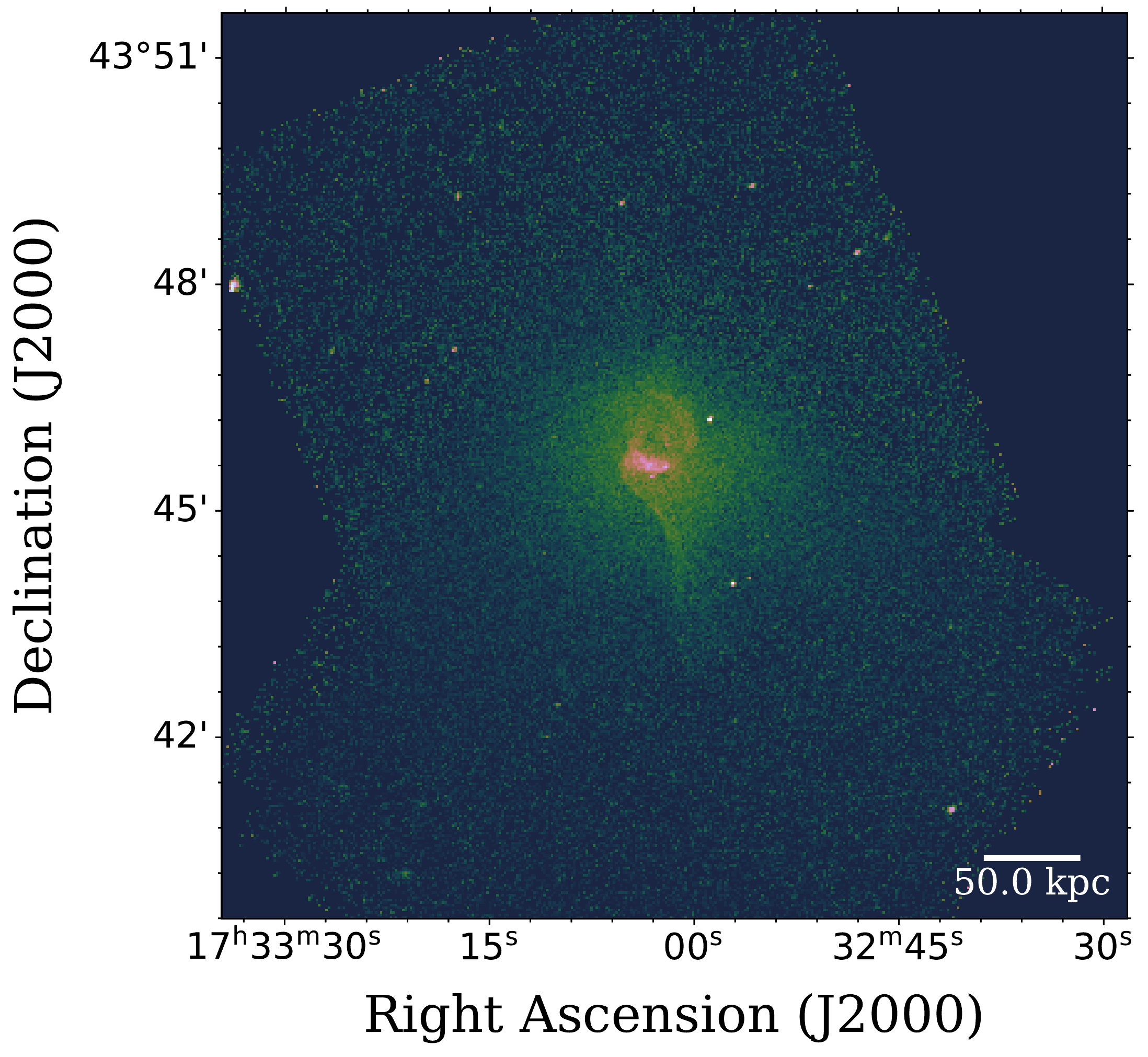} \\
\includegraphics[width=54mm]{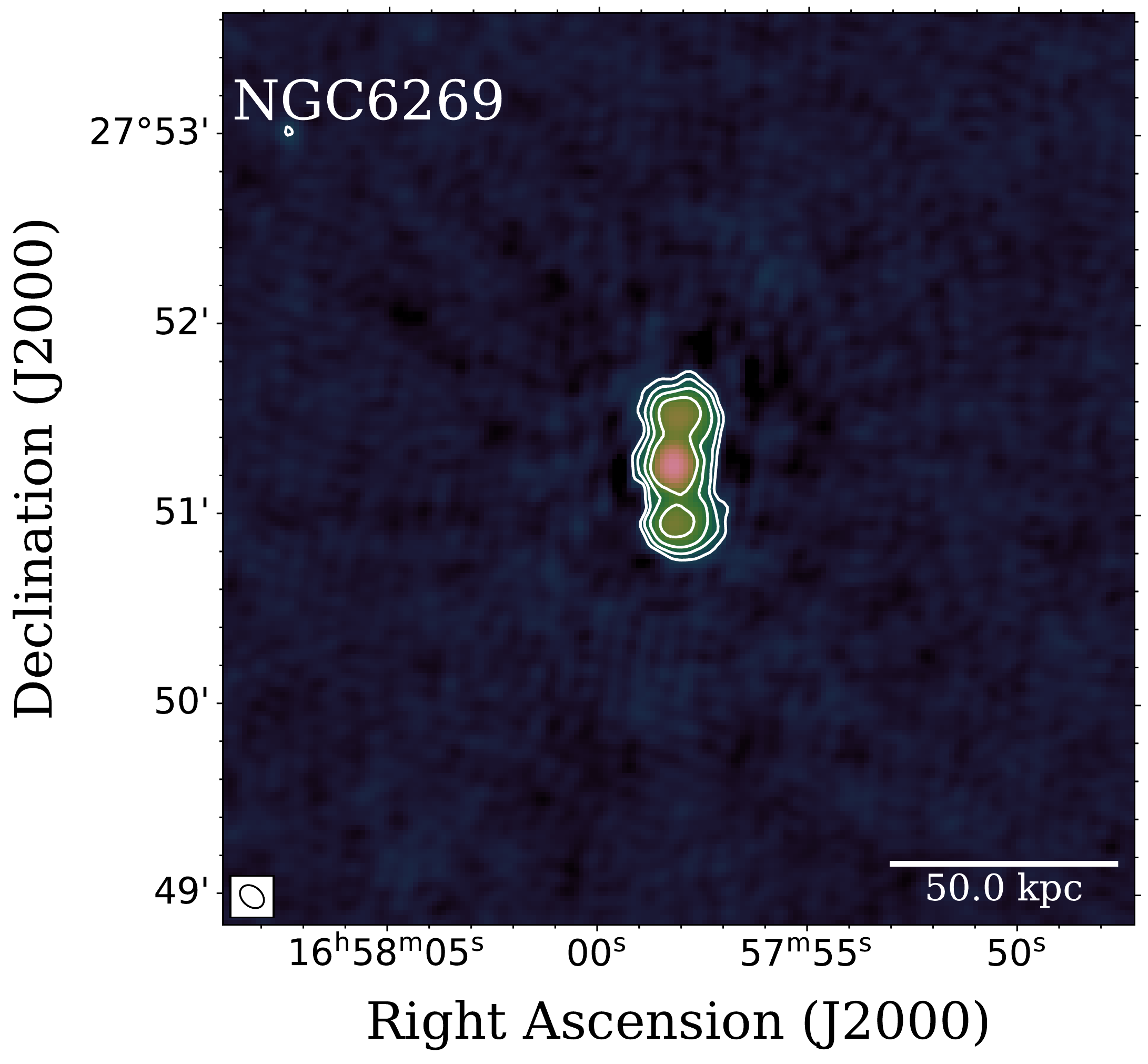} &
\includegraphics[width=54mm]{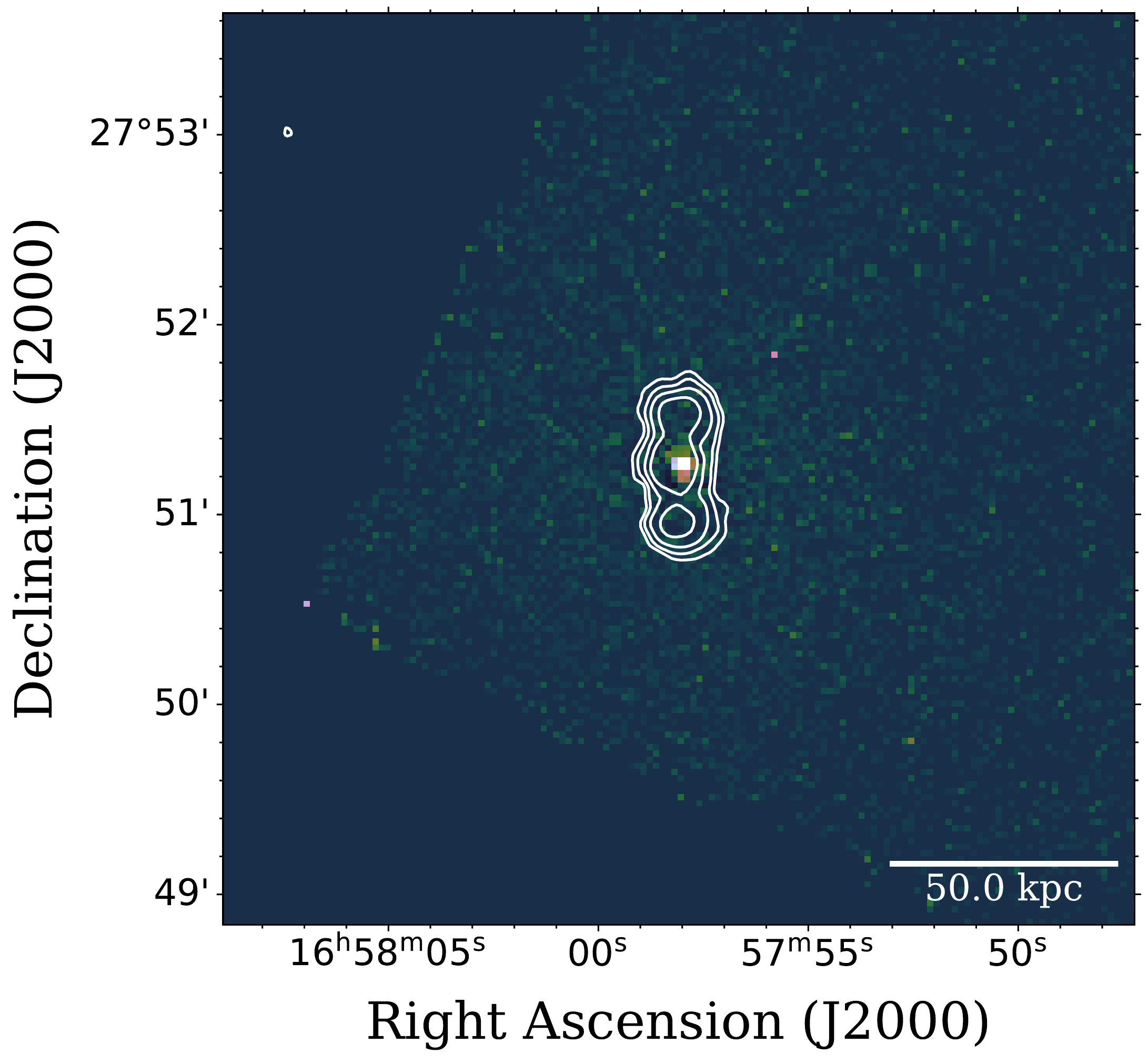}
\includegraphics[width=54mm]{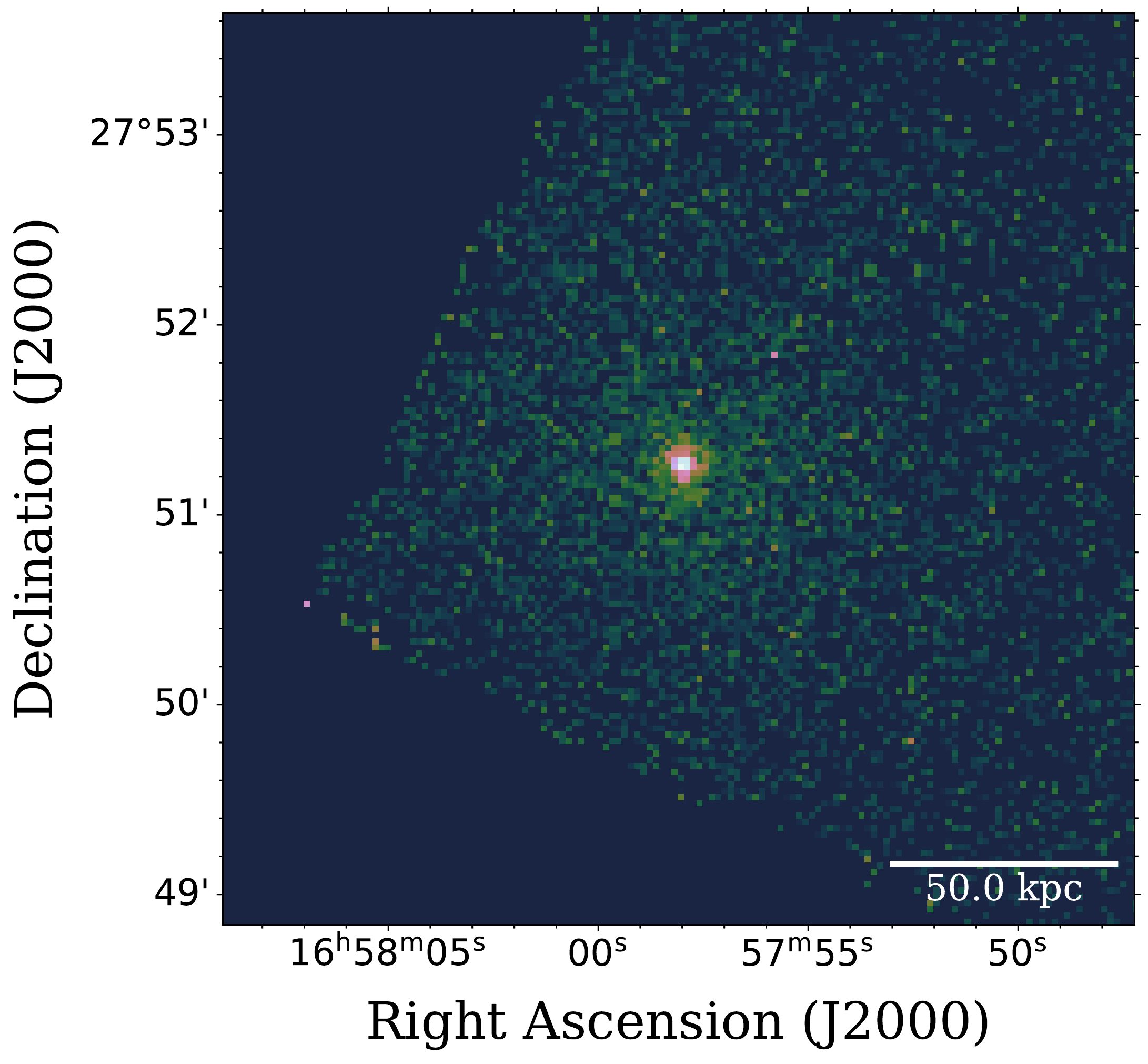} \\
\includegraphics[width=54mm]{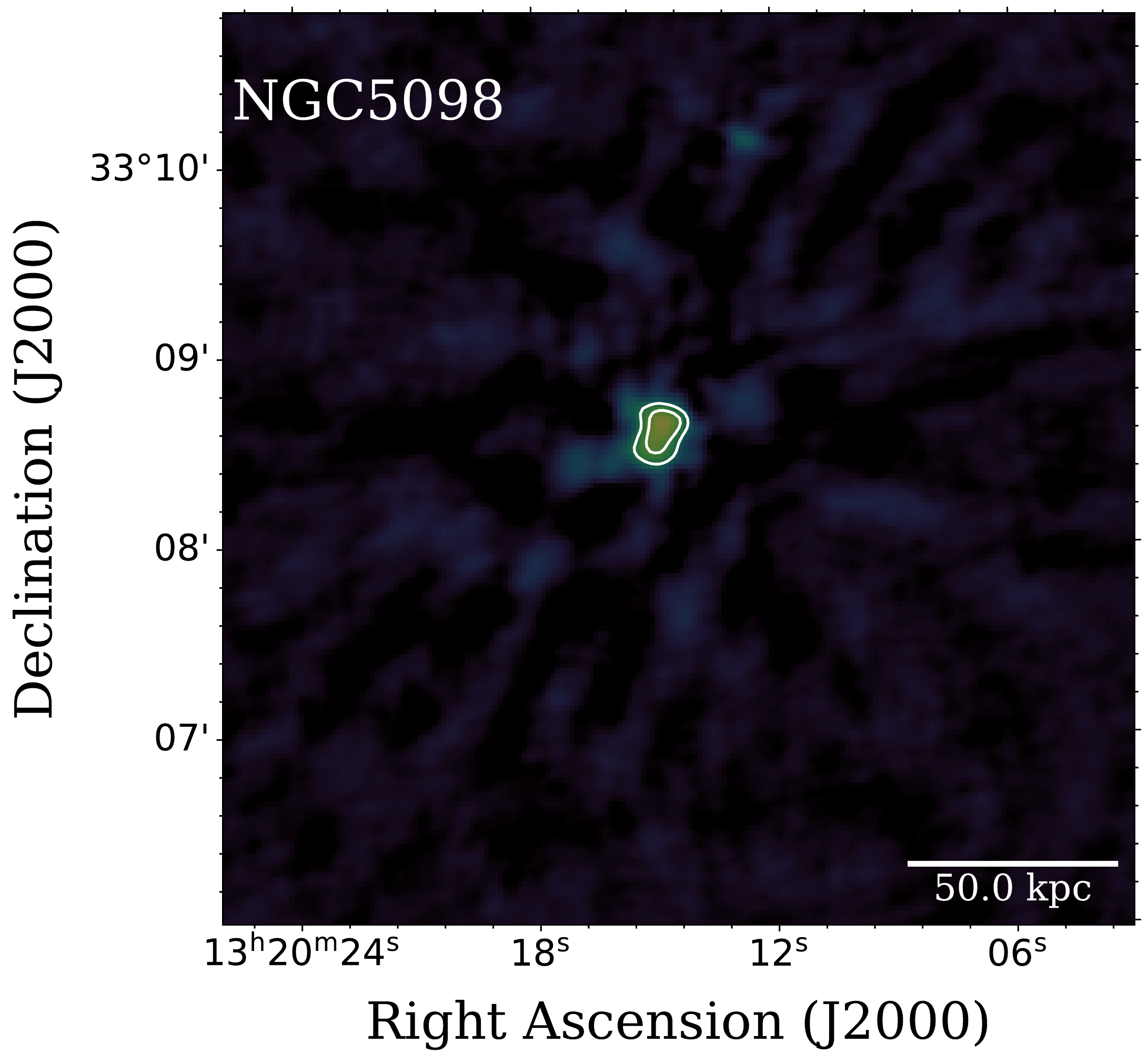} &
\includegraphics[width=54mm]{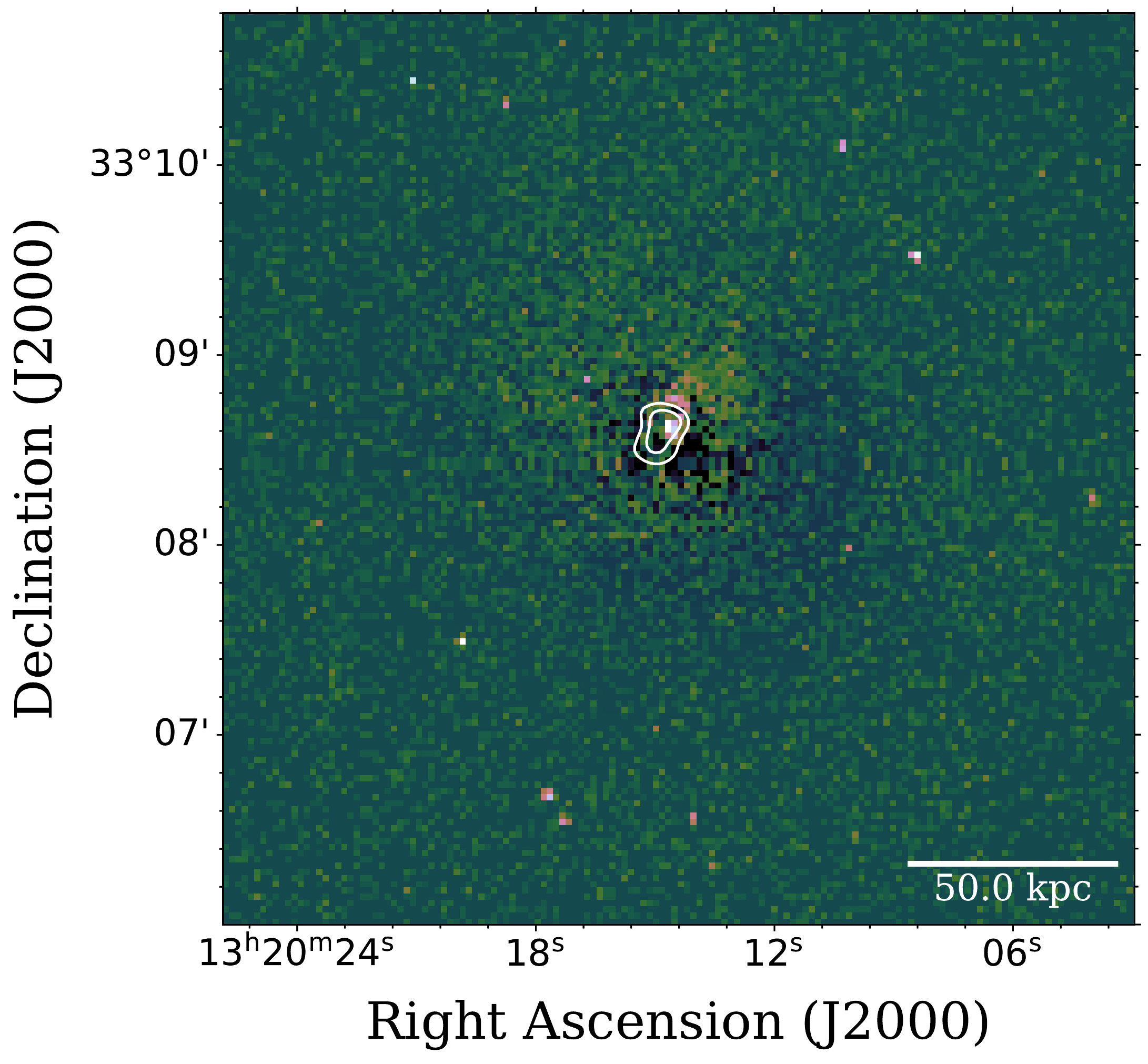}
\includegraphics[width=54mm]{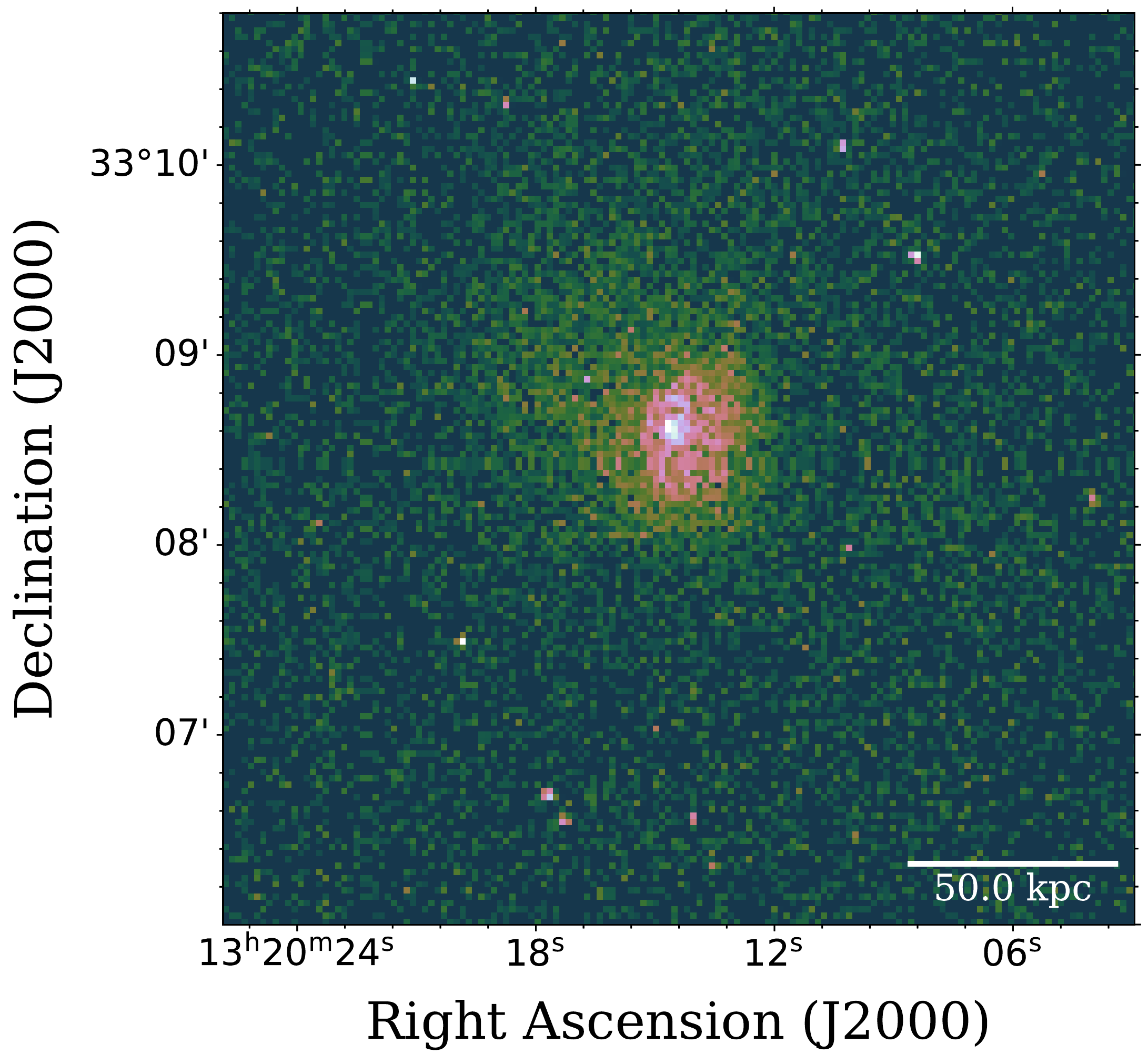} \\ \end{tabular}
\textbf{Figure}~1. --- continued (NGC 6338, IC1262, NGC 6269 and NGC 5098). For the LOFAR image, the first contour is at 0.00105~mJy beam$^{-1}$ (NGC 6338), 0.00105~mJy beam$^{-1}$ (IC1262), 0.0017~mJy beam$^{-1}$ (NGC 6269), 0.018~mJy beam$^{-1}$ (NGC 5098), and each contour increases by a factor of two. \\
\end{figure*}

\begin{figure*} \begin{tabular}{@{}cc}
\includegraphics[width=54mm]{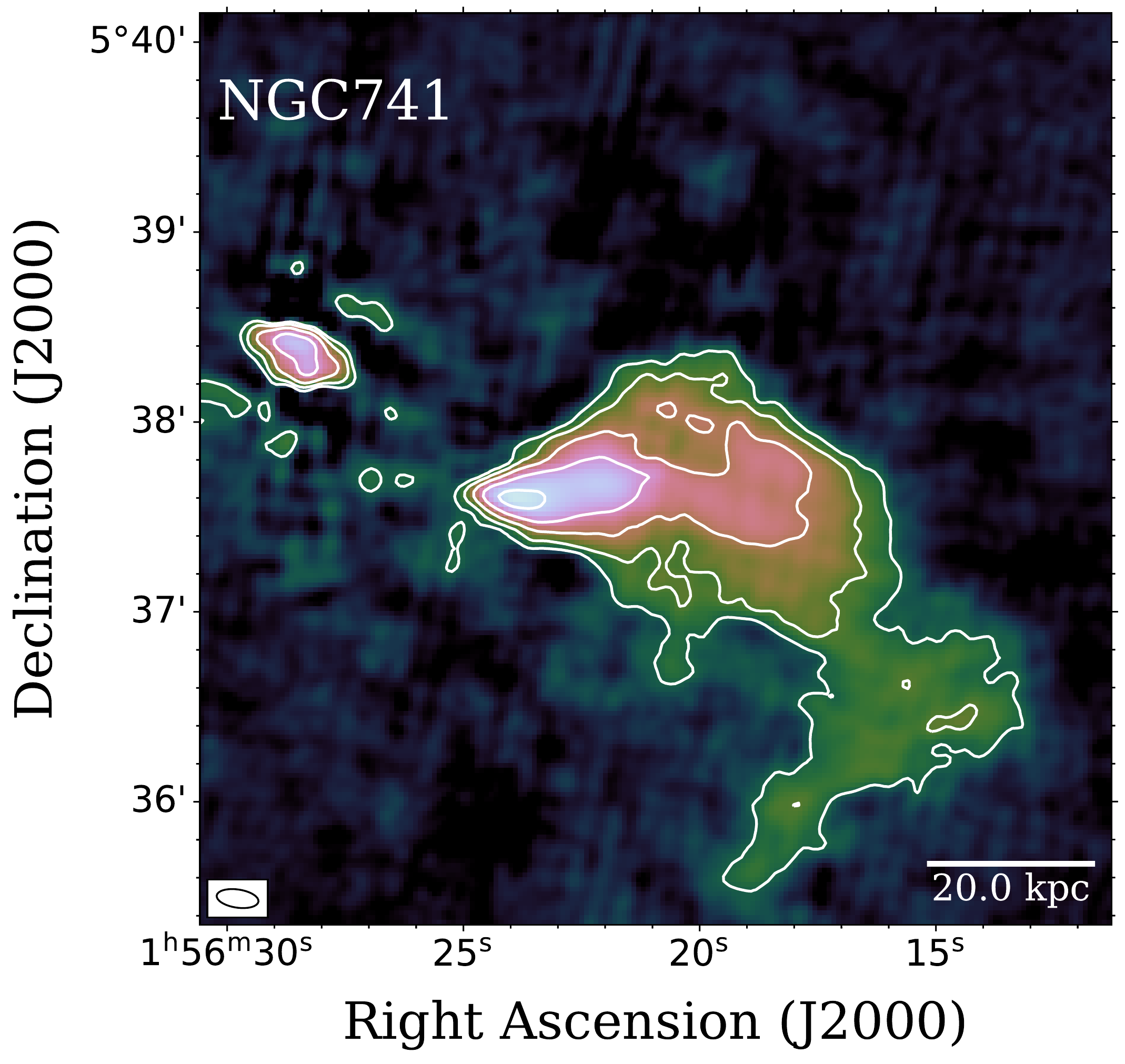} &
\includegraphics[width=54mm]{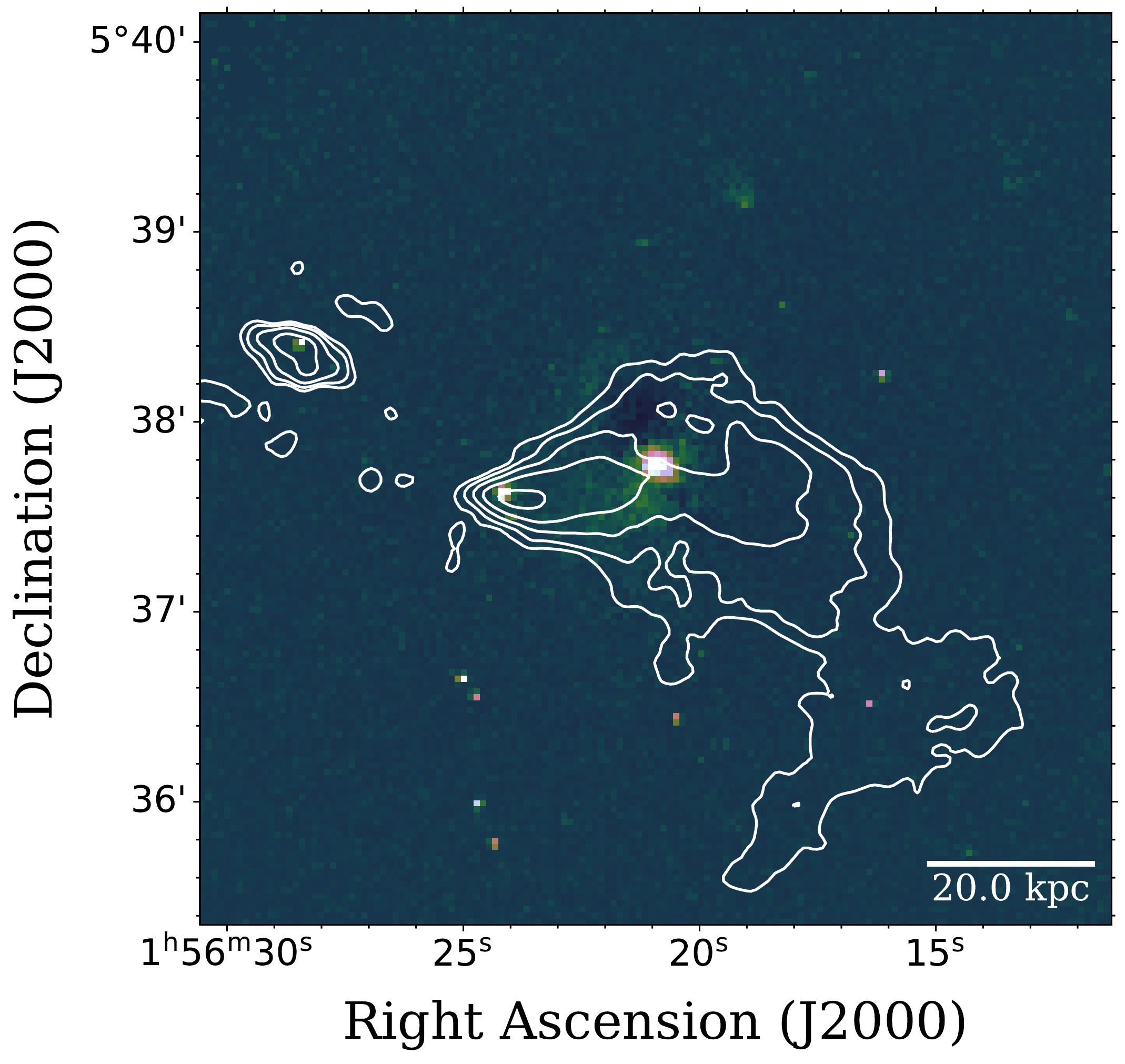}
\includegraphics[width=54mm]{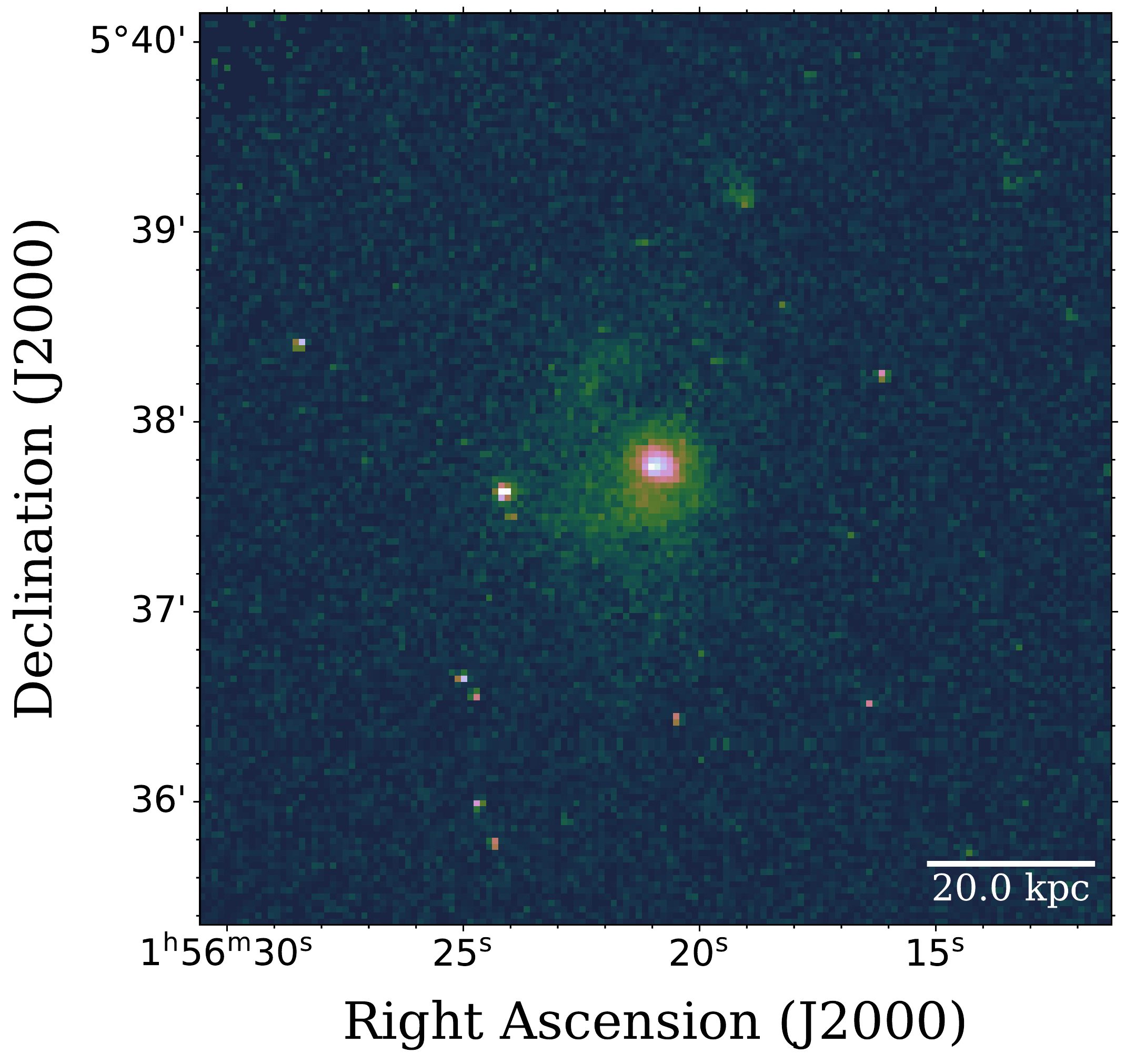}  \\
\includegraphics[width=54mm]{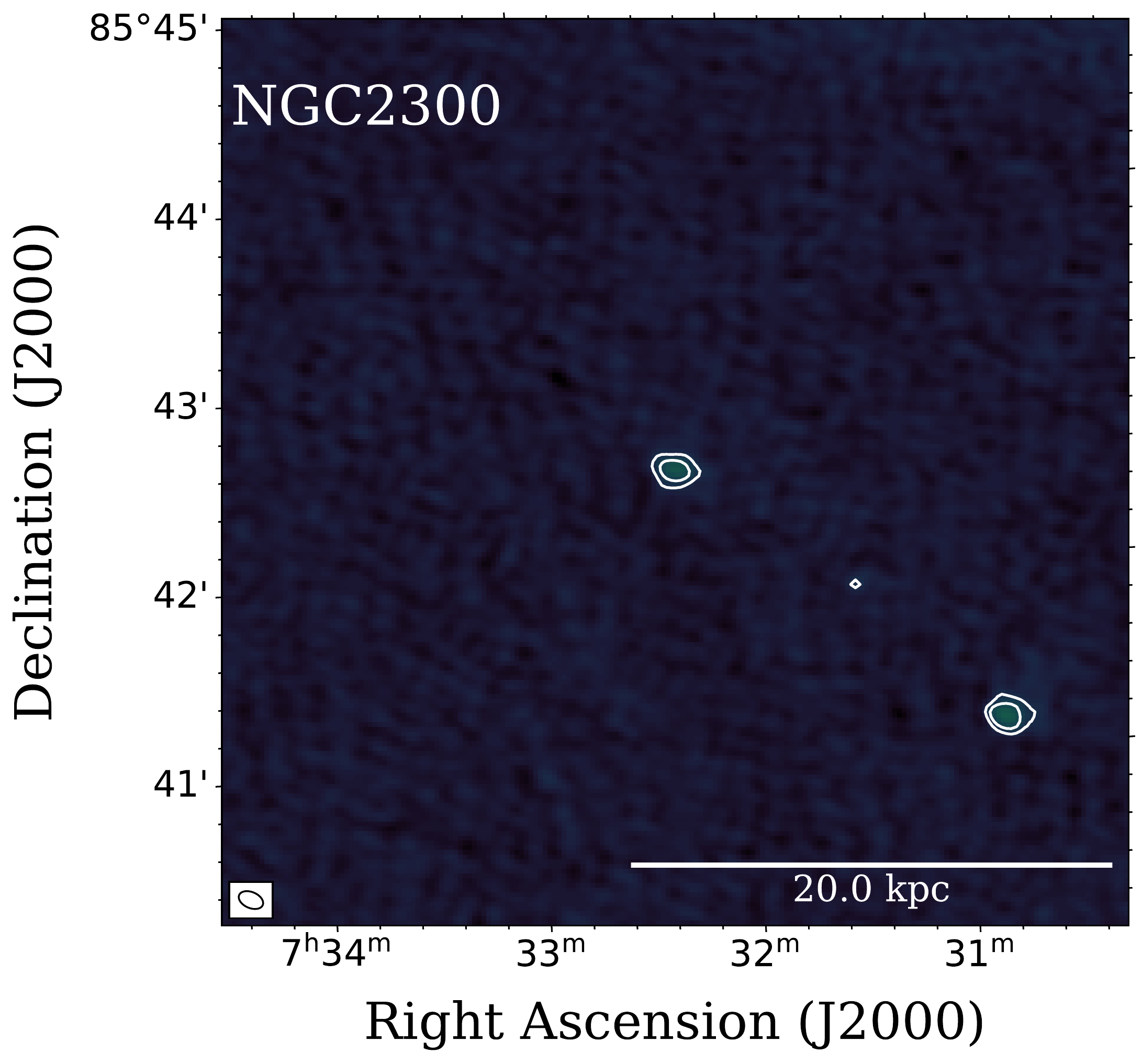} &
\includegraphics[width=54mm]{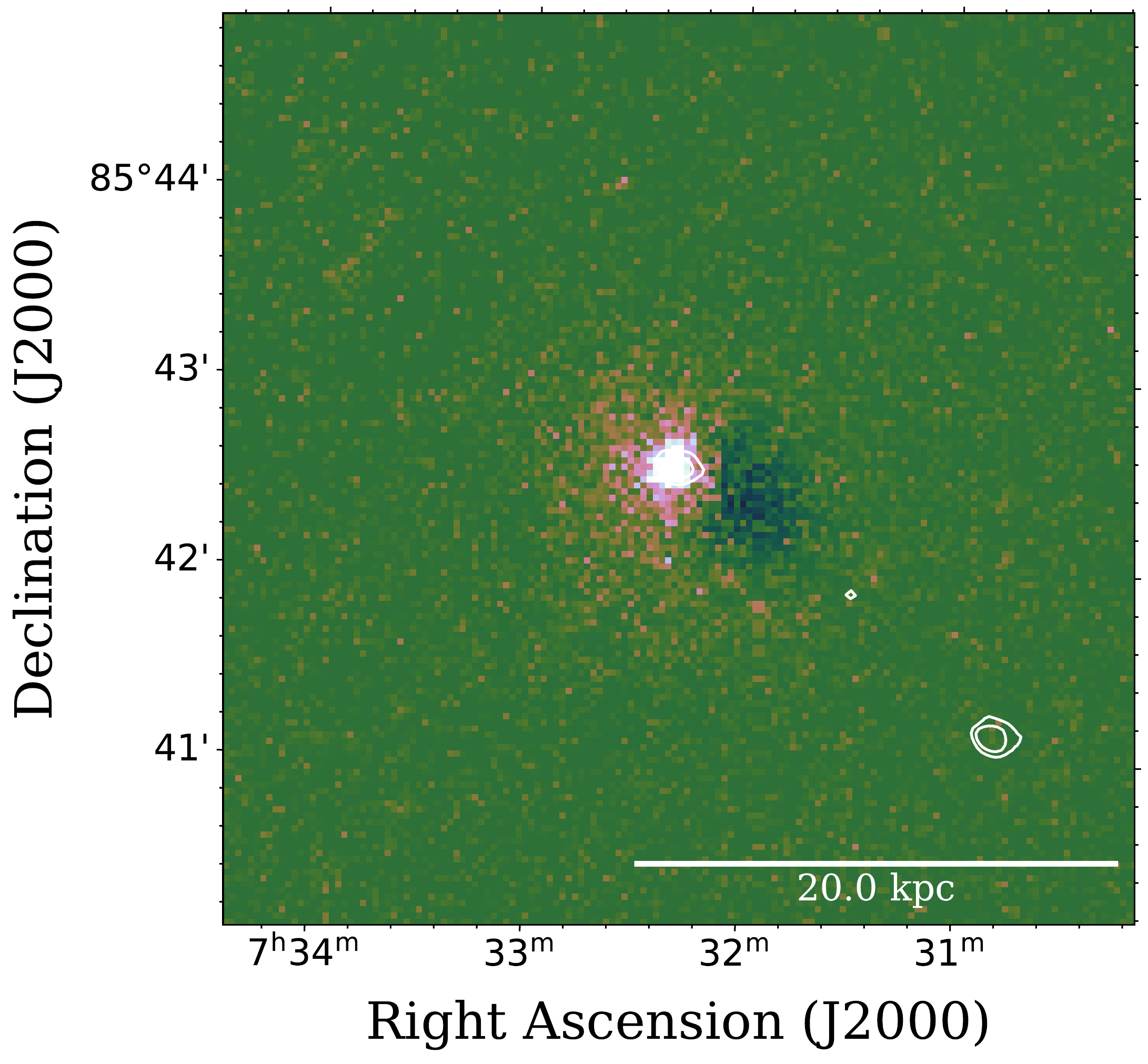}
\includegraphics[width=54mm]{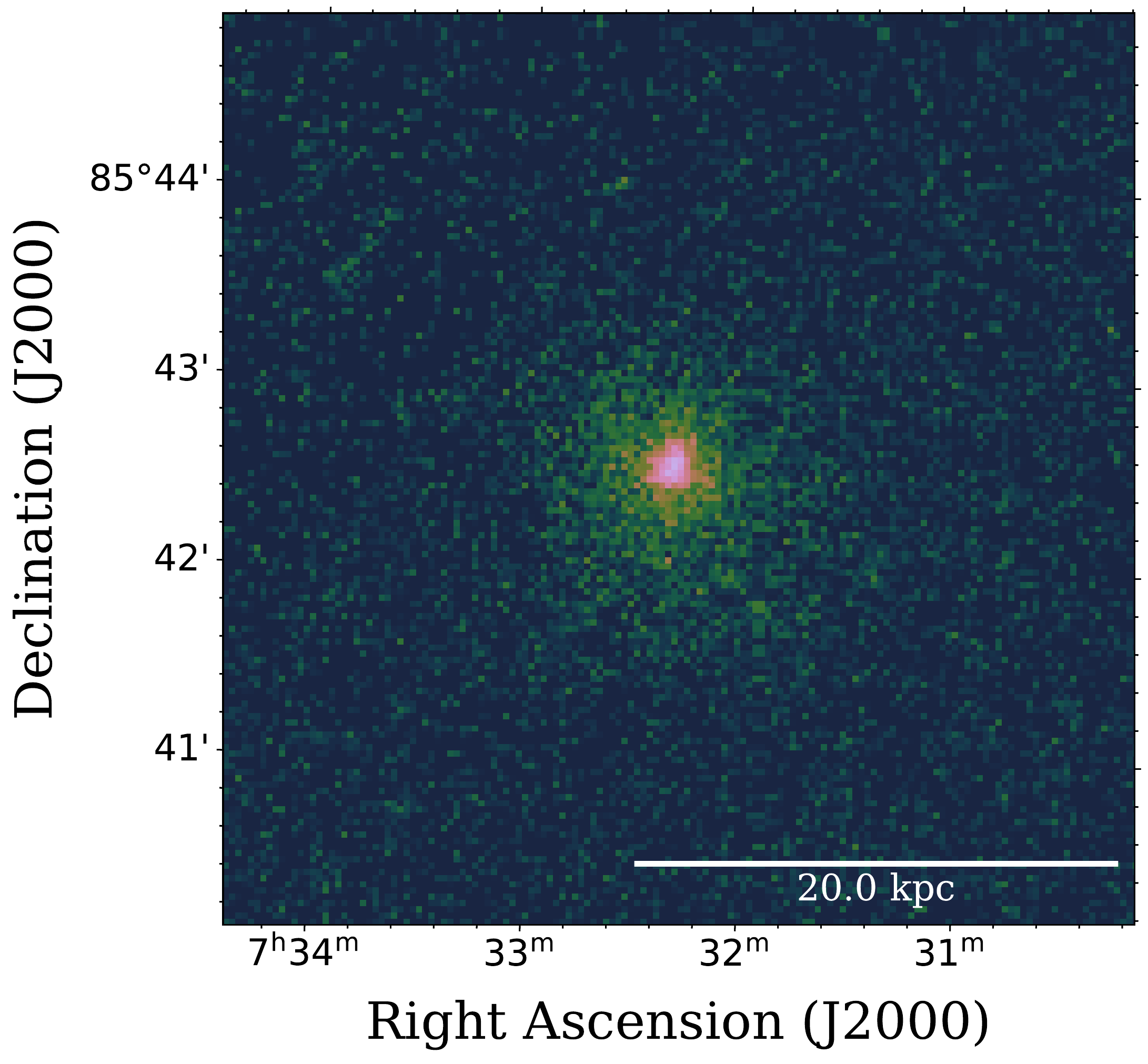}  \\
\includegraphics[width=54mm]{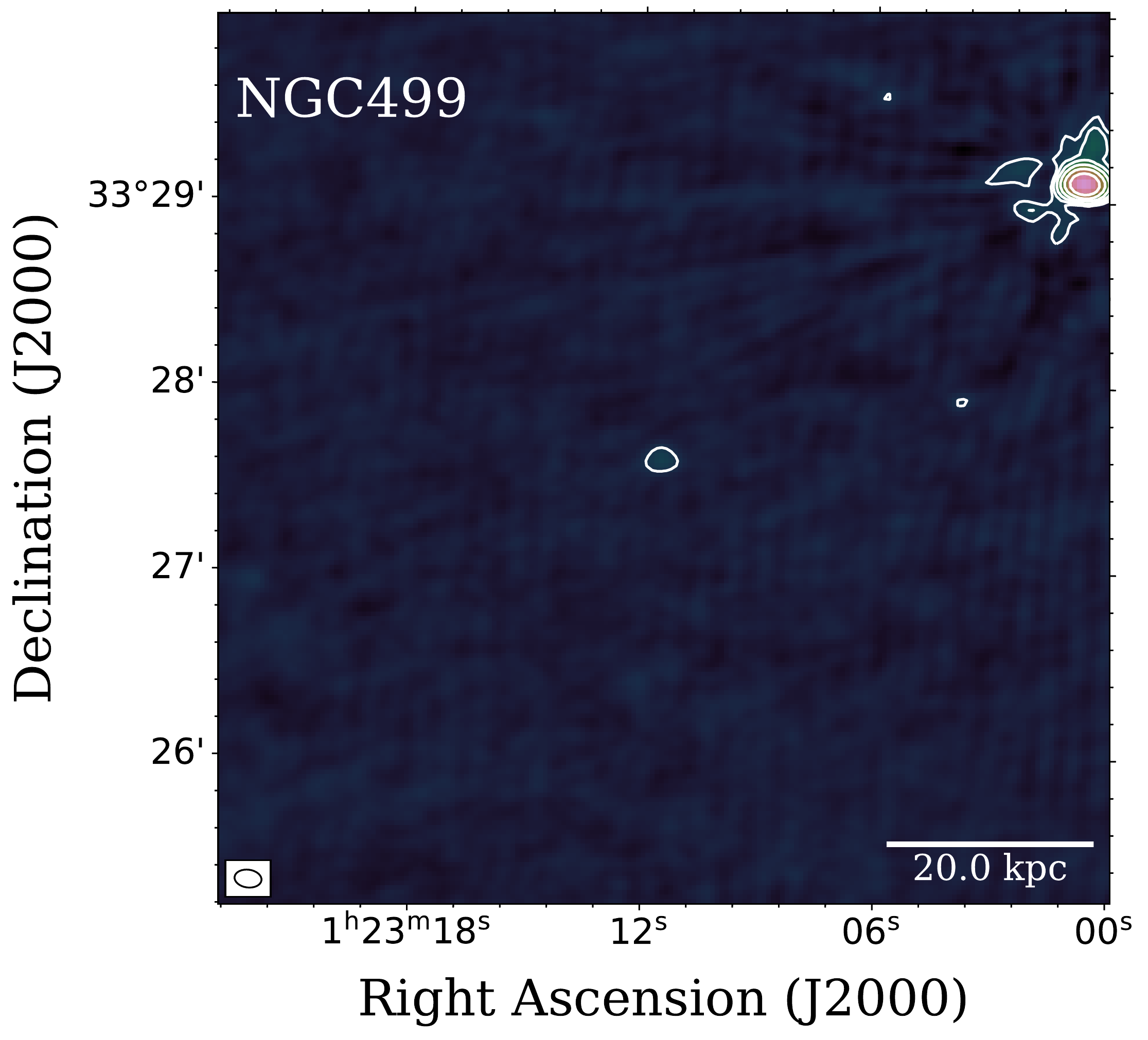} &
\includegraphics[width=54mm]{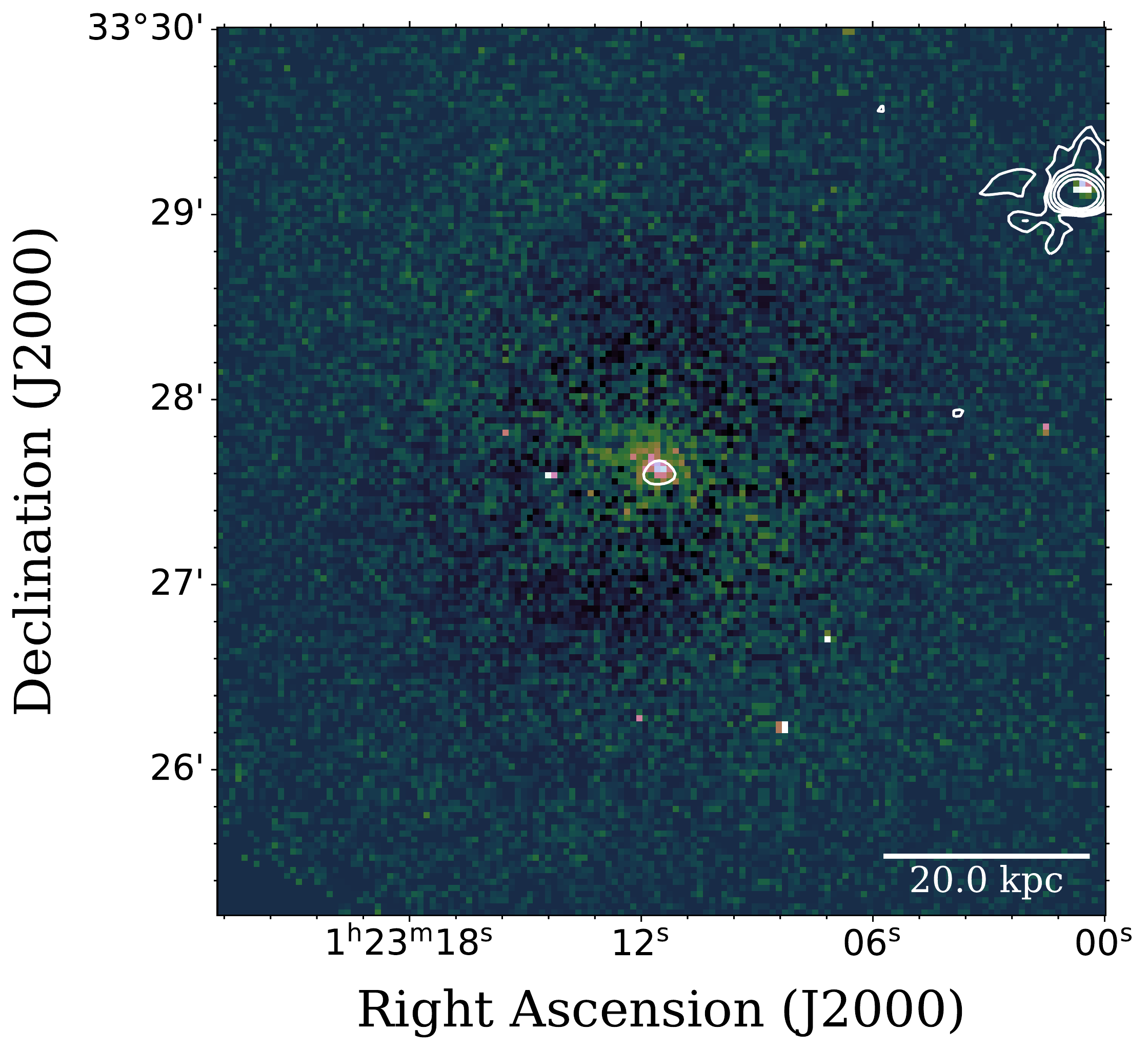}
\includegraphics[width=54mm]{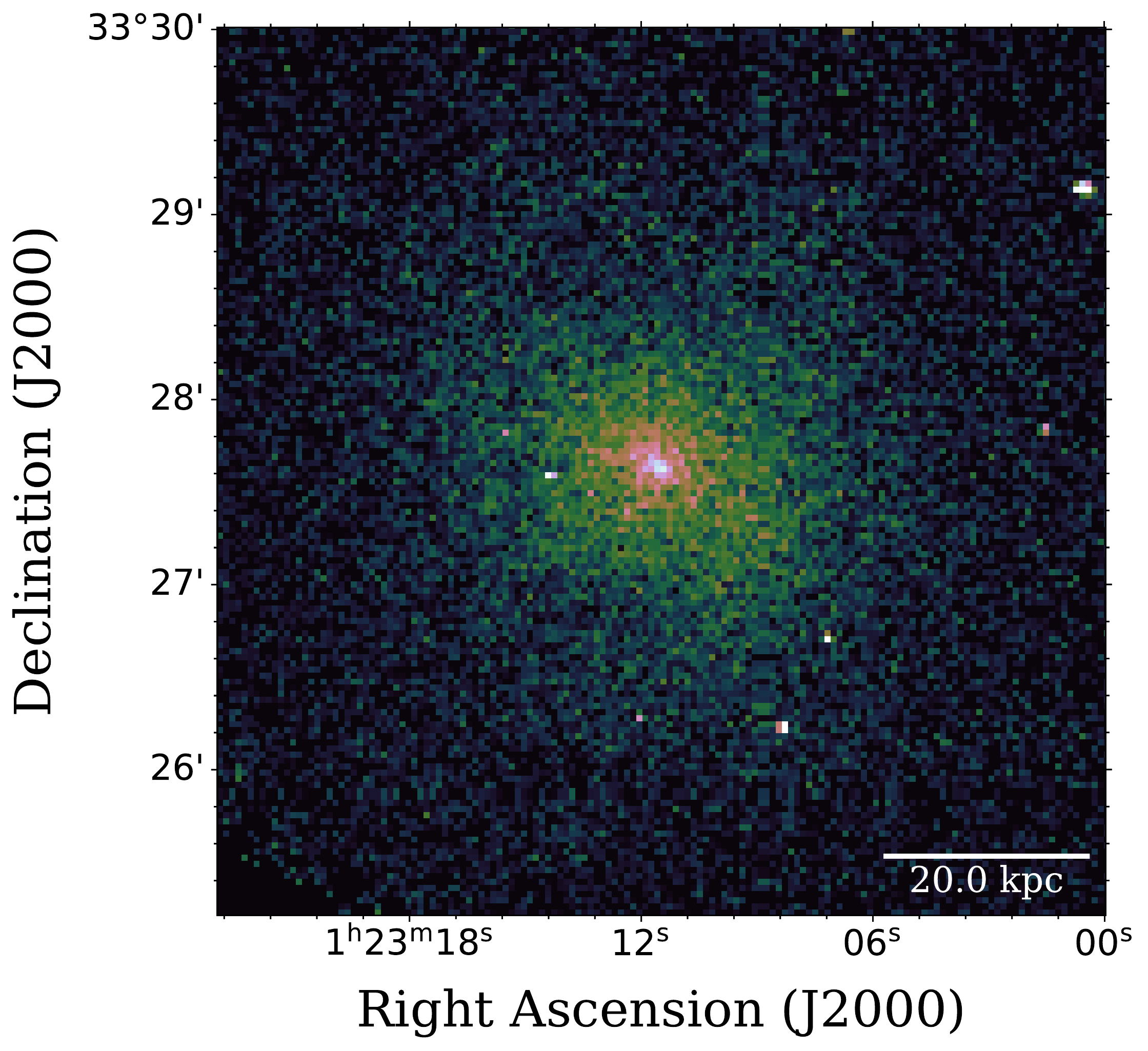} \\
\includegraphics[width=54mm]{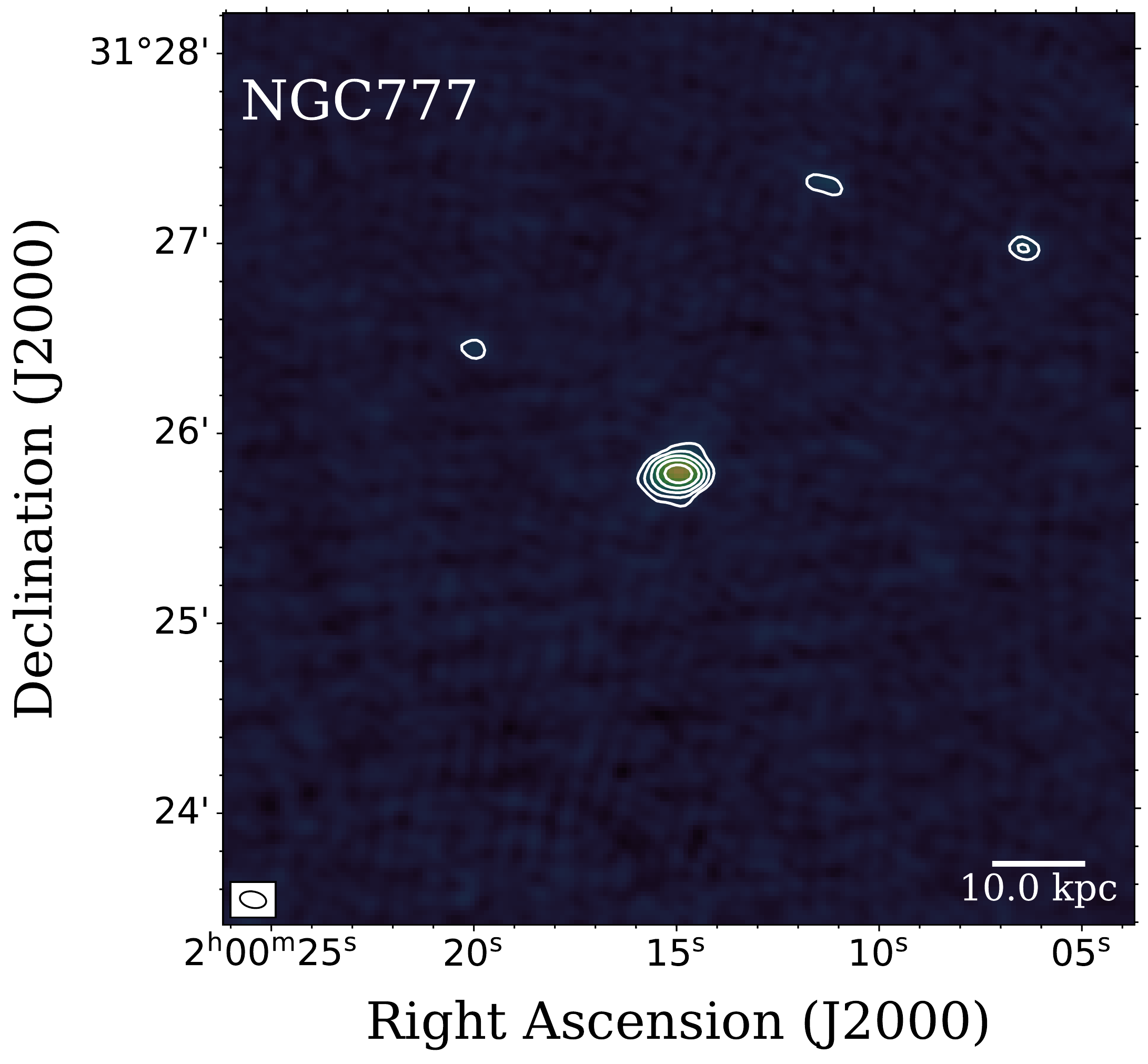} &
\includegraphics[width=54mm]{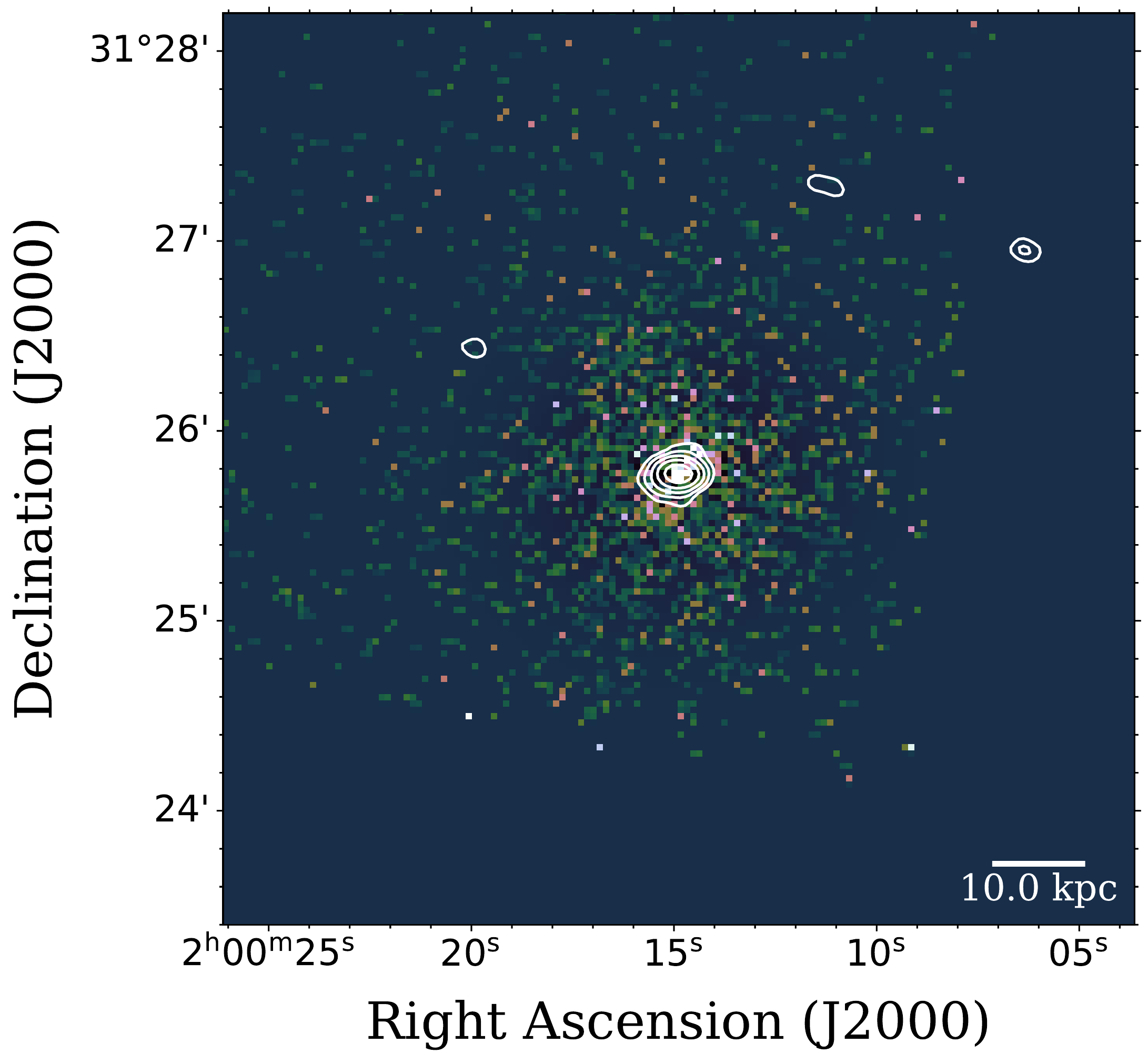}
\includegraphics[width=54mm]{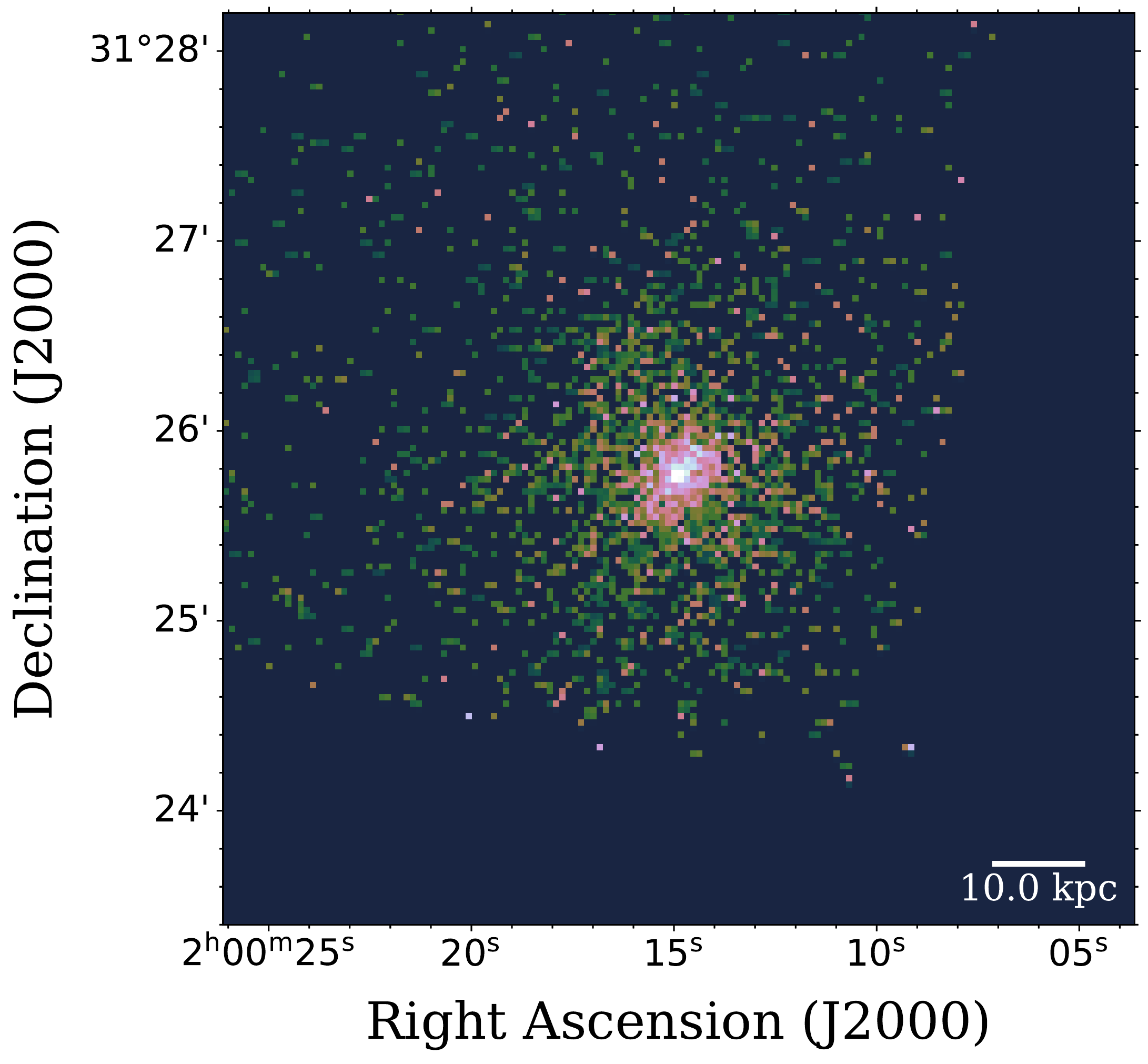} \\ \end{tabular}
\caption{\emph{Chandra} and LOFAR images for the groups and ellipticals  with putative  X-ray cavities which are not filled by radio emission, in the same order as in the tables (NGC 741, NGC 2300, NGC 499 and NGC 777 are shown above, and the others are shown in Figure 2-continued). However, NGC 3608 is not shown since no central radio source was detected in the LOFAR image. The panel organization  is  the same as in Figure~1. For the LOFAR image, the first contour is at 0.006~mJy beam$^{-1}$ (NGC 741), 0.00096~mJy beam$^{-1}$ (NGC 2300), 0.0021~mJy beam$^{-1}$ (NGC 499), 0.00087~mJy beam$^{-1}$ (NGC 777), and each contour increases by a factor of two.}
\label{F:images_2}
\end{figure*}

\begin{figure*} \begin{tabular}{@{}cc}
\includegraphics[width=54mm]{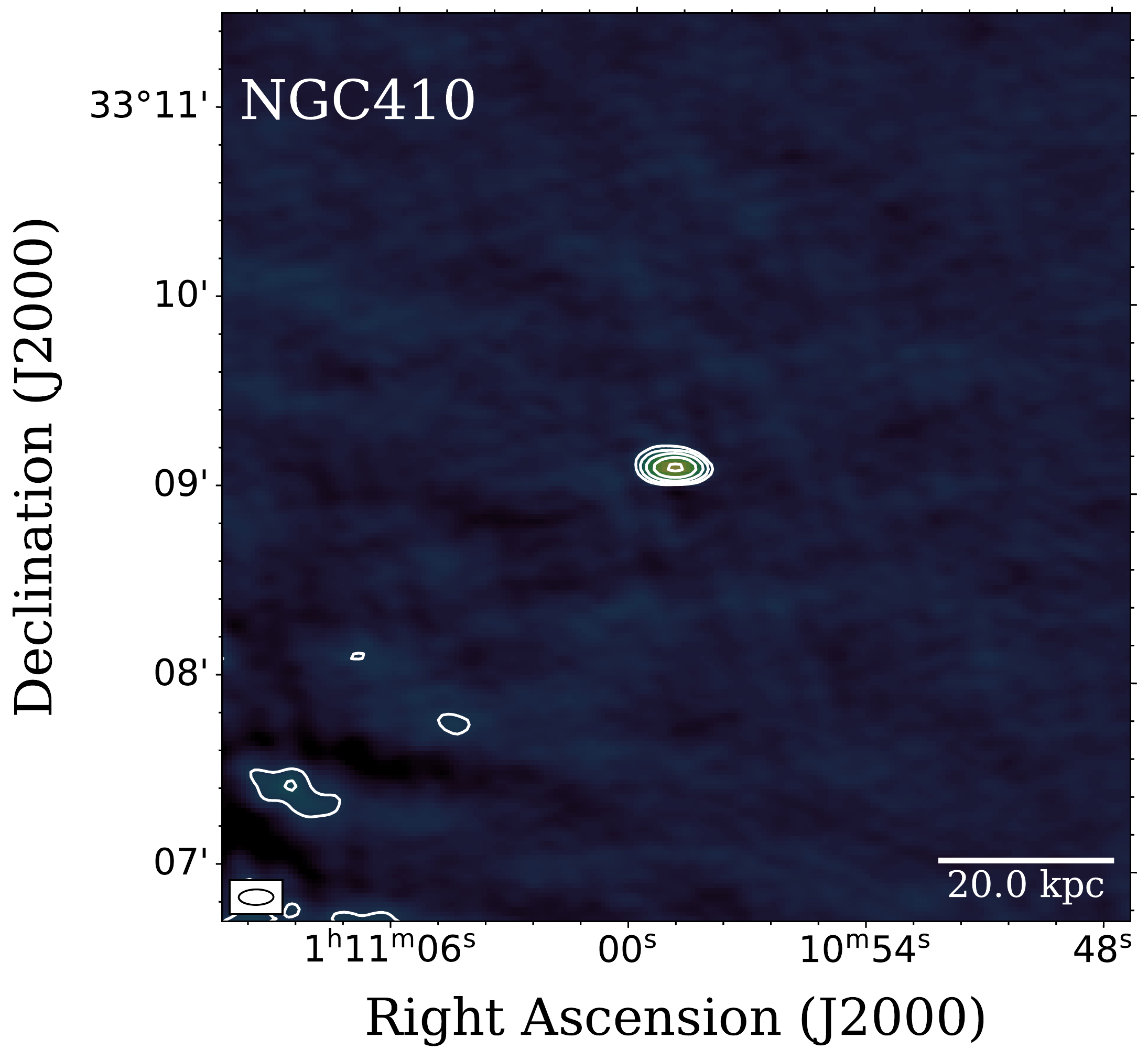} &
\includegraphics[width=54mm]{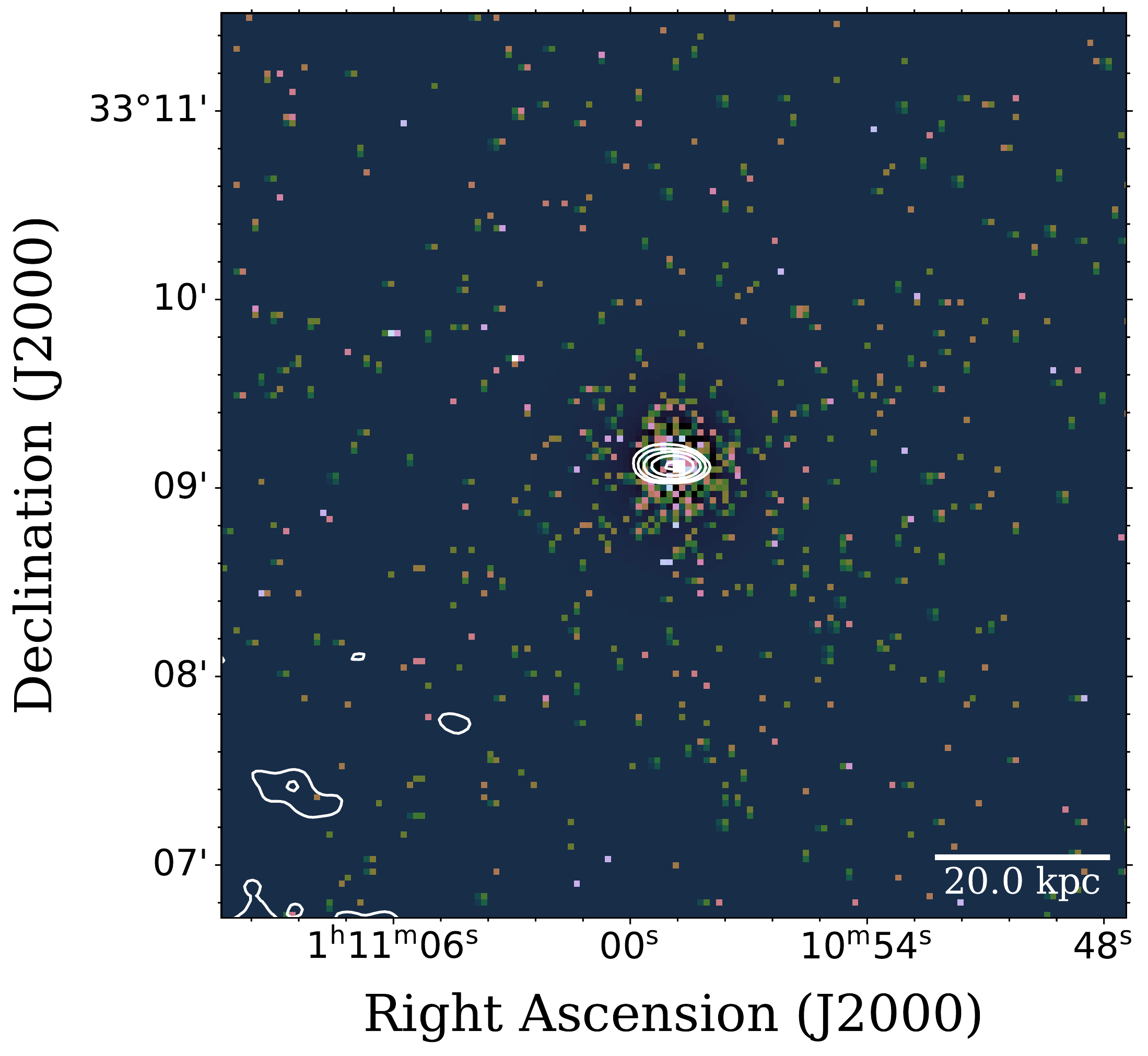}
\includegraphics[width=54mm]{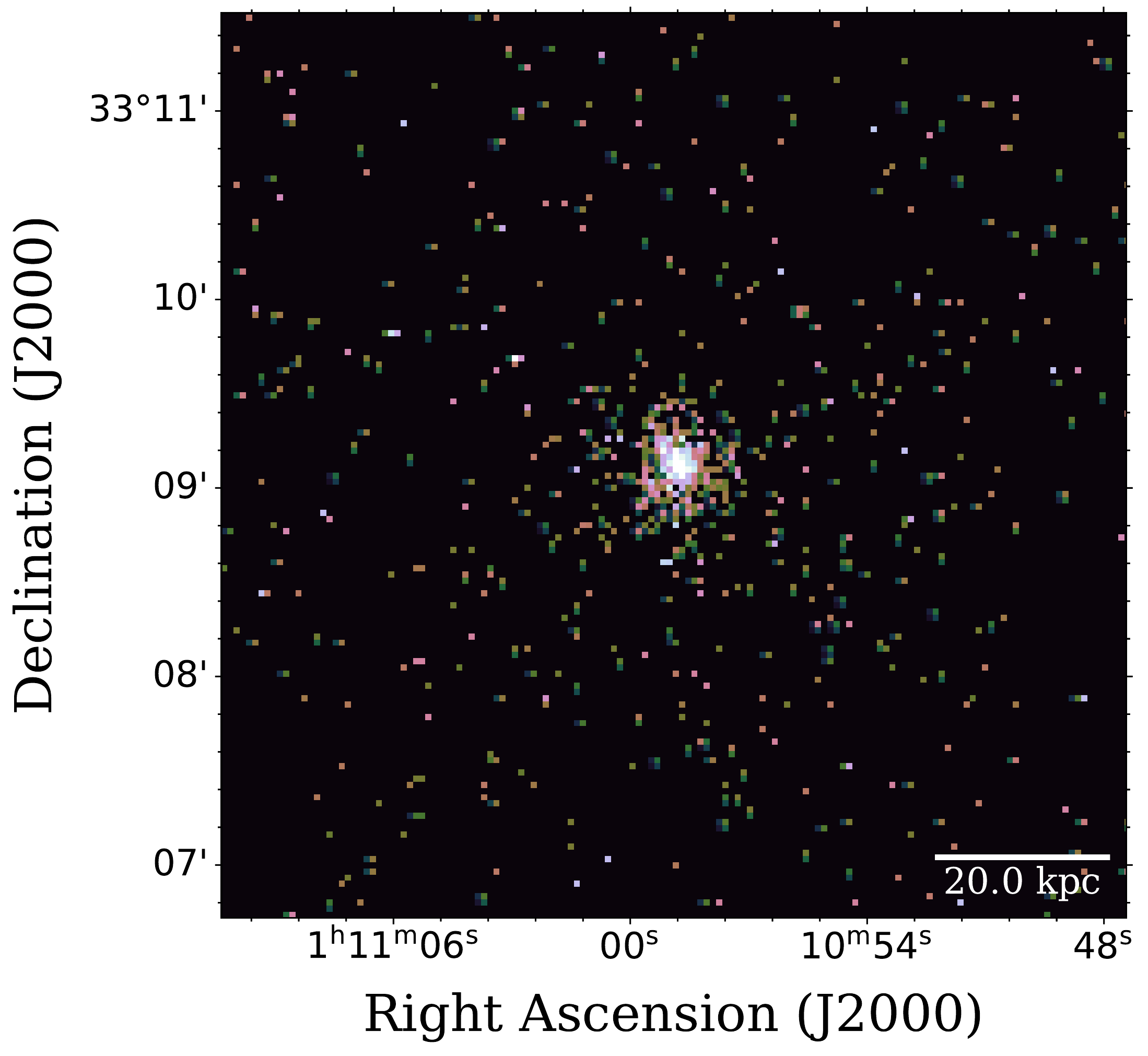} \\
\includegraphics[width=54mm]{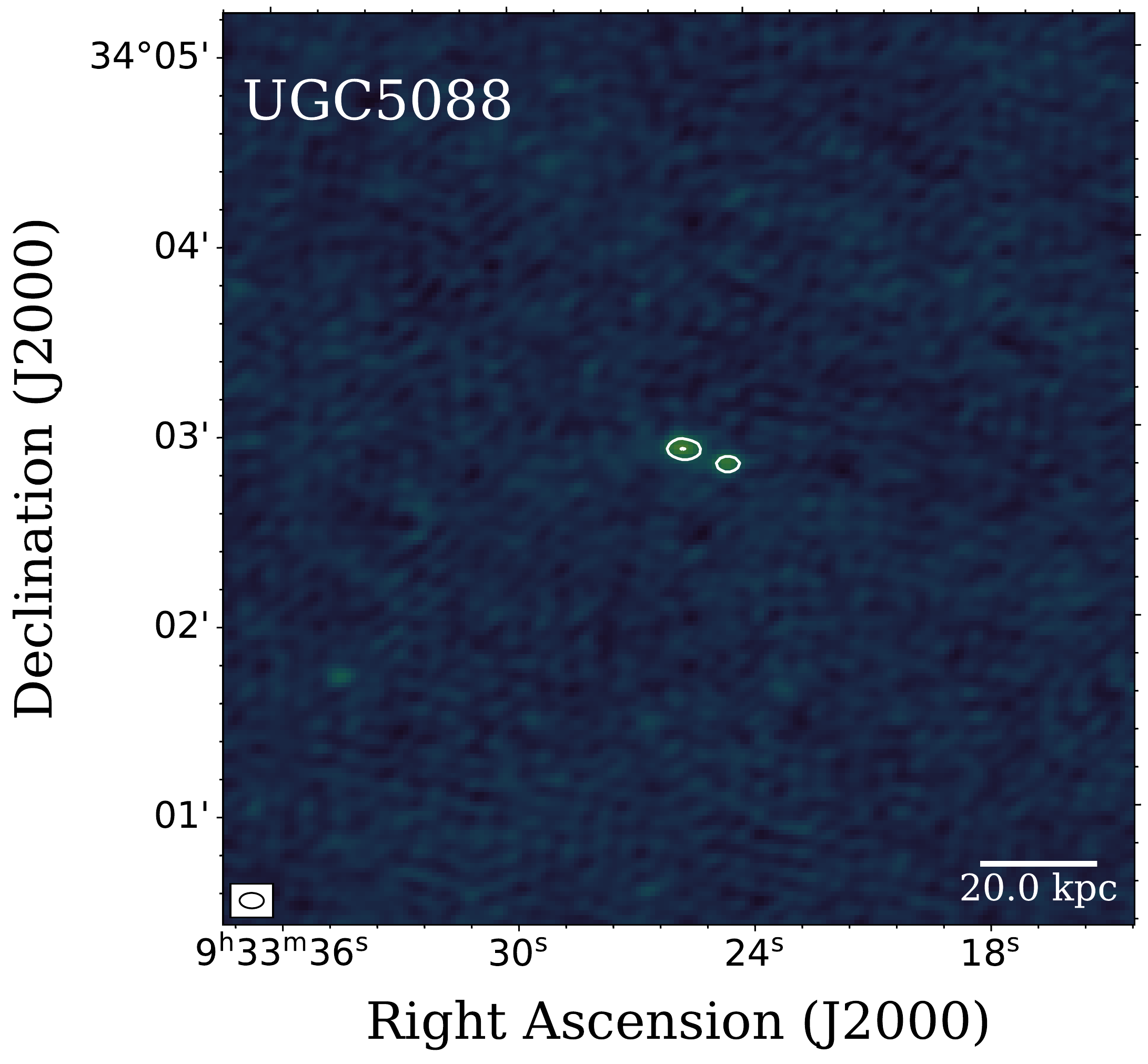} &
\includegraphics[width=54mm]{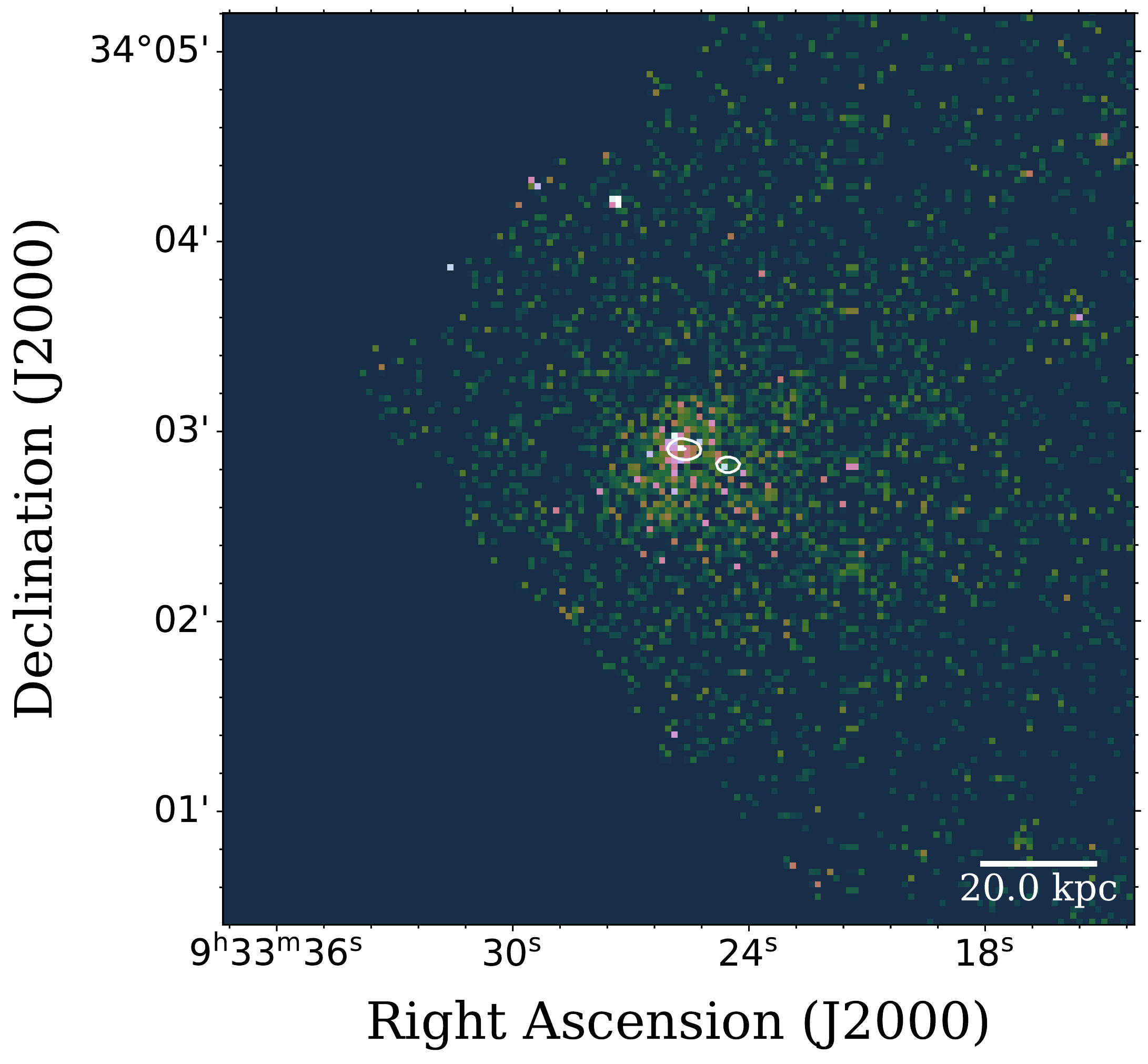}
\includegraphics[width=54mm]{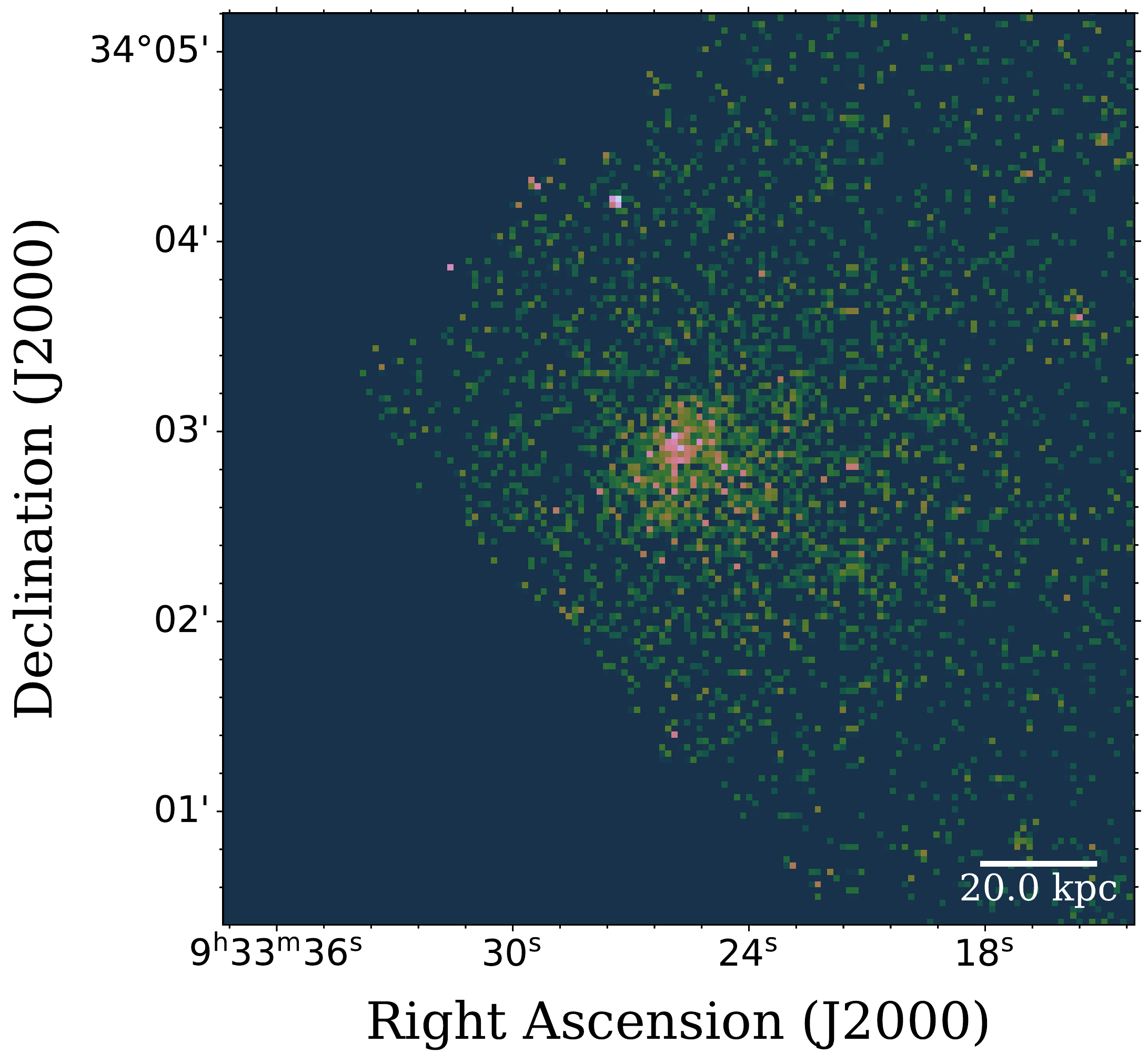} \\
\includegraphics[width=54mm]{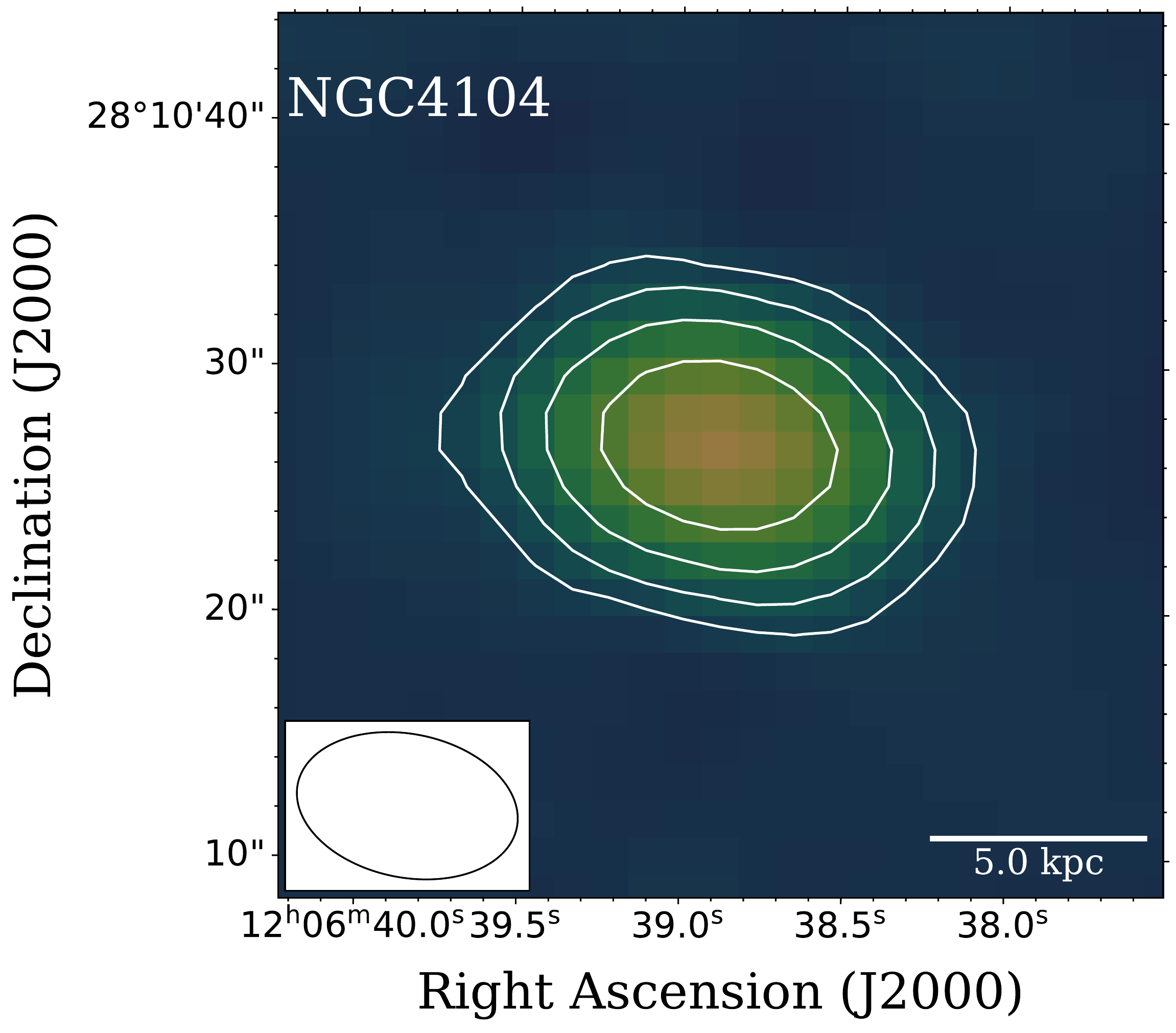} &
\includegraphics[width=54mm]{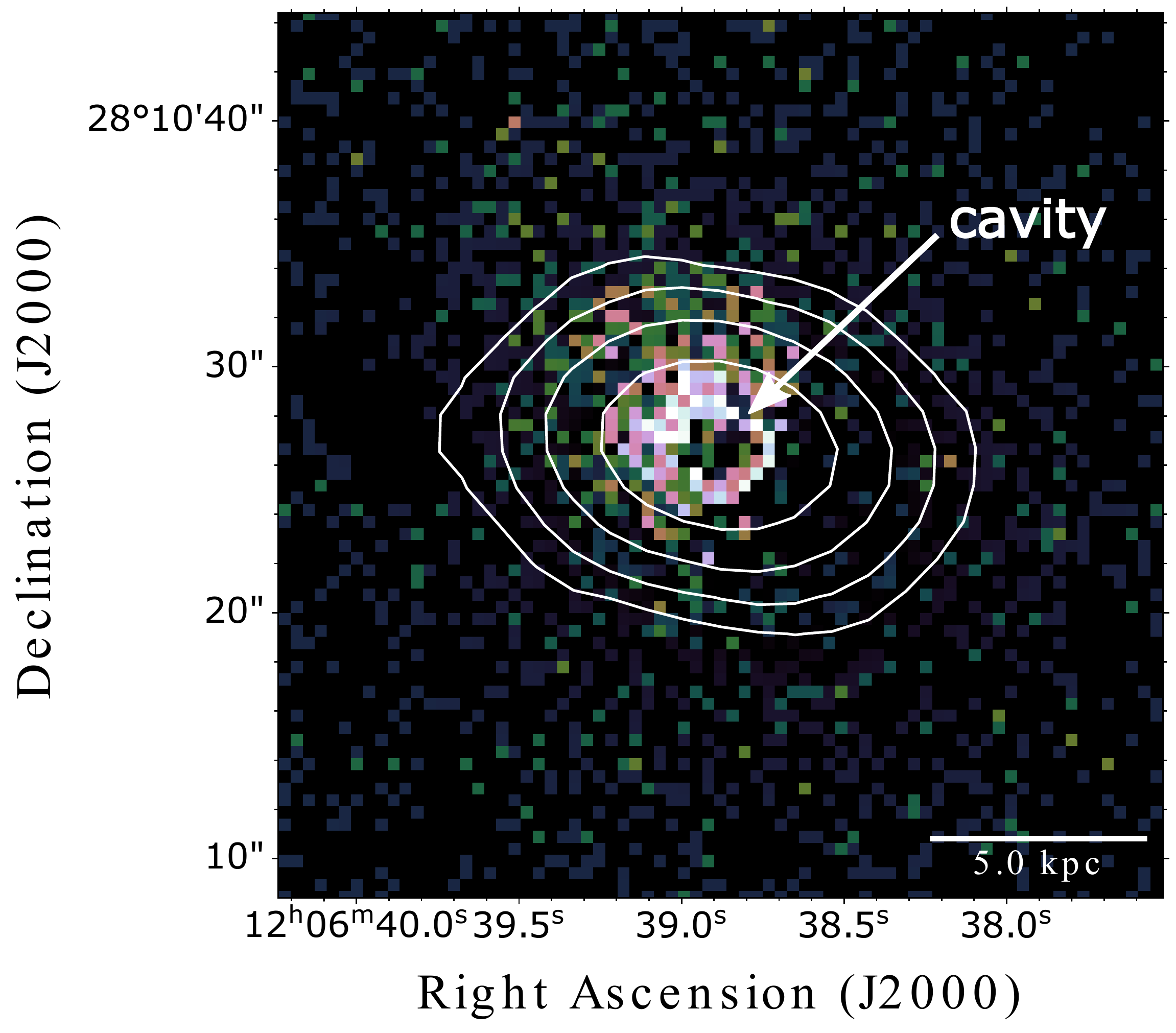}
\includegraphics[width=54mm]{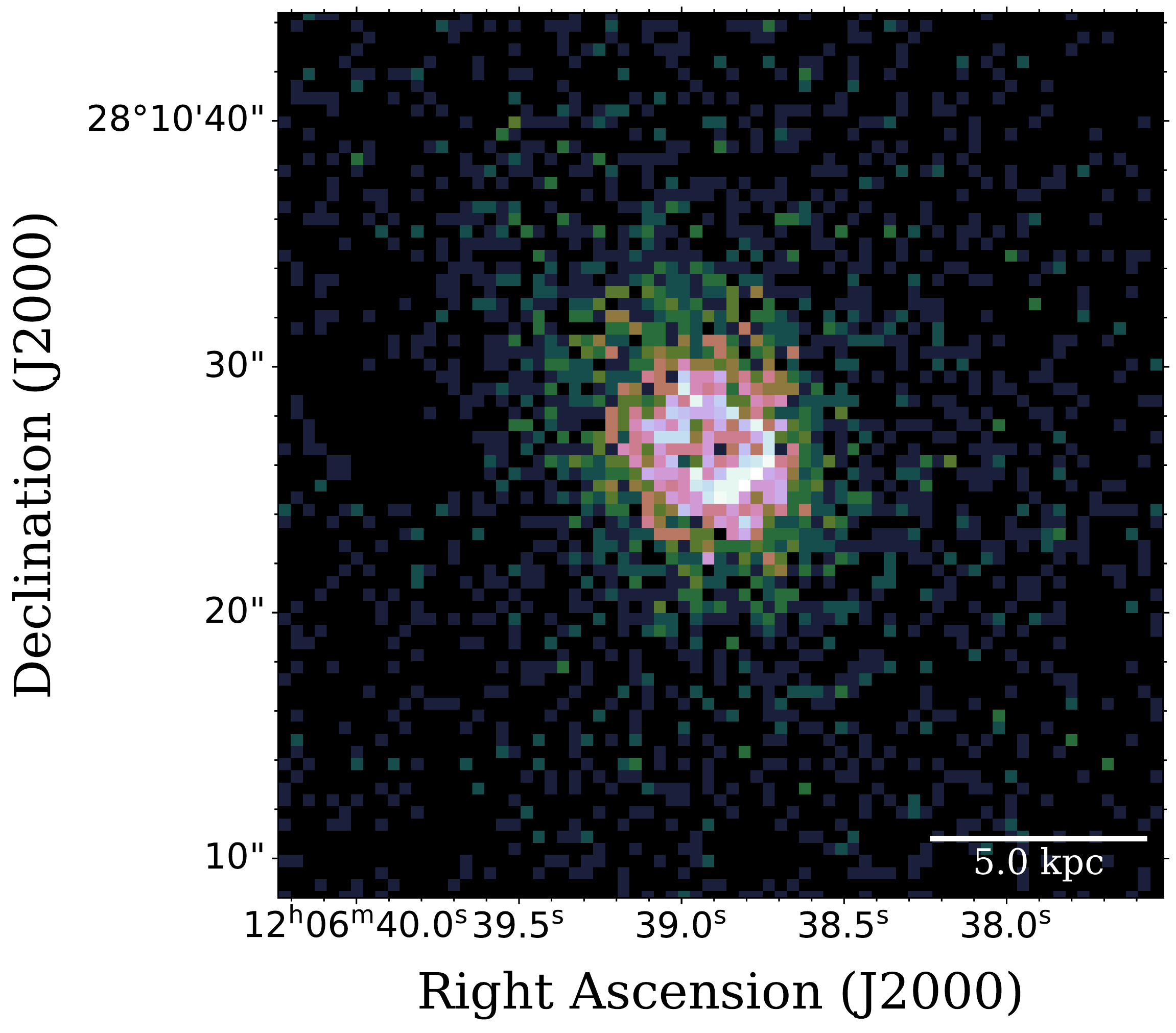}  \\
\includegraphics[width=54mm]{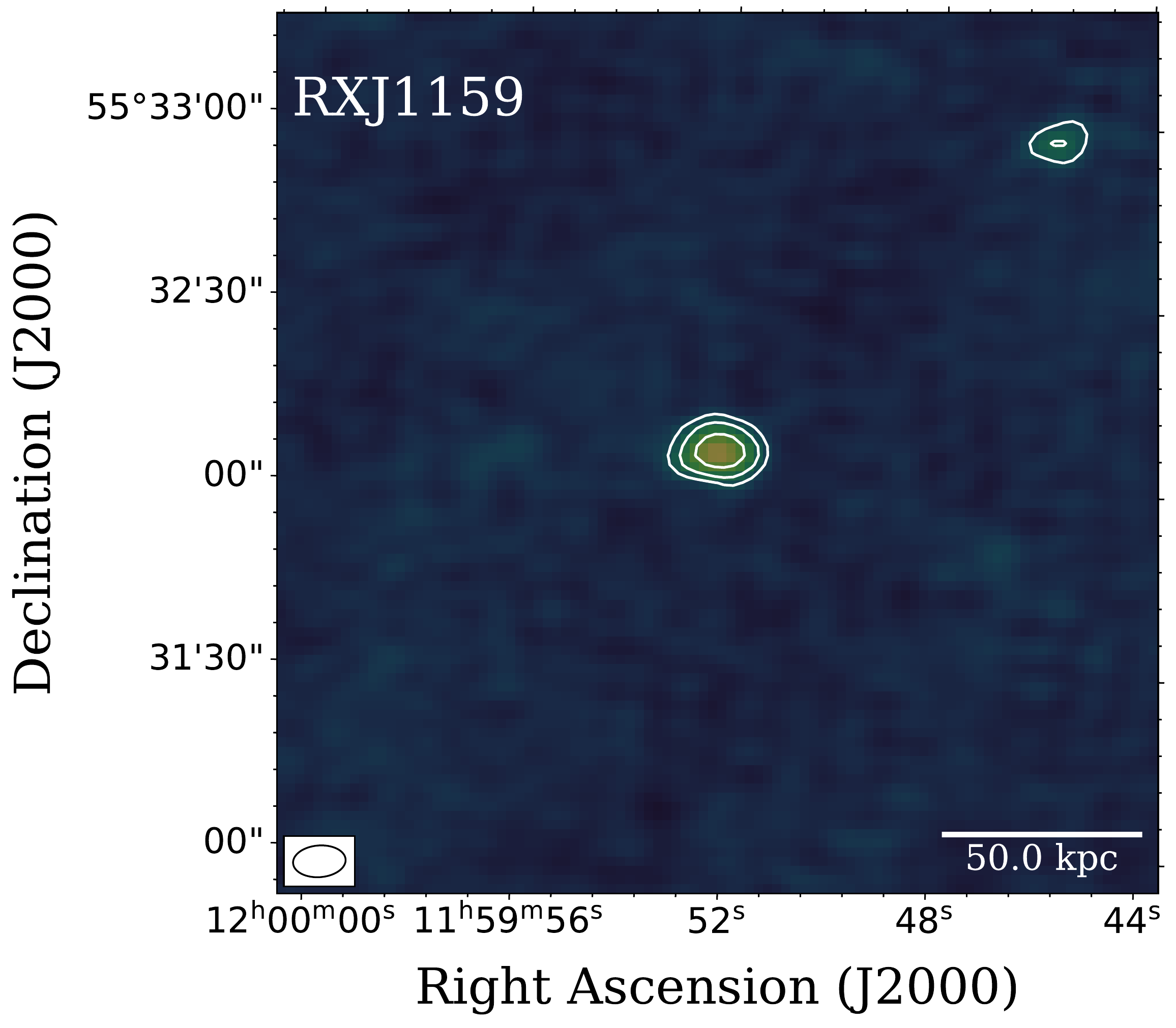} &
\includegraphics[width=54mm]{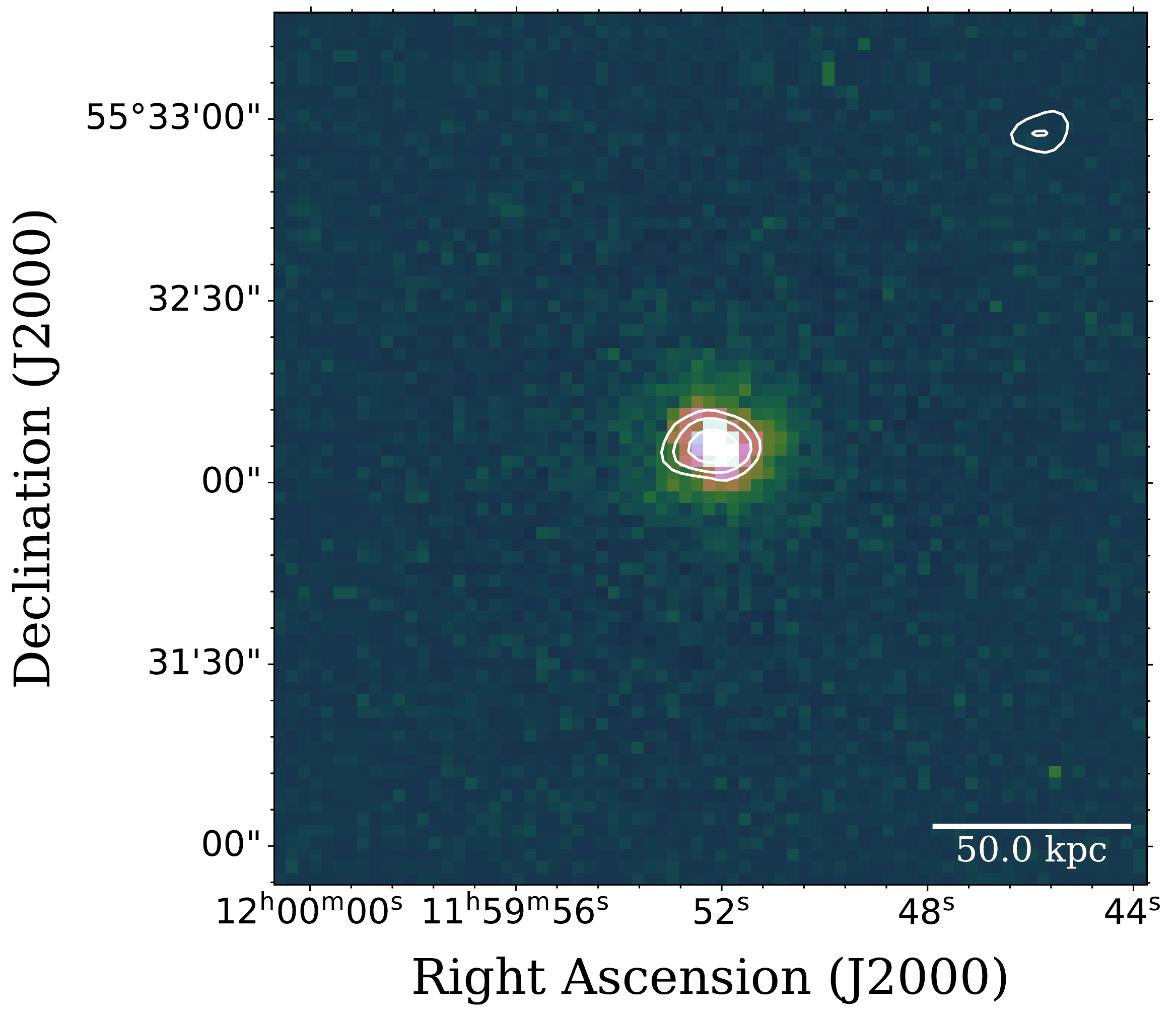}
\includegraphics[width=54mm]{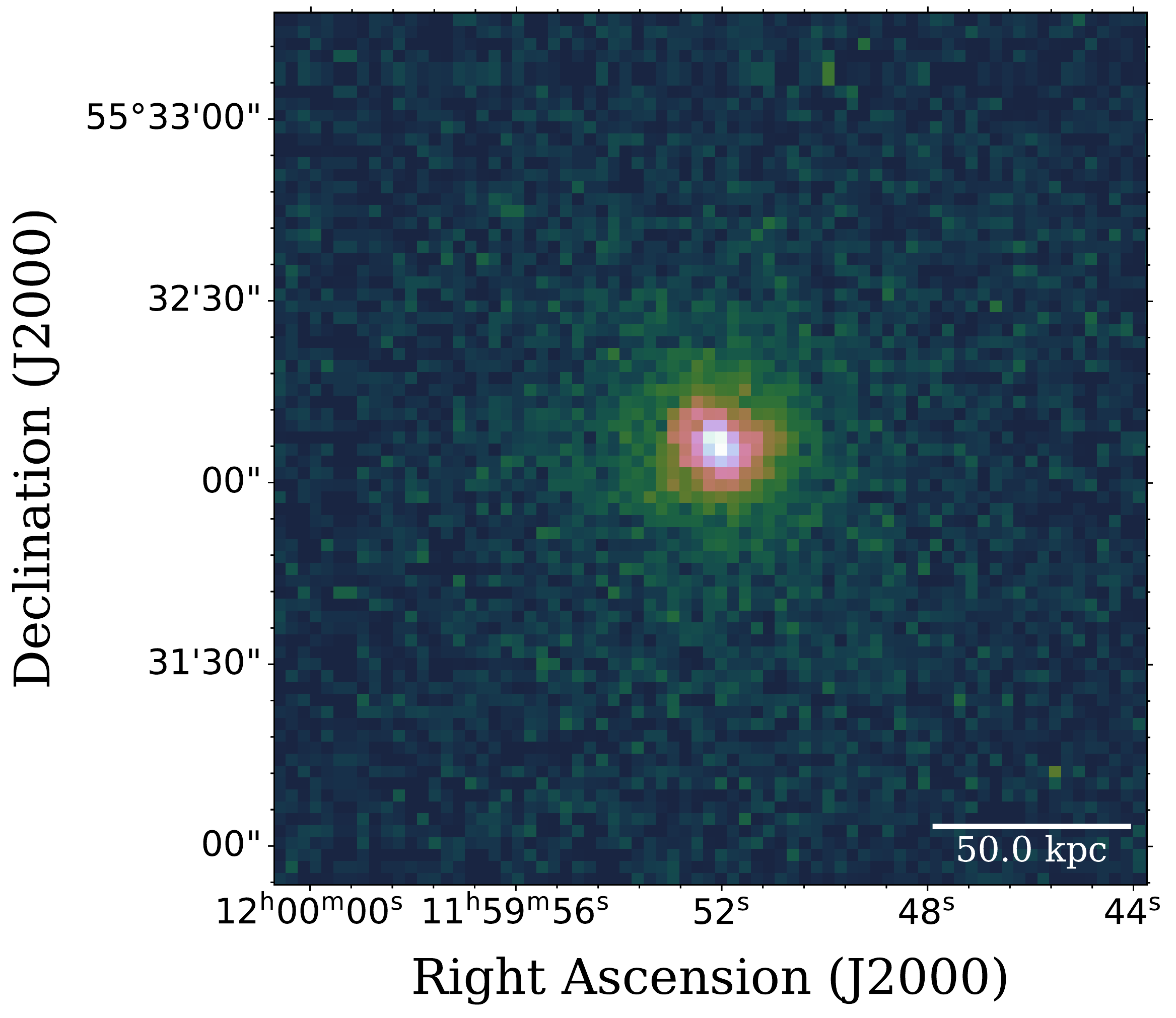} \\ \end{tabular}
\textbf{Figure}~2. --- continued (NGC 410, UGC 5088, NGC 4104 and RX J1159.8+5531). For the LOFAR image, the first contour is at 0.0021~mJy beam$^{-1}$ (NGC 410), 0.0012~mJy beam$^{-1}$ (UGC 5088), 0.0012~mJy beam$^{-1}$ (NGC 4104), 0.0006~mJy beam$^{-1}$ (RX J1159.8+5531), and each contour increases by a factor of two. \\
\end{figure*}

\begin{figure*} \begin{tabular}{@{}cc}
\includegraphics[width=54mm]{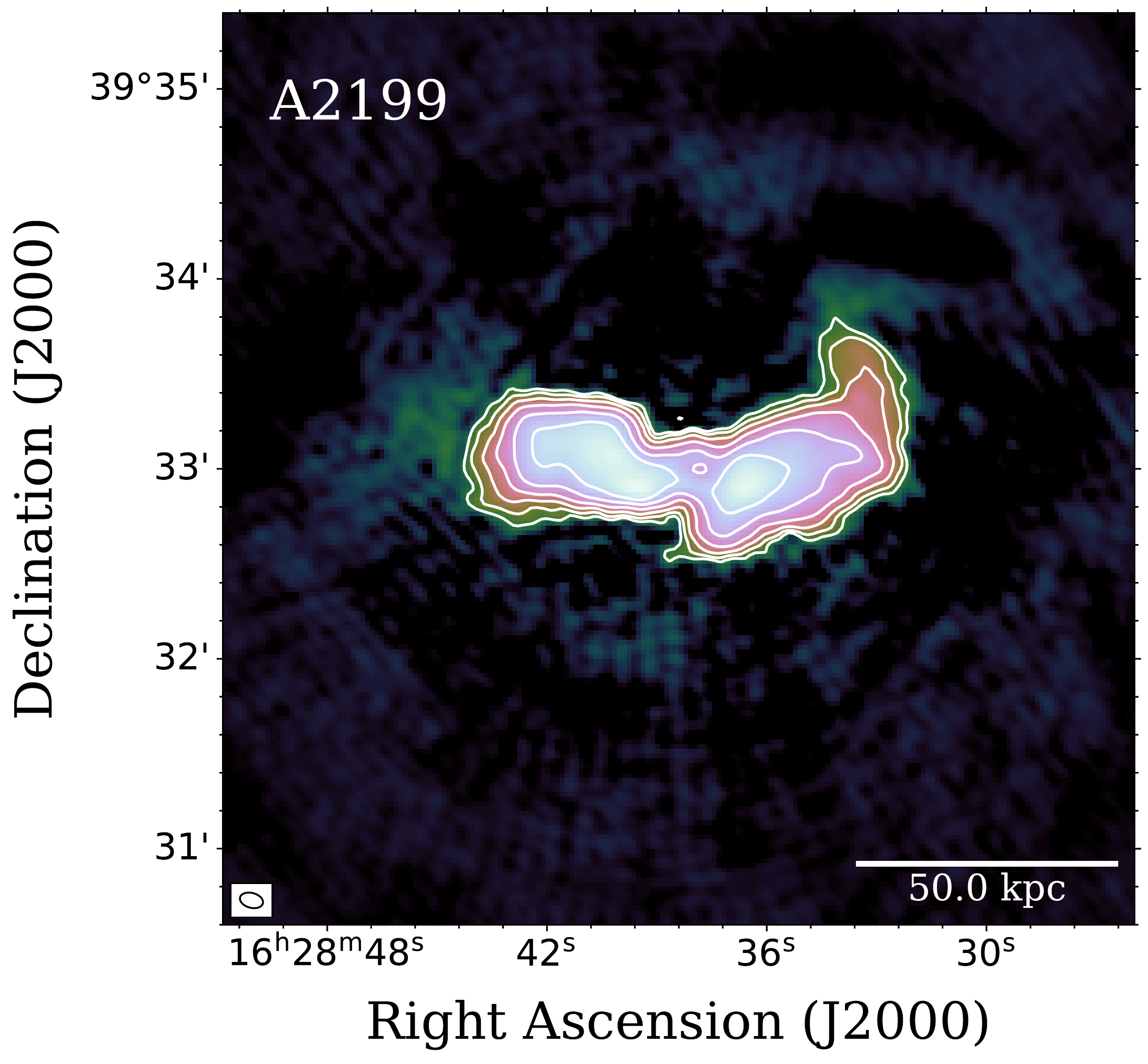}
\includegraphics[width=54mm]{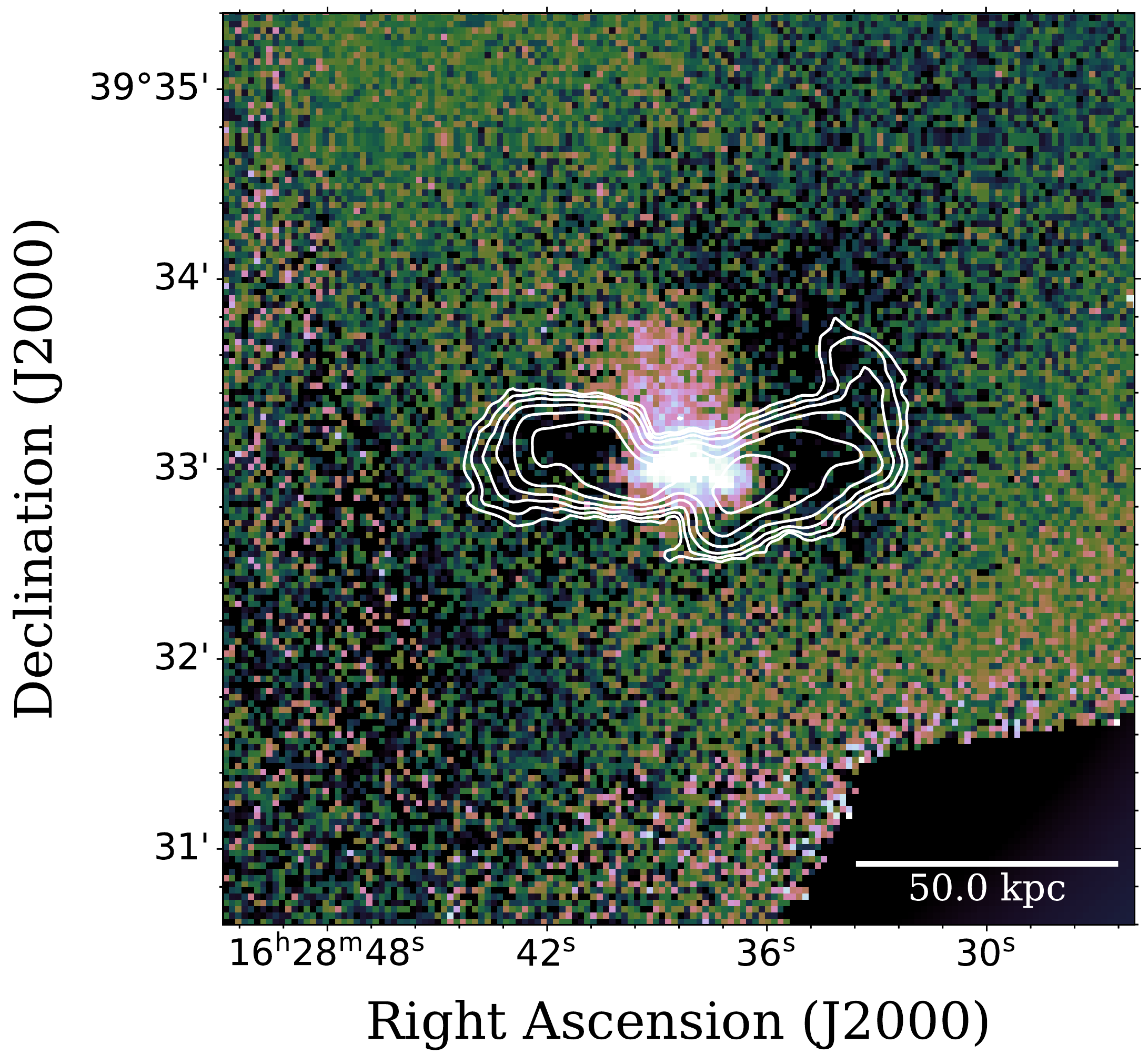}
\includegraphics[width=54mm]{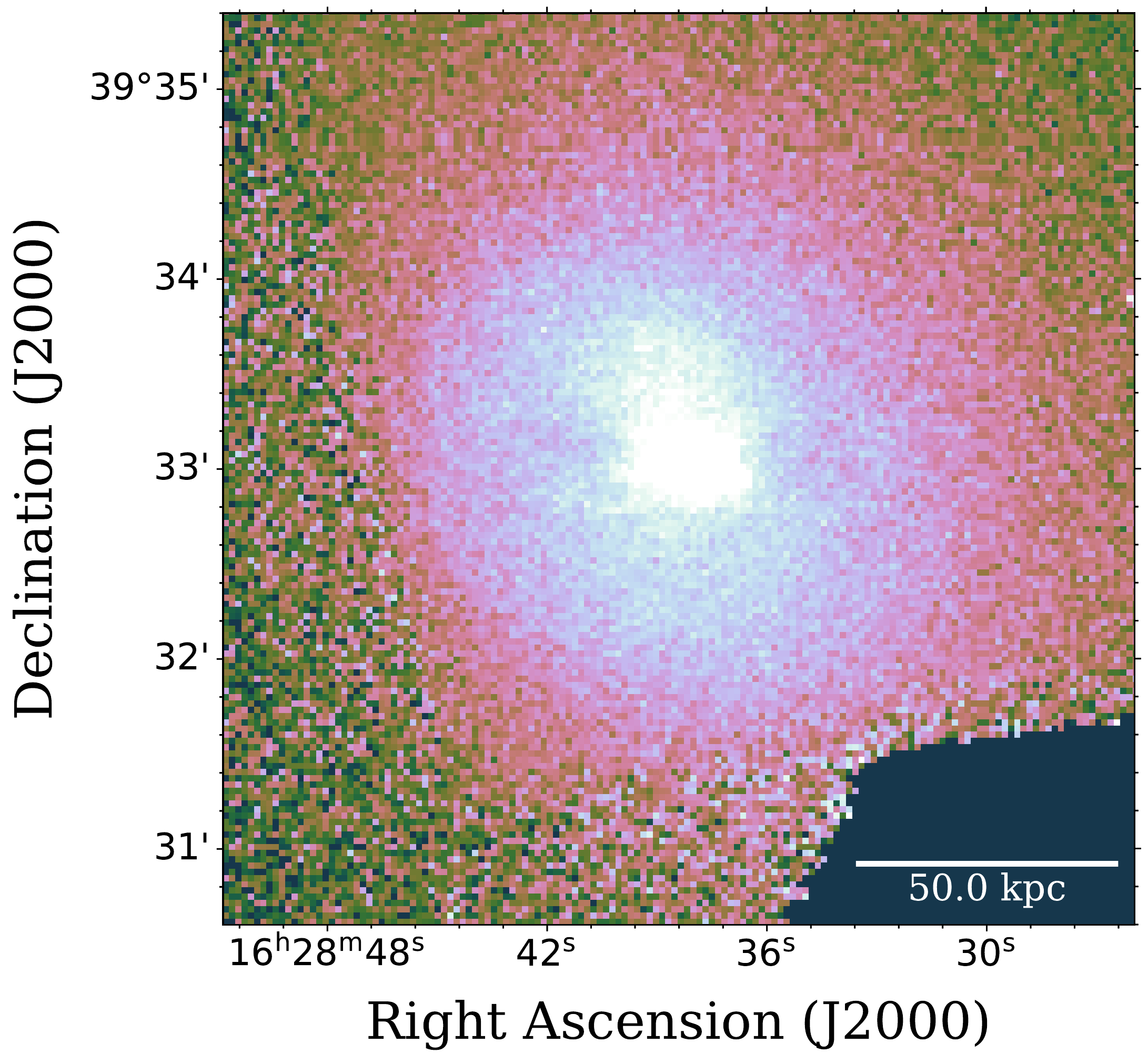} \\
\includegraphics[width=54mm]{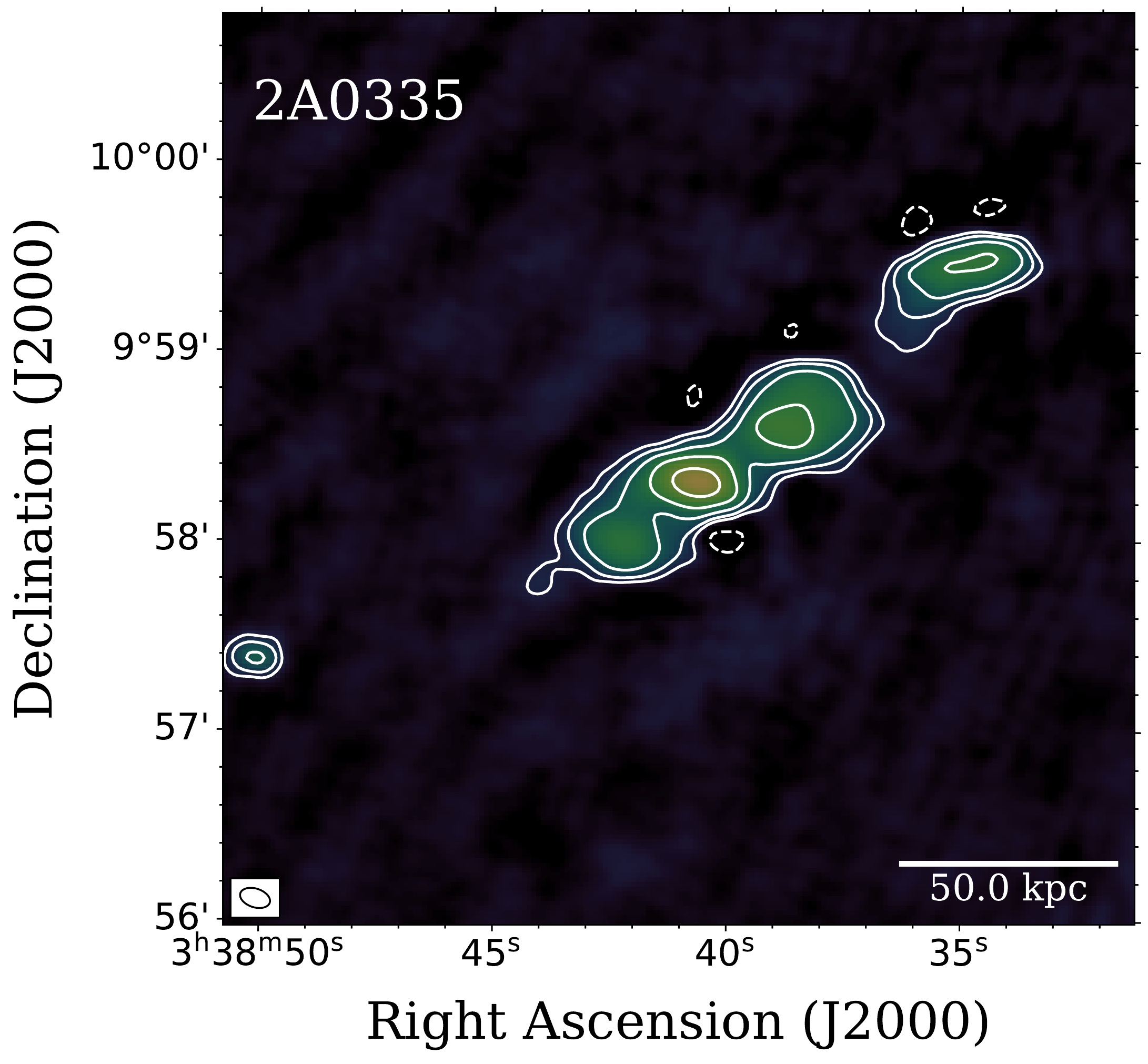}
\includegraphics[width=54mm]{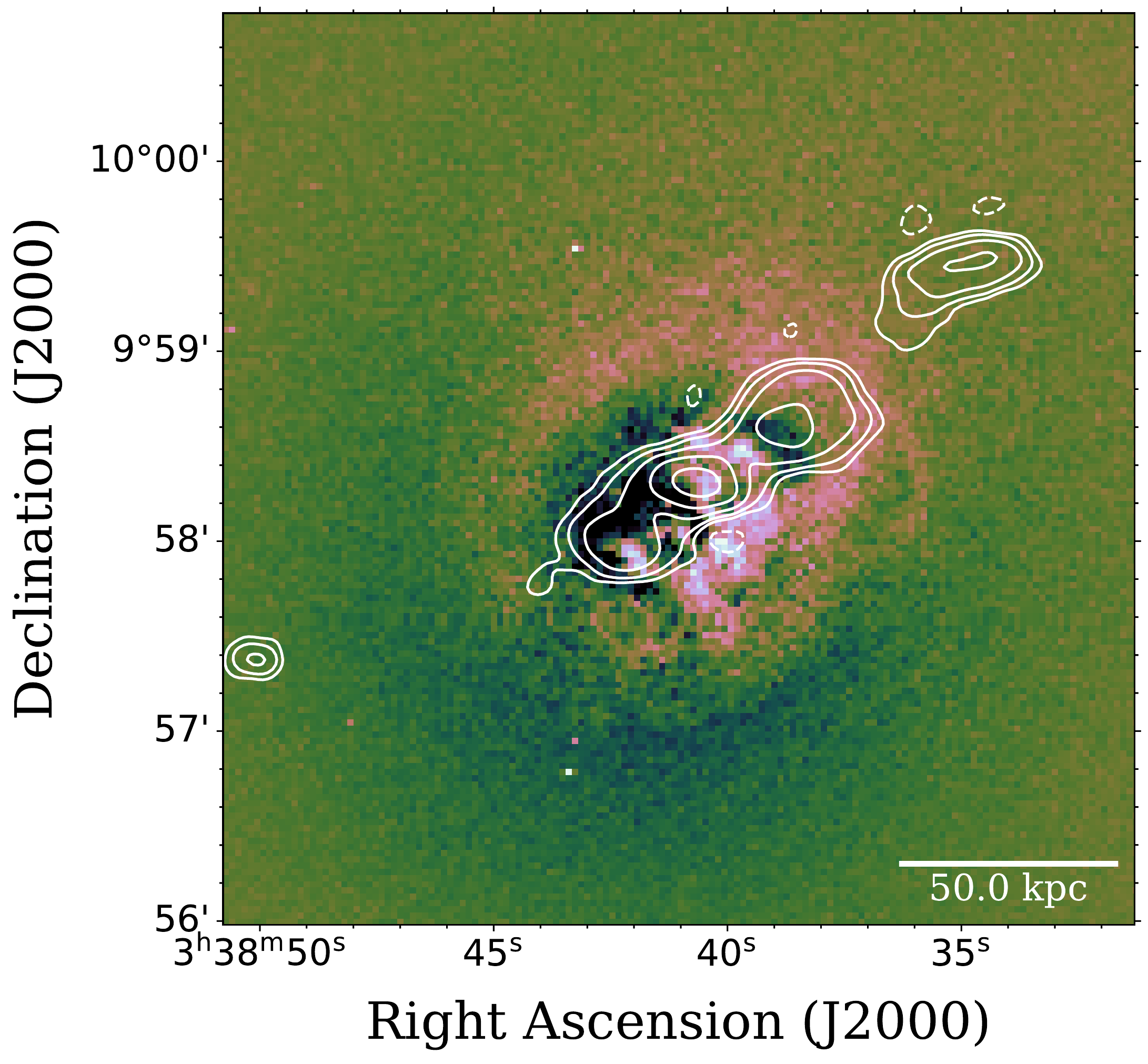}
\includegraphics[width=54mm]{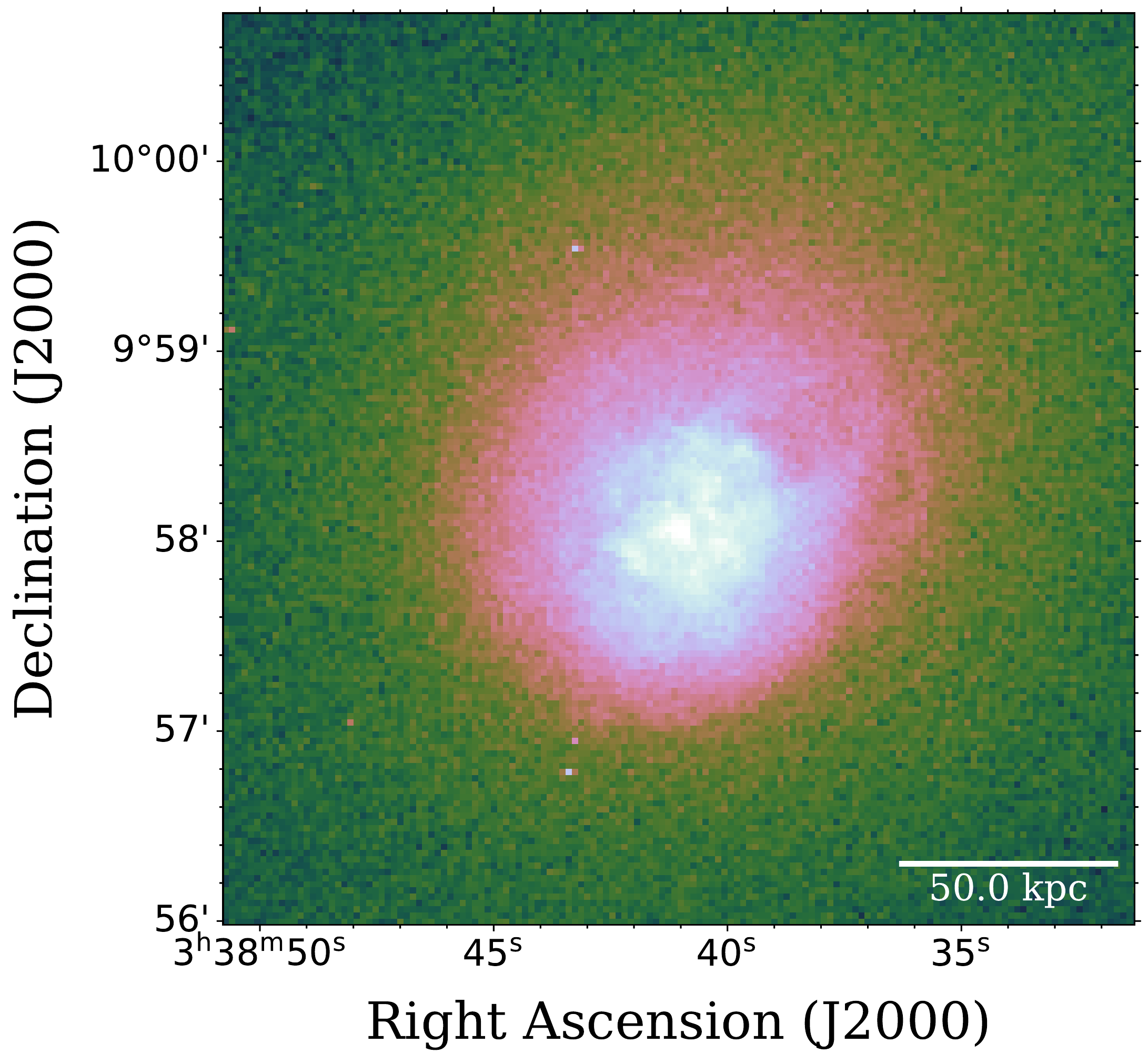}\\
\includegraphics[width=54mm]{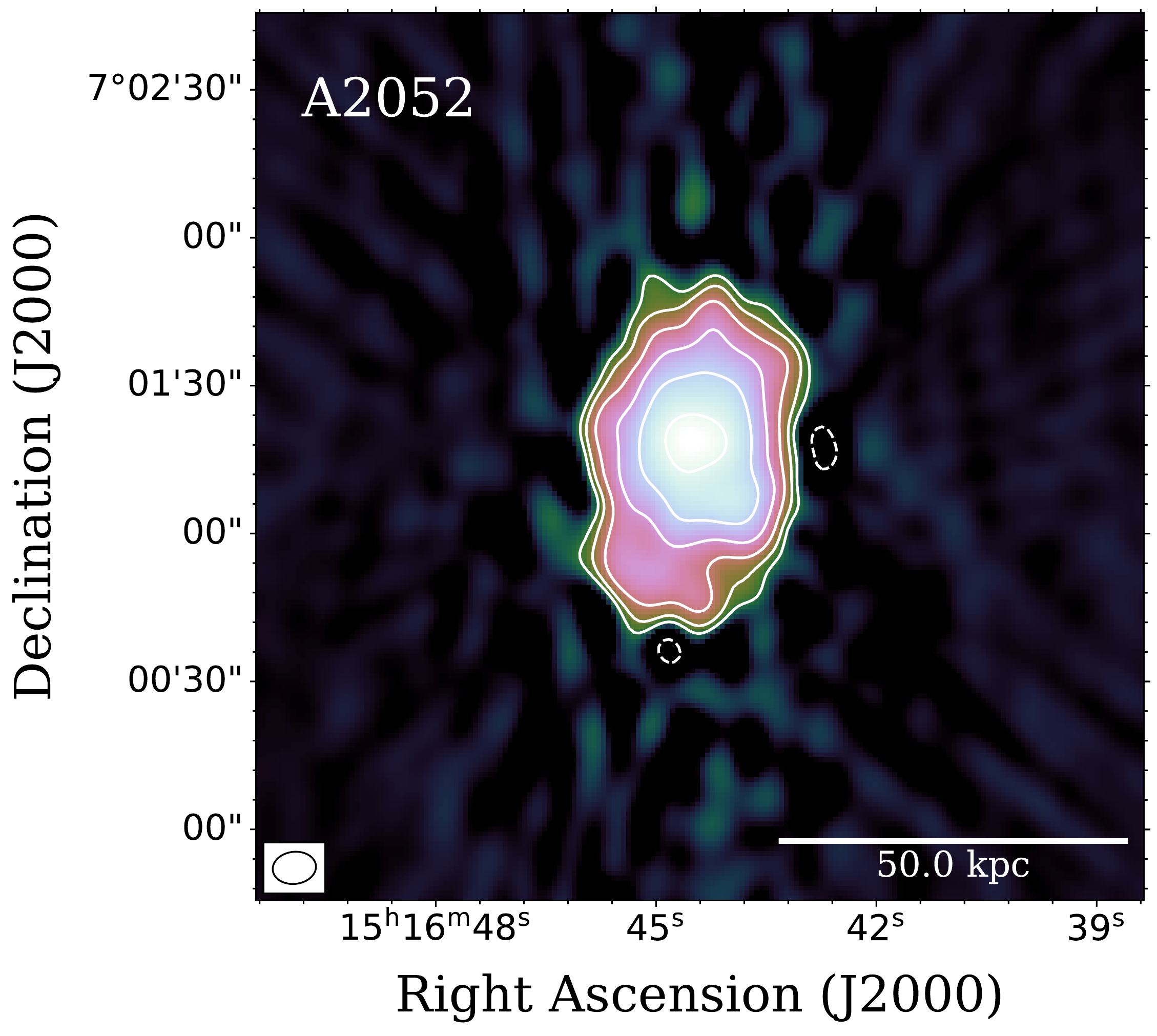}
\includegraphics[width=54mm]{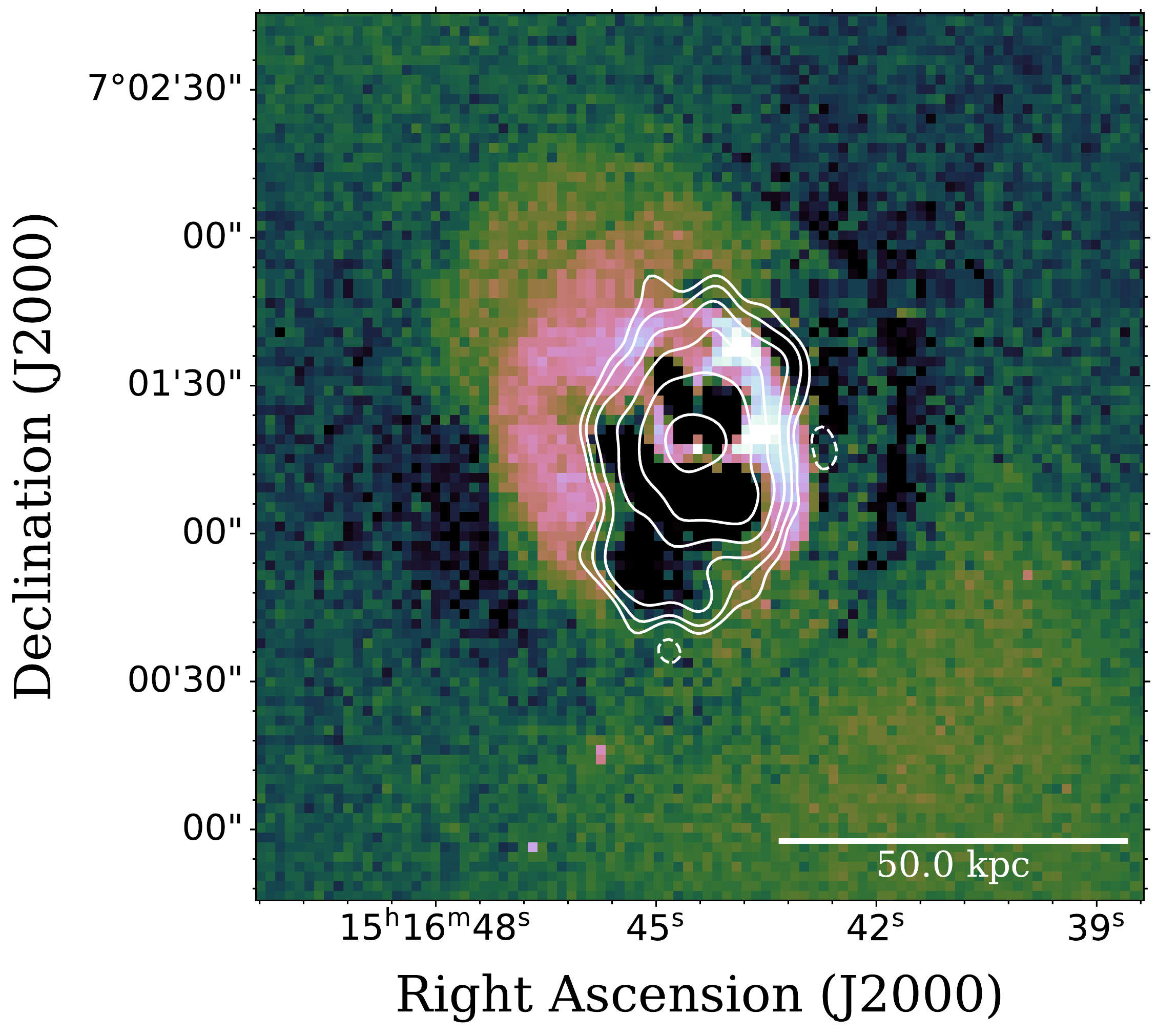}
\includegraphics[width=54mm]{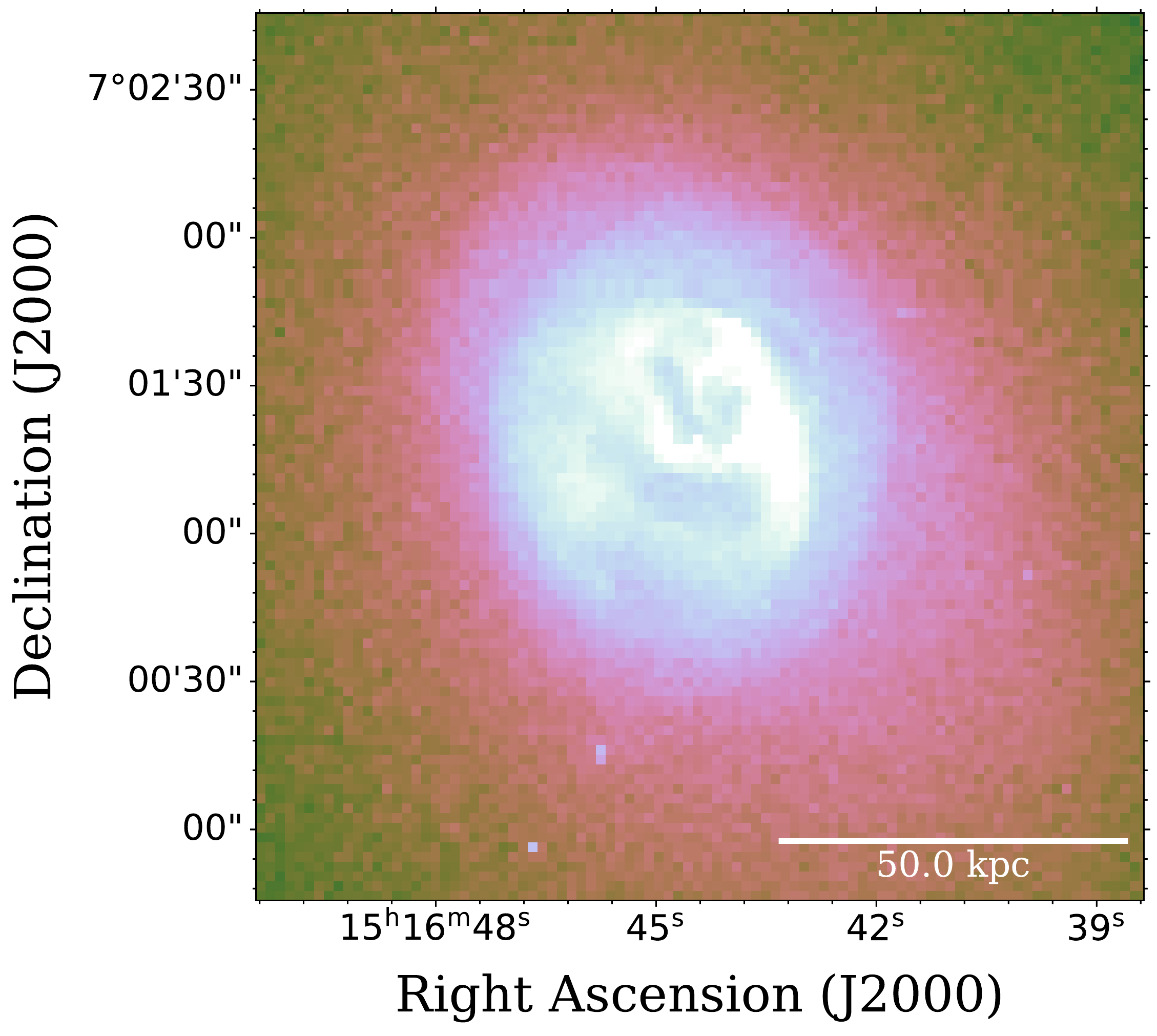} \\
\includegraphics[width=54mm]{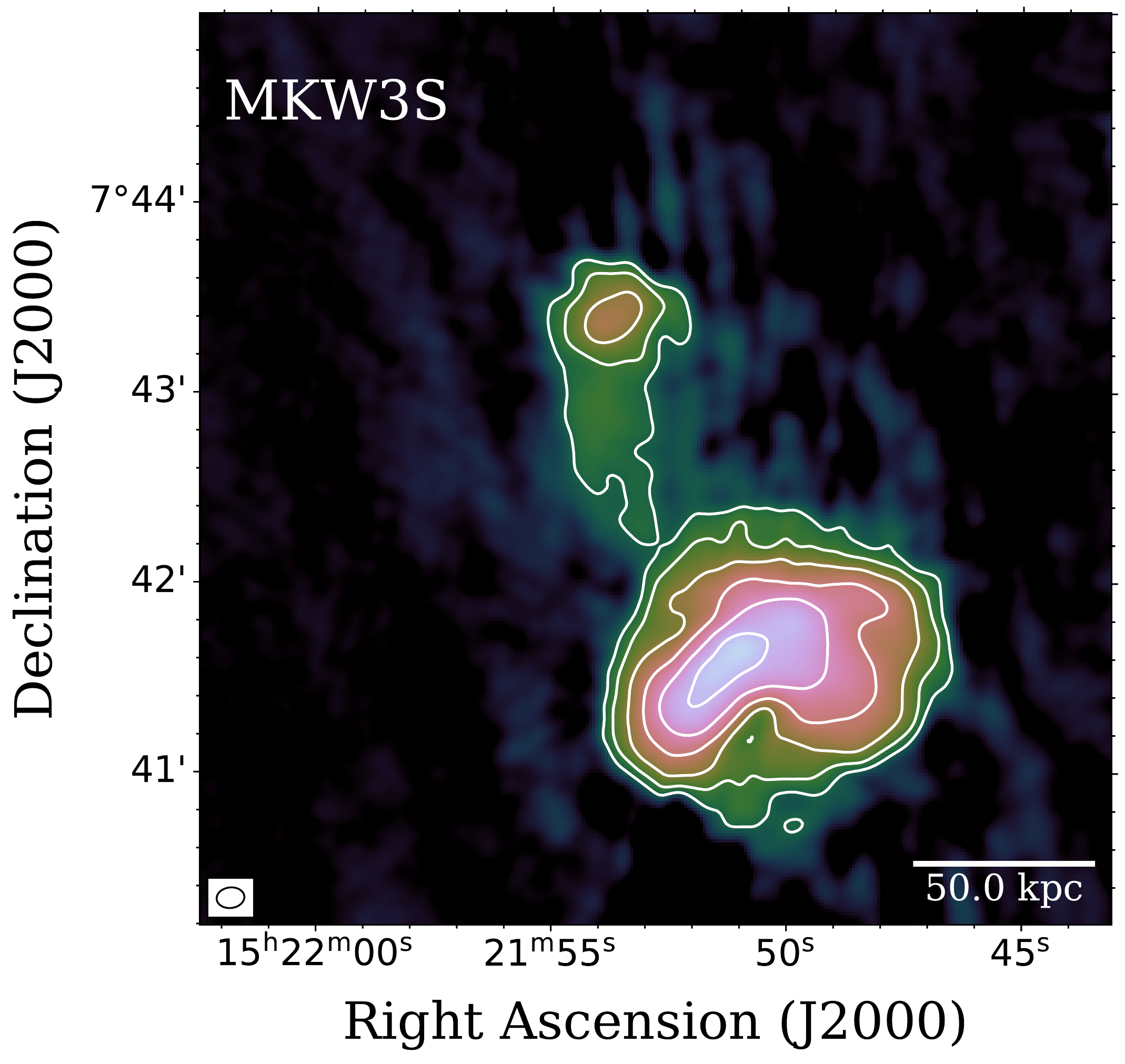}
\includegraphics[width=54mm]{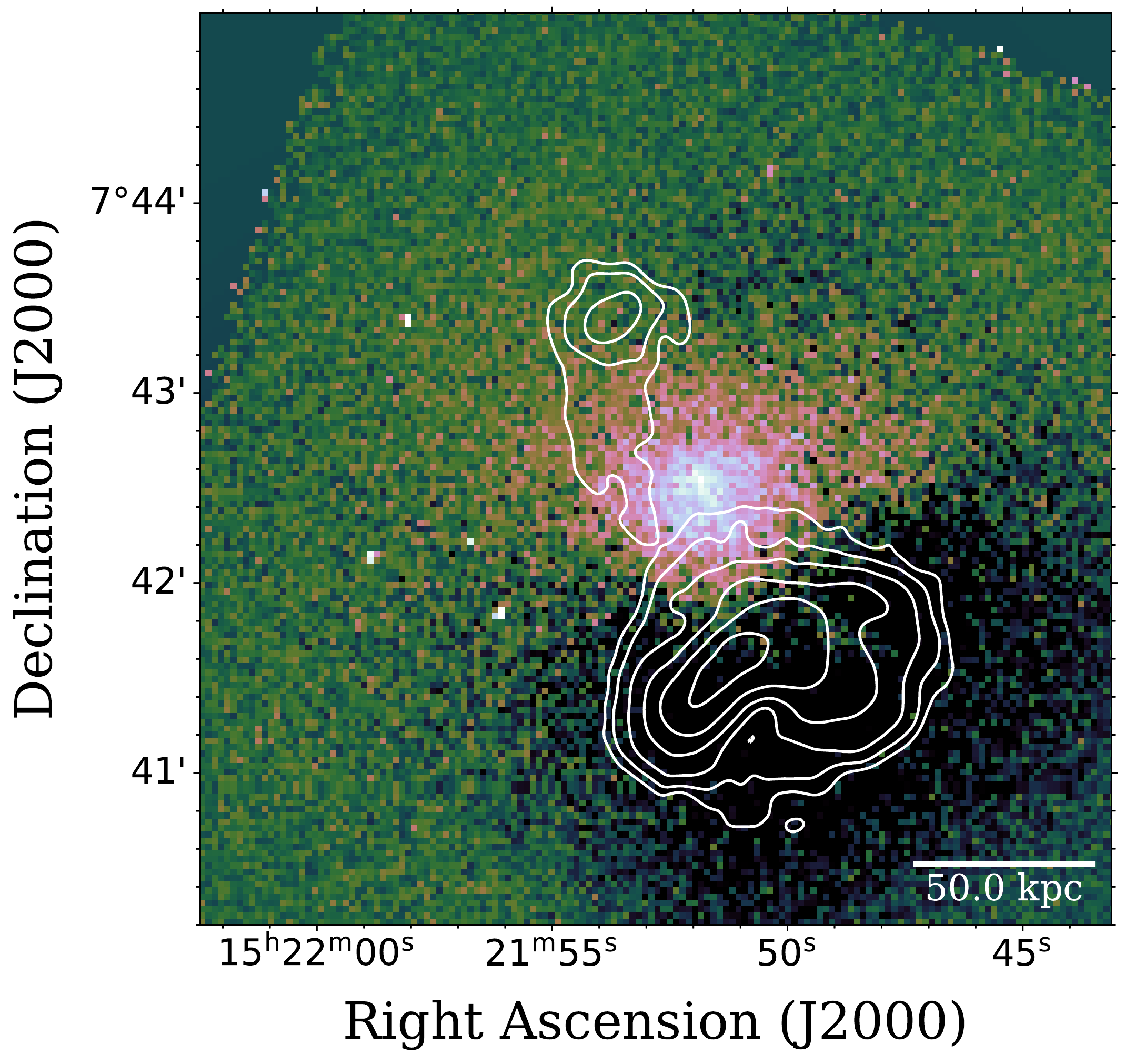}
\includegraphics[width=54mm]{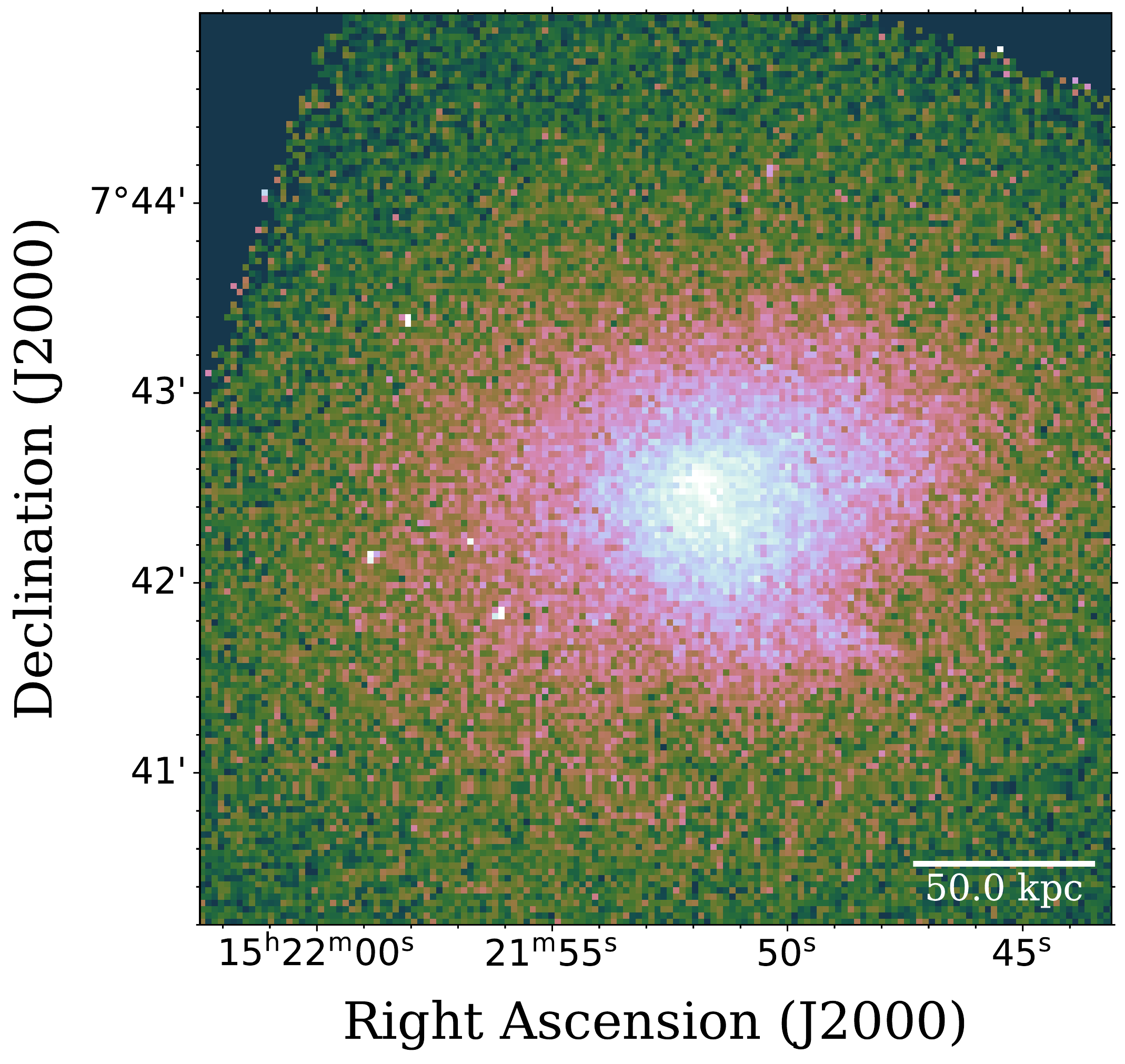} \\ \end{tabular}
\caption{ \emph{Chandra} and LOFAR images for the nearby clusters ($z<0.3$) with X-ray cavities shown in the same order as in Tables \ref{Xray_table}, \ref{LOFAR_table} and \ref{summary_table} (A2199, 2A 0335+096, A2052, and MKW3S are shown above, with the others shown in Figure 3-continued). The panel organization  is  the same as in Figure~1. For the LOFAR image, the first contour is at 0.0285~mJy beam$^{-1}$ (A2199), 0.0033~mJy beam$^{-1}$ (2A0335+096), 144~mJy beam$^{-1}$ (A2052), 0.018~mJy beam$^{-1}$ (MKW3S), and each contour increases by a factor of two.}
\label{F:images_3}
\end{figure*}

\begin{figure*} \begin{tabular}{@{}cc}
\includegraphics[width=54mm]{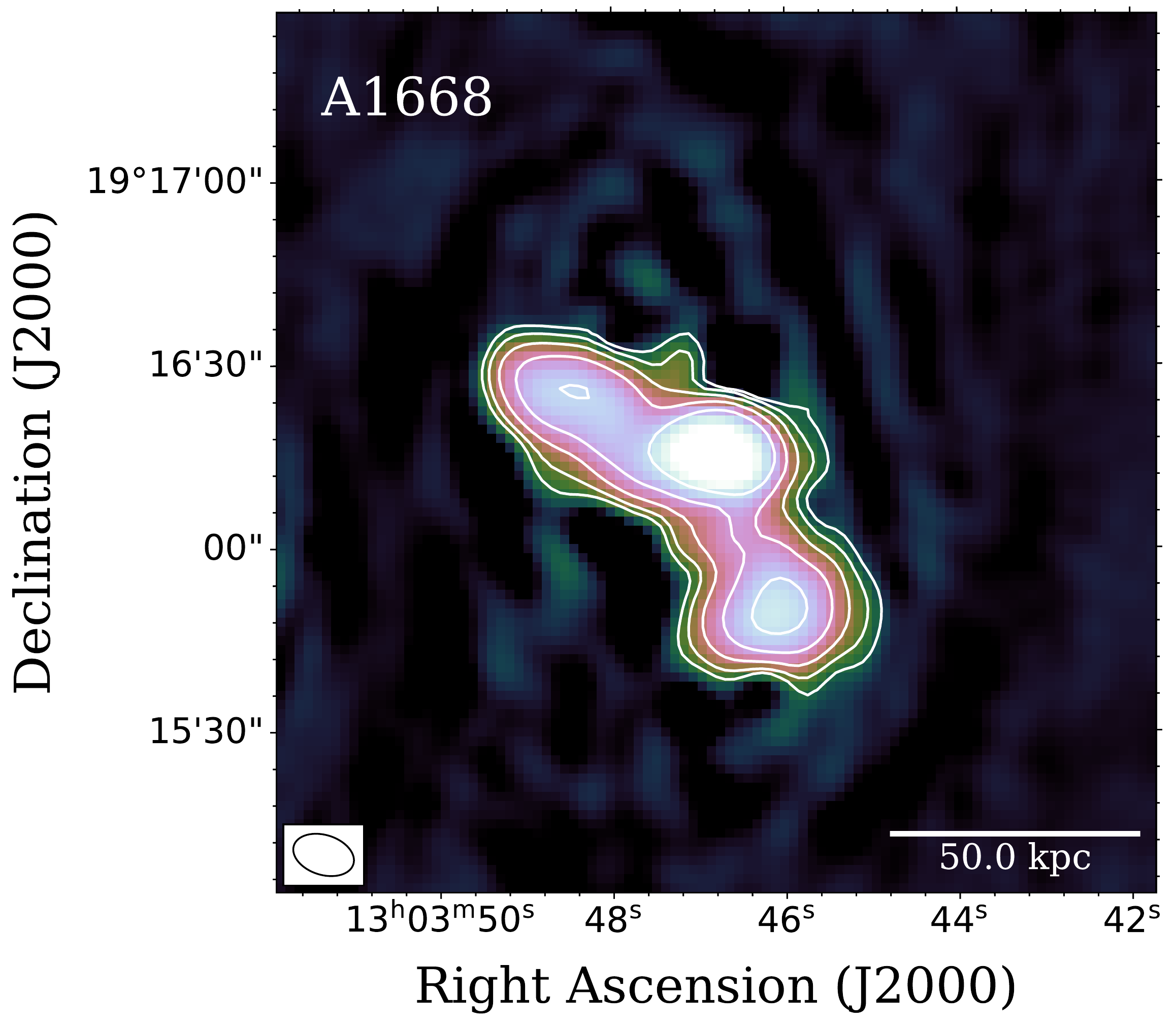} &
\includegraphics[width=54mm]{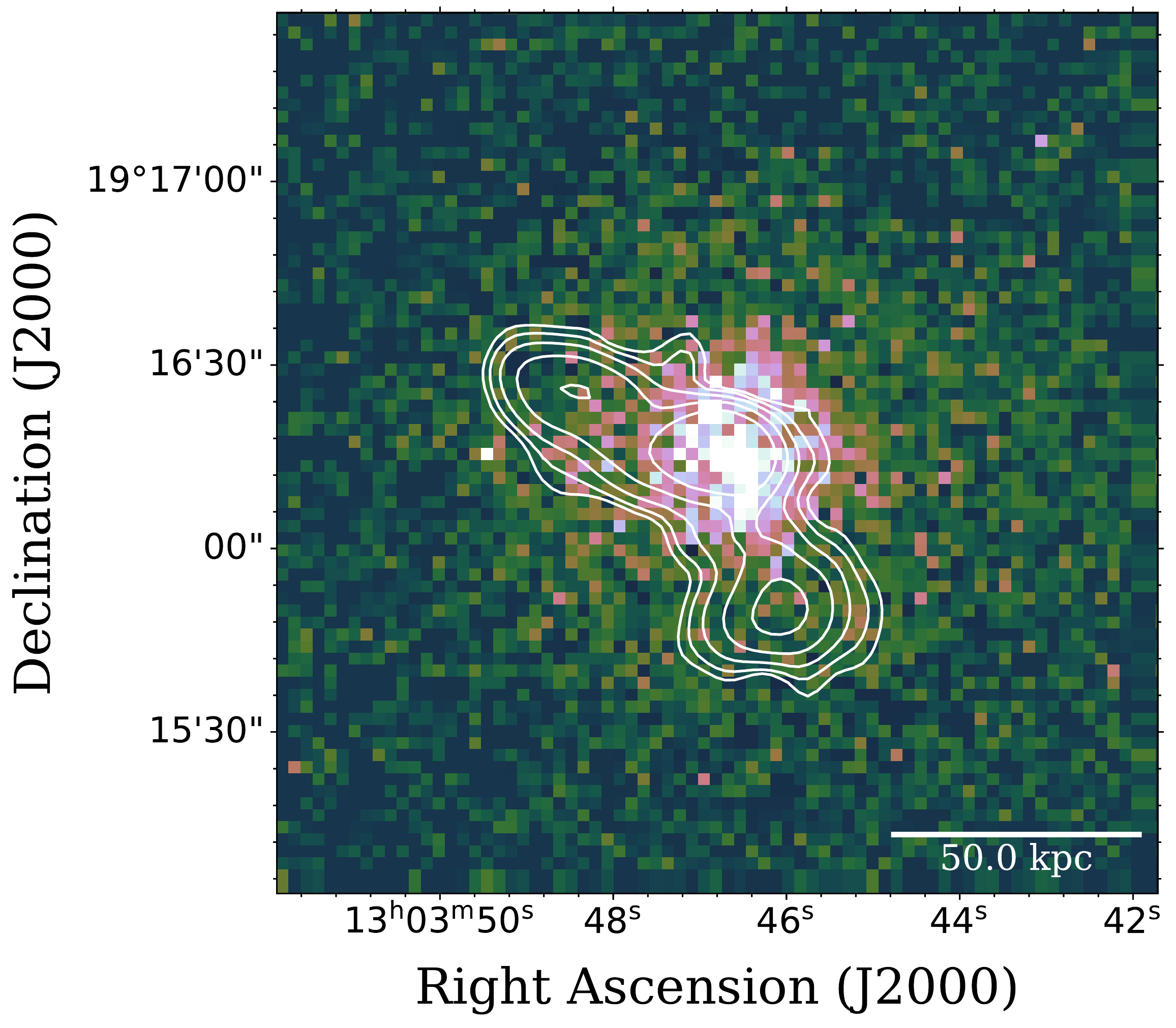}
\includegraphics[width=54mm]{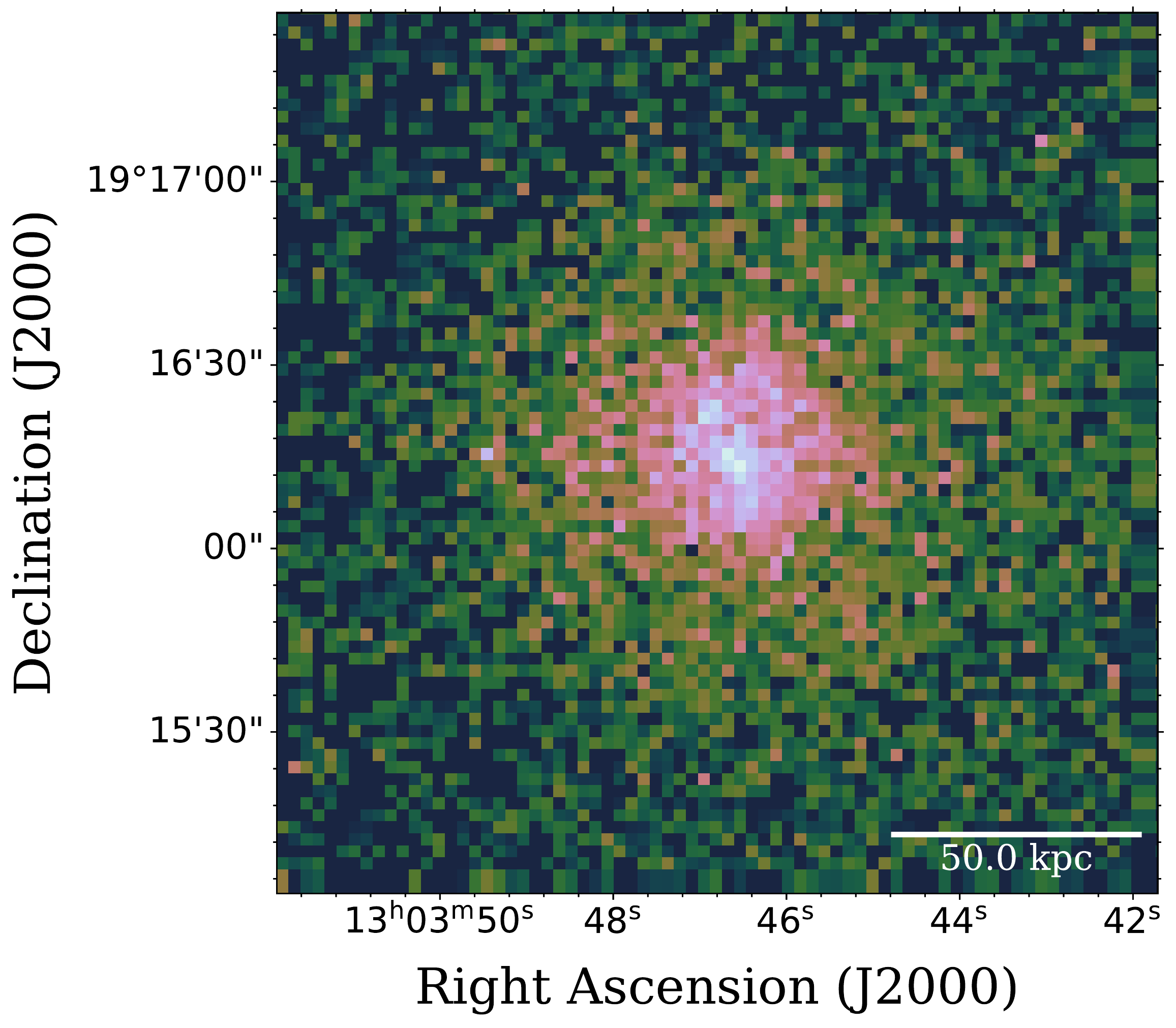} \\
\includegraphics[width=54mm]{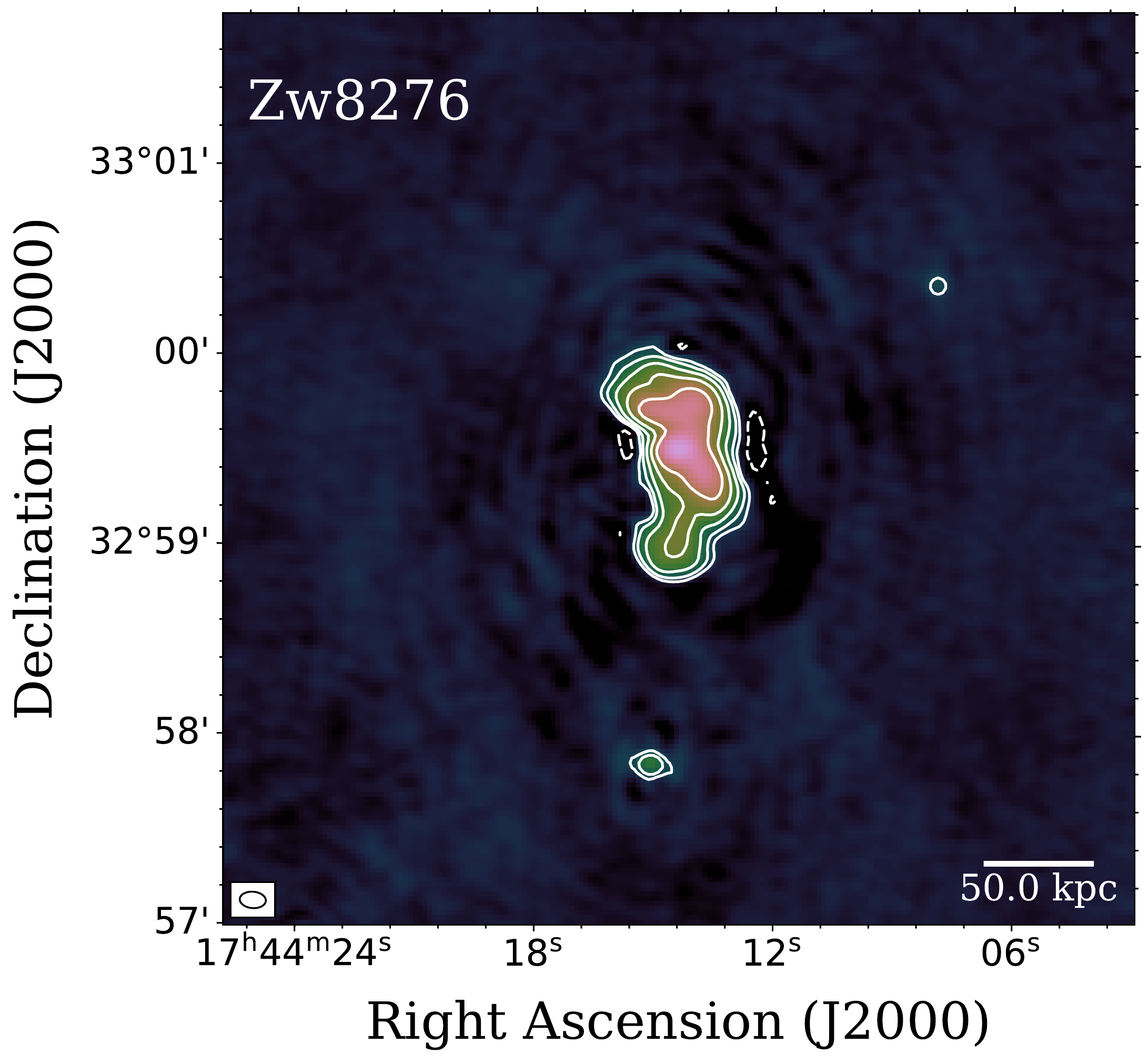} &
\includegraphics[width=54mm]{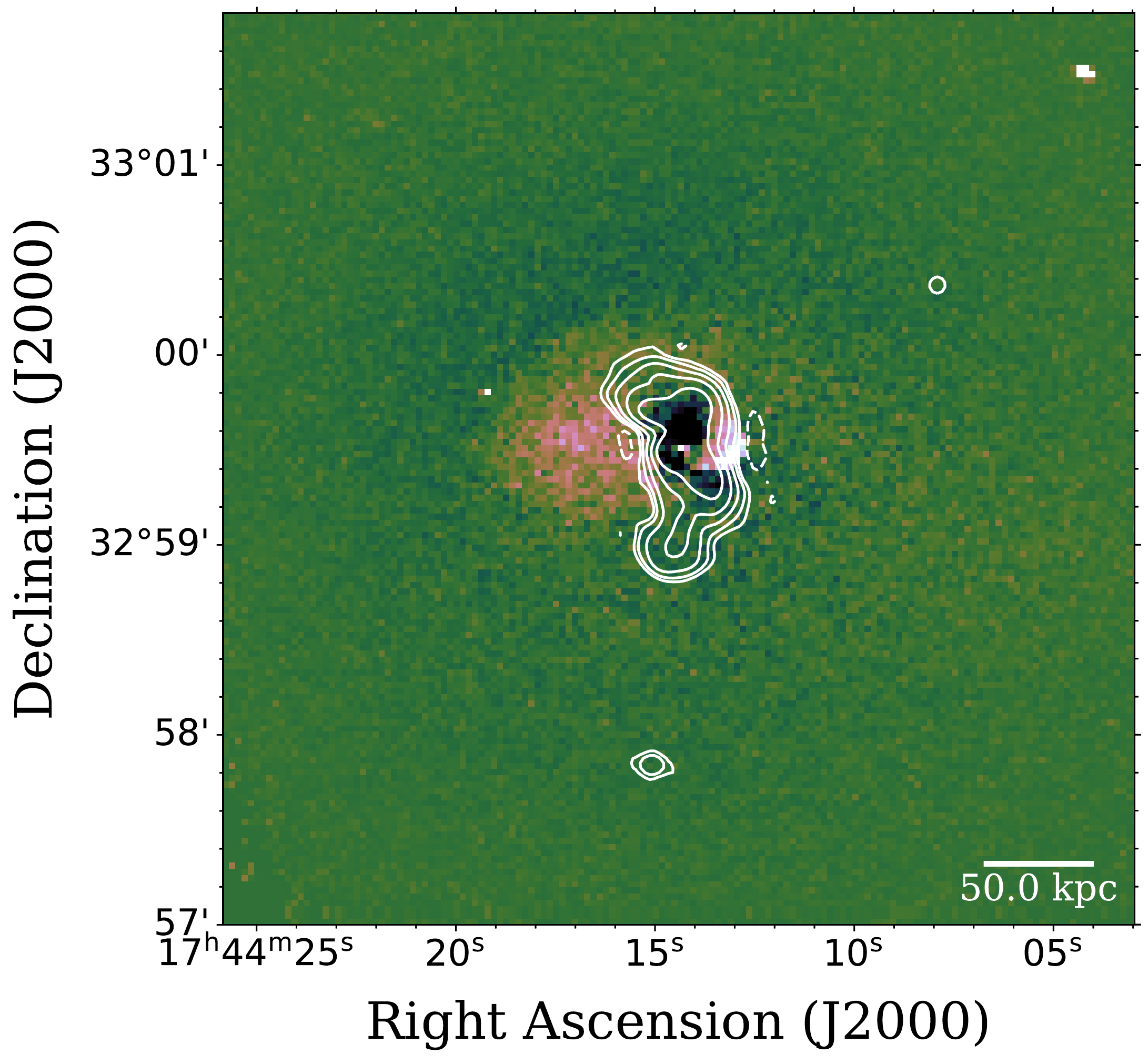}
\includegraphics[width=54mm]{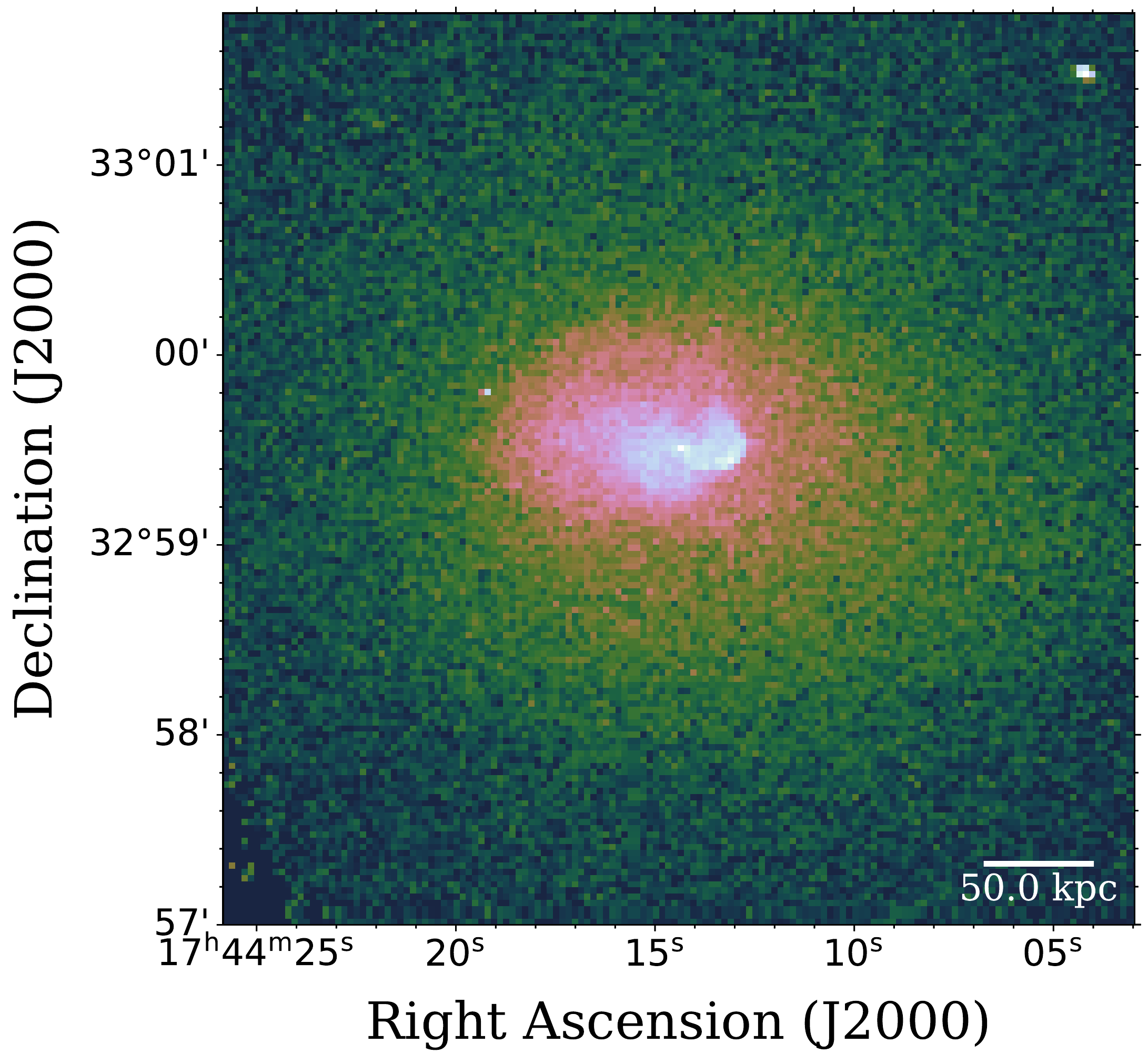} \\
\includegraphics[width=54mm]{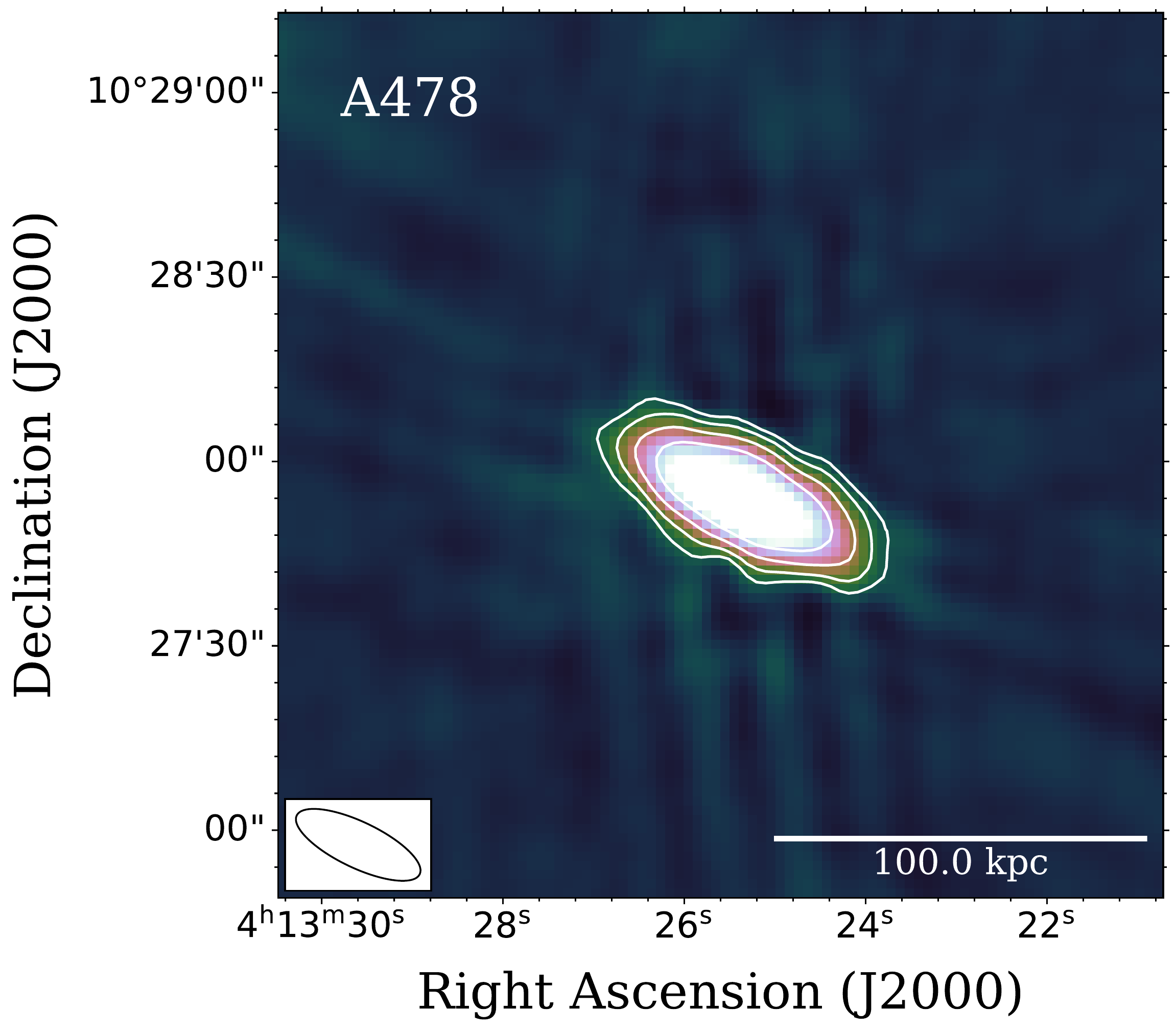} &
\includegraphics[width=54mm]{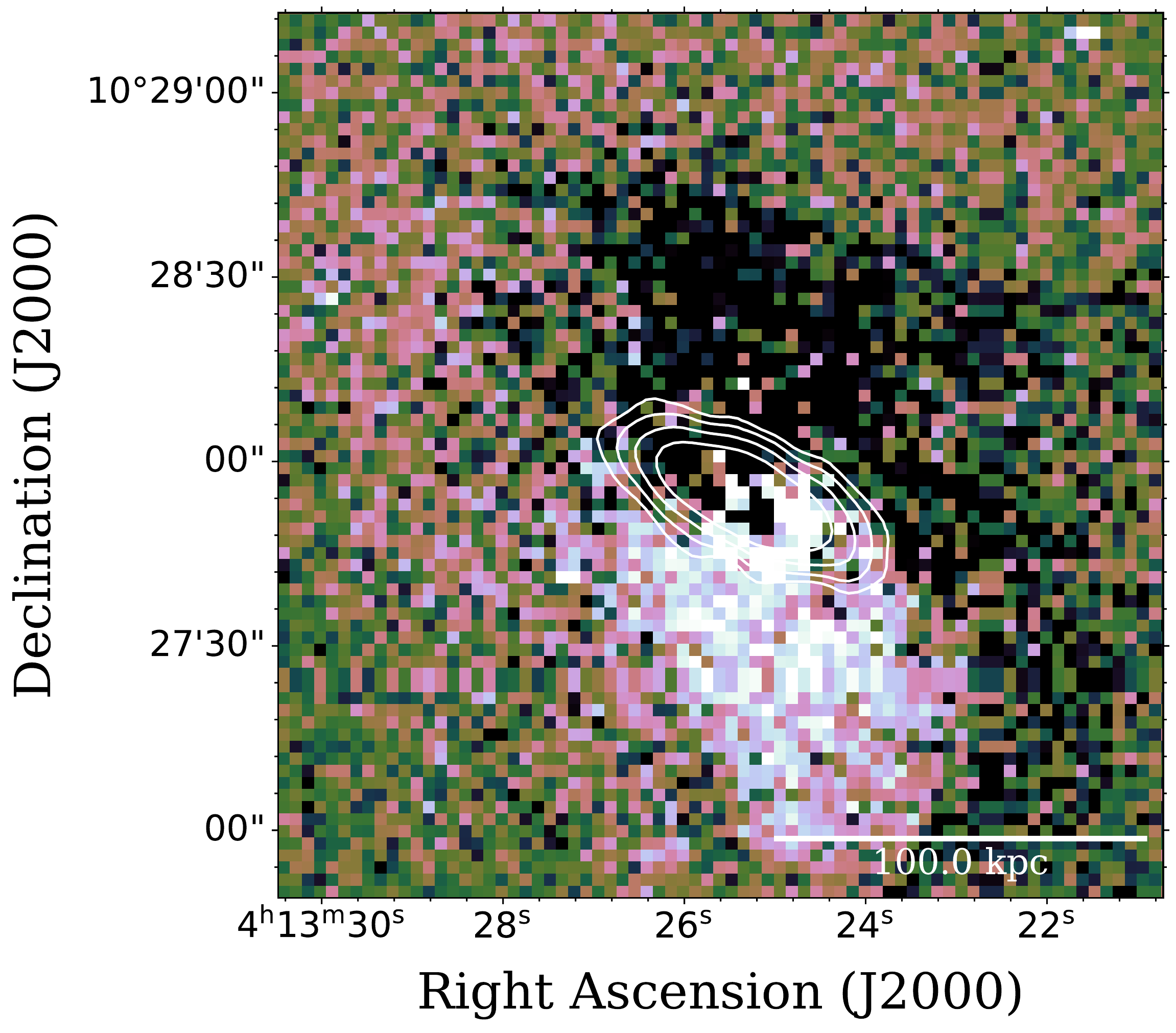}
\includegraphics[width=54mm]{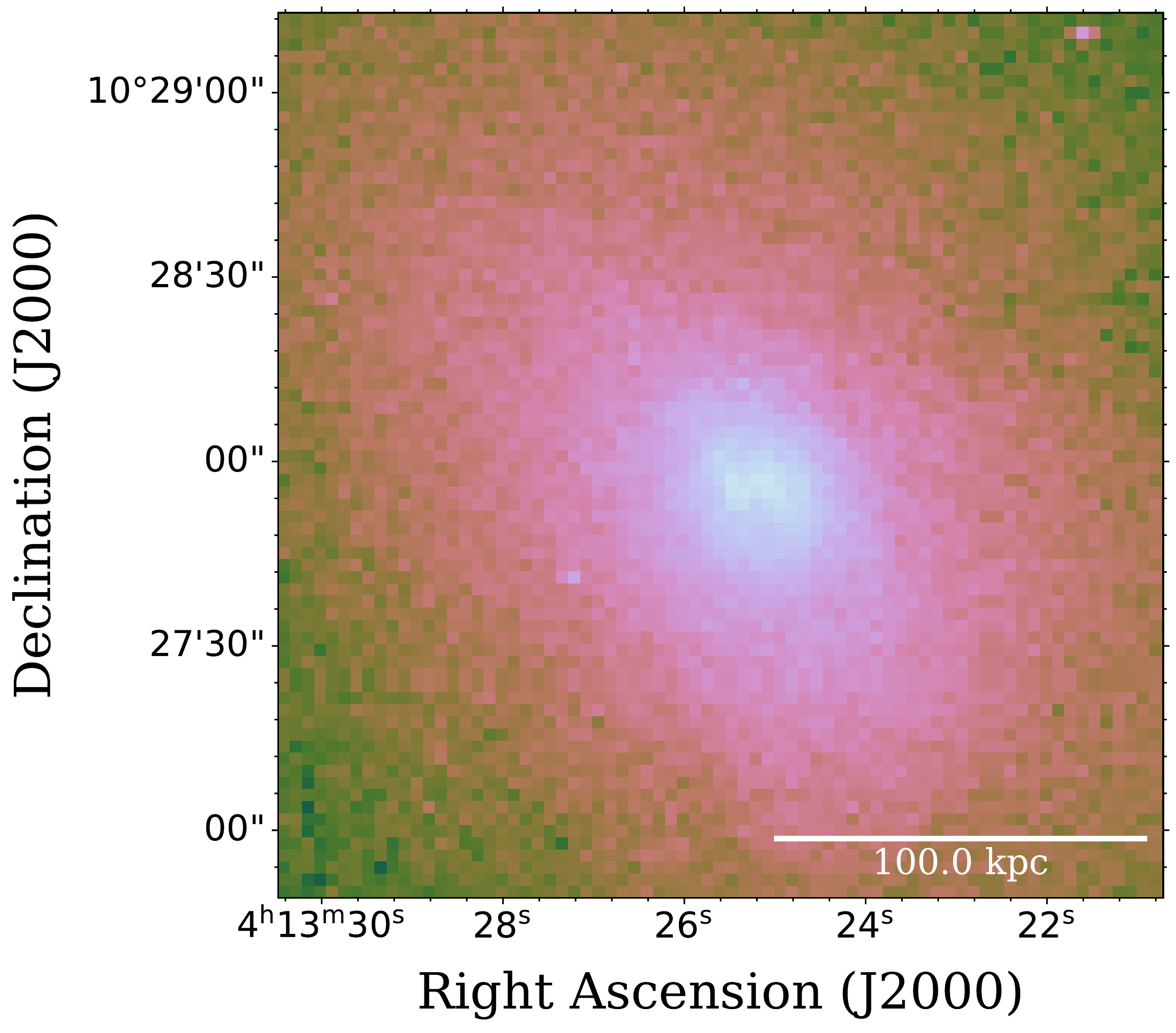}\\
\includegraphics[width=54mm]{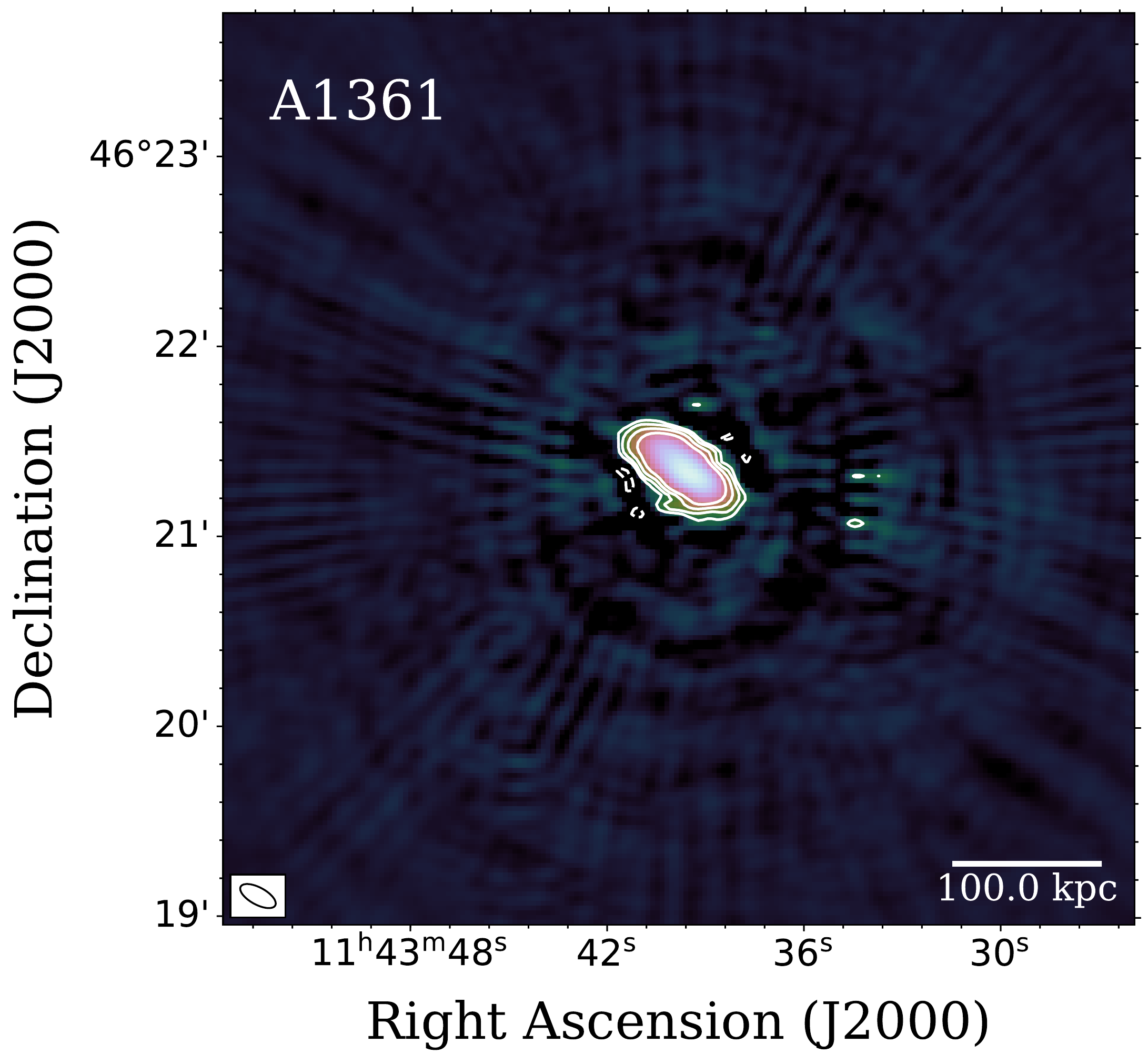} &
\includegraphics[width=54mm]{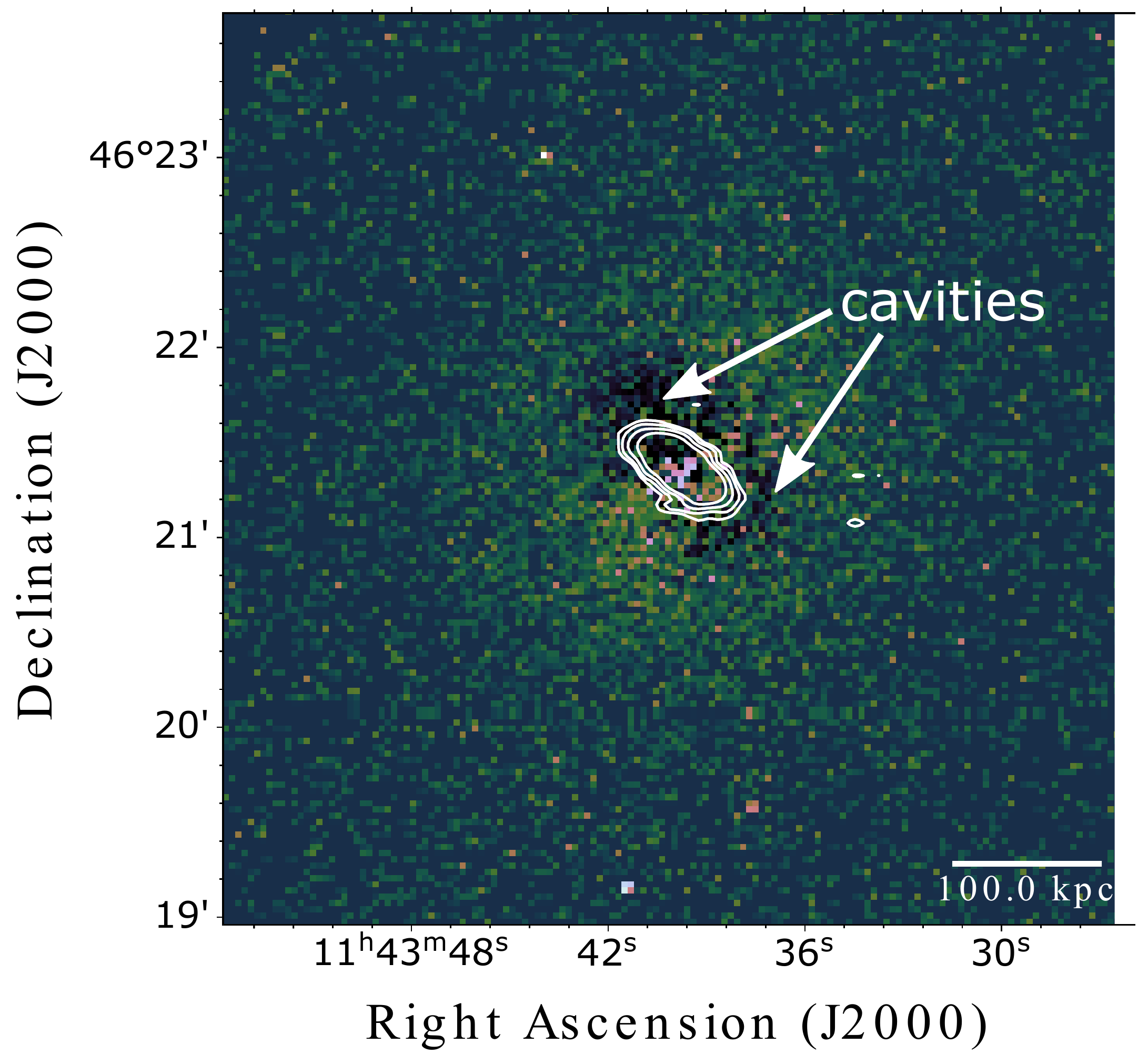}
\includegraphics[width=54mm]{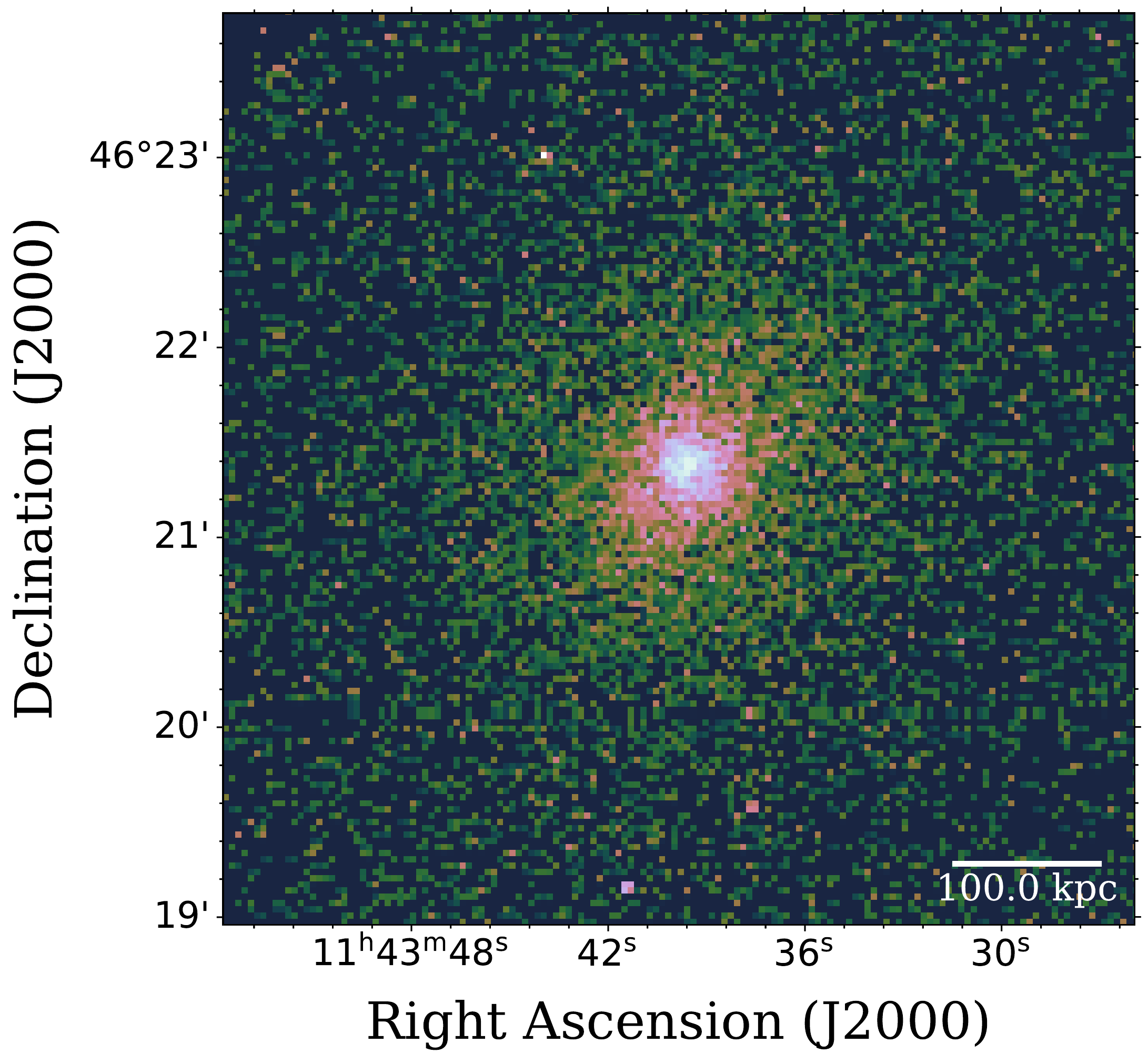} \\ \end{tabular}
\textbf{Figure}~3. --- continued (A1668, ZwCl 8276, A478, and A1361). The panel organization  is  the same as in Figure~1. For the LOFAR image, the first contour is at 0.006~mJy beam$^{-1}$ (A1668), 0.0021~mJy beam$^{-1}$ (ZwCl 8276), 0.0077~mJy beam$^{-1}$ (A478), 0.0587~mJy beam$^{-1}$ (A1361), and each contour increases by a factor of two. \\
\end{figure*}

\begin{figure*} \begin{tabular}{@{}cc}
\includegraphics[width=54mm]{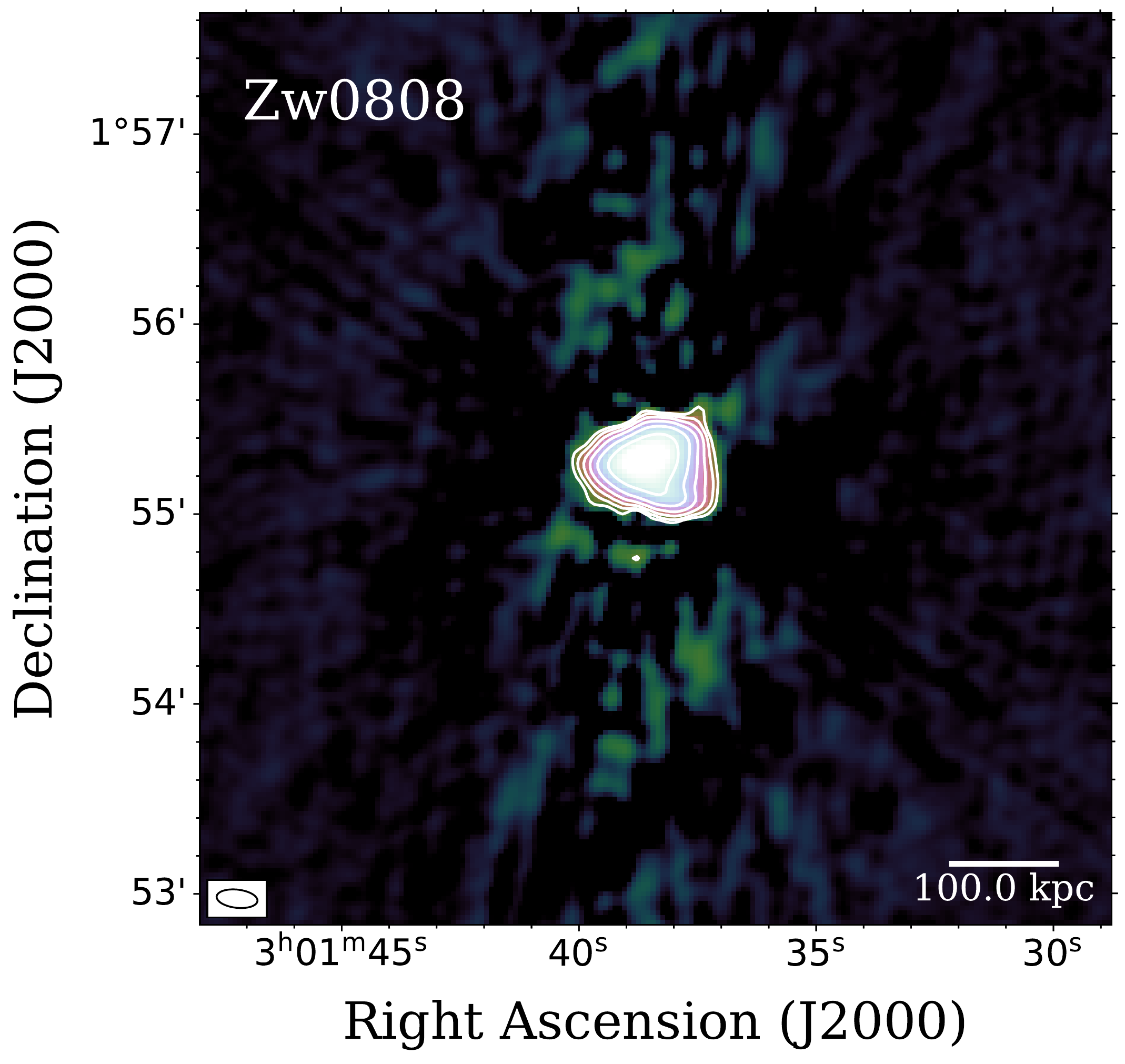} &
\includegraphics[width=54mm]{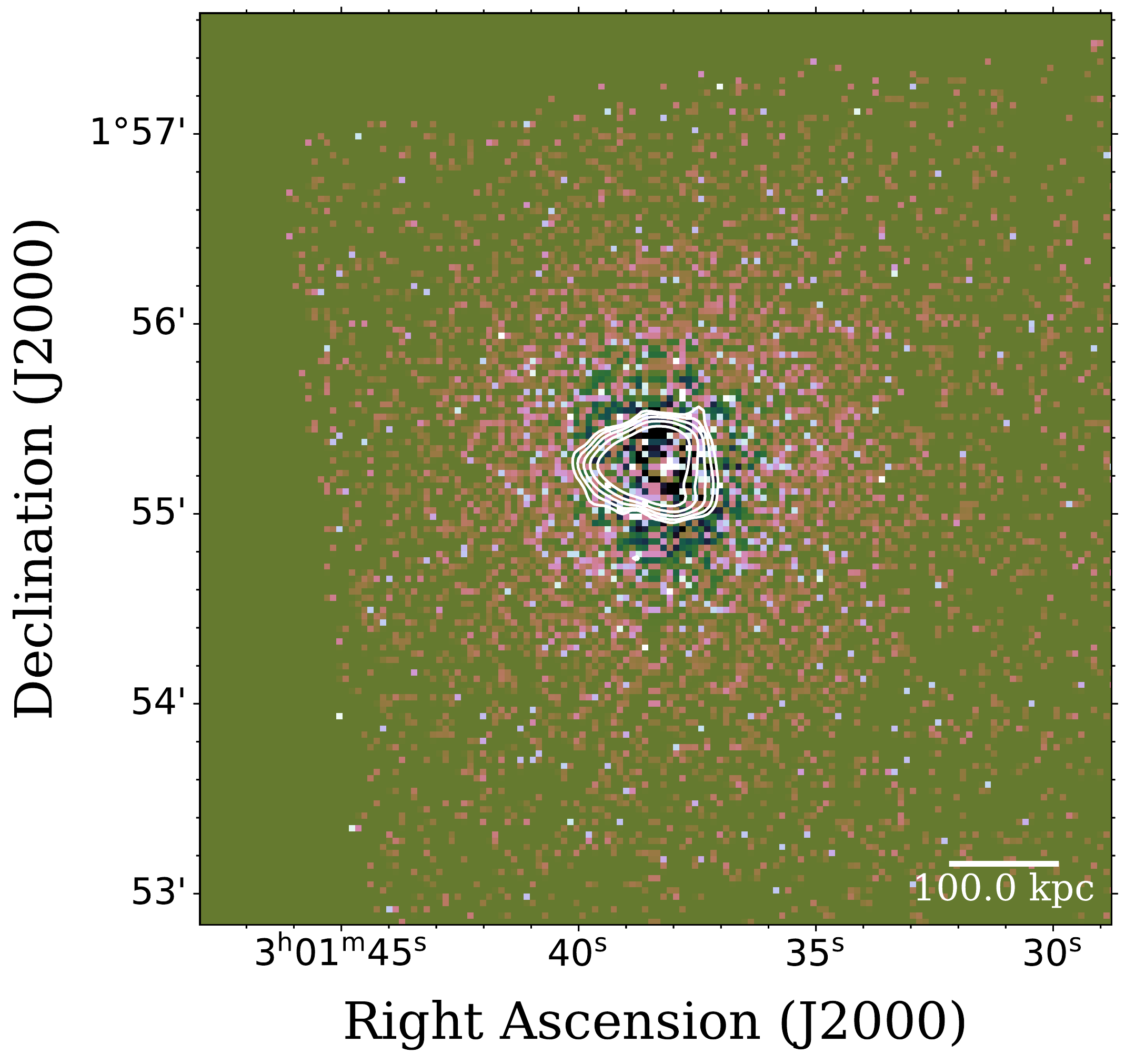}
\includegraphics[width=54mm]{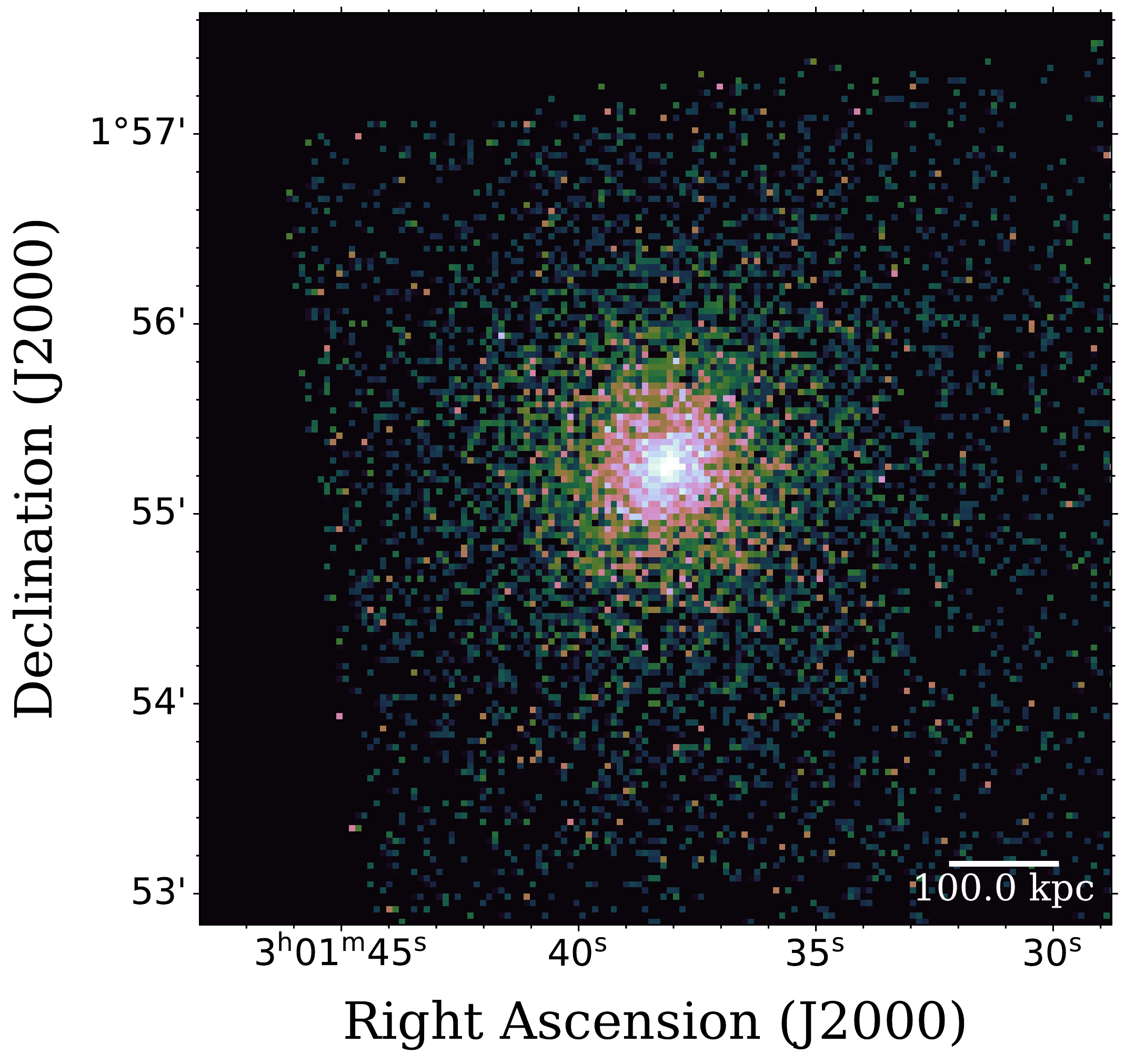} \\
\includegraphics[width=54mm]{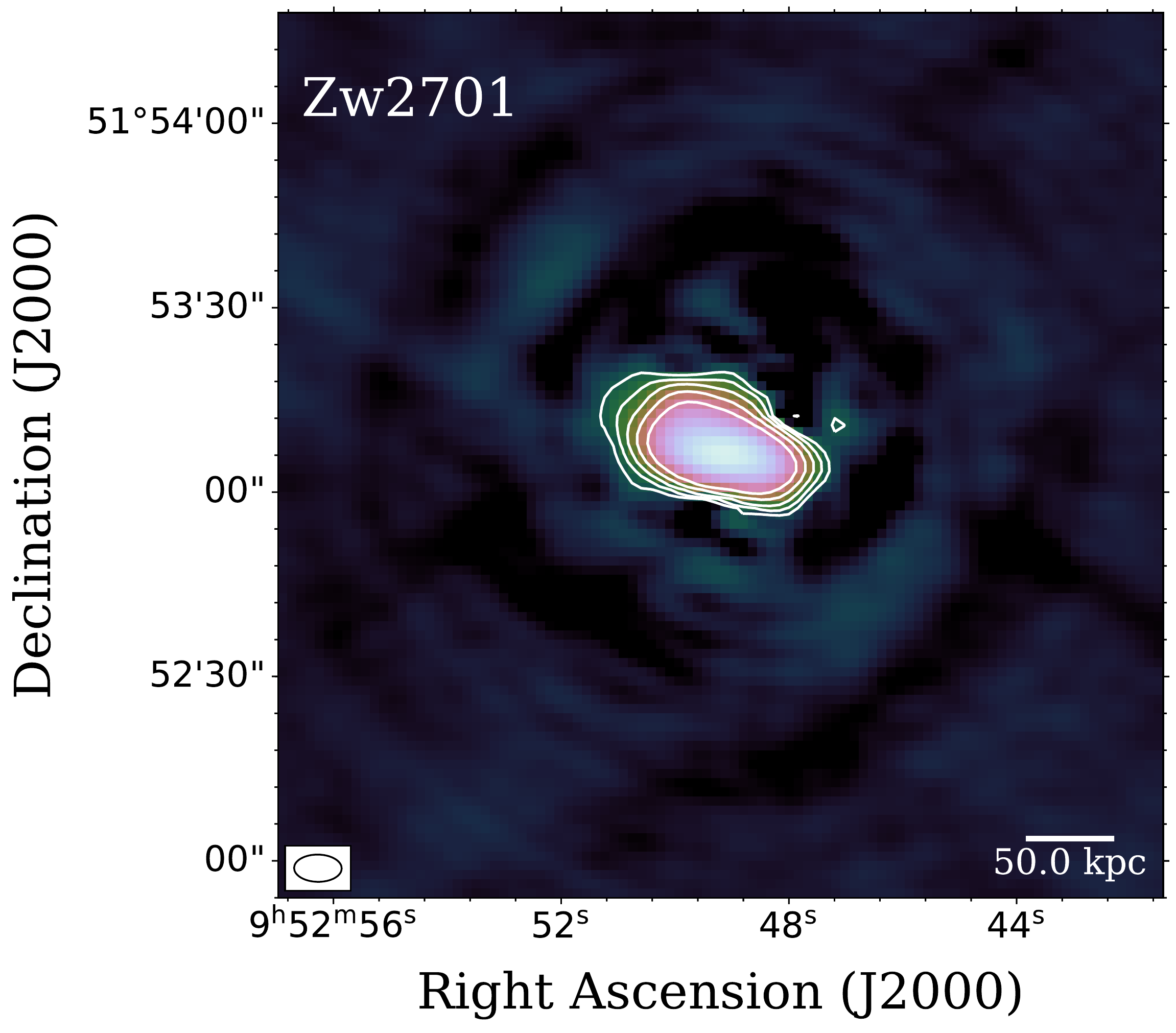} &
\includegraphics[width=54mm]{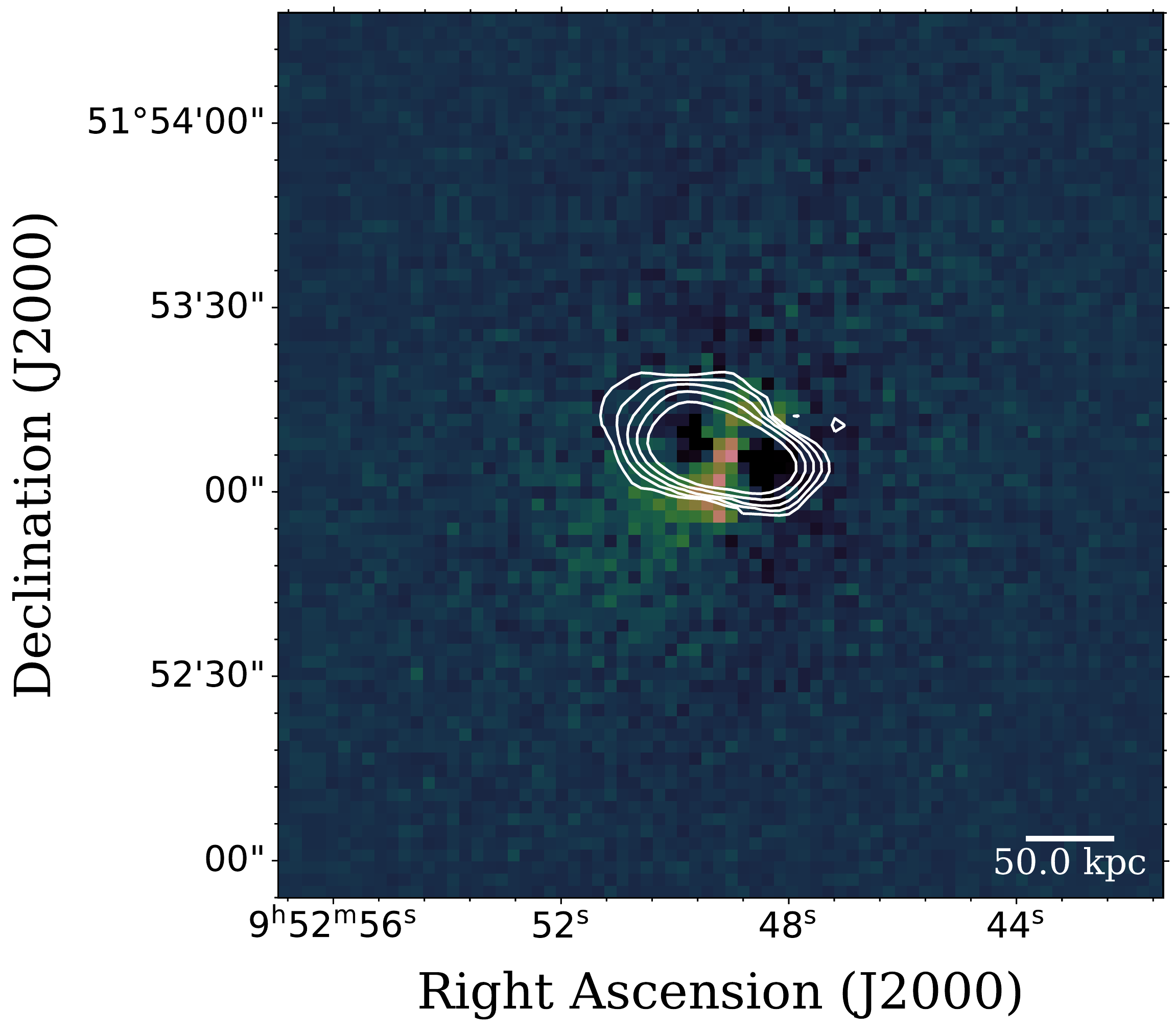}
\includegraphics[width=54mm]{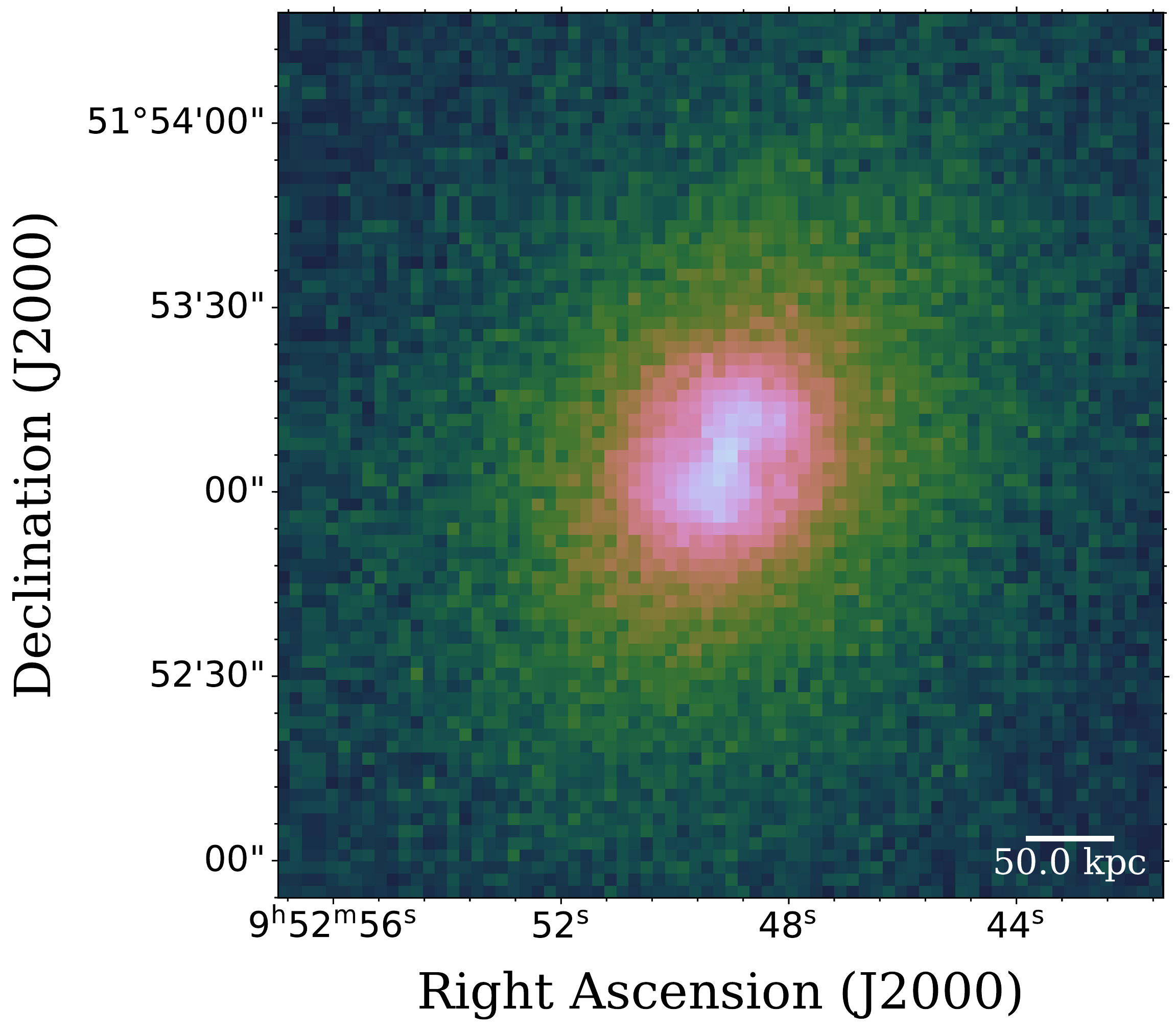} \\
\includegraphics[width=54mm]{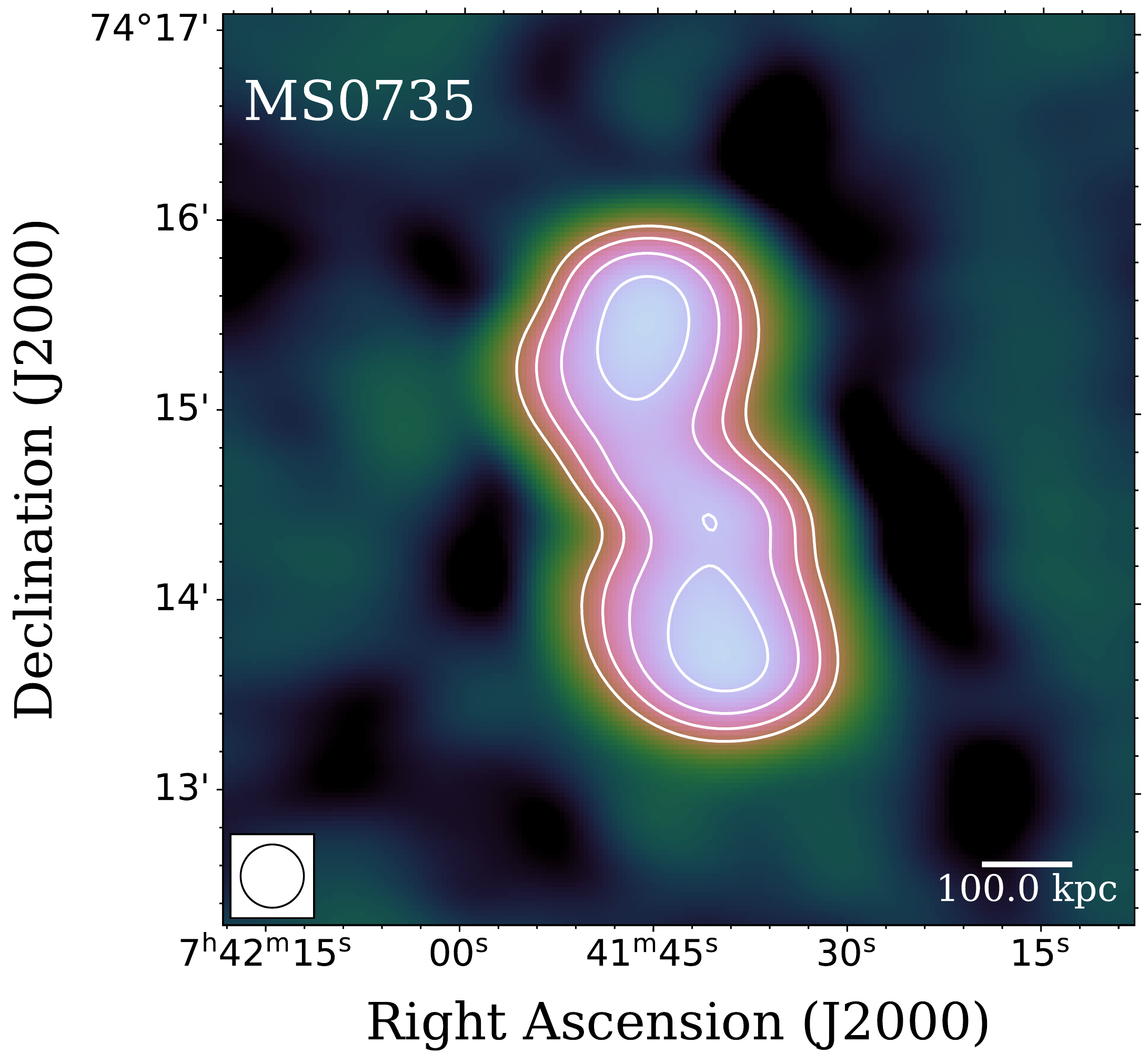} &
\includegraphics[width=54mm]{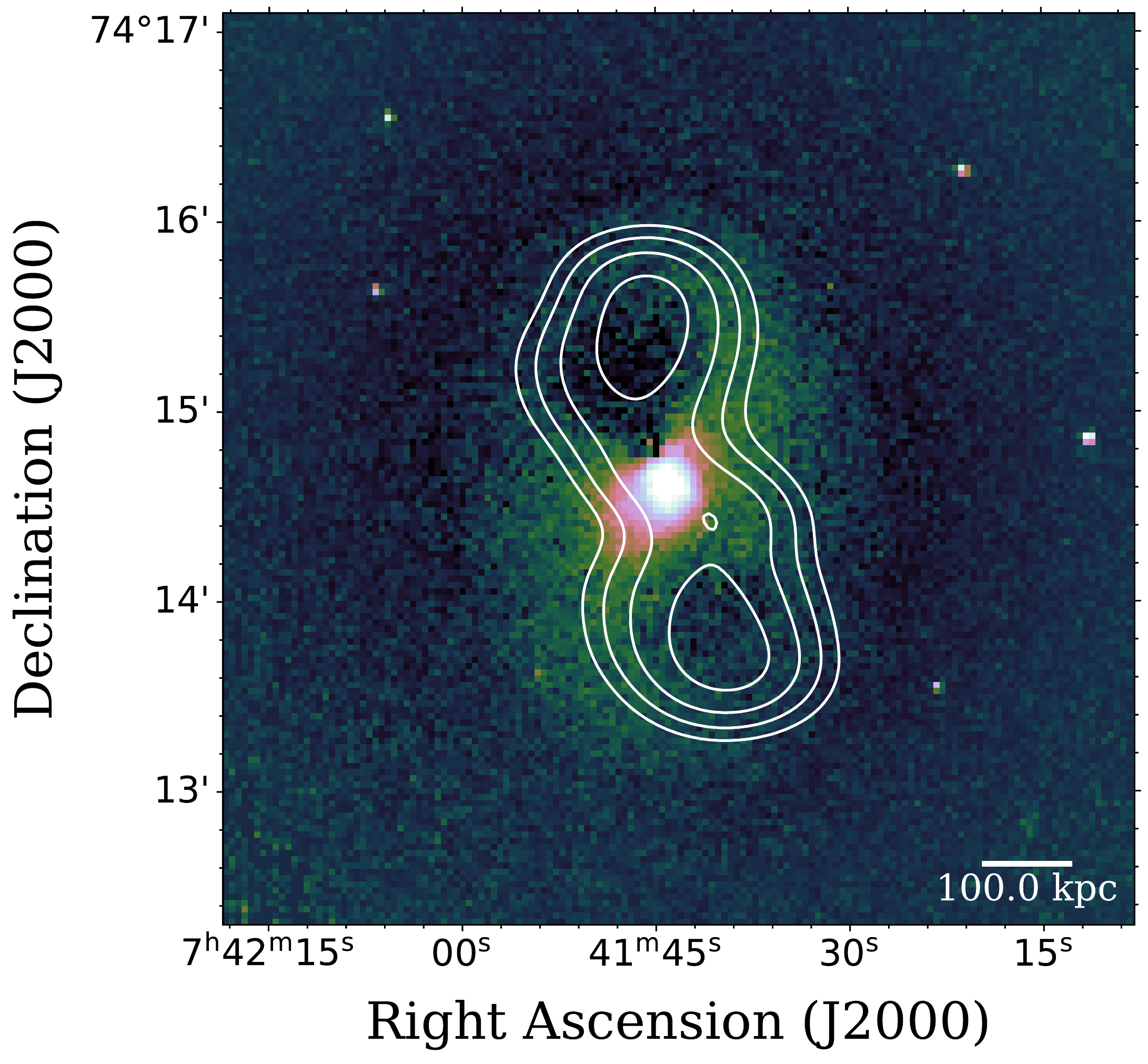}
\includegraphics[width=54mm]{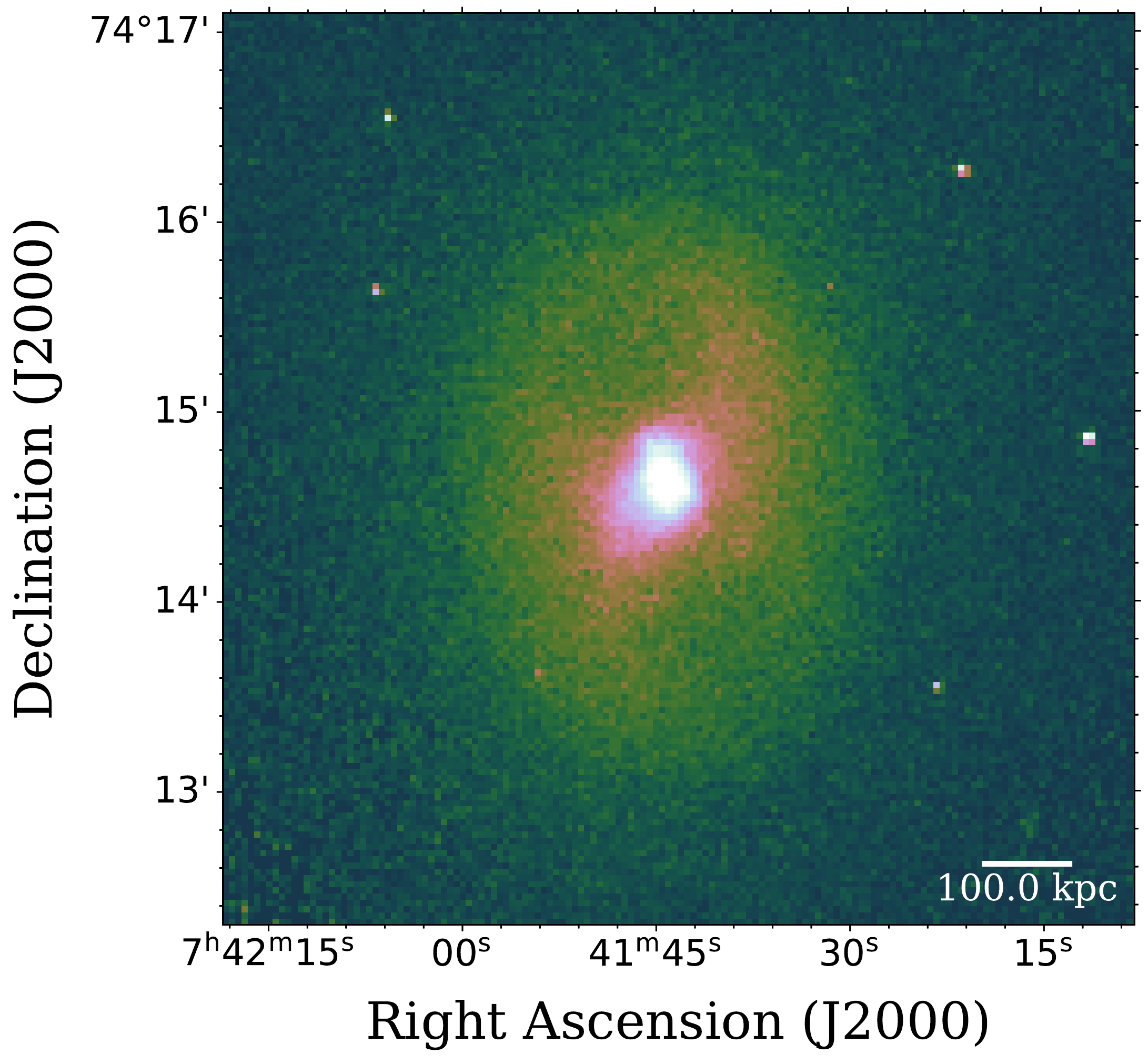} \\
\includegraphics[width=54mm]{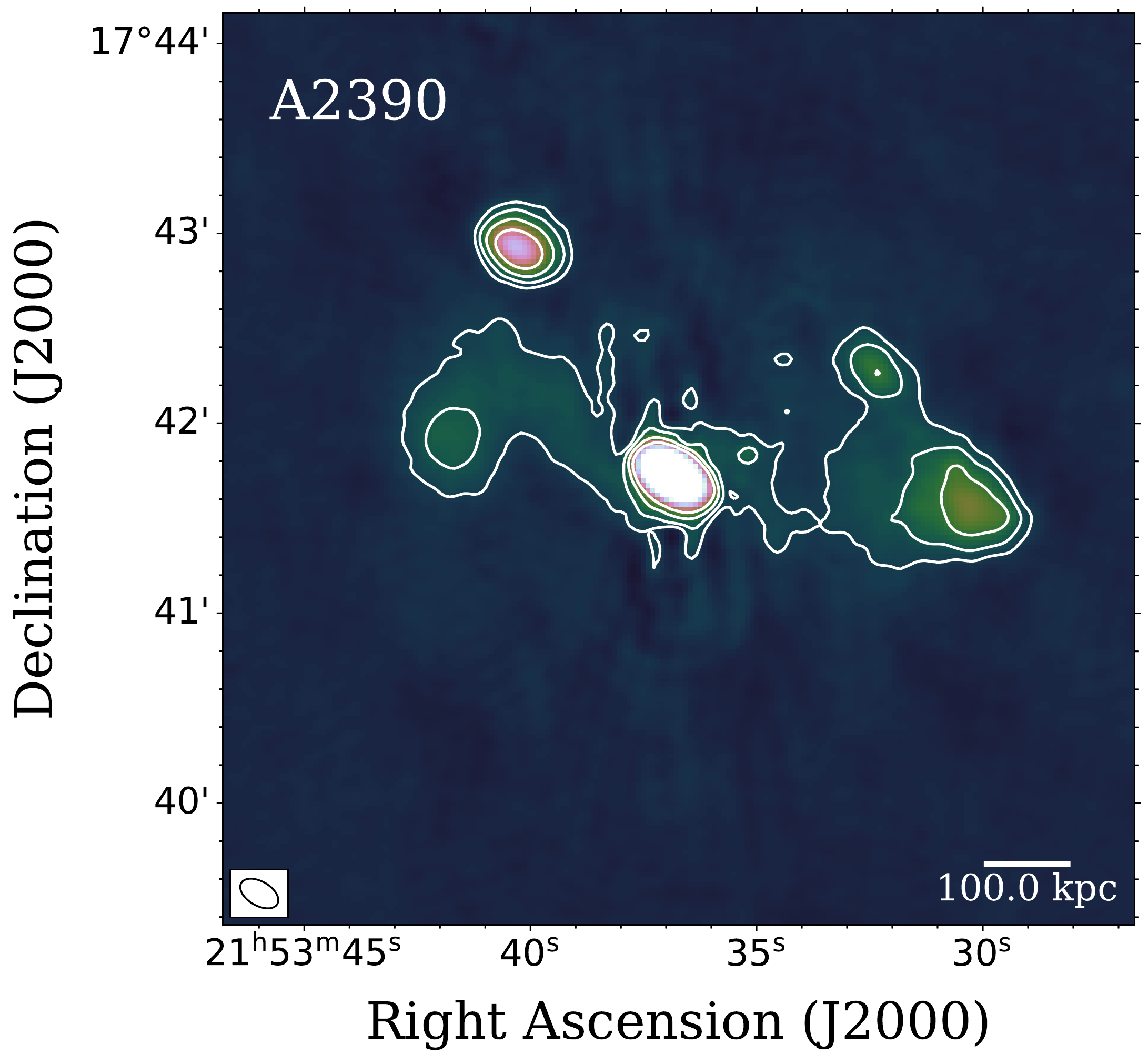} &
\includegraphics[width=54mm]{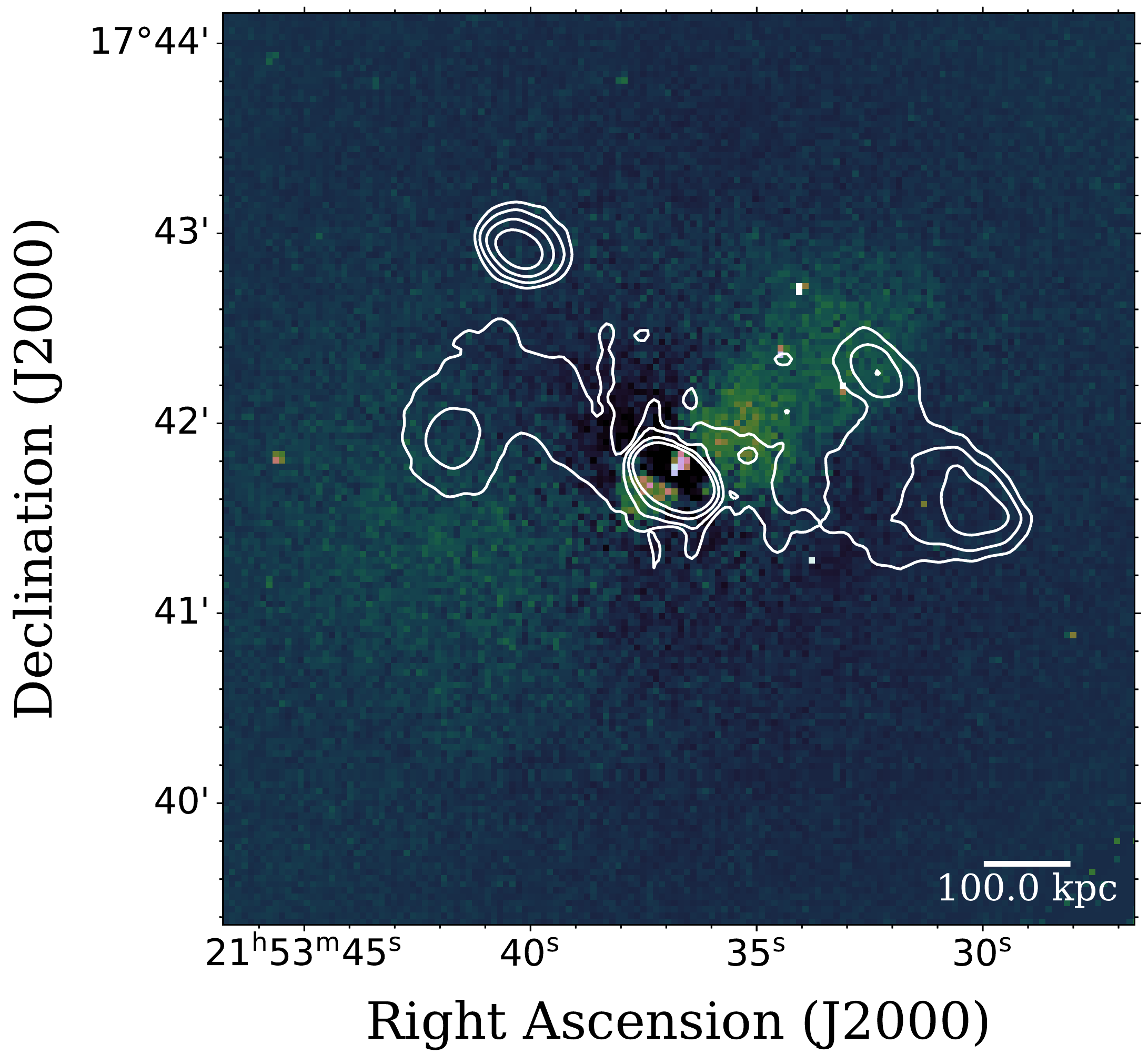}
\includegraphics[width=54mm]{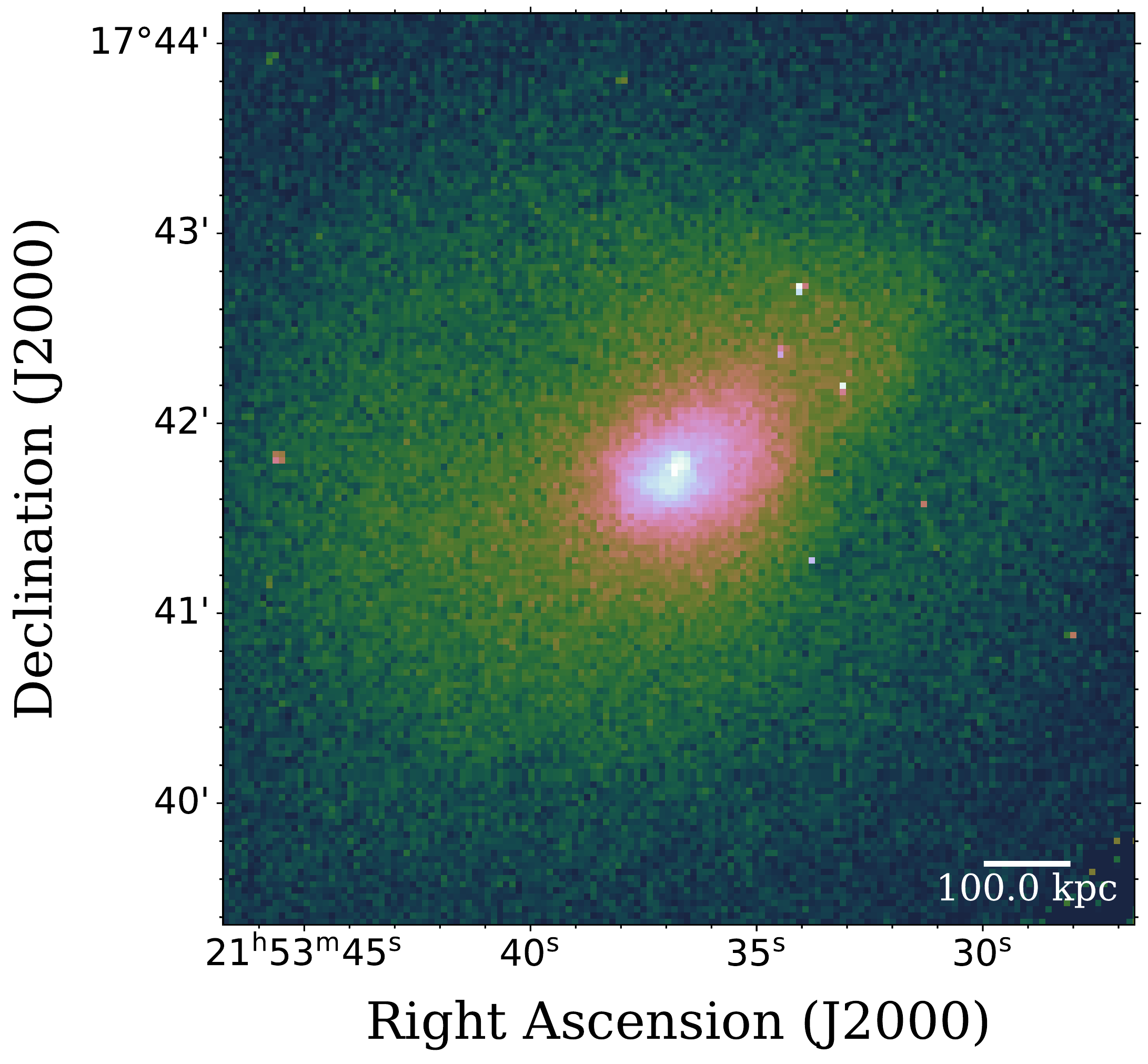} \\ \\ \end{tabular}
\textbf{Figure}~3. --- continued (ZwCl 0808, ZwCl 2701, MS 0735.6+7421, and A2390).  For the LOFAR image, the first contour is at 0.036~mJy beam$^{-1}$ (ZwCl 0808), 0.0039~mJy beam$^{-1}$ (ZwCl 2701), 0.046~mJy beam$^{-1}$ (MS 0735.6+7421), 0.00357~mJy beam$^{-1}$ (A2390), and each contour increases by a factor of two. \\
\end{figure*}

\begin{figure*} \begin{tabular}{@{}cc}
\includegraphics[width=54mm]{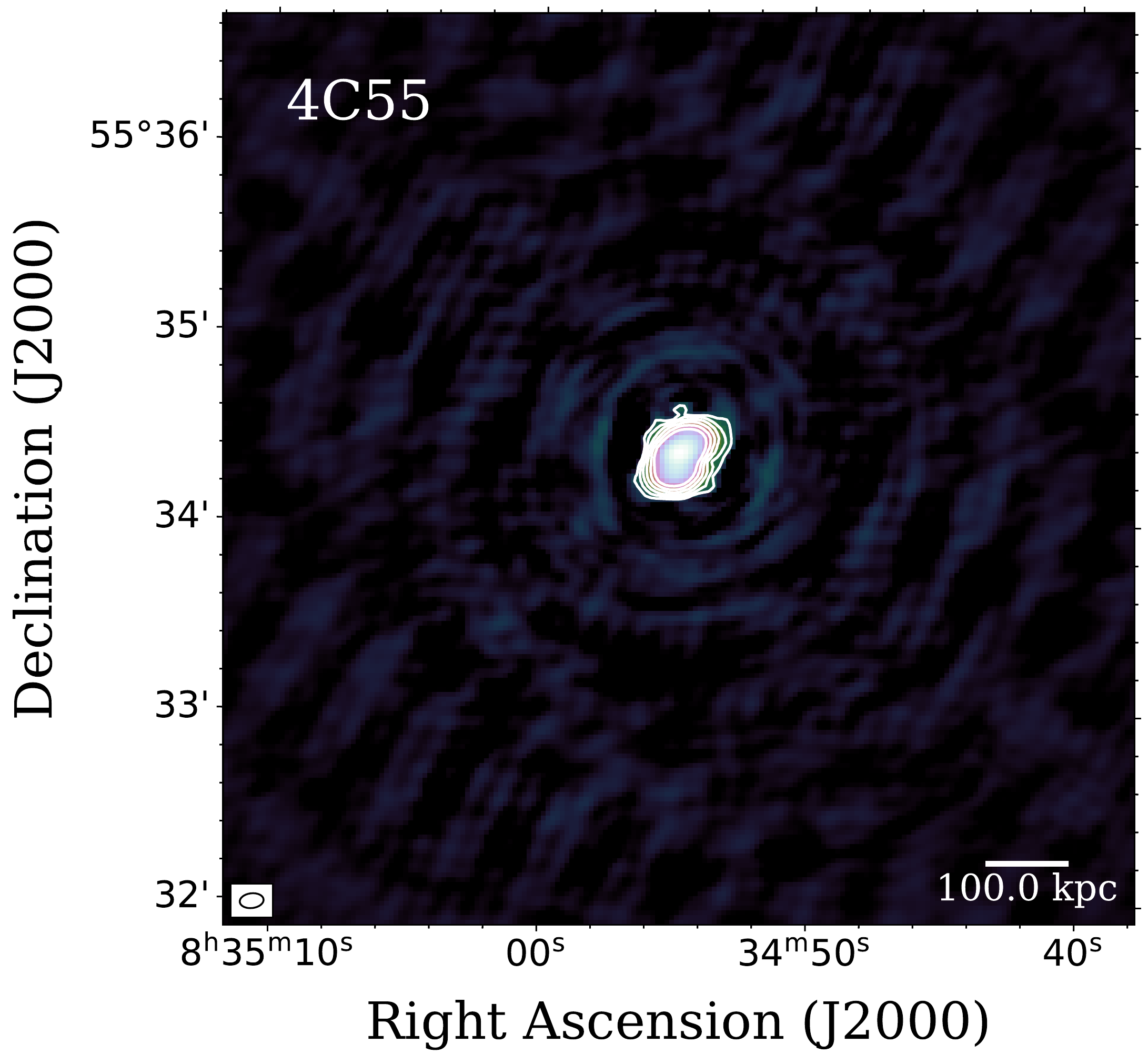} &
\includegraphics[width=54mm]{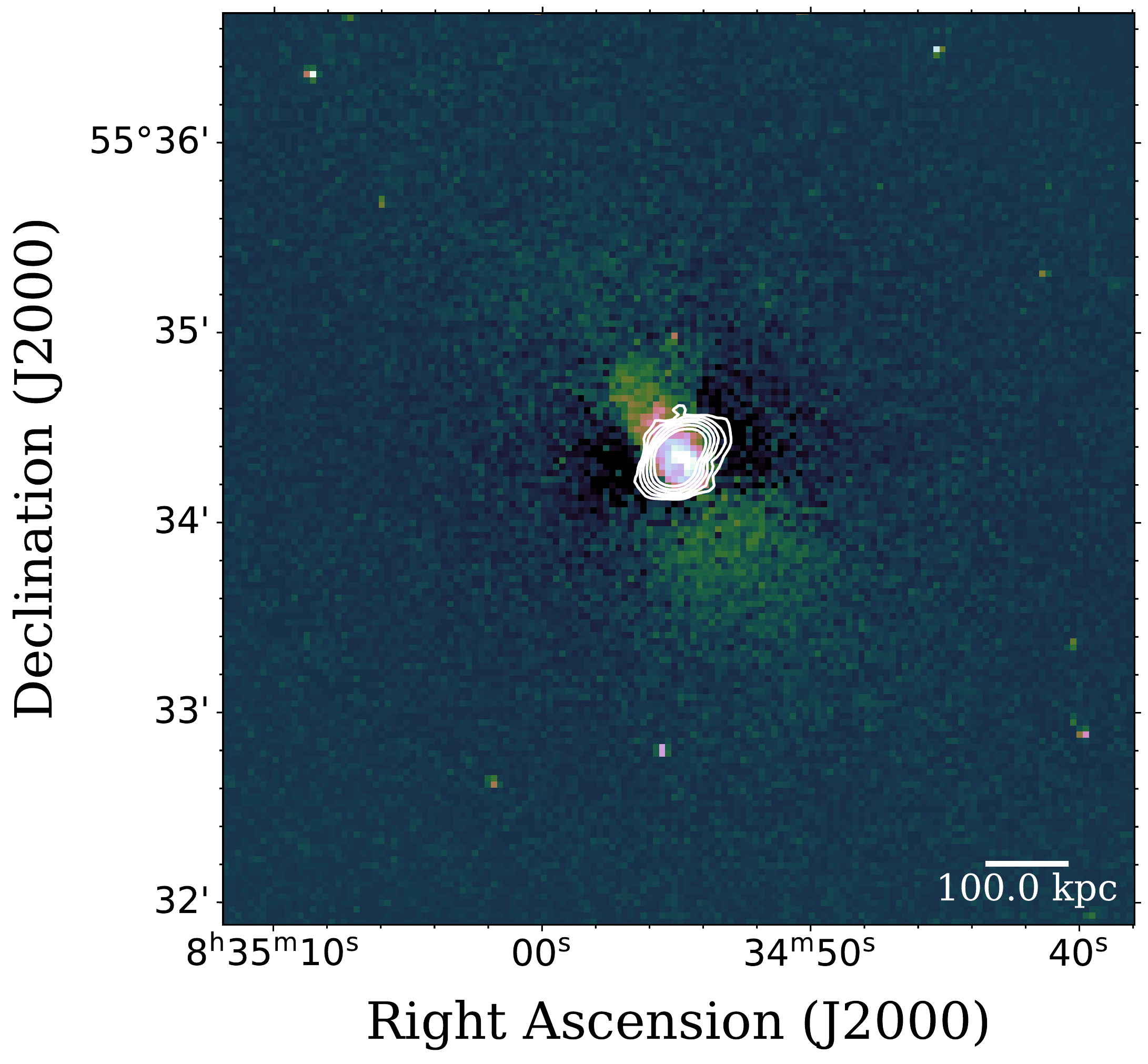}
\includegraphics[width=54mm]{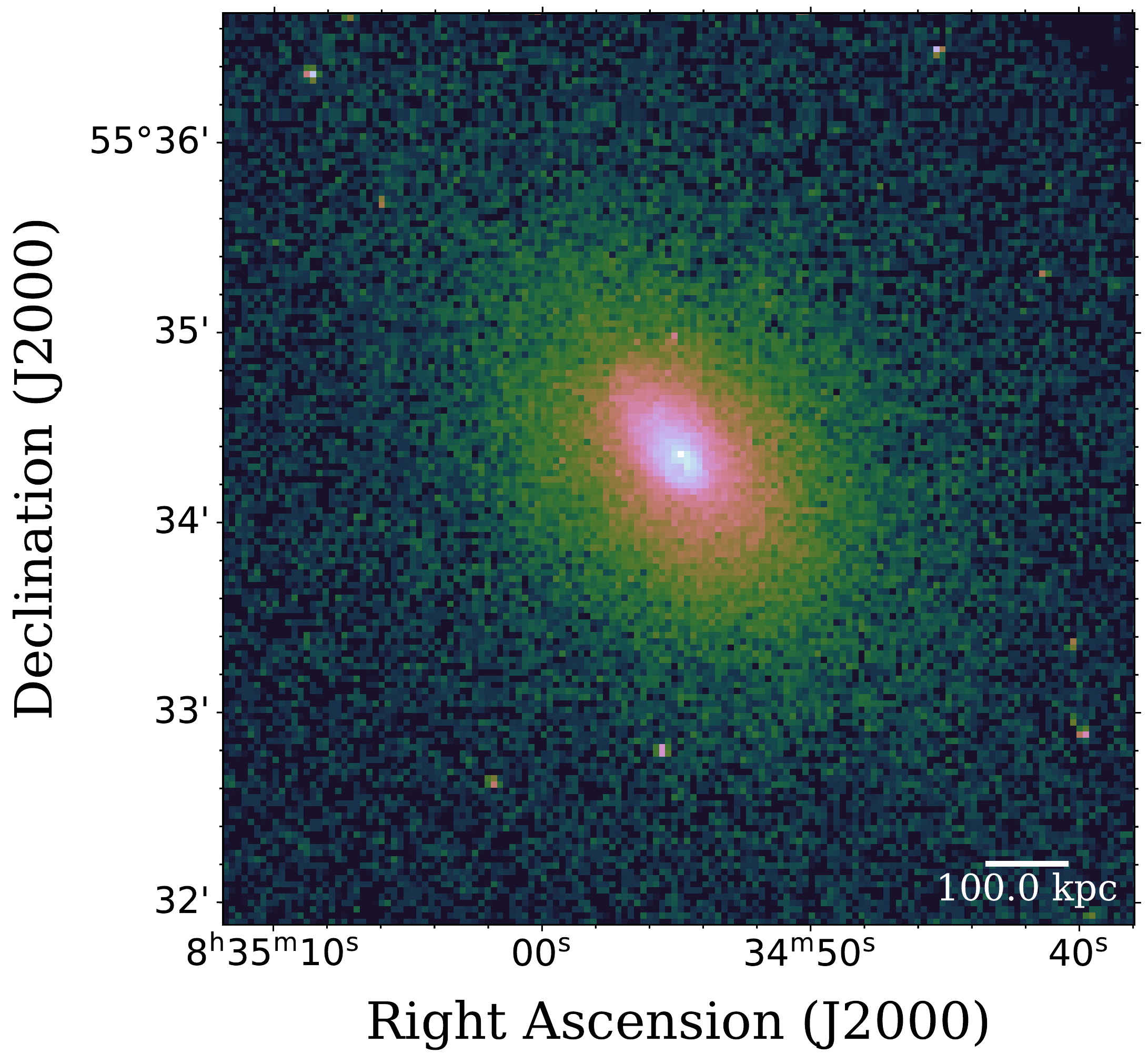} \\ \end{tabular}
\textbf{Figure}~3
. --- continued (4C+55.16).  For the LOFAR image, the first contour is at 0.012~mJy/beam and each contour increases by a factor of two. \\
\end{figure*}

\begin{figure*} \begin{tabular}{@{}cc}
\includegraphics[width=54mm]{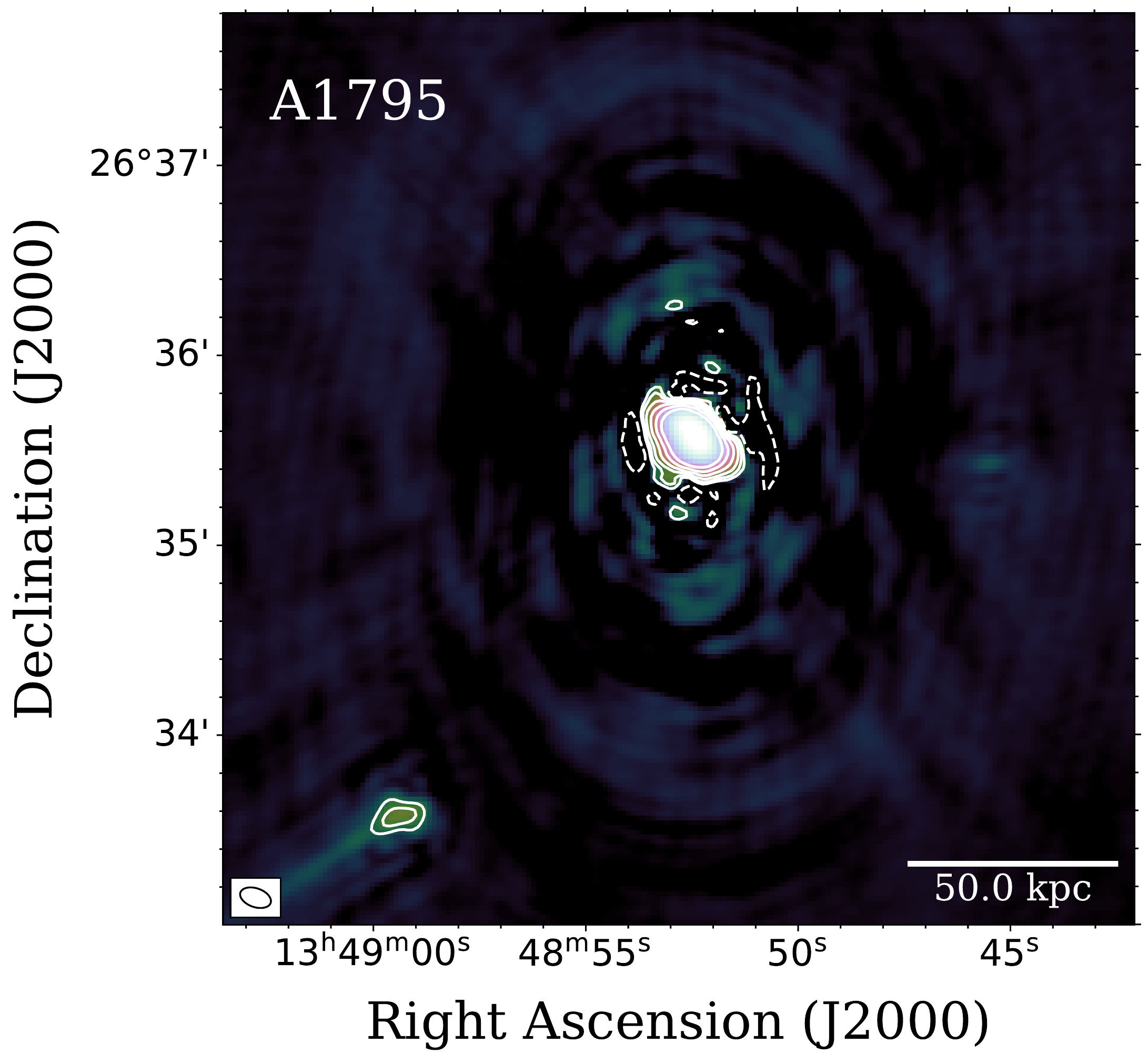} &
\includegraphics[width=54mm]{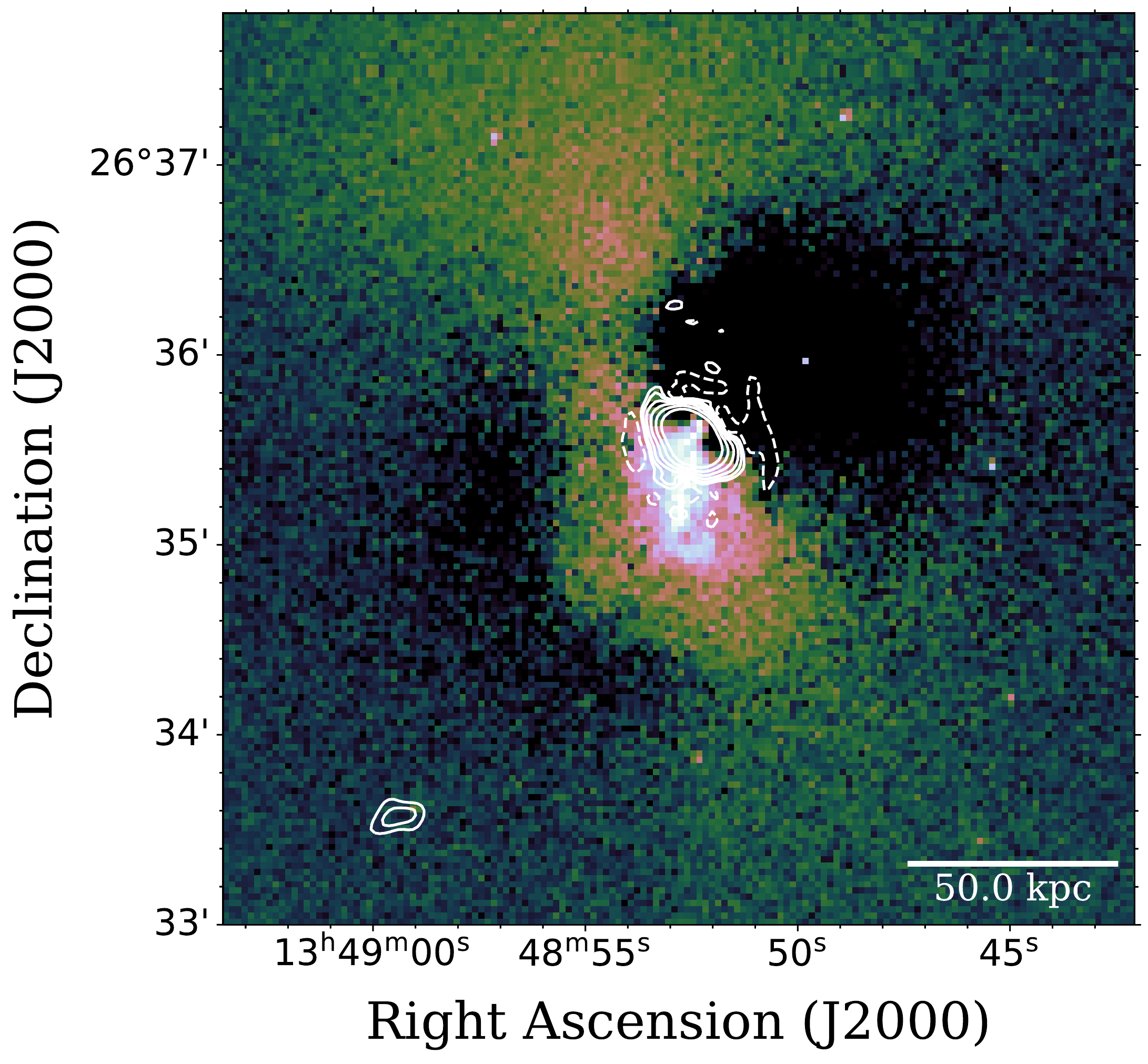}
\includegraphics[width=54mm]{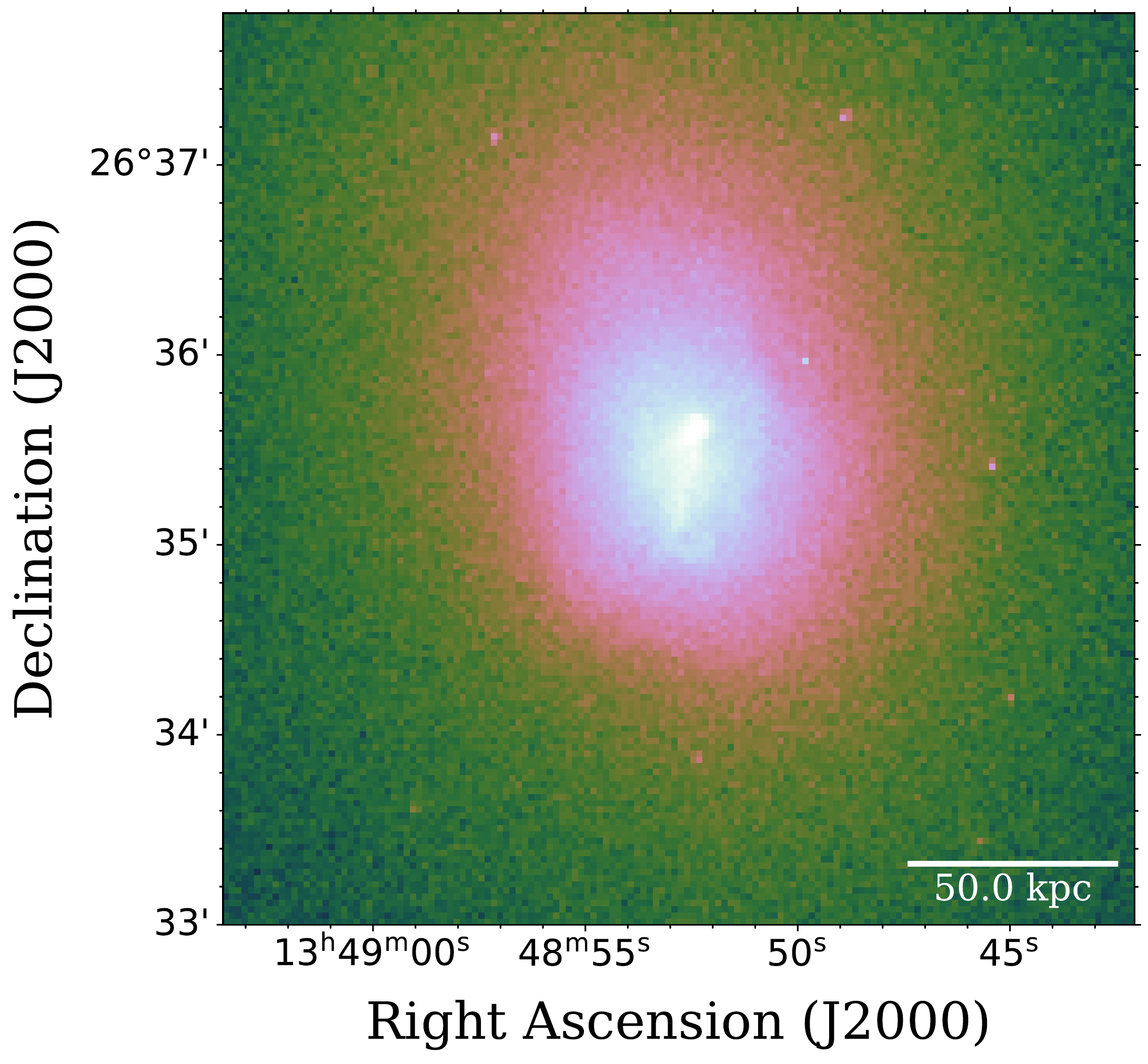} \\
\includegraphics[width=54mm]{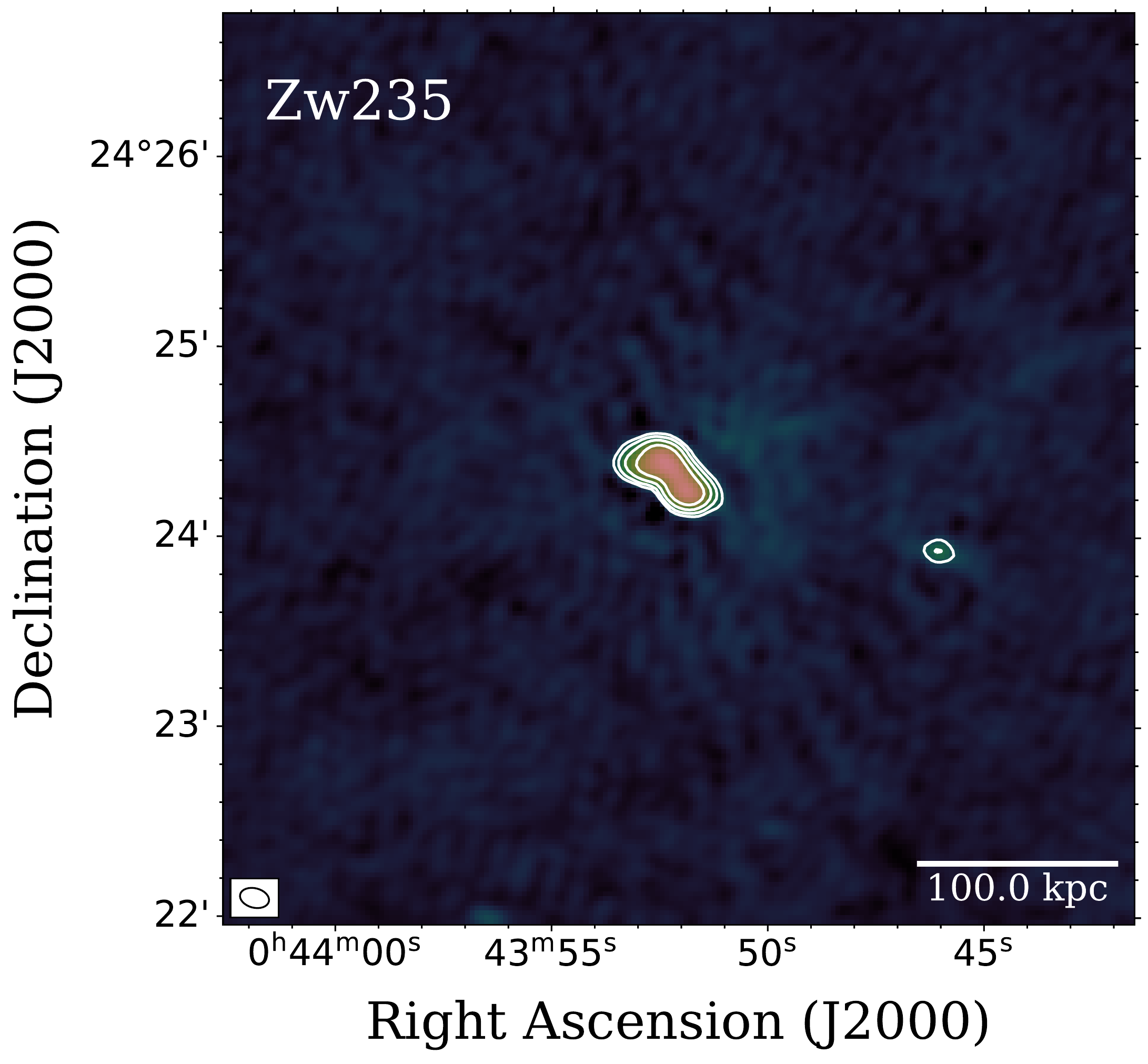} &
\includegraphics[width=54mm]{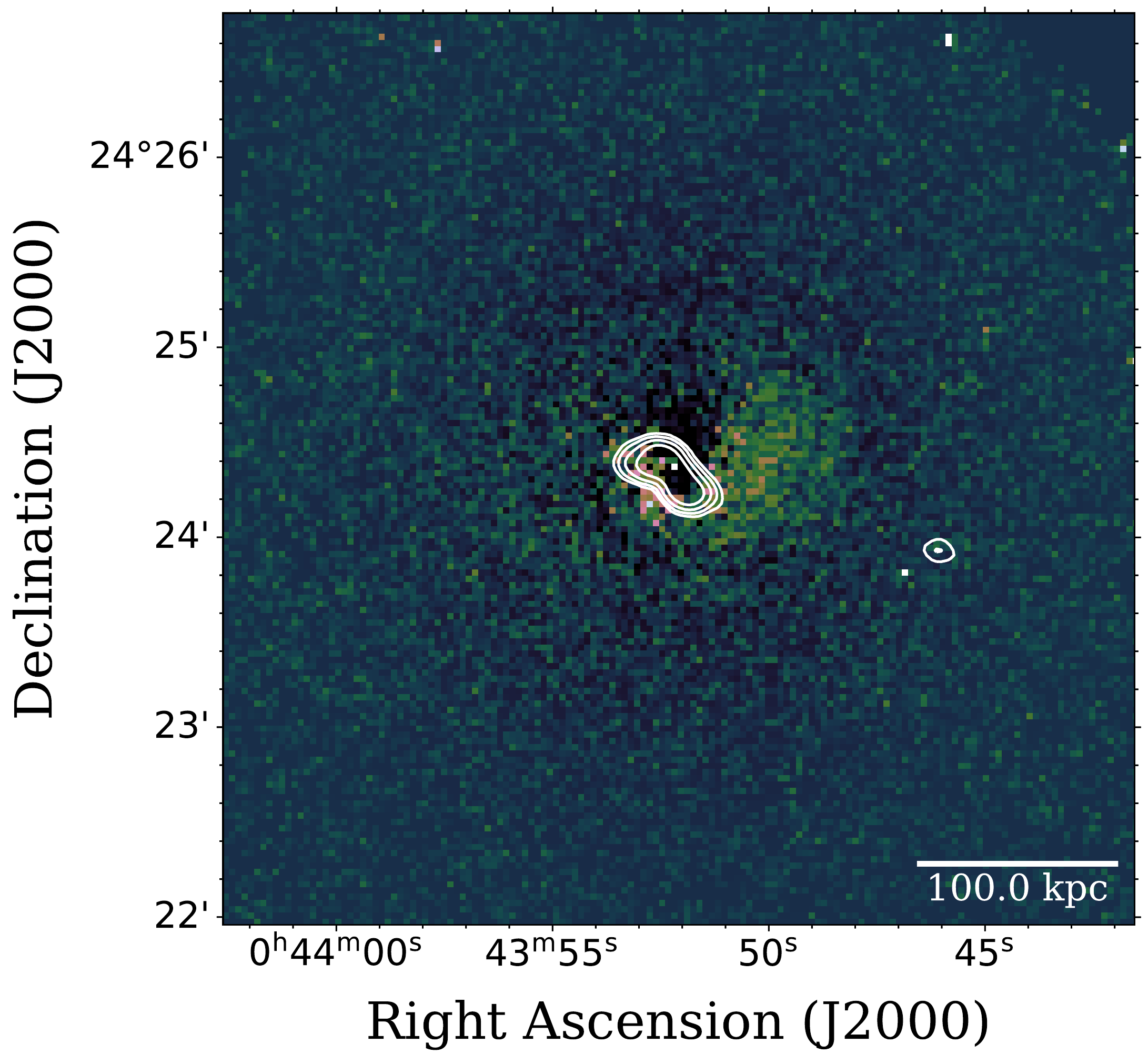}
\includegraphics[width=54mm]{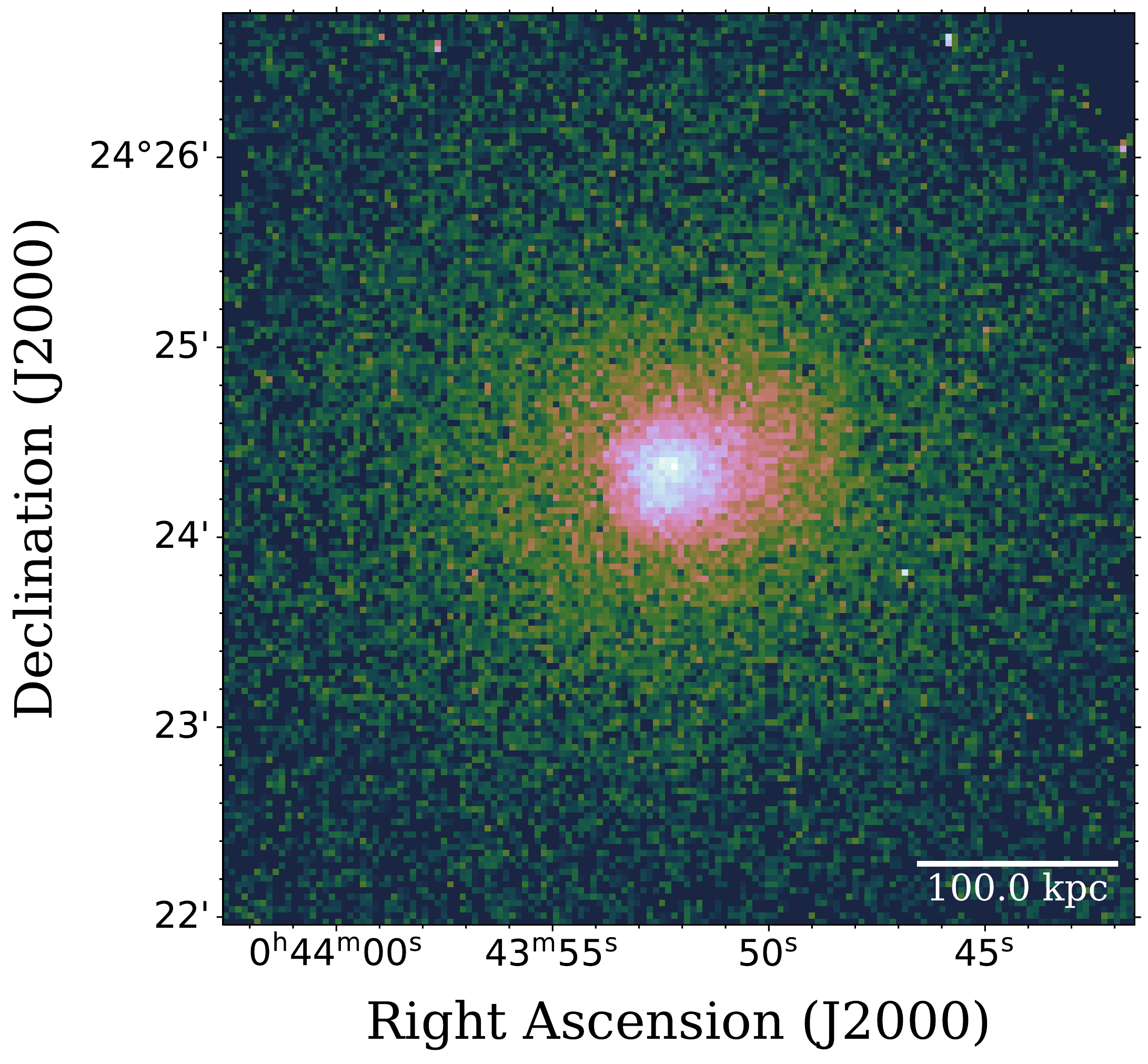} \\
\includegraphics[width=54mm]{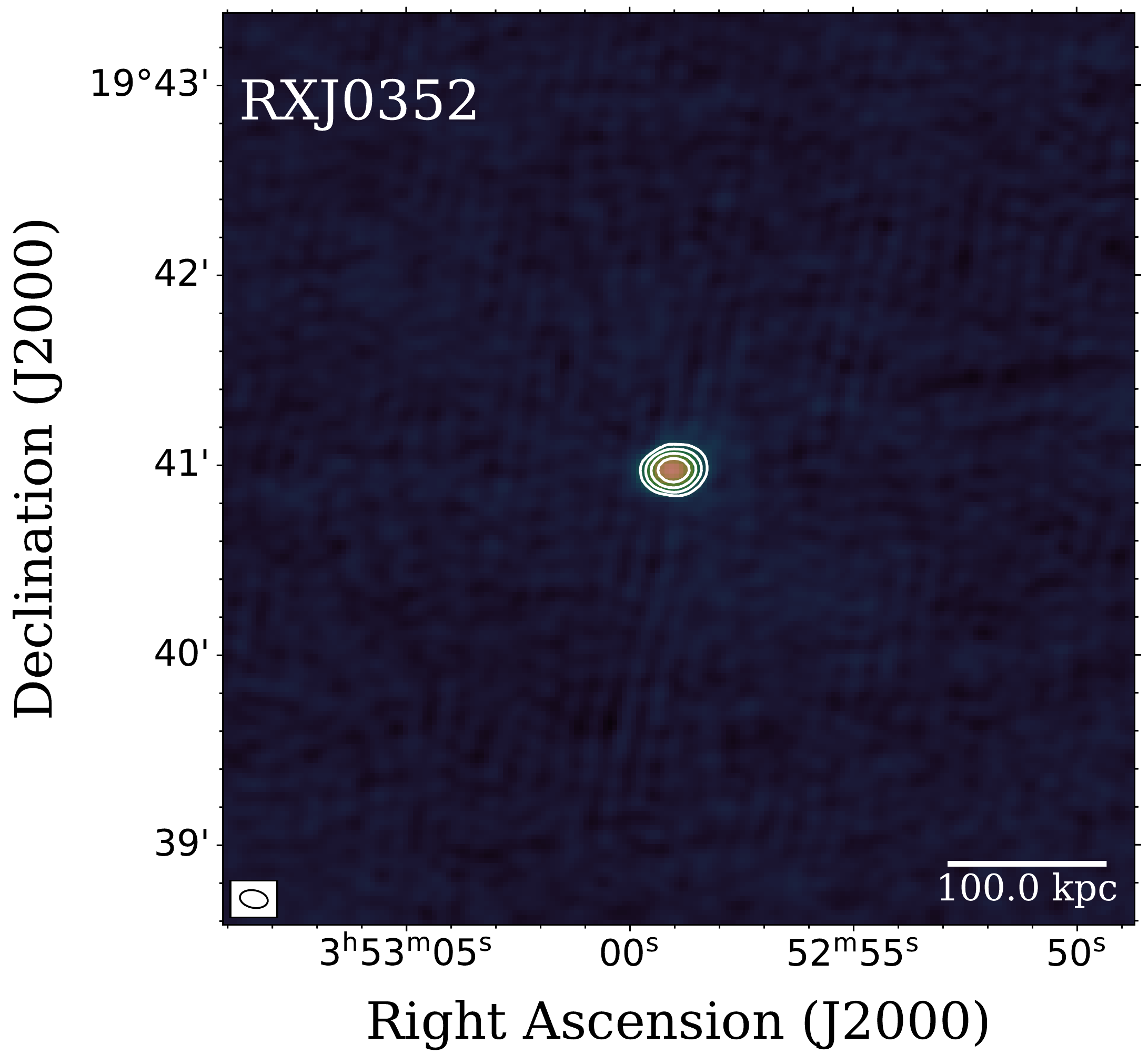} &
\includegraphics[width=54mm]{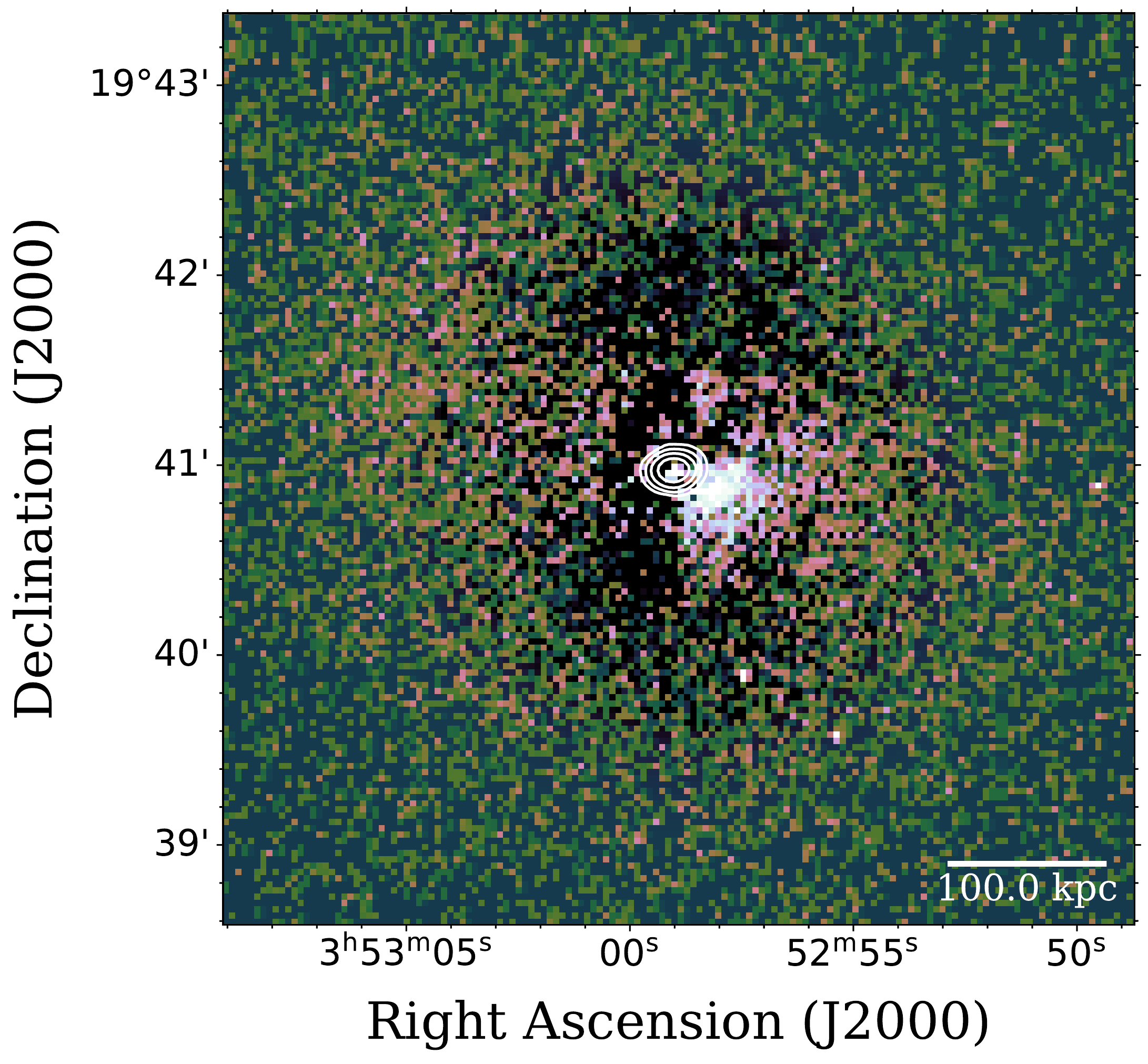}
\includegraphics[width=54mm]{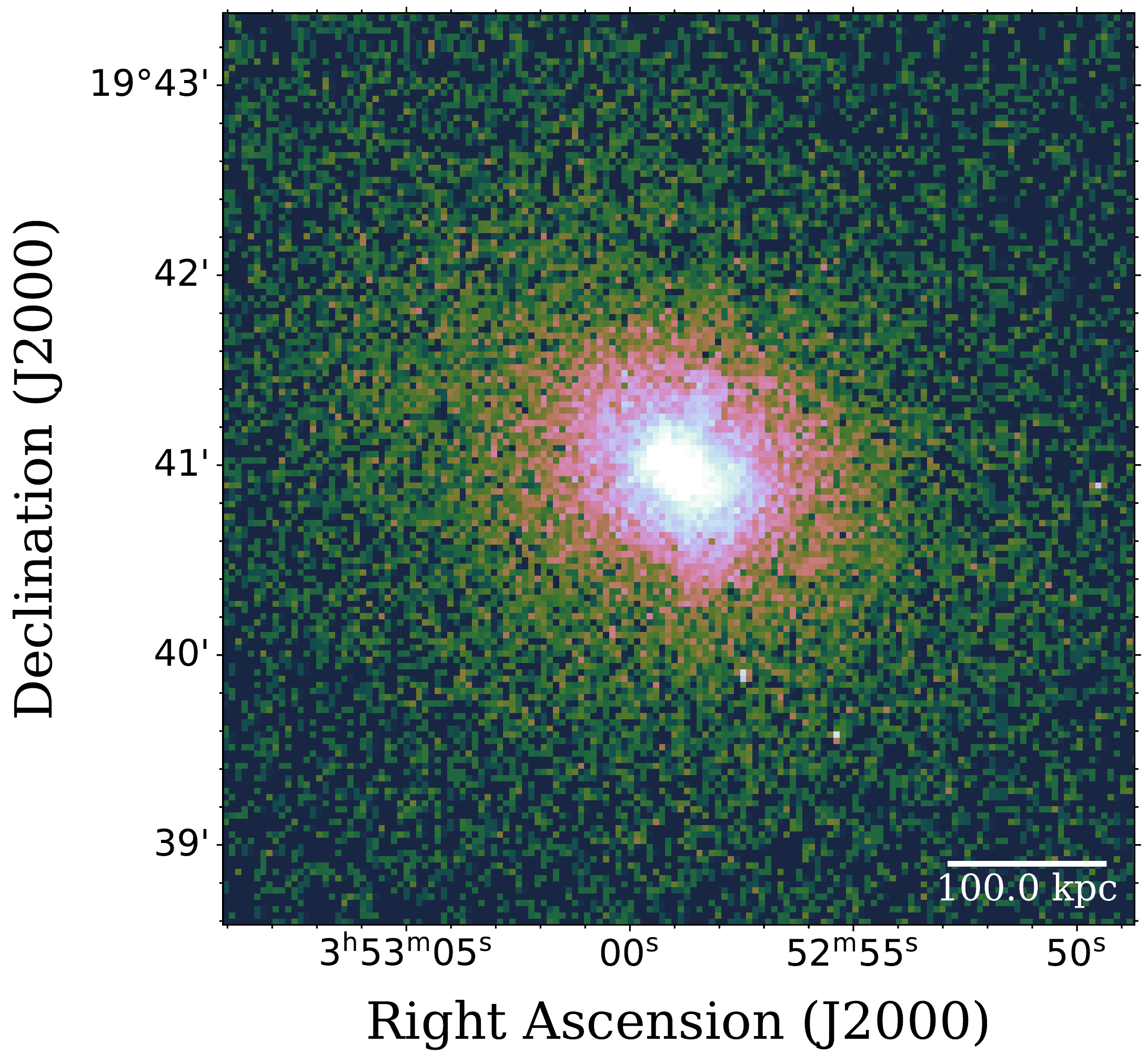} \\\end{tabular}
\label{F:images_4}
\caption{ \emph{Chandra} and LOFAR images for the nearby clusters ($z<0.3$) with X-ray cavities which are not filled with low-frequency radio emission
in the same order as in Tables \ref{Xray_table}, \ref{LOFAR_table} and \ref{summary_table}. The panel organization  is  the same as in Figure 1. Shown here are A1795, ZwCl 0235, and RX J0352.9+1941, with the others shown in Figure 4-continued.  For the LOFAR image, the first contour is at 0.018~mJy beam$^{-1}$ (A1795; the dashed contour is at -0.018~mJy beam$^{-1}$), 0.00267~mJy~beam$^{-1}$ (ZwCl 0235), 0.00267~mJy beam$^{-1}$ (RX J0352.9+1941), and each contour increases by a factor of two.}
\end{figure*}

\begin{figure*} \begin{tabular}{@{}cc}
\includegraphics[width=54mm]{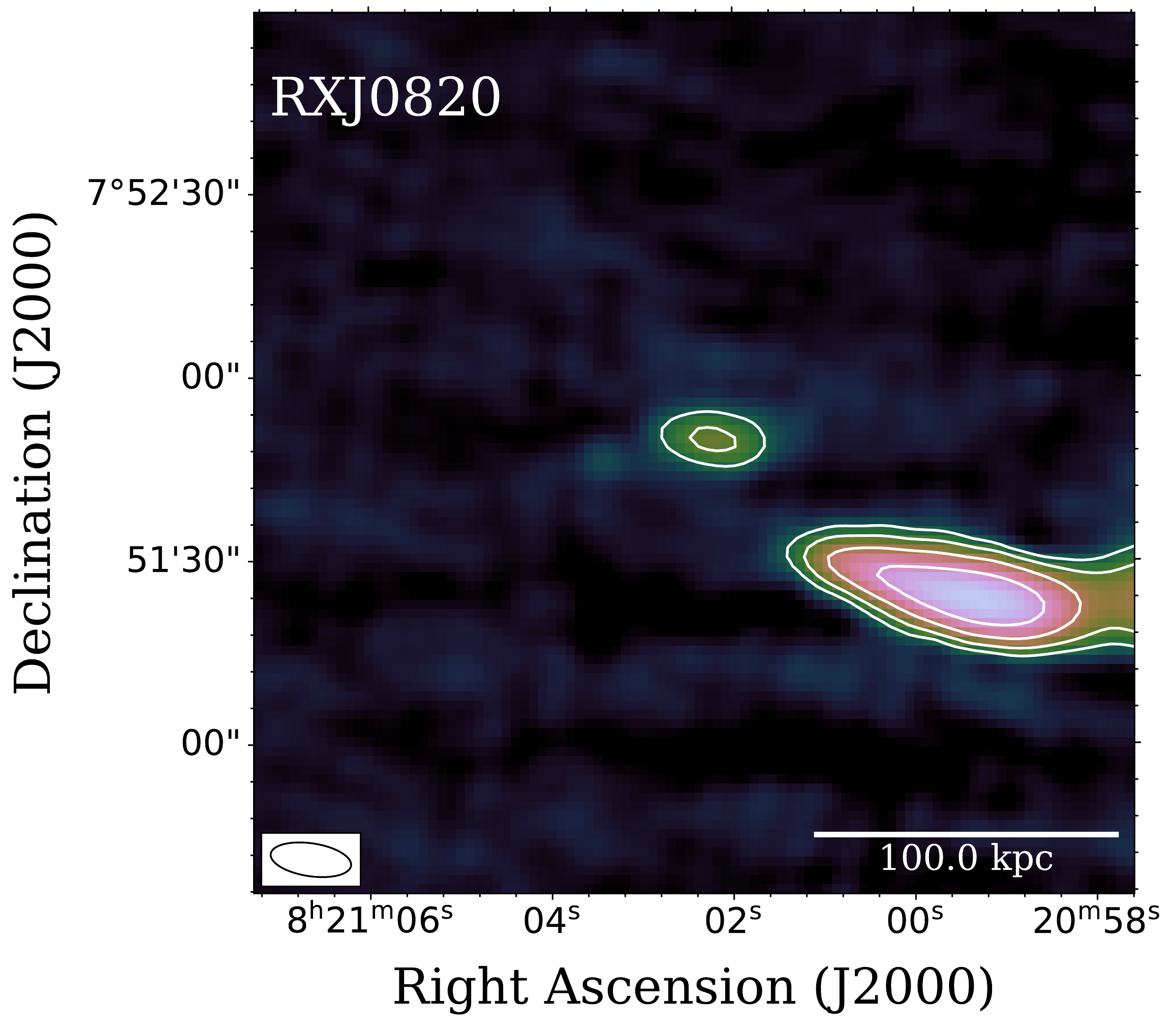} &
\includegraphics[width=54mm]{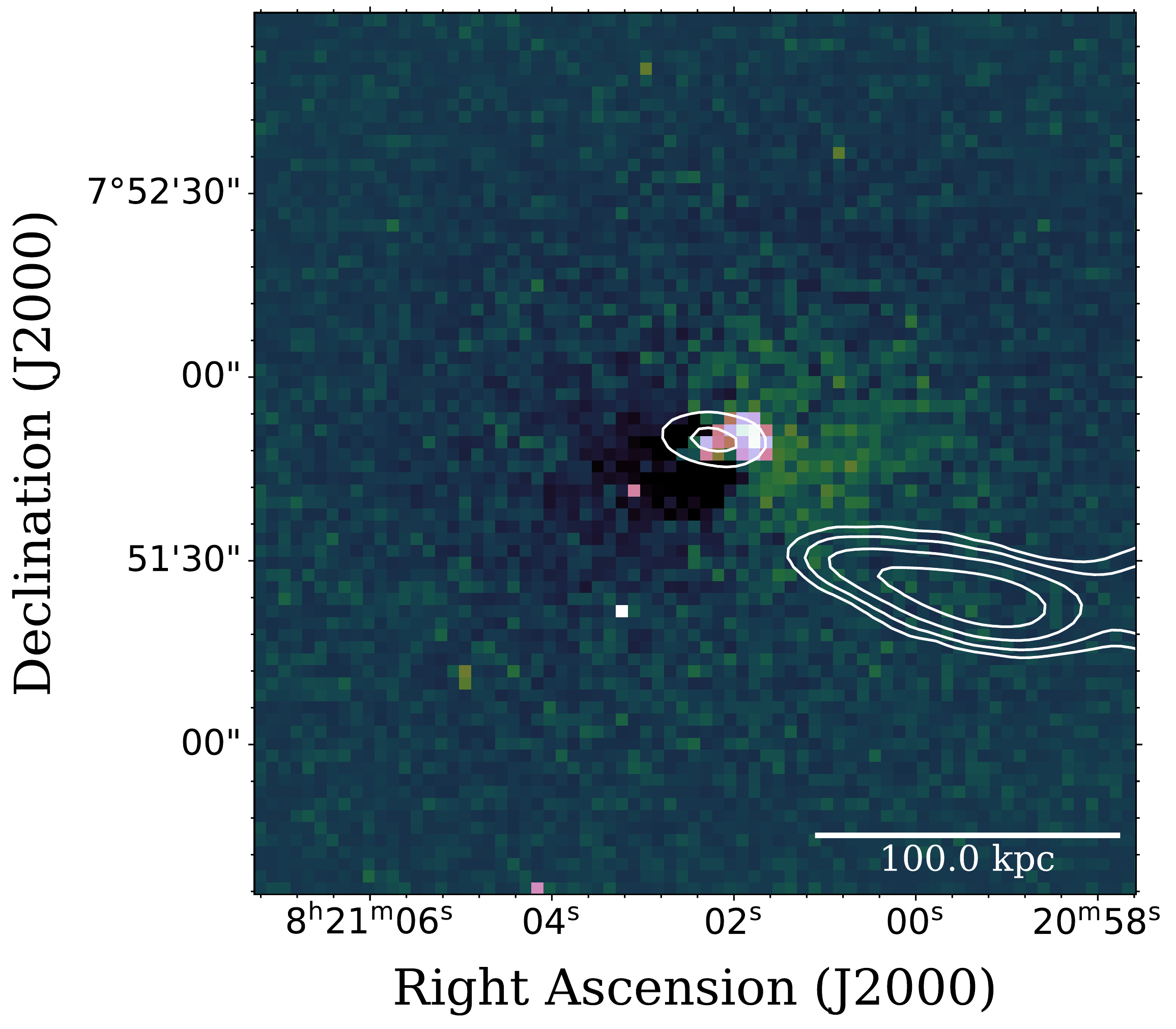}
\includegraphics[width=54mm]{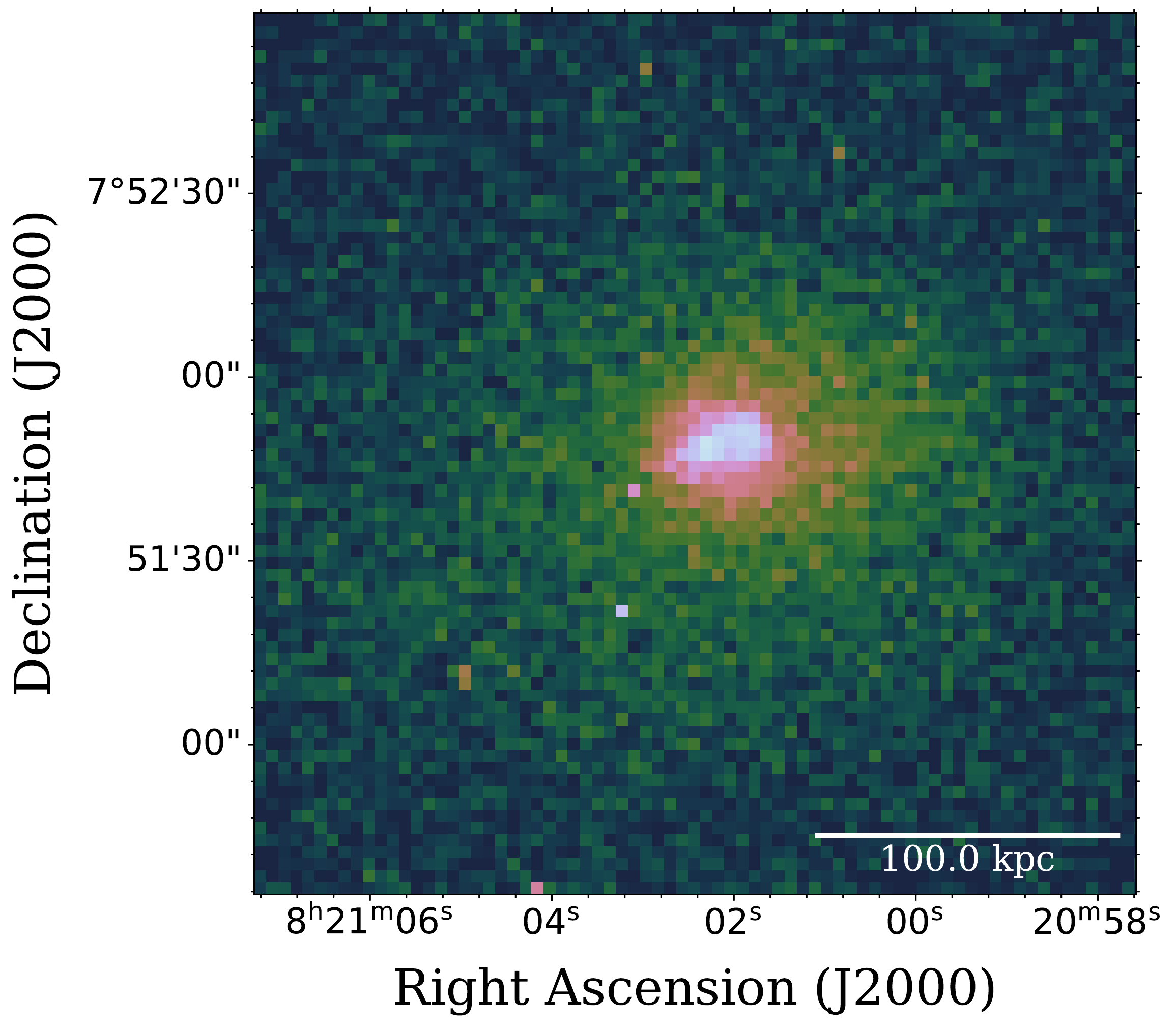} \\
\includegraphics[width=54mm]{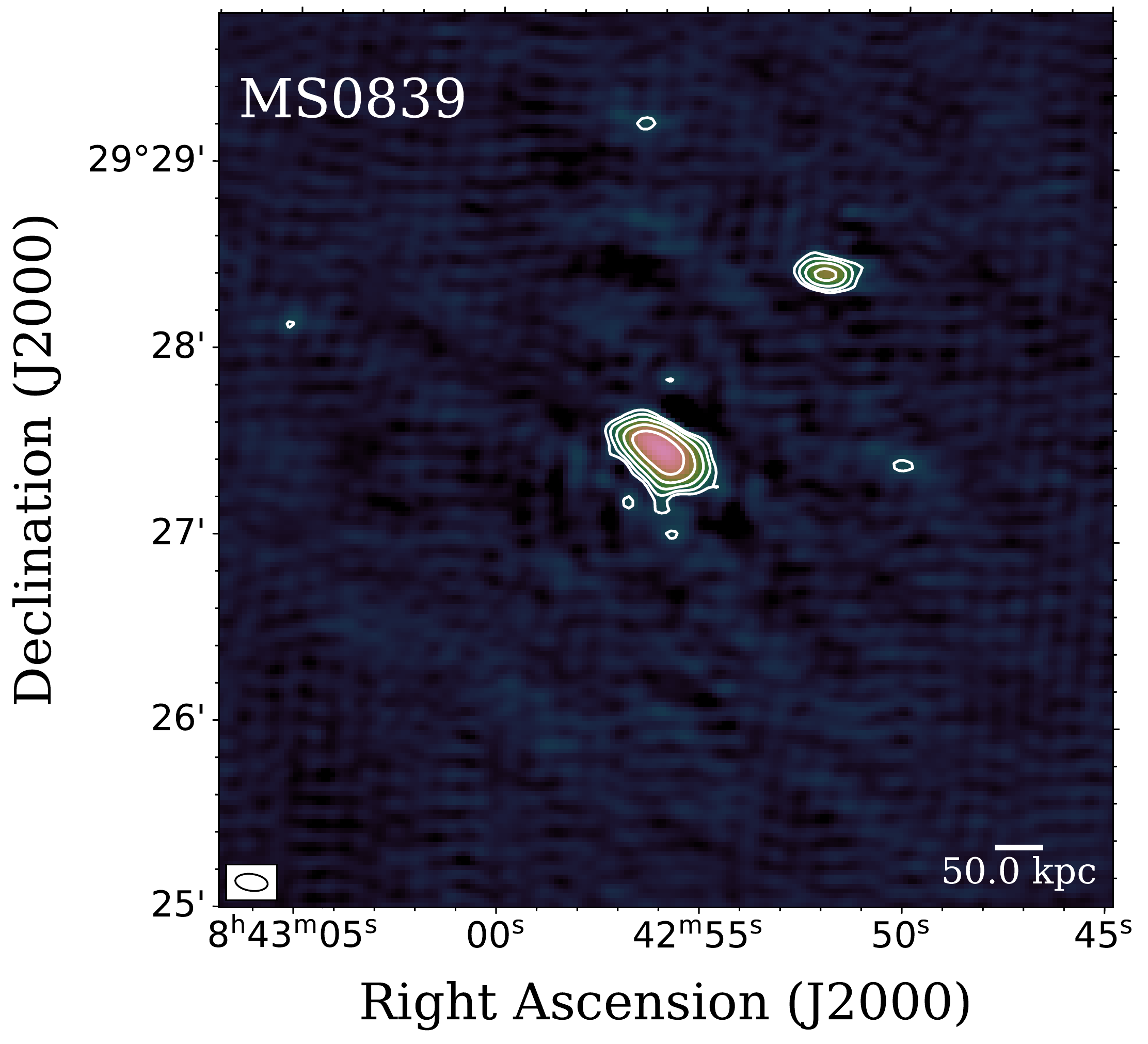} &
\includegraphics[width=54mm]{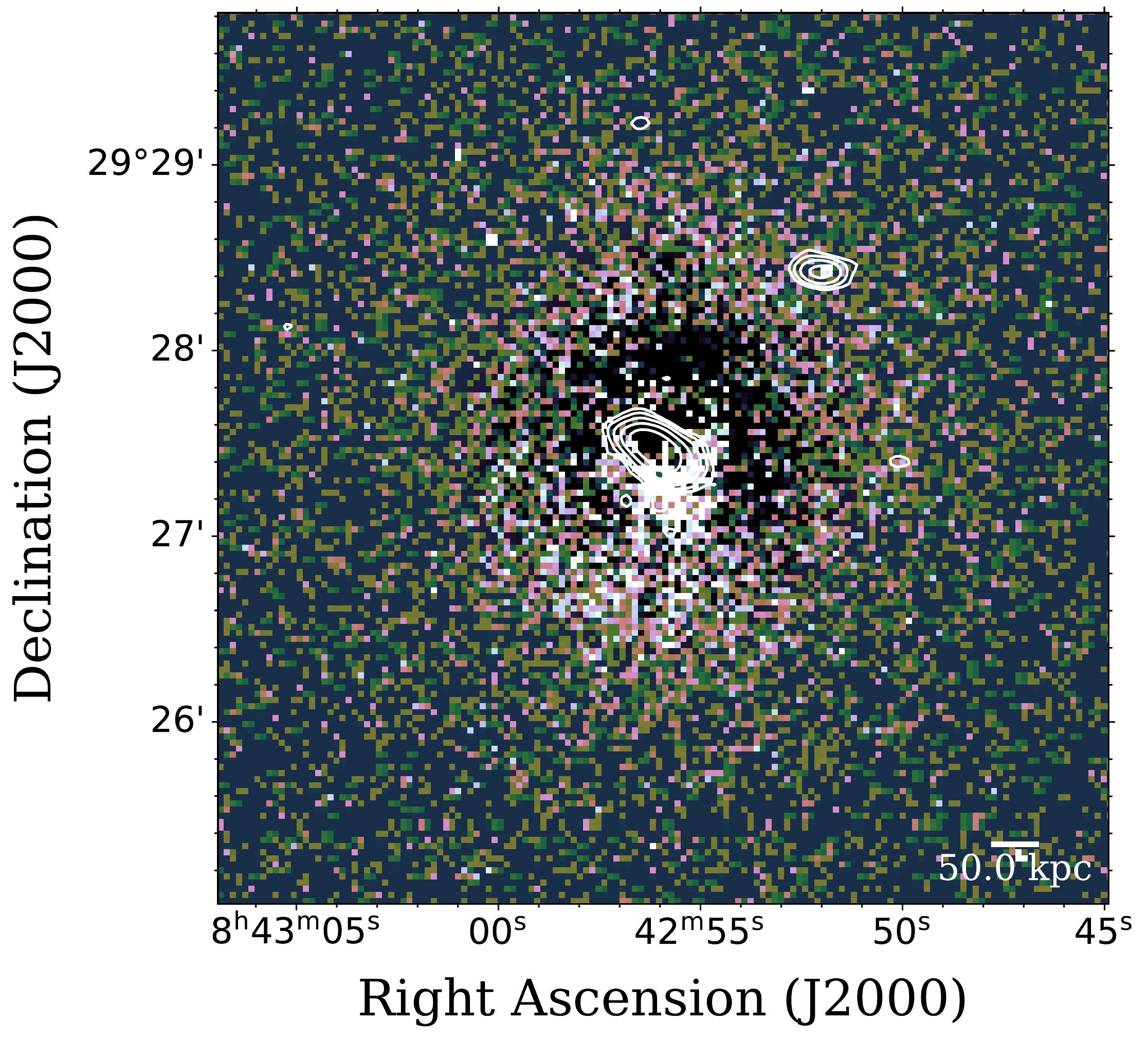}
\includegraphics[width=54mm]{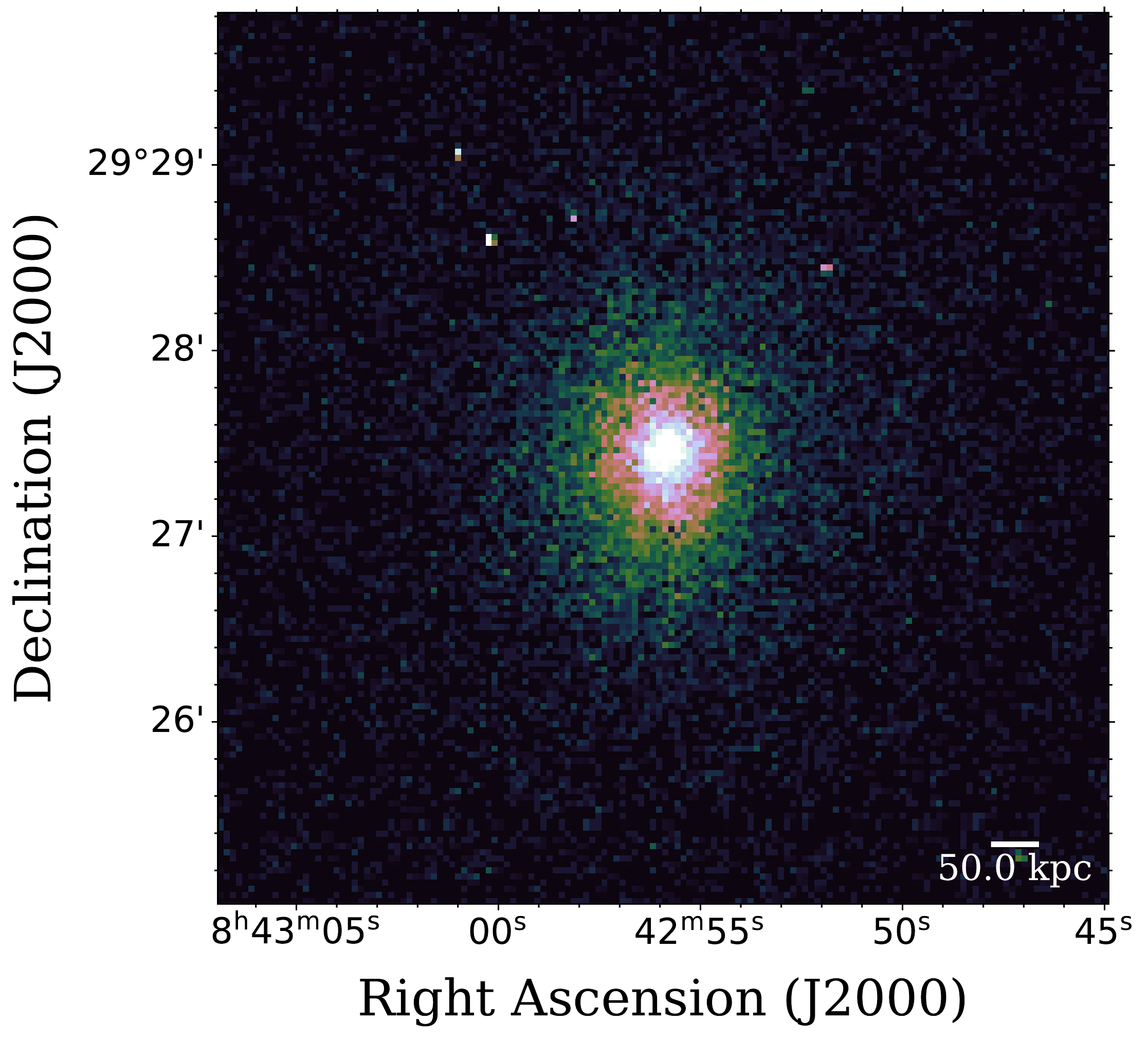} \\
\includegraphics[width=54mm]{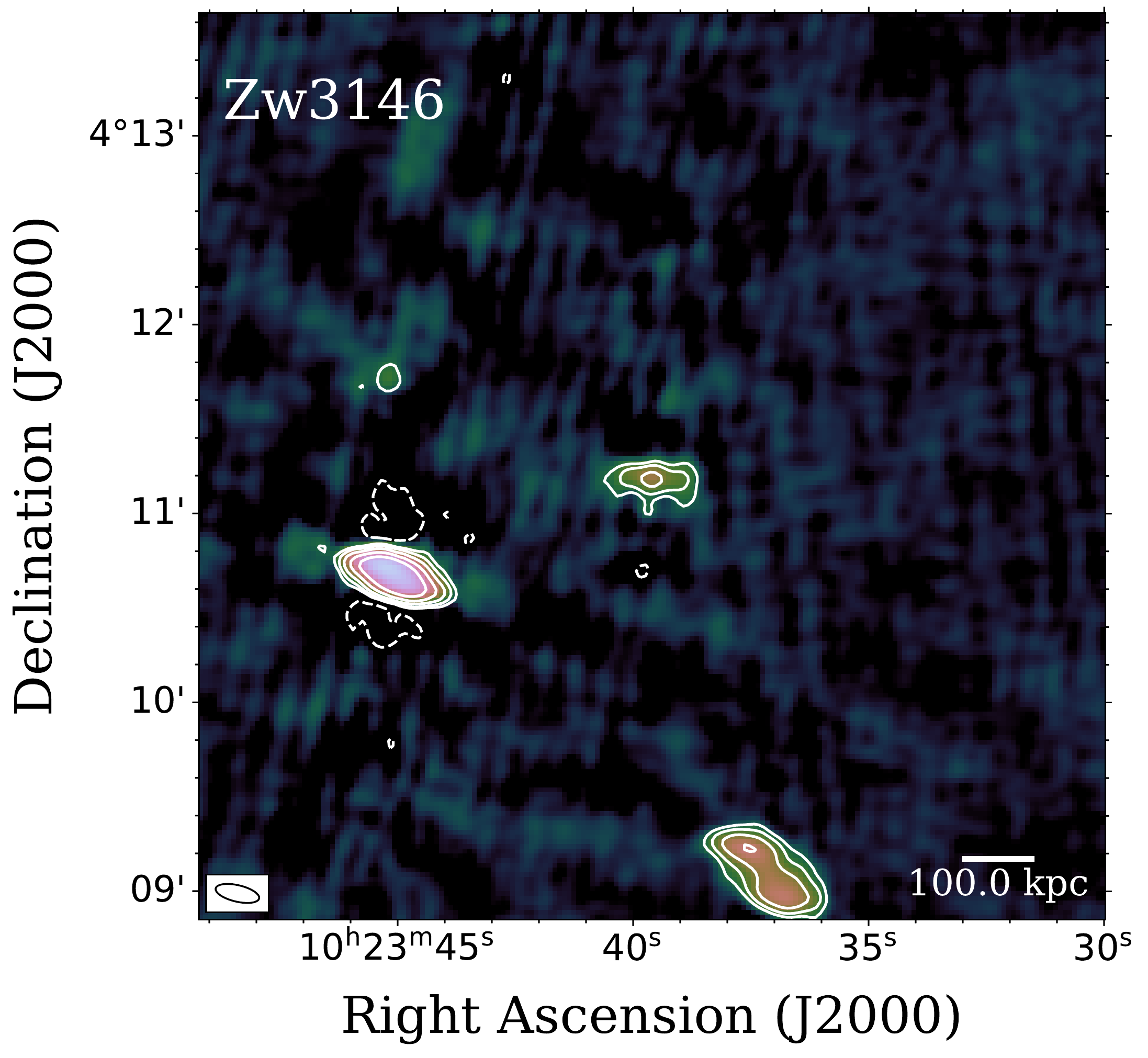} &
\includegraphics[width=54mm]{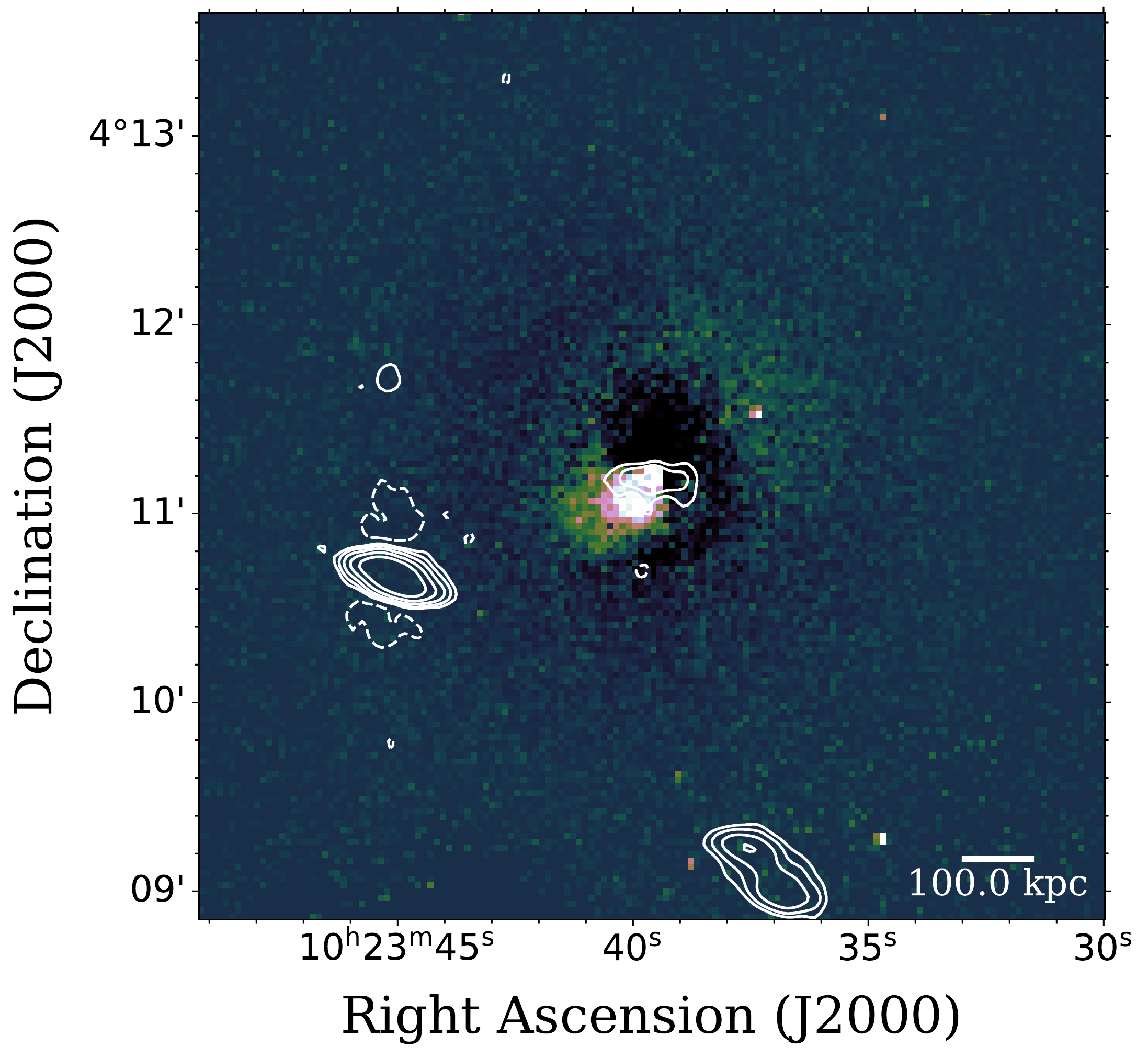}
\includegraphics[width=54mm]{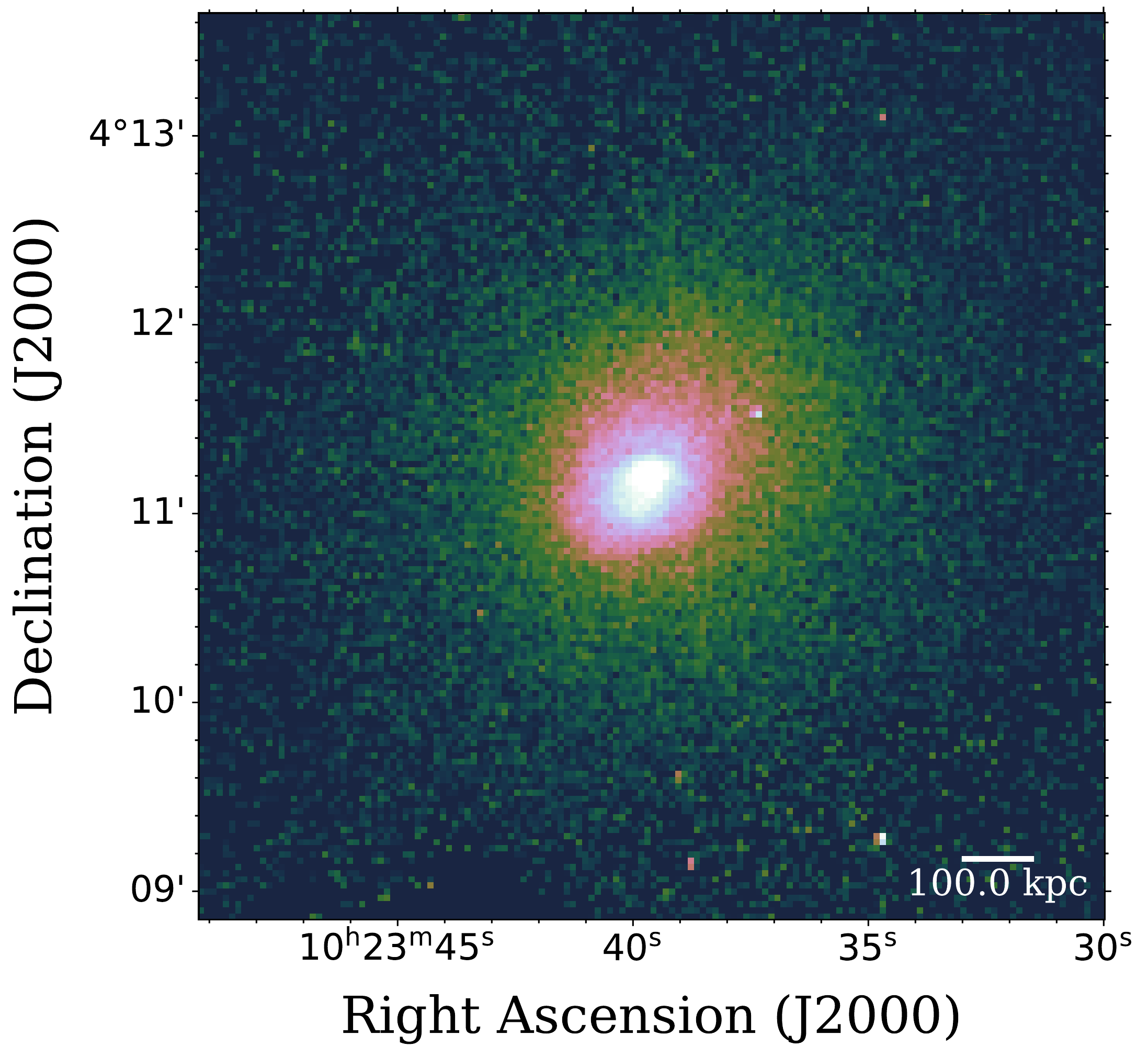} \\ \end{tabular}
\textbf{Figure}~4. --- continued (RX J0820.9+0752, MS 0839.9+2938 and ZwCl 3146). For the LOFAR image, the first contour is at 0.006~mJy beam$^{-1}$ (RX J0820.9+0752), 0.00225~mJy beam$^{-1}$ (MS 0839.9+2938), 0.0057~mJy beam$^{-1}$ (ZwCl 3146, the dashed contour is at -0.0057~mJy beam$^{-1}$), and each contour increases by a factor of two. \\
\end{figure*}

\begin{figure*} \begin{tabular}{@{}cc}
\includegraphics[width=54mm]{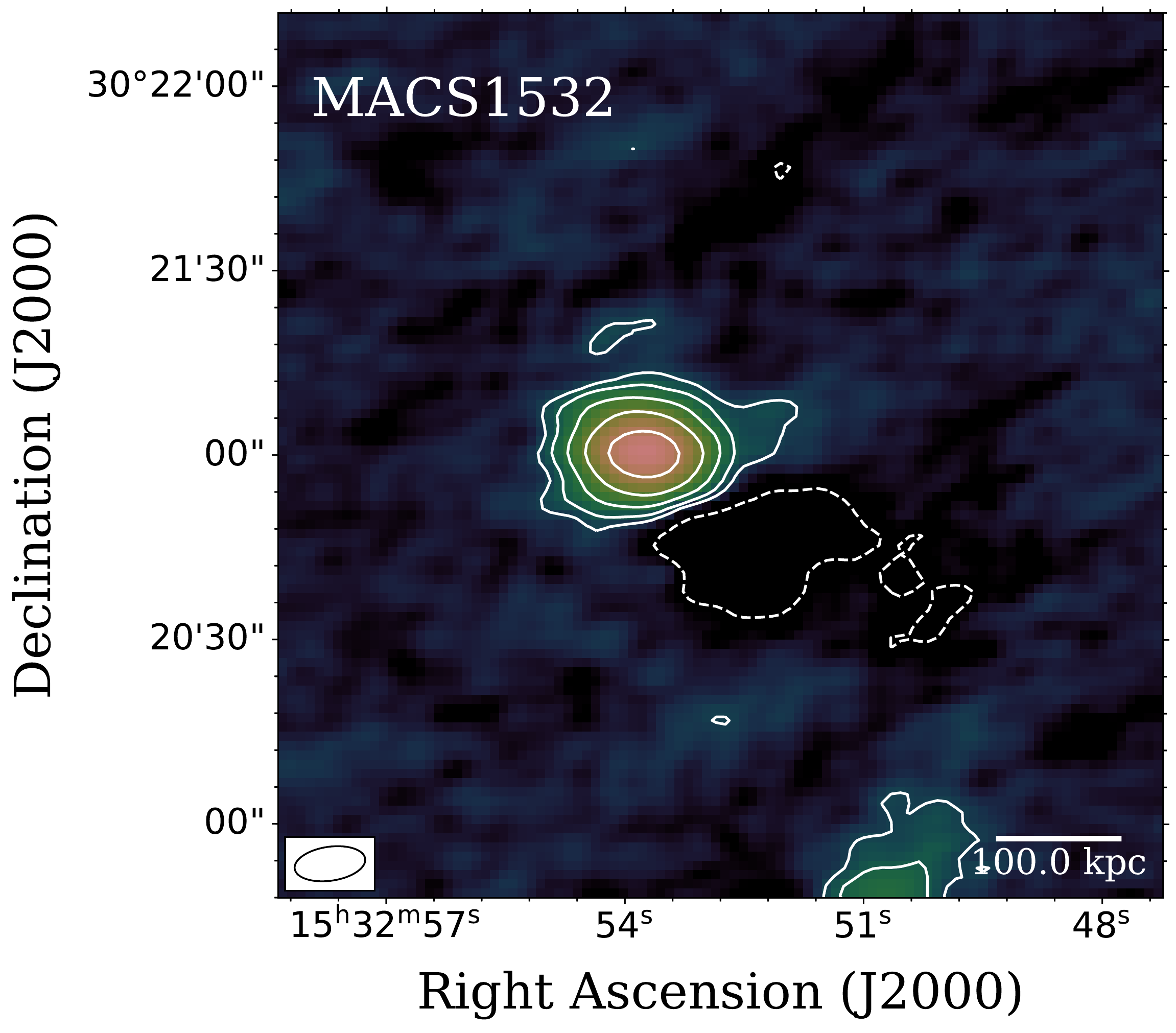} &
\includegraphics[width=54mm]{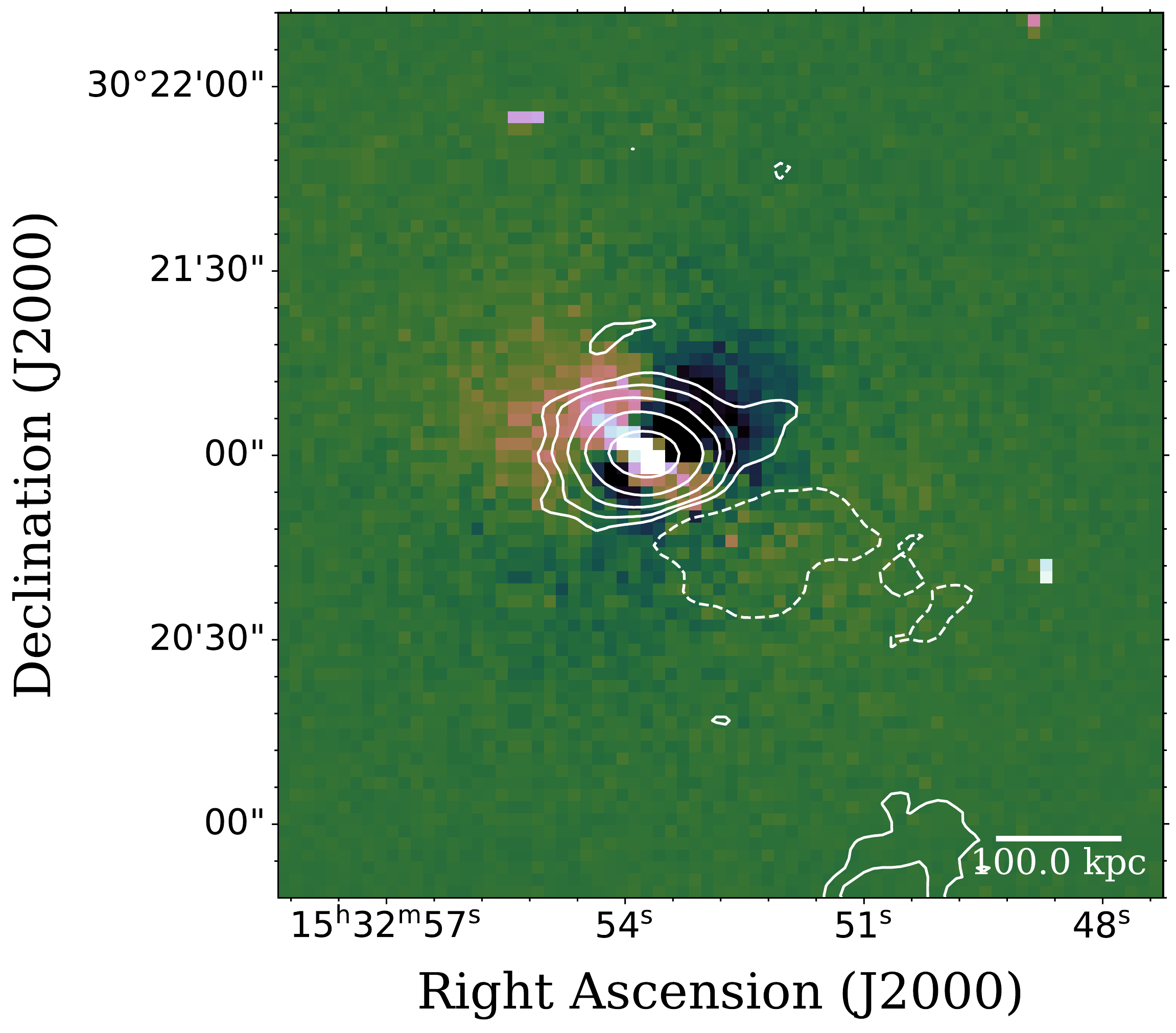}
\includegraphics[width=54mm]{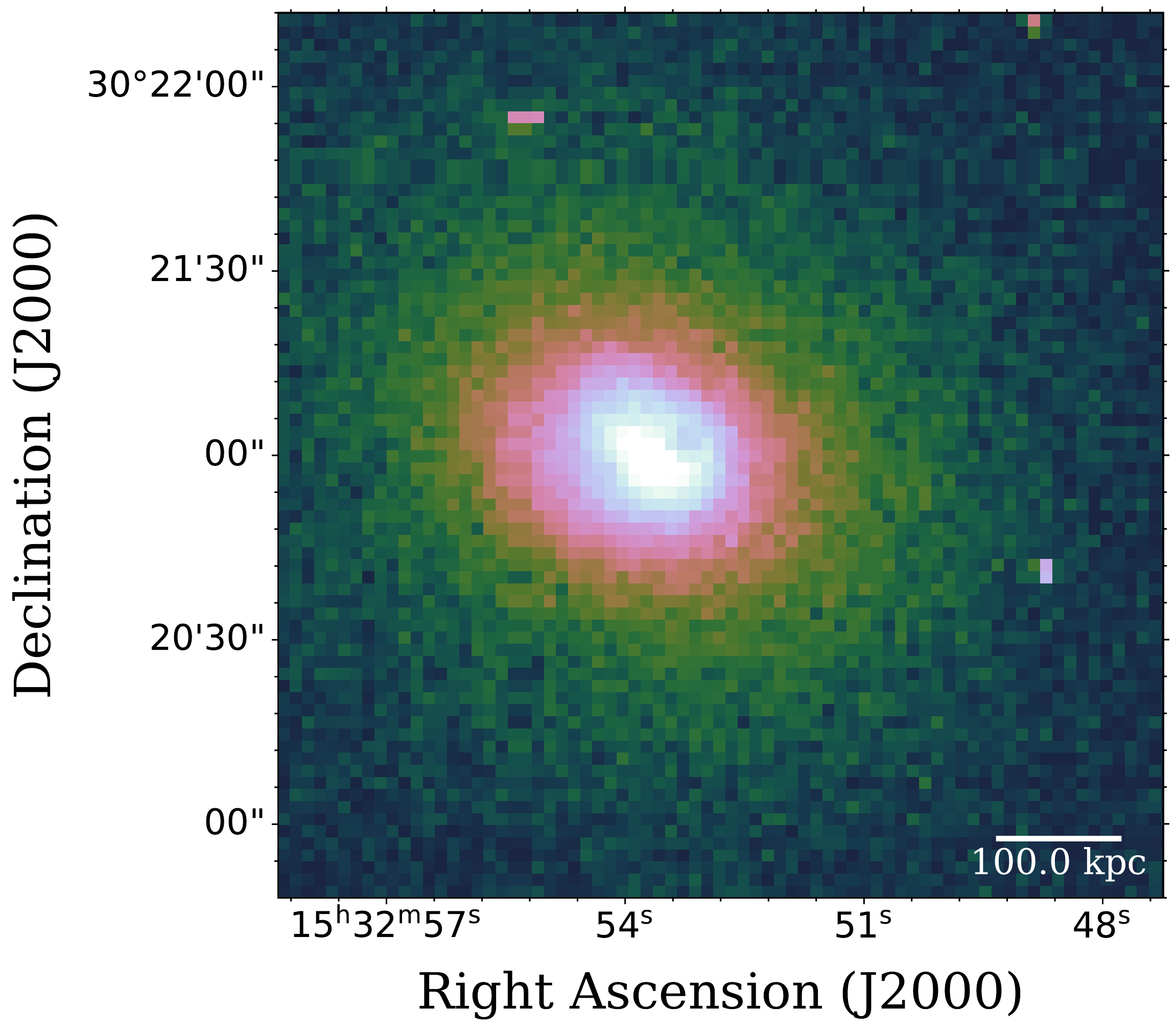}  \\ \end{tabular}
\caption{\emph{Chandra} and LOFAR images for the high redshift clusters ($z>0.3$) with X-ray cavities that are filled with low-frequency radio emission, in the same order as in the tables (with MACS J1532.9+3021 shown above, and the others shown in  Figure 5-continued). The panel organization  is the same as in Figure~1. For the LOFAR image, the first contour is at 0.0021~mJy beam$^{-1}$ (the dashed contour is at -0.0021~mJy beam$^{-1}$), and each contour increases by a factor of two. }
\label{F:images_5}
\end{figure*}

\begin{figure*} \begin{tabular}{@{}cc}
\includegraphics[width=54mm]{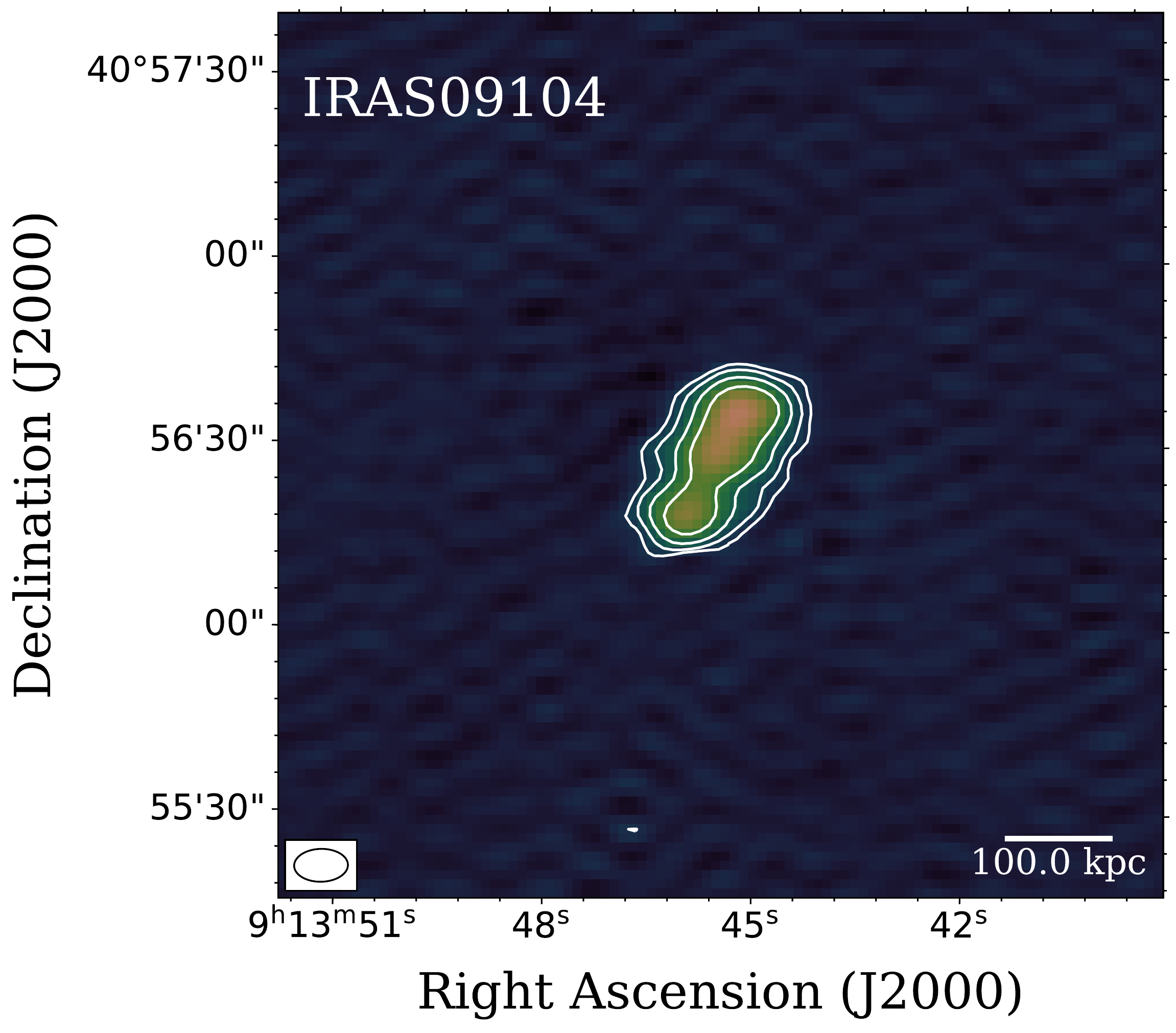} &
\includegraphics[width=54mm]{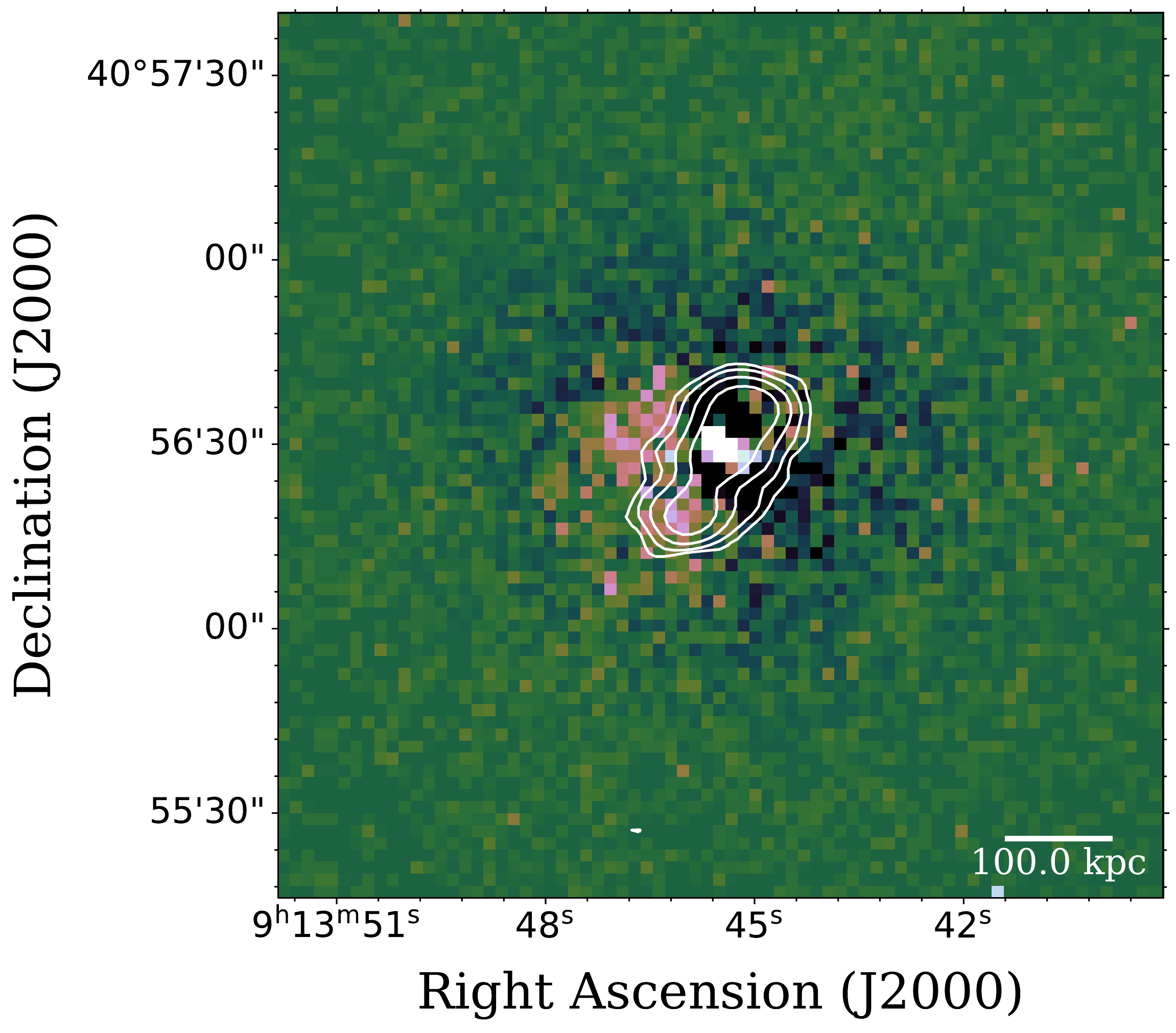}
\includegraphics[width=54mm]{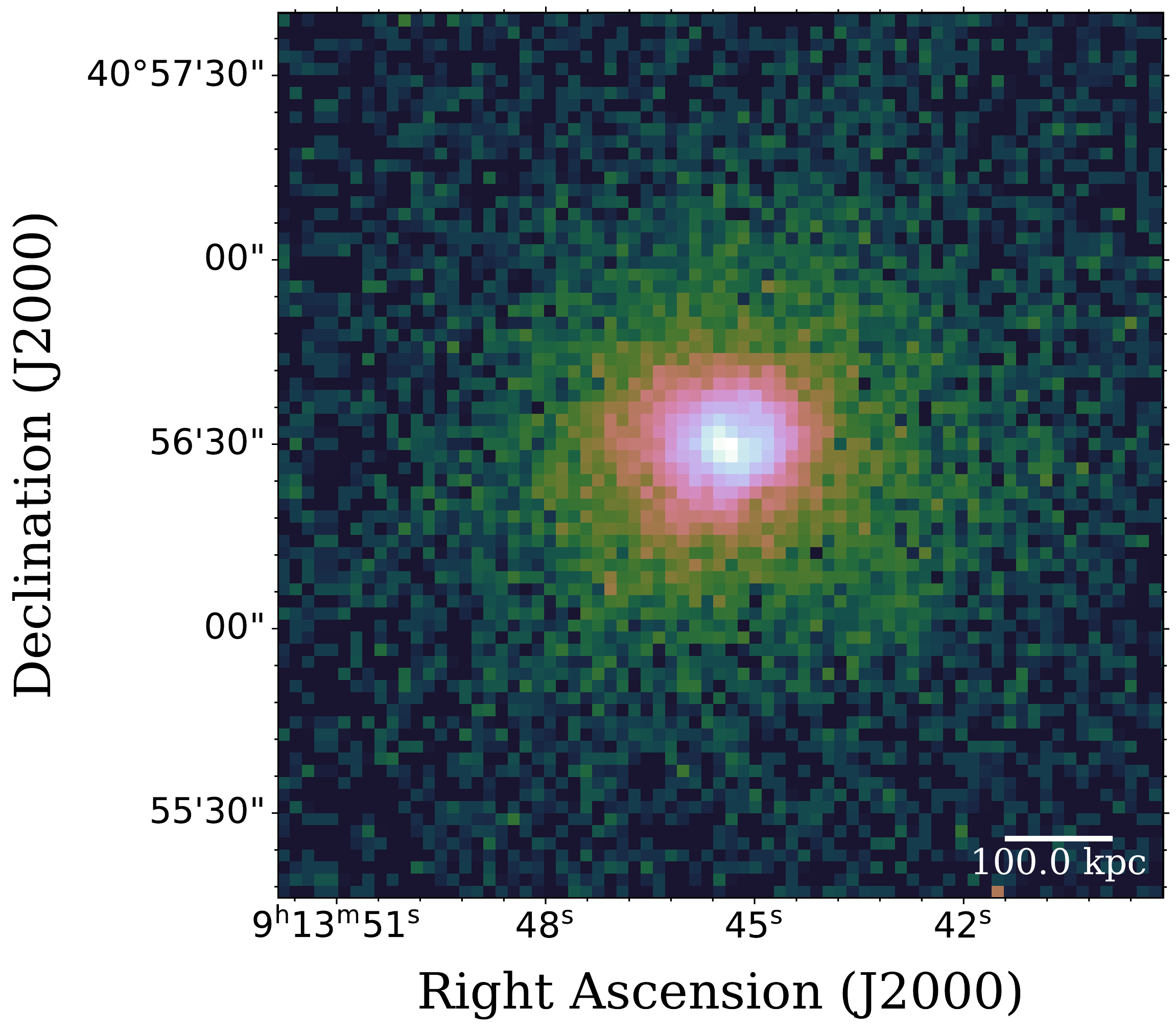}  \\
\includegraphics[width=54mm]{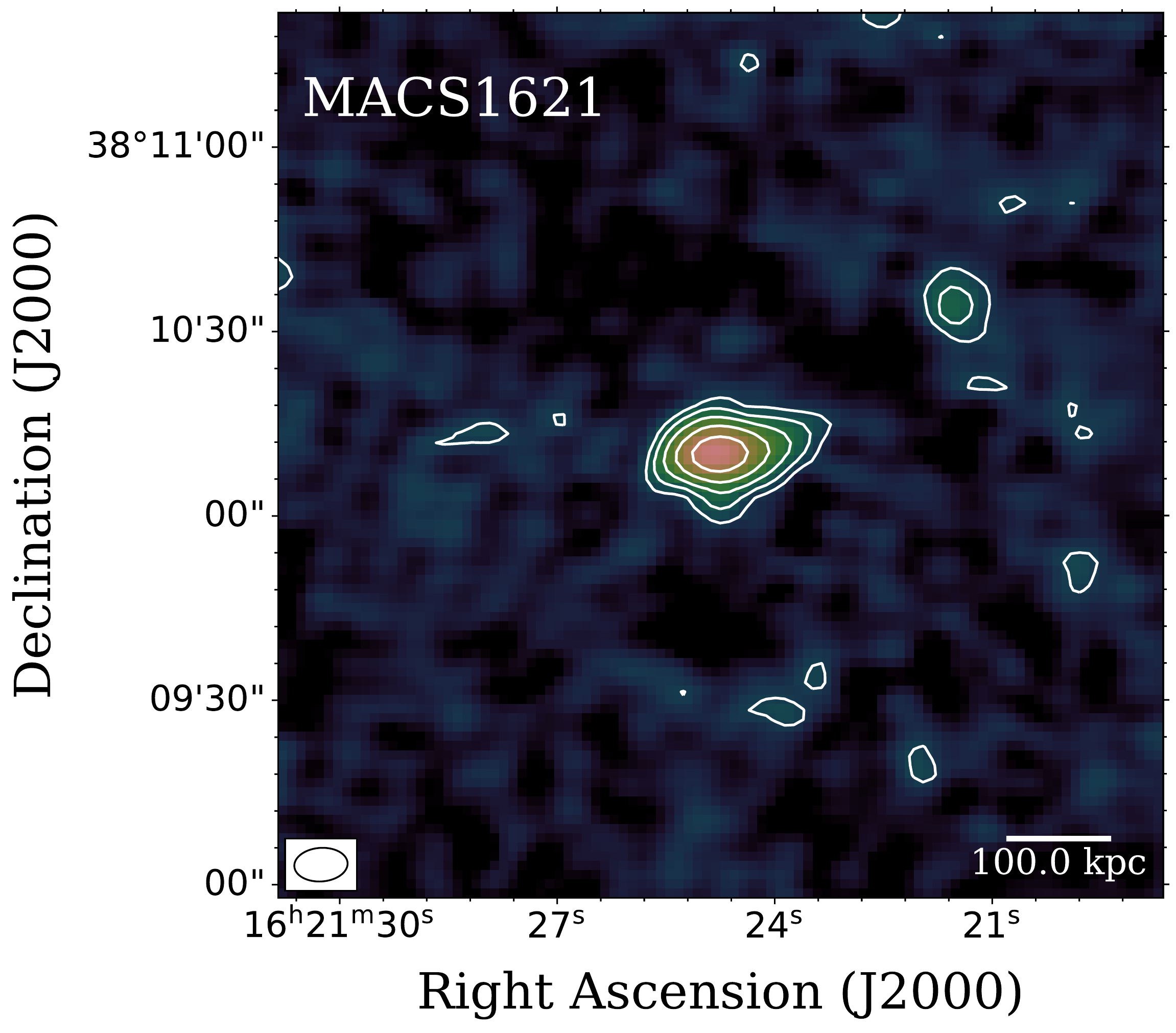} &
\includegraphics[width=54mm]{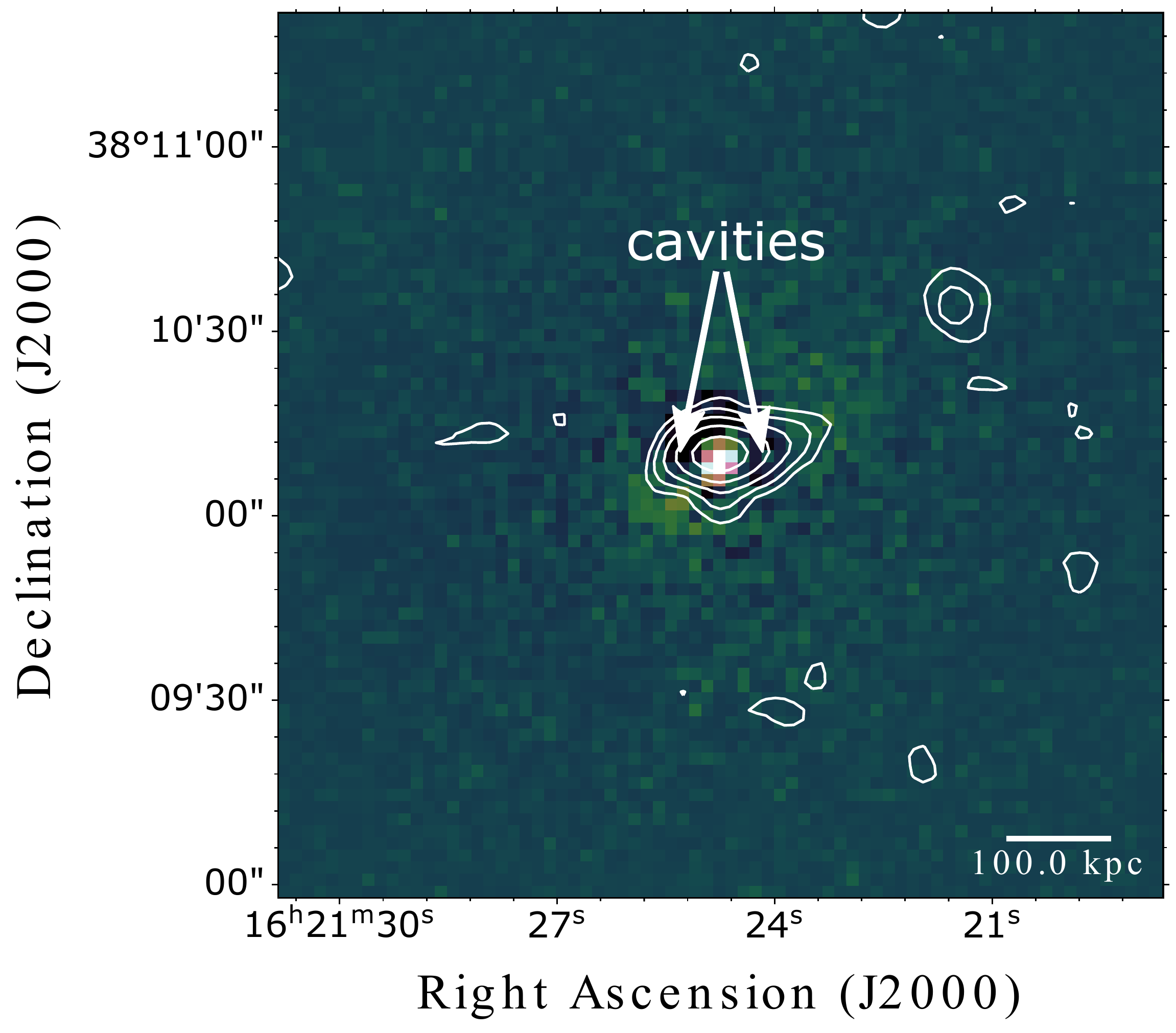}
\includegraphics[width=54mm]{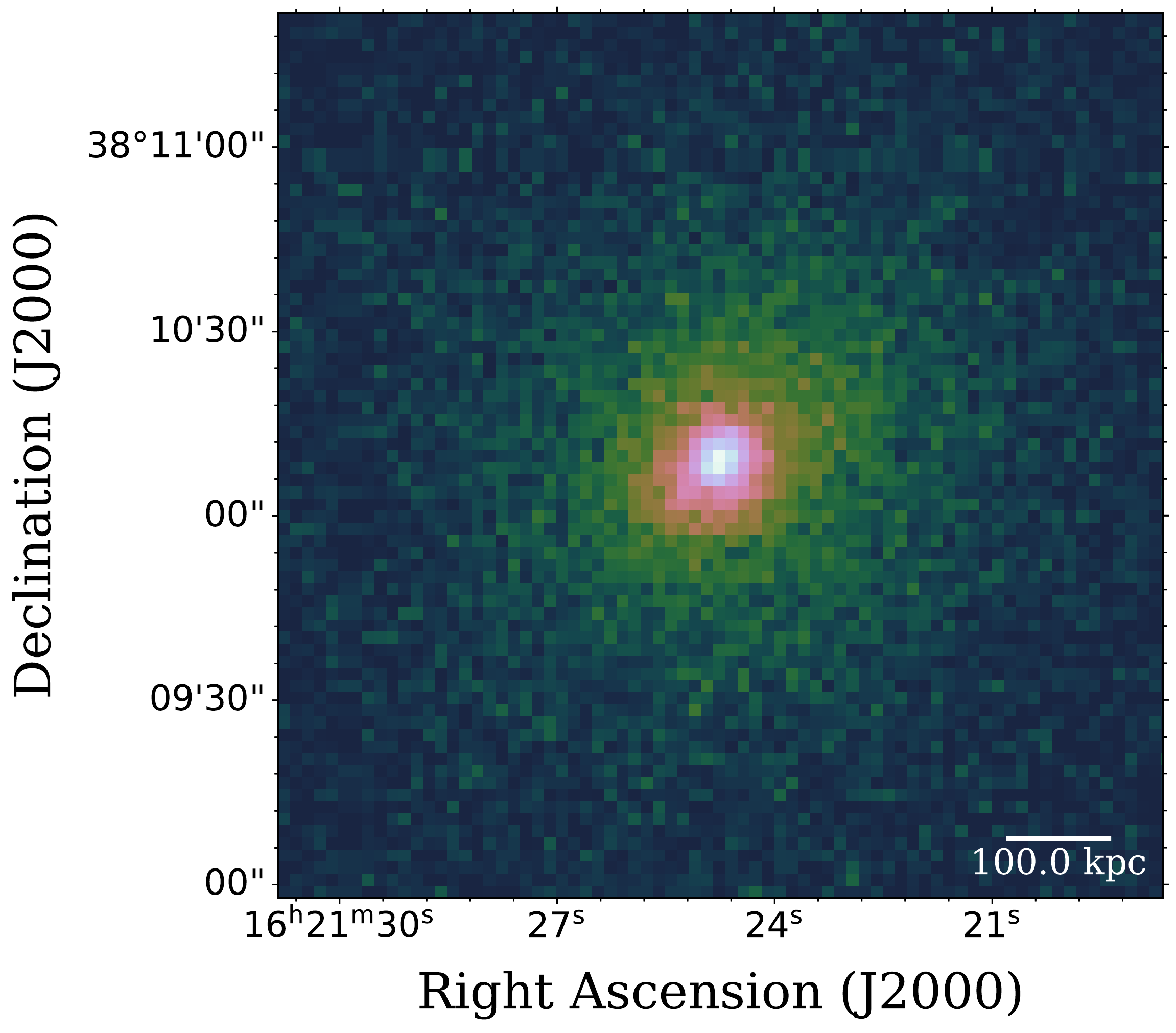} \\ \end{tabular}
\textbf{Figure}~5. --- continued (IRAS 09104+4109 and MACS J1621.3+3810). For the LOFAR image, the first contour is at 0.00267~mJy beam$^{-1}$ (IRAS 09104+4109), 0.00195~mJy beam$^{-1}$ (MACS J1621.3+3810), and each contour increases by a factor of two. \\
\end{figure*}

\begin{figure*} \begin{tabular}{@{}cc}
\includegraphics[width=54mm]{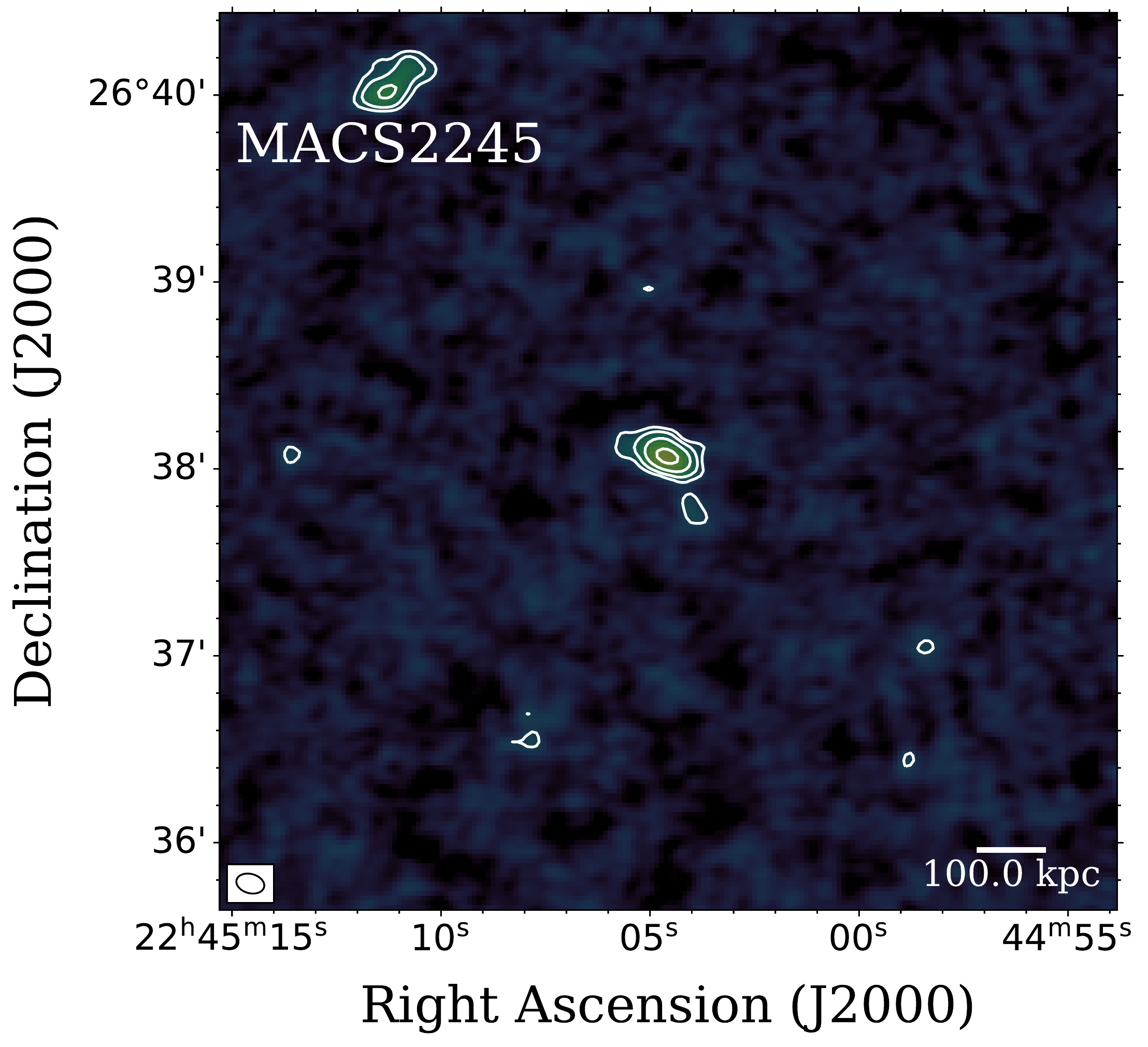} &
\includegraphics[width=54mm]{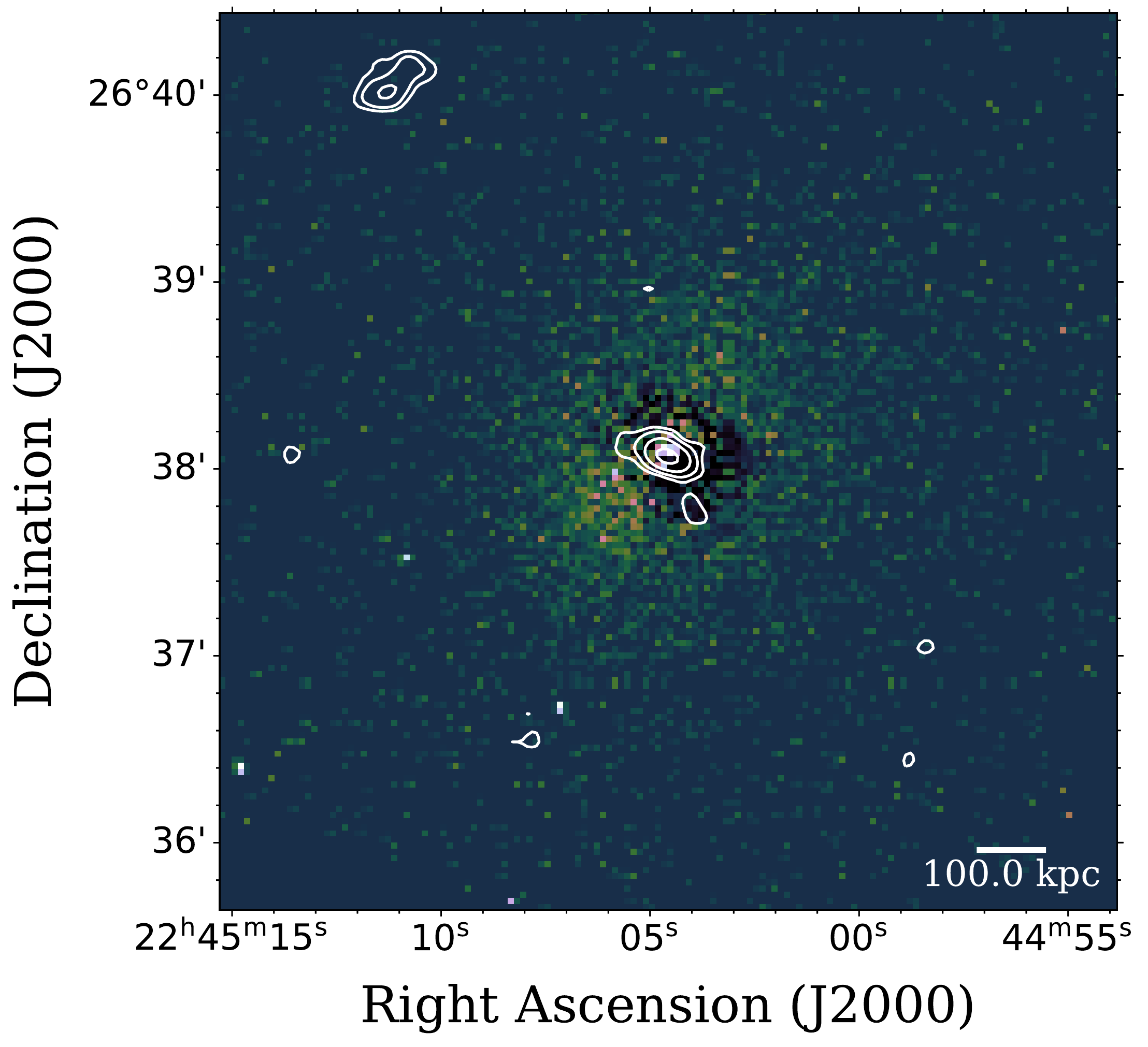}
\includegraphics[width=54mm]{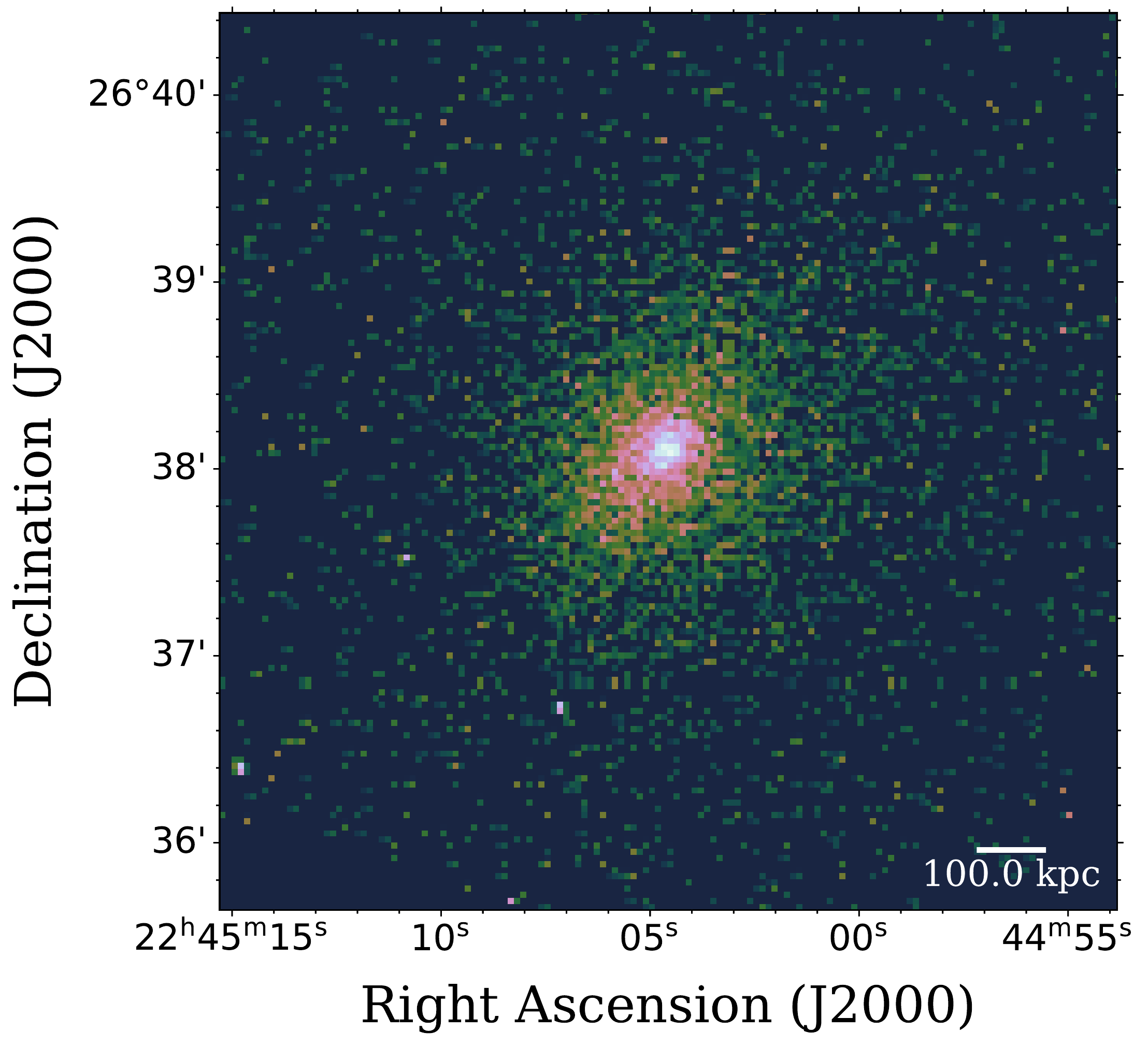} \\
\includegraphics[width=54mm]{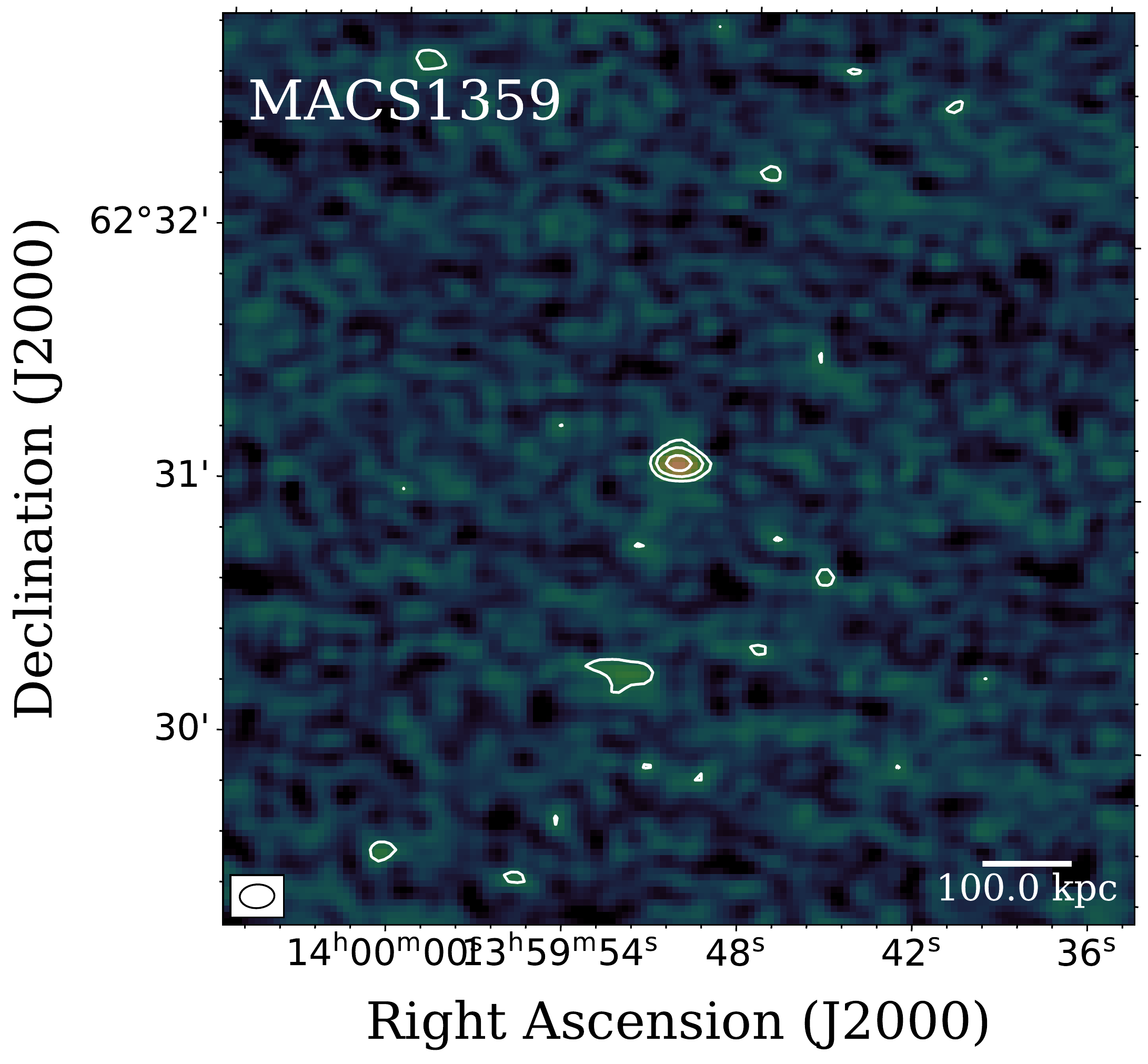} &
\includegraphics[width=54mm]{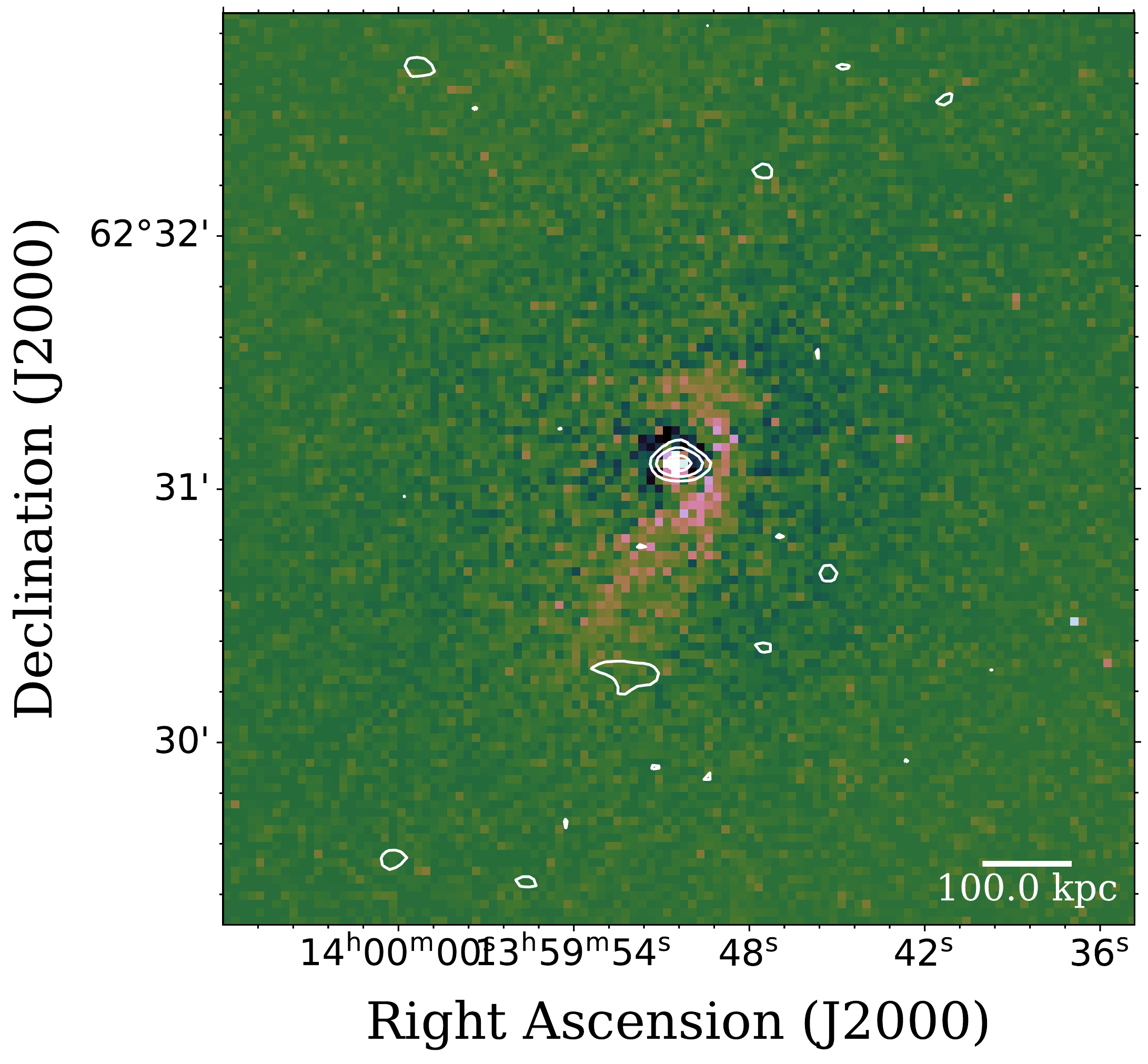}
\includegraphics[width=54mm]{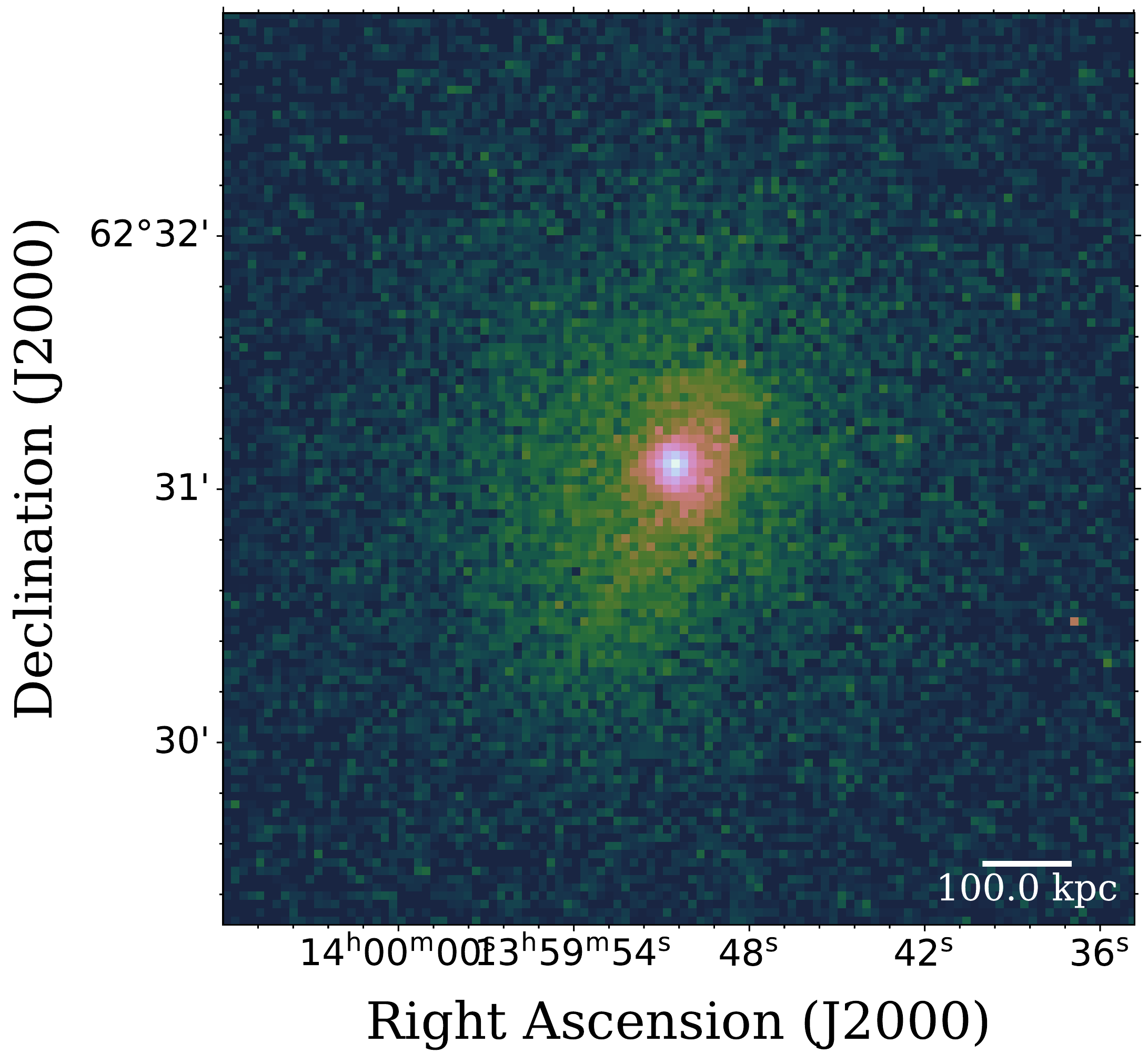}  \\ \end{tabular}
\caption{\emph{Chandra} and LOFAR images for the high redshift clusters ($z>0.3$) with  putative X-ray cavities which are not filled by radio emission, in the same order as in the tables (with MACS J2245.0+2637 and MACS J1359.8+6231 shown above, and the others shown in Figure 6-continued). The panel organization is the same as in Figure~1. For the LOFAR image, the first contour is at 0.00177~mJy beam$^{-1}$ (MACS J2245.0+2637), 0.00096~mJy beam$^{-1}$ (MACS J1359.8+6231), and each contour increases by a factor of two.}
\label{F:images_6}
\end{figure*}

\begin{figure*} \begin{tabular}{@{}cc}
\includegraphics[width=54mm]{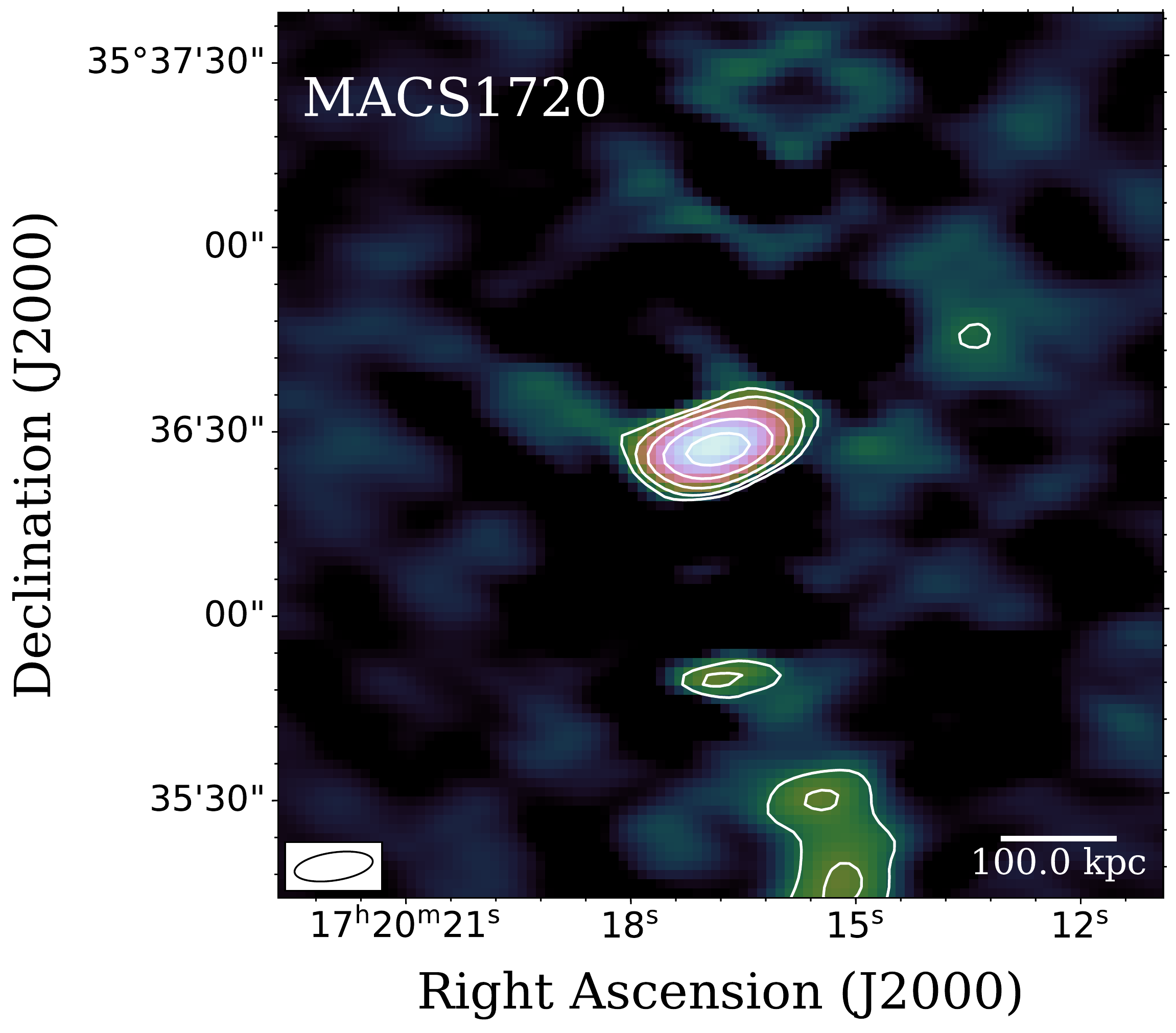} &
\includegraphics[width=54mm]{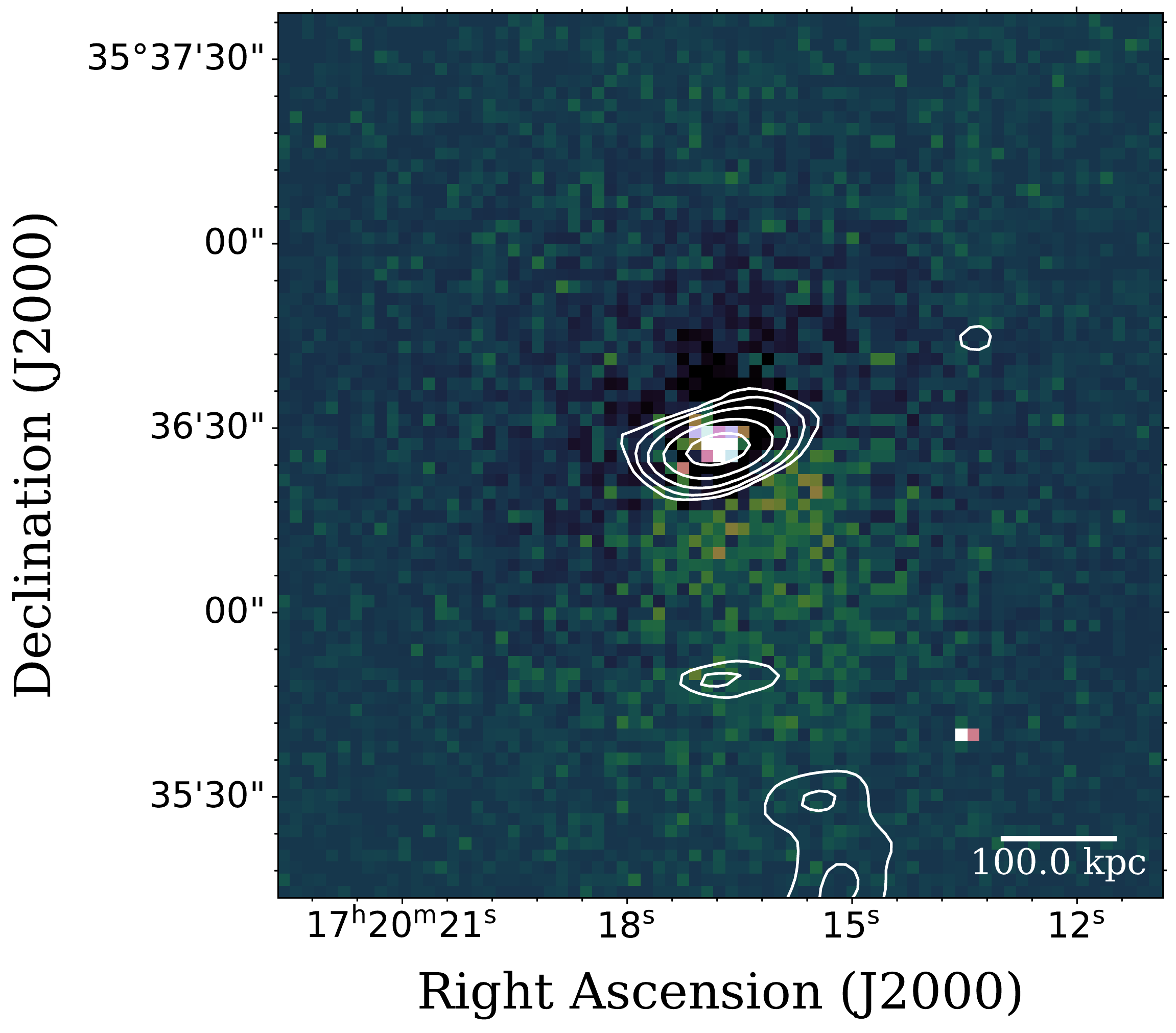}
\includegraphics[width=54mm]{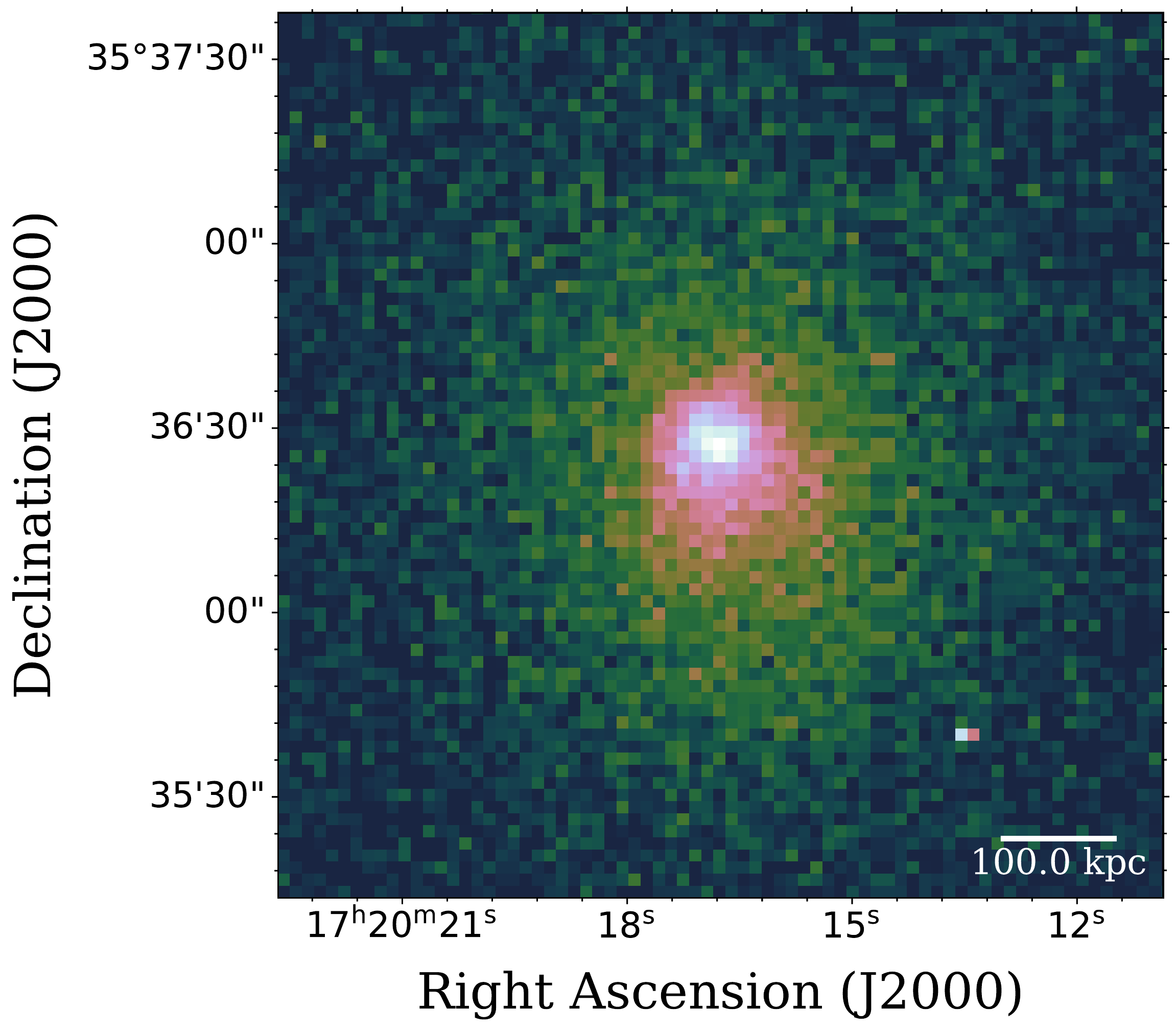} \\ \end{tabular}
\textbf{Figure}~6. --- continued (MACS J1720.2+3536). For the LOFAR image, the first contour is at 0.006~mJy beam$^{-1}$, and each contour increases by a factor of two. \\
\end{figure*}

\begin{figure*} \begin{tabular}{@{}cc}
\includegraphics[width=84mm]{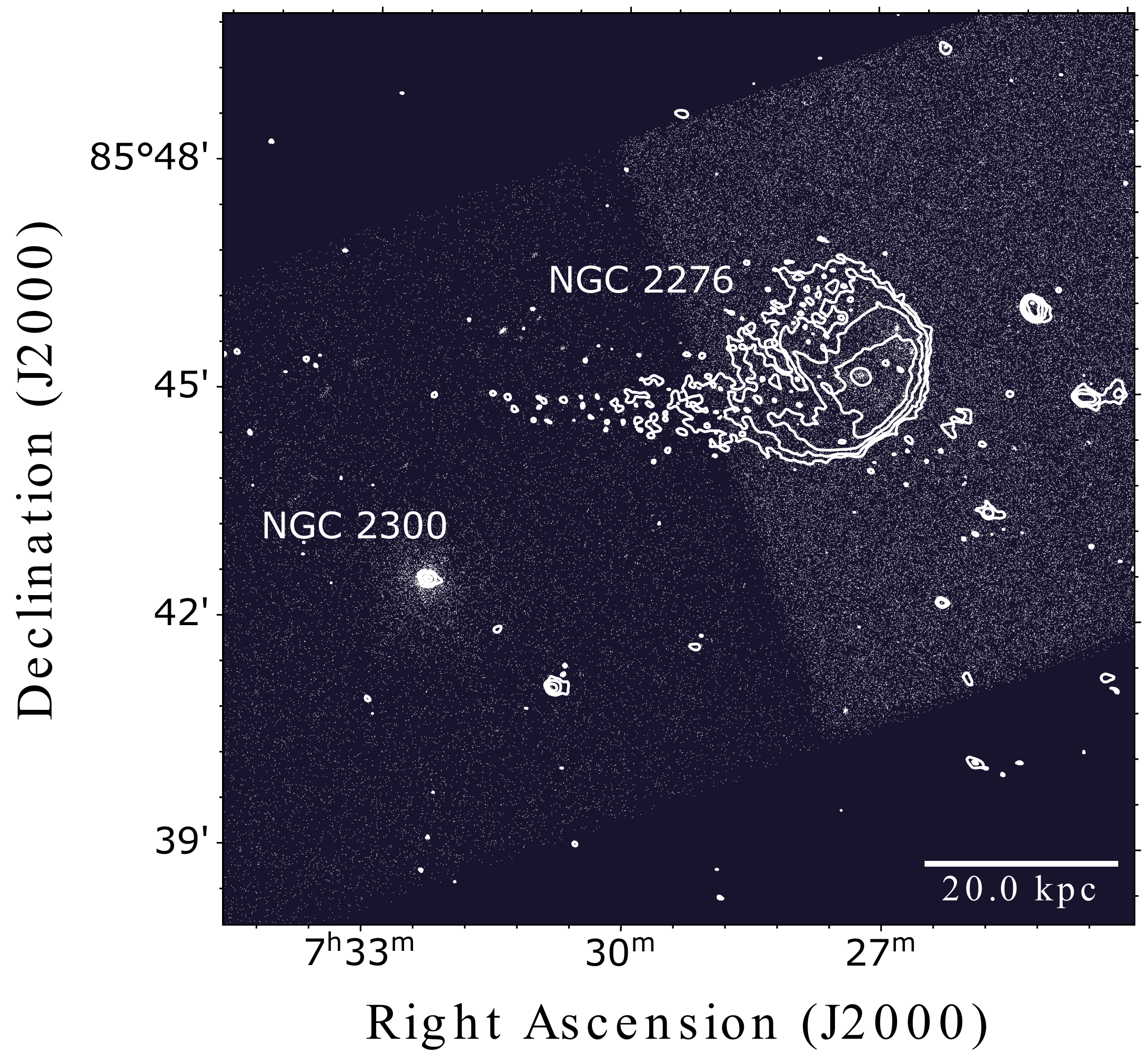}  &
\includegraphics[width=84mm]{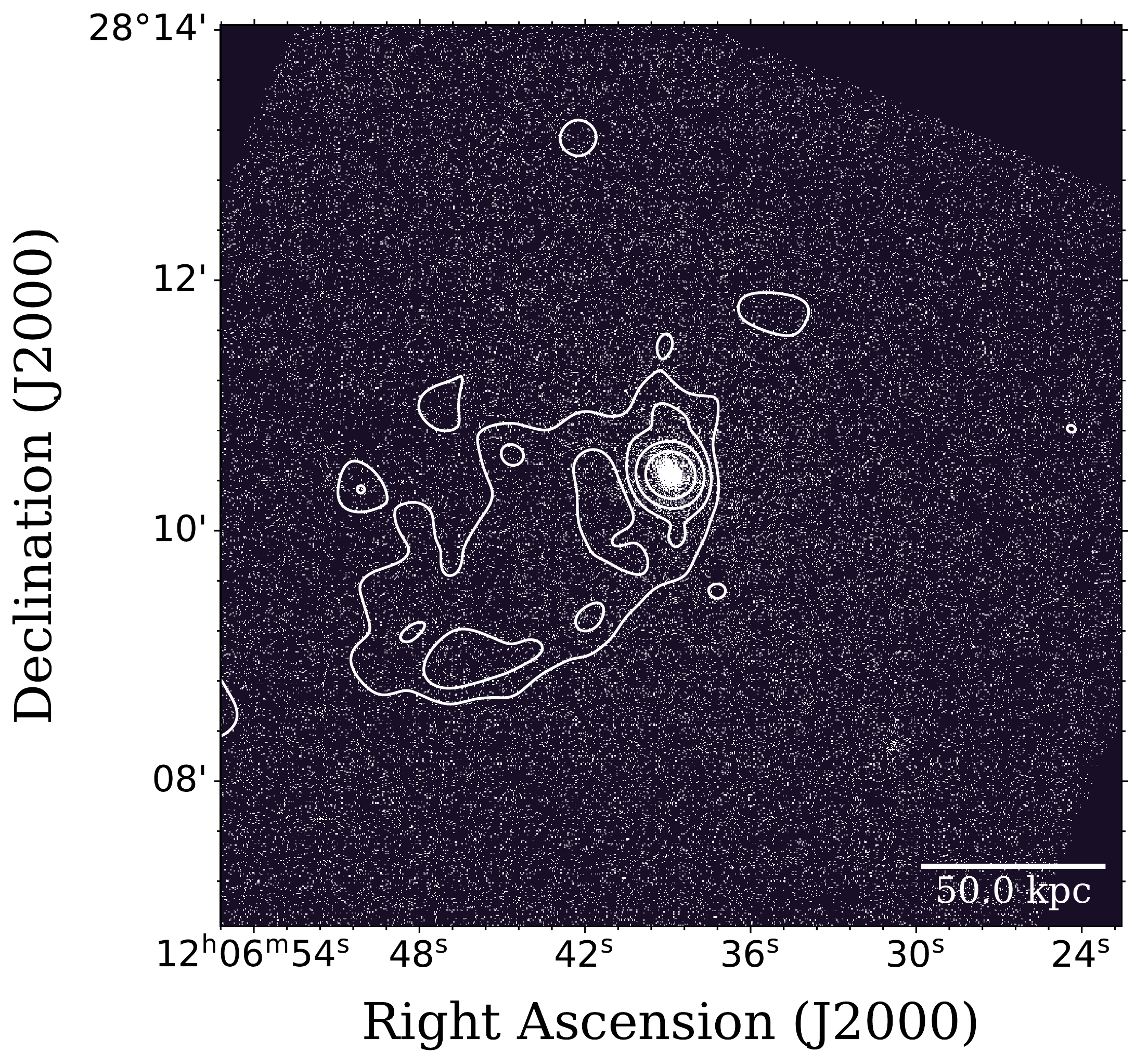} \\
\end{tabular}
\caption{\emph{Left panel}: Overlay of LOFAR contours on the  \emph{Chandra} X-ray image, showing  together NGC 2300 and NGC 2276.  For the LOFAR image, the first contour is at 0.00075~mJy beam$^{-1}$, and each contour increases by a factor of two. \emph{Right panel}: Overlay of LOFAR contours of the   diffuse emission on the  \emph{Chandra} X-ray image in NGC 4104, where the first contour is at 0.0002~mJy beam$^{-1}$, and each contour increases by a factor of two.}
\label{F:images_7}
\end{figure*}

\begin{table*}
\small
\begin{minipage}{145mm}
\caption{\emph{Chandra} observations}
\label{Xray_table}
\scalebox{0.95}{
\begin{tabular}{@{}lccccccc}
\hline
&          & \multicolumn{2}{c}{X-Ray Core (J2000)} &  & Int.$^b$     & Cavities  \\
System$^a$ & $z$   & RA & DEC & Obs. ID &     (ks) & (Ref.)  \\
 \hline
NGC 5846                    & 0.00571   & 15 06 29.25    & +01 36 21.26  & 788, 7923                   & 89.0
& (8,13,23,29) \\
NGC 5813                    & 0.00653   & 15 01 11.27    & +01 42 07.06  & 5907, 9517, 12951, 12952, 12953,   & 588
& (8,29,37) \\
                            &           &                &               & 13246, 13247, 13253, 13255 &     \\
NGC 193*                   & 0.01472    & 00 39 18.57    & +03 19 52.03  & 4053, 11389                 & 95.8
& (7,8,29) \\
\textbf{A262*}              & 0.016     & 01 52 46.20     & +36 09 11.80   & 2215, 7921                  & 138.4
& (5,9) \\
NGC 6338*                   & 0.027     & 17 15 22.90     & +57 24 38.70   & 4194, 18892, 18893, 19934, 19935, 19937,  & 295
& (8,31,33,48) \\
                            &           &                &               &  20089, 20104, 20112, 20113, 20117 & &    \\
IC1262*                     & 0.03265   & 17 33 03.44    & +43 45 34.59  & 2018, 6949, 7321, 7322      & 134.7 &
(10,34) \\
NGC 6269*                   & 0.03480   & 16 57 58.08    & +27 51 15.85  & 4972                        & 37.3
& (1,8,29) \\
NGC 5098*                    & 0.03789   & 13 20 14.73    & +33 08 36.05  & 6941                        & 36.5
& (10,36) \\
 \begin{tikzpicture}
\draw [thick,dash pattern={on 7pt off 2pt on 1pt off 3pt}] (0,0) -- (3,0);
\end{tikzpicture}\\
NGC 741                    & 0.01855   & 01 56 20.97     & +05 37 44.26   & 2223, 17198, 18718          & 170
& (10,19,20,40) \\
NGC 3608                    & 0.00409   & 11 16 59.34     & +18 08 51.90  & 2073                        & 32.8
& (8) \\
NGC 2300                    & 0.00635   & 07 32 18.86     &  +85 42 32.26  &    4968, 15648           & 51.1
& (10) \\
NGC 499*                      & 0.01467   & 01 23 11.51     & +33 27 36.33 & 10523, 10865, 10866, 10867 & 38.5 & \ldots  \\
NGC 777                     & 0.01673   & 02 00 14.90    & +31 25 44.95  & 5001                        & 9.0
& (8) \\
NGC 410                     & 0.01766 &  01 10 58.92 & +33 09 06.76 & 5897 & 2.6 & \ldots  \\
UGC 5088                    & 0.02693   &  09 33 25.69   & +34 02 53.46  & 3227                         & 28.4
& (10) \\
NGC 4104*                    & 0.0282   & 12 06 38.88  & +28 10 24.76 & 6339  & 36.0 & (41) \\
RX J1159.8+5531           & 0.081     & 11 59 52.23     & +55 32 06.68 & 4964                        & 64.8
& (10) \\
 \hline
\textbf{A2199}             & 0.030 & 16 28 38.20  & +39 33 04.94  & 10748, 10803, 10804, 10805  & 118.8 &
(2,11,21,28,32)\\
\textbf{2A0335+096}        & 0.035 & 03 38 40.90  & +09 58 04.62  &  919, 7939, 9792 & 100
 & (2,25,38) \\
\textbf{A2052}             & 0.035 & 15 16 44.46   & +07 01 17.88  & 5807, 10477, 10478, 10479, 10480    & 612
 & (2,3,4,6,11) \\
                           &       &               &               &   10879,  10914, 10915, 10916,10917 &         \\
\textbf{MKW3S}             & 0.045 & 15 21 51.80    & +07 42 31.0   & 900                           &
 & (2,11,24) \\
A1668                      & 0.0643 & 13 03 46.60 & +19 16 12.20 & 12877 & 9.7 & \ldots  \\
ZwCl 8276*                 & 0.0757 & 17 44 14.45   & +32 59 29.31  & 8267, 11708                  & 52.1 &
(14) \\
\textbf{A478}              & 0.081  & 04 13 25.35    &  +10 27 54.70  &  1669, 6102   & 52.4
& (2,12,43) \\
A1361                      & 0.117  & 11 43 39.76   & +46 21 21.21  & 3369                         & 1.0
 & (41,49) \\
ZwCl 0808*                  & 0.169 &  03 01 38.19   &  +01 55 14.98   & 12253                      & 16.6 &  \ldots           \\
\textbf{ZwCl 2701*}         & 0.214   & 09 52 49.25  & +51 53 05.32   & 3195, 12903                   & 117.5
 & (35,44) \\
\textbf{MS 0735.6+7421}     & 0.216   & 07 41 44.66  & +74 14 36.85   & 4197, 10468, 10469, 10470, 10471,    & 465.7
& (15,26,27,45) \\
                            &         &              &                &  10822, 10918, 10922          & &          \\
 A2390                      & 0.234   & 21 53 36.85  & +17 41 42.35   &   500, 4193                   & 89.5 &
(39,41,42) \\
 4C+55.16                   &  0.241   & 08 34 54.90  &  +55 34 20.90   &   1645, 4940                  &  75.8 &
(16,35) \\
   \begin{tikzpicture}
\draw [thick,dash pattern={on 7pt off 2pt on 1pt off 3pt}] (0,0) -- (3,0);
\end{tikzpicture}\\
\textbf{A1795}              & 0.063   & 13 48 52.30   & +26 35 36.78   & 493, 3666, 5286, 5287, 5288, 5289, 5290,      & 292
& (2,12,22,47) \\
                            &         &              &                & 6160, 6163, 10900, 12026, 12027,  13108,     & &         \\
                            &         &              &                &  13109, 13110, 13111, 13113, 14270, 14271  &  &        \\
ZwCl 0235*                  & 0.083   &  00 43 52.20  & +24 24 22.0    & 11735                         & 19.6
& (41) \\
RX J0352.9+1941            & 0.109   & 03 52 59.02  & +19 40 59.44   & 10466                         & 27.2
& (41) \\
RX J0820.9+0752             & 0.11087 & 08 21 02.30   & +07 51 46.39   & 17194, 17563                   & 64.4
& (46) \\
 MS 0839.9+2938*             & 0.194    & 08 42 55.90   & +29 27 26.90   & 2224                          & 26.7
 & (41)  \\
 \textbf{ZwCl 3146*}         & 0.291    & 10 23 39.57  & +04 11 12.92  & 909, 9371                     & 74.1
& (35) \\
    \hline
MACS J1532.9+3021            & 0.363     & 15 32 53.74   & +30 20 58.50    & 1649, 1665, 14009           & 104.6
& (17,18) \\
IRAS 09104+4109*            & 0.442     & 09 13 45.49   & +40 56 27.92   & 10445                       & 68.9
& (17,30) \\
MACS J1621.3+3810           & 0.465     & 16 21 24.75   & +38 10 07.58   & 3254, 6109, 6172, 9379, 10785    & 123
& (41,49) \\
 \begin{tikzpicture}
\draw [thick,dash pattern={on 7pt off 2pt on 1pt off 3pt}] (0,0) -- (3,0);
\end{tikzpicture}\\
MACS J2245.0+2637           & 0.301     & 22 45 04.54   & +26 38 04.45   & 3287                        & 11.8
& (17) \\
MACS J1359.8+6231*          & 0.330     & 13 59 50.51    & +62 31 05.58    &   516, 7714    & 29.3
& (17) \\
MACS J1720.2+3536           & 0.3913    & 17 20 16.90    &   +35 36 28.85   &  3280, 6107, 7718          & 53.1
& (17) \\
 \hline
\end{tabular}
}

References:
(1) \citet{bald09b};
(2) \citet{birz04};
(3) \citet{blan01};
(4)  \citet{blan03};
(5) \citet{blan04b};
(6)  \citet{blan11};
(7)  \citet{bogd14};
(8) \citet{cava10};
(9) \citet{clar09};
(10) \citet{dong10};
(11)  \citet{dunn04};
(12) \citet{dunn05};
(13) \citet{dunn10};
(14) \citet{etto13};
(15) \citet{gitt07};
(16) \citet{hlav11};
(17) \citet{hlav12};
(18) \citet{hlav13b};
(19) \citet{jeth07};
(20) \citet{jeth08};
(21) \citet{john02};
(22) \citet{koko18};
(23) \citet{mach11};
(24) \citet{mazz02};
(25) \citet{mazz03};
(26) \citet{mcna05};
(27) \citet{mcna09};
(28)  \citet{nuls13};
(29) \citet{osul11};
(30) \citet{osul12};
(31)  \citet{osul19};
(32)  \citet{owen98};
(33)  \citet{pand12};
(34)  \citet{pand19};
(35) \citet{raff06};
(36) \citet{rand09};
(37) \citet{rand15};
(38) \citet{sand09a};
(39) \citet{savi19};
(40) \citet{sche17};
(41) \citet{shin16};
(42)   \citep{sonk15};
(43) \citet{sun03};
(44) \citet{vags16};
(45) \citet{vant14};
(46) \citet{vant19};
(47) \citet{walk14n};
(48) \citet{wang19};
(49) this work.

$^a$ The systems in bold are from B08 sample. The asterisk marks systems with alternative names, as in Table \ref{LOFAR_table}.  \\
$^b$ Total integration time on source, after reprocessing.  \\
\end{minipage} \end{table*}

\clearpage

\begin{table*}
\small
\begin{minipage}{145mm}
\caption{143 MHz LOFAR observations}
\label{LOFAR_table}
\scalebox{0.95}{
\begin{tabular}{@{}lcccccc}
\hline
&  &        &  Total Flux Density & Rms noise & Resolution &   Radio   \\
System$^a$  & $z$  & Obs. date$^b$    & (Jy)  & ($\mu$Jy beam$^{-1}$) &  ($arcsec \times arcsec$) & (Ref.)  \\
\hline
NGC 5846                   & 0.00571  & 01-09-2018* & 0.091 $\pm$ 0.017  & 780    & 14.07 $\times$ 5.93  &
(10,21) \\
NGC 5813                   & 0.00653  & 01-09-2018* & 0.066 $\pm$ 0.013  & 430    &  14.06  $\times$ 5.9  &
(10) \\
NGC 193*                   & 0.01472  & 06-06-2019* & 6.16 $\pm$ 0.93    & 360    &  13.42    $\times$ 5.76 &
(10,20) \\
\textbf{A262*}             & 0.016    & 23-05-2014  & 0.576 $\pm$ 0.090  & 585    & 14.41 $\times$ 10.76 &
(2,5,7,16,29) \\
NGC 6338*                  & 0.027    & 21-12-2017  & 0.160 $\pm$ 0.025  & 138    & 13.66 $\times$ 8.68  &
(16,25,38) \\
IC 1262*                   & 0.03265  & 13-06-2018  & 5.61 $\pm$ 0.85    & 180    & 7.86 $\times$ 4.52   &
(16,28,32) \\
NGC 6269*                  & 0.03480  & 25-01-2017  & 0.272 $\pm$ 0.042  & 250    & 8.48 $\times$ 6.21   &
(1,9,10,16) \\
NGC 5098*                 & 0.03789  & 10-05-2018  & 0.186 $\pm$ 0.033   & 733    & 7.87 $\times$ 5.17   &
(16,31) \\
  \begin{tikzpicture}
\draw [thick,dash pattern={on 7pt off 2pt on 1pt off 3pt}] (0,0) -- (3,0);
\end{tikzpicture}\\
NGC 741                   & 0.01855  & 09-06-2019* & 3.90 $\pm$ 0.60 & 513 & 13.33 $\times$ 5.73  &
(10,17,36) \\
NGC 3608                   & 0.00409   & 11-01-2018  & $\ldots$           & 636    & 9.32 $\times$ 5.49    & \ldots   \\
NGC 2300                   & 0.00635  & 26-07-2017 & 0.006 $\pm$ 0.002  & 198 & 8.16 $\times$ 5.09 & \ldots  \\
NGC 499*                   & 0.01467  & 28-10-2016 & 0.046 $\pm$ 0.009  & 585 & 8.75 $\times$ 5.76 & \ldots  \\
NGC 777                    & 0.01673  & 24-10-2016  & 0.032 $\pm$ 0.006  & 246    & 8.42 $\times$ 5.13   & \ldots    \\
NGC 410                    & 0.01766  &  08-08-2016  & 0.039 $\pm$ 0.007 & 520 & 11.05 $\times$ 4.91 & \ldots \\
UGC 5088                   & 0.02693  & 01-09-2018 & 0.003 $\pm$ 0.001 & 140 & 5.59 $\times$ 3.27  & \ldots  \\
NGC 4104*                  & 0.0282   & 04-04.2017 & 0.018  $\pm$ 0.004 & 184 & 9.09 $\times$ 5.84 & (16) \\
RX J1159.8+5531           & 0.081     & 15-02-2015  & 0.004 $\pm$ 0.001  & 83     & 8.6 $\times$ 5.18     & \ldots  \\
\hline
\textbf{A2199}             & 0.030   & 24-03-2016  & 53.97 $\pm$ 8.10  &  850   & 7.47 $\times$ 4.69  &
(2,4,13,27) \\
\textbf{2A0335+096}        & 0.035   & 04-01-2018  & 0.852 $\pm$ 0.135  &  1100  & 9.79 $\times$ 5.91  &
(2,30,33,34) \\
\textbf{A2052}             & 0.035   & 14-08-2014  & 58.53 $\pm$ 8.82   & 4925   & 8.80 $\times$ 6.51  &
(2,3,40) \\
\textbf{MKW3S}             & 0.045   & 14-08-2014  & 21.24 $\pm$ 3.20   & 726    & 8.80 $\times$ 6.51  &
(2,9,23) \\
A1668                      & 0.0643  & 05-04-2019  & 1.83 $\pm$ 0.44    & 286    & 10.21 $\times$ 6.52  &
(16) \\
ZwCl 8276*                 & 0.0757  & 30-05-2018  & 0.90 $\pm$ 0.14    & 468    & 8.29 $\times$ 5.20  &
(11,16) \\
\textbf{A478}              & 0.081   &  \ldots     &  \ldots            &  \ldots  &     \ldots        &
(2,11,16,35) \\
A1361                      & 0.117   & 15-06-2014  & 5.22 $\pm$ 0.8   & 1030   & 12.46 $\times$ 5.41 &
(16,26) \\
ZwCl 0808*                   & 0.169 & 17-07-2018* & 11.64 $\pm$ 1.78  & 4070 & 13.0 $\times$ 5.75  &
(16) \\
\textbf{ZwCl 2701*}        & 0.214   & 20-02-2018  & 1.34 $\pm$ 0.20    & 204    & 7.75 $\times$ 4.43  &
(2,37) \\
\textbf{MS 0735.6+7421}    & 0.216   & 18-05-2013  & 4.33 $\pm$ 0.65    & 404    & 9.65 $\times$ 8.23  &
(1,22) \\
A2390                      & 0.234   & \ldots & \ldots & \ldots & \ldots &
(16,35) \\
 4C+55.16                  &  0.241   & 15-06-2019  &   11.74   $\pm$ 1.77  & 1350 & 8.92  $\times$ 2.91 &
 (6,39) \\
    \begin{tikzpicture}
\draw [thick,dash pattern={on 7pt off 2pt on 1pt off 3pt}] (0,0) -- (3,0);
\end{tikzpicture}\\
\textbf{A1795}             & 0.063   & 03-05-2014  & 6.45 $\pm$ 0.98    & 1610   & 10.14 $\times$ 5.93 &
(2,8,11) \\
ZwCl 0235*                 & 0.083   & 28-09-2018  & 0.151 $\pm$ 0.024  & 387    & 9.34 $\times$ 6.15  &
(16) \\
RX J0352.9+1941           & 0.109   & 03-07-2018  & 0.068 $\pm$ 0.011  & 157    & 8.91 $\times$ 5.53  &
(16)\\
RX J0820.9+0752            & 0.11087 & 08-29-2018  &  0.021  $\pm$ 0.004 & 590 &  13.29   $\times$ 5.38         &
(16) \\
MS 0839.9+2938             & 0.194   & 18-02-2016  & 0.221 $\pm$ 0.035  & 331    & 10.51 $\times$ 5.40 &
(12) \\
\textbf{ZwCl 3146*}        & 0.291   & 08-06-2018* & 0.055 $\pm$ 0.009  & 765    & 14.20 $\times$ 5.23 &
(2,11,19) \\
  \hline
MACS J1532.9+3021           & 0.363   & 17-08-2018  & 0.097 $\pm$ 0.017  & 496    & 11.59 $\times$ 5.53 &
(6,11,15,16,18) \\
IRAS 09104+4109*           & 0.442   & 01-02-2018  & 0.205 $\pm$ 0.033  & 582    & 8.76 $\times$ 5.35  &
(6,14,24) \\
MACS J1621.3+3810          & 0.465   & 20-08-2015  & 0.082 $\pm$ 0.015  & 707    & 8.68  $\times$ 5.47 & (6)  \\
\begin{tikzpicture}
\draw [thick,dash pattern={on 7pt off 2pt on 1pt off 3pt}] (0,0) -- (3,0);
\end{tikzpicture}\\
MACS J2245.0+2637          & 0.301   & 14-07-2016  & 0.026 $\pm$ 0.006  & 594    & 9.15 $\times$ 6.19  & \ldots  \\
MACS J1359.8+6231*         & 0.330   & 29-09-2018   & 0.003 $\pm$  0.002    &   310    &  8.23 $\times$  5.6   & \ldots  \\
MACS J1720.2+3536          & 0.3913  & 04-08-2018  & 0.172 $\pm$ 0.030  &   1004    &  7.54  $\times$ 4.9    & \ldots   \\
 \hline
\end{tabular}
}

References:
(1) \citet{bald09b};
(2)  \citet{birz08};
(3)  \citet{blan11};
(4) \citet{burn83};
(5) \citet{clar09};
(6)  \citet{edge03};
(7) \citet{fant87};
(8) \citet{ge93};
(9) \citet{giac07};
(10) \citet{giac11b};
(11) \citet{giac14};
(12) \citet{giac17};
(13) \citet{giov98};
(14) \citet{hine93};
(15) \citet{hlav13b};
(16) \citet{hoga15};
(17)   \citet{jeth08};
(18) \citet{kale13};
(19) \citet{kale15b};
(20)   \citet{lain11};
(21) \citet{mach11};
(22) \citet{mcna05};
(23) \citet{mazz02};
(24) \citet{osul12};
(25)  \citet{osul19};
(26) \citet{owen97};
(27) \citet{owen98};
(28)  \citet{pand19};
(29) \citet{parm86};
(30) \citet{patn88};
(31) \citet{rand09};
(32) \citet{rudn09};
(33) \citet{sand09a};
(34) \citet{sara95b};
(35) \citet{savi19};
(36) \citet{sche17};
(37) \citet{vags16};
(38) \citet{wang19};
(39) \citet{xu95};
(40) \citet{zhao93}.

$^a$  For A478 and A2390, see \citet{savi19} for details of the LOFAR observations.
The asterisk marks systems with alternative names:
ZwCl 8276 (ZwCl 1742.1+3306);
ZwCl 0808 (ZwCl 0258.9+0142);
ZwCl 2701 (ZwCl 0949.6+5207); ZwCl 0235 (ZwCl 0040.8+2404); ZwCl 3146 (ZwCl 1021.0+0426);
   IRAS 09104+4109 (RX J0913.7+4056);
MACS J1359.8+6231 (ZwCl 1358.1+6245, MS 1358.4+6245);
NGC 193 (UGC408);
A262 (NGC 708); NGC 6338 (RX J1715.3+5725); IC 1262 (RX J1733.0+4345); NGC 5098 (RXC J1320.2+3308);
NGC 6269 (AWM 5, RX J1657.8+2751);
NGC 499 (RX J0123.2+3327); NGC 4104 (RX J1206.6+2810). \\
$^b$ The integration time is 8 h, except for those with an asterisk for which it is 4 h.\\
\end{minipage} \end{table*}

\clearpage

\begin{table*}
\small
\begin{minipage}{145mm}
\caption{Summary}
\label{summary_table}
\scalebox{0.95}{
\begin{tabular}{@{}lccccc}
\hline
&  &          Sloshing/ & H$\alpha$ filam./   & Central-radio source  &   Radio-filled    \\
System  & $z$      & Cold fronts(Ref.)$^a$  &  Mol. gas$^b$ (Ref.)  &  with lobes$^c$ (yes/no) & cavities$^d$ (yes/no)  \\
 \hline
NGC 5846                   & 0.00571  & (17,21)  & (19,21,35,40)     & yes & yes   \\
NGC 5813                   & 0.00653  & \ldots & (19,40)     &  yes & yes  \\
NGC 193*                   & 0.01472  & \ldots & (1)      &  yes & yes  \\
\textbf{A262*}             & 0.016    & (18)  & (4,9,34)     & yes* & yes  \\
NGC 6338*                  & 0.027    & (39) & (31)    & yes* & yes   \\
IC 1262*                   & 0.03265  & (32)  &  (4)      & yes & yes    \\
NGC 6269*                  & 0.03480  & \ldots  & no (4)     & yes & ?   \\
NGC 5098*                  & 0.03789  & (33)  & (4)    & yes & yes    \\
  \begin{tikzpicture}
\draw [thick,dash pattern={on 7pt off 2pt on 1pt off 3pt}] (0,0) -- (3,0);
\end{tikzpicture}\\
NGC 741                   & 0.01855  & \ldots & no (20)  & yes & no  \\
NGC 3608                   & 0.00409   & \ldots  & no (42)        & no & no      \\
NGC 2300                   & 0.00635  & \ldots & no (20)   & no & no   \\
NGC 499*                   & 0.01467  & \ldots &  (20)  & no & no \\
NGC 777                    & 0.01673  & \ldots  & no (20)     & no & no     \\
NGC 410                    & 0.01766  & \ldots &  (20)   & no & no \\
UGC 5088                   & 0.02693  & \ldots & \ldots  & no & no    \\
NGC 4104*                  & 0.0282   & \ldots &  (4) &  no & no \\
RX J1159.8+5531           & 0.081     & \ldots  & \ldots      & no & no       \\
\hline
\textbf{A2199}             & 0.030   & (18,28)  & (27)   &  yes   & yes   \\
\textbf{2A0335+096}        & 0.035   & (18,22)  & (7,37)    & yes* & yes   \\
\textbf{A2052}             & 0.035   & (3)  & (19,24)      & yes & yes   \\
\textbf{MKW3S}             & 0.045   &  \ldots & (34,41)      & yes & yes   \\
A1668                      & 0.0643  & \ldots  & (19)       & yes* & ?   \\
ZwCl 8276*                 & 0.0757  & (12)  &  (4,9)       & yes* & yes   \\
\textbf{A478}              & 0.081   &  \ldots     &  (19,24)    &   yes & yes         \\
A1361                      & 0.117   &    \ldots & (4)      & yes* & yes   \\
ZwCl 0808*                   & 0.169 & \ldots & (4)   & yes* & ?   \\
\textbf{ZwCl 2701*}        & 0.214   & \ldots & (4)      & yes* & yes  \\
\textbf{MS 0735.6+7421}    & 0.216   & \ldots  & (26)       & yes & yes   \\
A2390                      & 0.234   & \ldots & (19)   & yes* & yes  \\
4C+55.16                  &  0.241   & \ldots  &   (9)   & yes & yes  \\
    \begin{tikzpicture}
\draw [thick,dash pattern={on 7pt off 2pt on 1pt off 3pt}] (0,0) -- (3,0);
\end{tikzpicture}\\
\textbf{A1795}             & 0.063   & (11,18) & (5,23,24,25,27,36)       & yes* & no  \\
ZwCl 0235*                 & 0.083   & \ldots & (4,34)     & yes* & no     \\
RX J0352.9+1941           & 0.109   & \ldots & (19)     & no & no   \\
RX J0820.9+0752            & 0.11087 & \ldots  &  (2,19,34,38)  &  no & no    \\
MS 0839.9+2938             & 0.194   & \ldots  & (6)      & yes & no  \\
\textbf{ZwCl 3146*}        & 0.291   & (14,15) & (27,29)      & yes* & no \\
  \hline
MACS J1532.9+3021           & 0.363   & (16)  & (8,13)      & yes & yes  \\
IRAS 09104+4109*           & 0.442   &  (30) & (30)      & yes & yes   \\
MACS J1621.3+3810          & 0.465   & \ldots  & (10)    & yes* & yes   \\
\begin{tikzpicture}
\draw [thick,dash pattern={on 7pt off 2pt on 1pt off 3pt}] (0,0) -- (3,0);
\end{tikzpicture}\\
MACS J2245.0+2637          & 0.301   & \ldots  & \ldots      & ?* & no    \\
MACS J1359.8+6231*         & 0.330   & \ldots   & (6)        &  no & no     \\
MACS J1720.2+3536          & 0.3913  & \ldots & (8,13)     &  no & no      \\
 \hline
\end{tabular}
}

References:
(1)  \citet{baby19};
(2)  \citet{baye02};
(3) \citet{blan11};
(4) \citet{craw99};
(5) \citet{craw05};
(6)  \citet{dona92};
(7)  \citet{dona07};
(8)  \citet{dona15};
(9) \citet{edge02};
(10)  \citet{edge03};
(11) \citet{ehle15};
(12) \citet{etto13};
(13) \citet{foga15};
(14) \citet{form02};
(15) \citet{form02b};
(16) \citet{hlav13b};
(17) \citet{gast13};
(18) \citet{ghiz10};
(19)  \citet{hame16};
(20)  \citet{lakh18};
(21)  \citet{mach11};
(22) \citet{mazz03};
(23) \citet{mcdo09};
(24) \citet{mcdo10};
(25) \citet{mcdo14};
(26) \citet{mcna09};
(27) \citet{mitt15};
(28)  \citet{nuls13};
(29) \citet{odea10};
(30) \citet{osul12};
(31) \citet{pand12};
(32) \citet{pand19};
(33) \citet{rand09};
(34) \citet{salo03};
(35) \citet{temi18};
(36) \citet{trem15};
(37) \citet{vant16};
(38) \citet{vant19};
(39) \citet{wang19};
(40) \citet{wern14};
(41) \citet{whit97};
(42) \citet{youn11}.

$^a$ Indicates the presence of sloshing and/or cold fronts using information available in the literature. \\
$^b$ Indicates the presence of H$\alpha$ filaments, and/or molecular gas using information available in the literature.\\
$^c$ Indicates the presence of a central radio source with resolved lobes (marked with `yes'), and point source only emission or unresolved sources (marked with `no'), and `?' for the sources with possible hints of resolved extended emission (radio lobes).
The asterisk marks the systems for which the LOFAR data
are a significant improvement over the previous observations and strengthen the evidence for AGN feedback in those systems.\\
$^d$ Systems for which the X-ray cavities are filled with lobe radio emission are marked with  `yes' and those without such emission with `no'. The uncertain systems are marked with `?'.\\
\end{minipage} \end{table*}

\clearpage

\section*{Acknowledgements}

The scientific results reported in this article are based on data obtained with the International LOFAR Telescope (ILT), and archive data from \emph{Chandra} Data Archive and XMM-\emph{Newton} archive. LOFAR \citep{vanH13} is the Low Frequency Array designed and constructed by ASTRON. It has observing, data processing, and data storage facilities in several countries, that are owned by various parties (each with their own funding sources), and that are collectively operated by the ILT foundation under a joint scientific policy. The ILT resources have benefitted from the following recent major funding sources: CNRS-INSU, Observatoire de Paris and Université d'Orléans, France; BMBF, MIWF-NRW, MPG, Germany; Science Foundation Ireland (SFI), Department of Business, Enterprise and Innovation (DBEI), Ireland; NWO, The Netherlands; The Science and Technology Facilities Council, UK. The LOFAR reduction was done using the PREFACTOR and FACTOR packages. The authors thank Federica Savini for providing LOFAR images of A478 and A2390.   ACE acknowledges support from STFC grant ST/P00541/1. HR acknowledges support from the ERC Advanced Investigator programme NewClusters 321271. AB acknowledges support from the VIDI research programme with project number 639.042.729, which is financed by the Netherlands Organisation for Scientific Research (NWO). We acknowledge the ''Multiphase AGN Feeding and Feedback'' workshop held in Sexten, Italy, on July 2018,  for stimulating discussion. The authors thank the referee for the careful reading of the draft and the comments which improved and clarified the paper.

\bibliographystyle{mn2e}
\bibliography{/Users/Laura/Documents/Bibliography/master_references}

\end{document}